\documentstyle[12pt,twoside]{report}

\renewcommand{\theequation}{\thesection.\arabic{equation}}

\newlength{\extraspace}
\newlength{\extraspaces}
\newcounter{dummy}

\newcommand{\baa}{
\addtocounter{equation}{1}
\setcounter{dummy}{\value{equation}}
\setcounter{equation}{0}
\renewcommand{\theequation}{\thesection.\arabic{dummy}\alph{equation}}
\begin{eqnarray}
\addtolength{\abovedisplayskip}{\extraspaces}
\addtolength{\belowdisplayskip}{\extraspaces}
\addtolength{\abovedisplayshortskip}{\extraspace}
\addtolength{\belowdisplayshortskip}{\extraspace}}

\newcommand{\eaa}{
\end{eqnarray}
\setcounter{equation}{\value{dummy}}
\renewcommand{\theequation}{\thesection.\arabic{equation}}}

\newcommand{\be}{\begin{equation}
\addtolength{\abovedisplayskip}{\extraspaces}
\addtolength{\belowdisplayskip}{\extraspaces}
\addtolength{\abovedisplayshortskip}{\extraspace}
\addtolength{\belowdisplayshortskip}{\extraspace}}
\newcommand{\ee}{\end{equation}}

\newcommand{\ba}{\begin{eqnarray}
\addtolength{\abovedisplayskip}{\extraspaces}
\addtolength{\belowdisplayskip}{\extraspaces}
\addtolength{\abovedisplayshortskip}{\extraspace}
\addtolength{\belowdisplayshortskip}{\extraspace}}
\newcommand{\ea}{\end{eqnarray}}

\newcommand{\bd}{\begin{displaymath}
\addtolength{\abovedisplayskip}{\extraspaces}
\addtolength{\belowdisplayskip}{\extraspaces}
\addtolength{\abovedisplayshortskip}{\extraspace}
\addtolength{\belowdisplayshortskip}{\extraspace}}
\newcommand{\ed}{\end{displaymath}}

\newcommand{\ban}{\begin{eqnarray*}
\addtolength{\abovedisplayskip}{\extraspaces}
\addtolength{\belowdisplayskip}{\extraspaces}
\addtolength{\abovedisplayshortskip}{\extraspace}
\addtolength{\belowdisplayshortskip}{\extraspace}}
\newcommand{\ean}{\end{eqnarray*}}

\newcommand{\nonu}{\nonumber \\[.5mm]}

\newcommand{\deel}[2]{{\textstyle{#1 \over #2}}}
\newcommand{\hf}{{\textstyle{1\over 2}}}
\newcommand{\hv}{{\textstyle{1\over 4}}}

\newcommand{\ie}{{\it i.e.}}

\newtheorem{lemma}{Lemma}
\newtheorem{theorem}{Theorem}
\newtheorem{thm}{Theorem}
\newtheorem{lmm}{Lemma}
\newtheorem{exam}{Example}
\newcommand{\bt}{\begin{thm}}
\newcommand{\et}{\end{thm}}
\newcommand{\bl}{\begin{lmm}}
\newcommand{\el}{\end{lmm}}
\newcommand{\bex}{\begin{exam}}
\newcommand{\eex}{\end{exam}}
\newcommand{\hj}{\hat{J}}
\newcommand{\whj}[1]{W(\hat{J}^{#1})}
\newcommand{\www}[4]{\deel{#1}{#2}\hj^{#3}\hj^{#4}}

\def\inbar{\,\vrule height1.5ex width.4pt depth0pt}
\font\rms=cmr12 at 12pt
\def\ce{\relax\ifmmode\mathchoice
{\hbox{$\inbar\kern-.3em{\rm C}$}}
{\hbox{$\inbar\kern-.3em{\rm C}$}}
{\lower.9pt\hbox{\rms $\inbar\kern-.3em{\rm C}$}}
{\lower1.2pt\hbox{\rms $\inbar\kern-.3em{\rm C}$}}
\else{$\inbar\kern-.3em{\rm C}$}\fi}
\font\cmss=cmss12 \font\cmsss=cmss12 at 12pt
\def\ze{\relax\ifmmode\mathchoice
{\hbox{\cmss Z\kern-.4em Z}}{\hbox{\cmss Z\kern-.4em Z}}
{\lower.9pt\hbox{\cmsss Z\kern-.4em Z}}
{\lower1.2pt\hbox{\cmsss Z\kern-.4em Z}}\else{\cmss Z\kern-.4em Z}\fi}

\newcommand{\dif}{\partial}

\newcommand{\tr}{\mbox{Tr}}

\newcommand{\actie}[1]{\deel{1}{2\pi}\int d^2z \, }

\newcommand{\mats}[9]{\left( \begin{array}{ccc}
				#1 & #2 & #3 \\
				#4 & #5 & #6 \\
				#7 & #8 & #9
                             \end{array} \right) }

\newcommand{\mat}[4]{\left( \begin{array}{cc}
				#1 & #2 \\
				#3 & #4
                             \end{array} \right) }

\newcommand{\np}[1]{Nucl. Phys. {\bf B#1}}
\newcommand{\cmp}[1]{Comm. Math. Phys. {\bf #1}}

\newcommand{\plb}[1]{Phys. Lett. {\bf B#1}}

\newcommand{\ad}[1]{\mbox{\rm ad}_{#1}}

\newcommand{\im}{\mbox{\rm im}}

\begin{document}


\thispagestyle{empty}
\vspace*{1cm}
\begin{center}
{\Huge Finite and infinite W algebras} \\[1cm]
{\LARGE and their applications} \\[10cm]
\end{center}
\vfill
\begin{center}
{\LARGE Tjark Tjin}
\end{center}
\vspace*{2cm}
\mbox{}
\newpage

\newpage
\thispagestyle{empty}

\vspace*{3cm}
\begin{center}
{\large Supervisor: Prof. dr. ir. F. A. Bais}

\vspace*{3cm}

\end{center}

\vspace*{3cm}

The various chapters of this thesis are based on the following papers:
\begin{itemize}
\item Chapter 3: T. Tjin, Finite $W$ algebras, {\em Phys. Lett} {\bf B292}
(1992)60; J. de Boer and T. Tjin,
Quantization and representation theory of finite $W$
algebras, {\em Comm. Math. Phys.} to appear.
\item Chapter 4: F. A. Bais, T. Tjin, P. van Driel, Covariantly coupled
chiral algebras, {\em Nucl. Phys.} {\bf B357} (1991) 632; J. de Boer and
T. Tjin, The relation between quantum $W$ algebras and Lie algebras,
{\em Comm. Math. Phys.} to appear;
F. A. Bais, T. Tjin, P. van Driel, J. de Boer, J. Goeree, $W$ algebras,
$W$ gravities and their moduli spaces, To be published in the proceedings
of the `workshop on Superstrings and related Topics', Trieste, Juli 1992.
\item Chapter 5: T. Tjin, P. van Driel, Coupled WZNW-Toda models and
Covariant KdV hierarchies, Preprint ITFA-91-04; T. Tjin, Finite $W$
symmetry in finite dimensional integrable systems, To be published in the
proceedings of the workshop on `Low dimensional Topology and Physics',
Isaac Newton Institute, Cambridge, 1992.
\end{itemize}

\vfill

\setcounter{page}{0}
\tableofcontents

\setcounter{chapter}{0}

\addtolength{\baselineskip}{.7mm}

\pagestyle{headings}

\chapter{General Introduction}
In this introductory
chapter we give a brief review of conformal field theory (CFT), define
the concept of a $W$
algebra and discuss some of their potential applications
to high energy physics,
condensed matter physics and solitary wave phenomena.
In 2 dimensions the conformal group is infinite dimensional and its Lie
algebra is the Virasoro algebra. The particular structure of
the representation theory of this algebra has lead to a partial
classification of CFT's but it has turned out that for a complete
classification it is necessary to consider {\em extensions} of conformal
symmetry. Such extensions are called $W$ algebras. In the first section
of this chapter we review some basic aspects of CFT and $W$ algebras in
order to set the stage. Having developed the
necessary background  we consider string theory which is the most
promising candidate for a unified theory of the four known forces of
nature and its particle spectrum. In this context CFT's arize as
perturbative groundstates around which one expands the theory.
As another promising application we consider some recent
work on the role played by CFT in the quantum Hall effect. The crucial
insight here is that (as first proposed by Laughlin) in the Hall state
the system behaves like an incompressible quantum fluid such that the
only nontrivial dynamics is concentrated around the boundary of the sample.
The boundary theory can then be described by a 2 dimensional CFT.
Next we discuss the result that 2 dimensional
lattice models, such as the Ising model,
at criticality correspond to  2 dimensional CFT's. This relation
is based on the fact that at criticality fluctuations in the system are
correlated over long distances (i.e. the correlation length goes to
infinity) which allows one to define a continuum limit in which the
lattice system corresponds
to a 2 dimensional quantum field theory. Due to the scale covariance
in critical lattice systems this field theory will by conformally
invariant and the critical exponents will be determined by the
representation theory of the $W$ algebra underlying the CFT.
In the last section of this chapter we discuss the role played
by $W$ algebras in the theory of integrable hierarchies with solitary wave
solutions. Here $W$ algebras appear in two distinct ways: first of all
through the second Hamiltonian structures and also through the so
called weak action angle structure which is intimately related to
the integrability of the hierarchy.

We hope that with this introductory
chapter we have provided the reader with sufficient motivation to
proceed to the more technical chapters in which we study $W$ algebras
in detail.

\section{Conformal Field Theory and W algebras}
The distinctive feature of the
conformal group in 2 dimensions is that it is, in contrast with other
dimensions, infinite dimensional. It turns out that this infinite
symmetry has dramatic consequences as was first shown in \cite{BPZ}.
In this section we shall give a brief account of this vast subject.
For more detailed reviews we refer the reader to \cite{CFT,Nahm}.

In what follows we shall always take spacetime to be
$M={\bf R}\times S^1$ with coordinates $(x^0,x^1)\equiv (t,x)$ and
metric $ds^2=dt^2-dx^2$. Here we think of space to be compactified, i.e
$x=x+2\pi$. A conformal transformation is a diffeomorphism (i.e. a smooth
and invertible map) from spacetime to itself that preserves the causal
structure (that is the lightlike lines are mapped to lighlike lines and
the orientation of time is preserved). The set of lightlike lines on $M$
is isomorphic to $S^1 \times S^1$ and from this it is easy to see that
the set of conformal diffeomorphisms is given by $Diff_c(M)=Diff(S^1)
\times Diff(S^1)$ where in light cone coordinates the first factor
corresponds to transformations $x^+ \mapsto \hat{x}^+(x^+)$ and the
second factor corresponds to $x^- \mapsto \hat{x}^-(x-)$. Note that
in these coordinates the metric has the form $ds^2=dx^+dx^-$ and
is form invariant (up to a scalar factor) under $Diff_c(M)$. Let
$\epsilon =\epsilon (x^{\pm})$ be a (small) function and let $A$ be
a smooth function of $x^{\pm}$, then one has
\be
A(x^{\pm}+\epsilon (x^{\pm}))=\mbox{exp}(\sum_n \epsilon_ne^{inx^{\pm}}
\frac{d}{dx^{\pm}})A(x^{\pm})
\ee
where $\epsilon_n$ are the Fourier coefficients of
$\epsilon (x^{\pm})$.
{}From this follows that the space of conformal Killing vectors is spanned by
\be \label{killing}
X^{\pm}_n=e^{inx_{\pm}}\frac{d}{dx_{\pm}}
\ee
which have the following commutation relations
\be    \label{confkil}
[X_n^{\pm},X_m^{\pm}]  =  -i(n-m)X_{n+m}^{\pm}
\ee

Now, let $\cal F$ be the `algebra of fields' of a certain
relativistic quantum field theory on $M$ with Hilbert space $\cal H$ (
i.e. $\cal F$ consists  of operator valued distributions $\phi (x)$
over $M$ such that $[\phi_1 (x),\phi_2 (y)]=0$ if $x$ and $y$
are spacelike separated. Furthermore the spectrum of the energy
momentum operators $P^{\mu}$  is concentrated in the closed forward
light cone $\bar{V}^+$ ($p^2\geq 0;\; p^0=E\geq 0$) and the Hilbert space
$\cal H$, which is an irreducible representation of $\cal F$, contains
precisely one state, called the physical vacuum, which is unvariant
under the unitary representation of the Poincare group that $\cal H$
carries). Physical operators like $P^{\mu}$ are usually unbounded
and can therefore not be defined on the whole Hilbert space. As usual
we therefore assume the existence of a common domain of definition
${\cal D} \subset {\cal H}$ for all fields which is dense in $\cal H$.

In a CFT the Hilbert space will carry a unitary projective representation
of the conformal group. This means that any element $f \in Diff_c(M)$
will be represented by a unitary operator $U(f)$ which is determined
up to a phase. If the theory is conformally invariant the representation
$U$ must give rise to a representation $R$ of the conformal group on the
field algebra $\cal F$ such that
\be \label{trans}
U(f)\phi (x)U(f)^{-1}=R_f(\phi (f(x)))
\ee
for any $f\in Diff_c(M)$ and $\phi \in \cal F$.
Equation (\ref{trans}) expresses the covariance of the theory under
conformal transformations.

Let us consider an infinitesimal conformal transformation generated
by a conformal Killing vector $X$ on $M$. of course it will be
represented on $\cal H$ by a Hermitean operator $T(X)$ which, due
to the projectivity of the representation, is determined up to
a constant. The fact that $T$ is a (projective) representation
of the Lie algebra of conformal Killing vectors is expressed by the
equation
\be
[T(X),T(X')]=T([X,X'])+c(X,X')\bf 1
\label{proj}
\ee
where $c(X,X')$ is a `central extension'. For an `ordinary'
representation this term
would have been absent. Defining  $L_n=T(X^+_n)$ and $\bar{L}_n=T(X^-_n)$
and using equations (\ref{confkil}) and (\ref{proj}) one finds
\ba
[L_n,L_m] & = & (n-m)L_{n+m}+\frac{c}{12}m(m^2-1)\delta_{n+m,0} \nonumber
\\
{} [\bar{L}_n,\bar{L}_m] & = & (n-m)\bar{L}_{n+m}+\frac{\bar{c}}{12}
m(m^2-1)\delta_{n+m,0} \nonumber \\
{}[\bar{L}_n,L_m] & = & 0 \label{vir}
\ea
where the central terms are fixed up to scalars $c,\bar{c}$ by the
Jacobi identities. In fact Lorentz invariance requires $c$ and $\bar{c}$
to be  equal. We conclude the Hilbert space of any CFT carries a
representation of $Vir \oplus Vir$.

{}From eq. (\ref{killing})
follows that $\frac{d}{dt}=X_0^++X_0^-$ while $\frac{d}{dx}=X_0^+-X_0^-$
which means that $L_0\pm \bar{L}_0$ are the generators of time and
spatial translations respectively in $\cal H$. Therefore $L_0+\bar{L}_0$
must be equal to the Hamiltonian $H\equiv P^0$ of the system while
$L_0-\bar{L}_0$ corresponds to the momentum operator $P \equiv P^1$.
Furthermore since space is compact we have $\phi (x+2\pi )=\phi (x)$
which means that the operator implementing spatial translations by
$2\pi$ given by $e^{2\pi iP}$ commutes with all fields in $\cal F$. Since
$\cal H$ is an irreducible representation of $\cal F$ we find by
Schur's lemma that $e^{2\pi iP}$ must be a multiple of the identity
operator, i.e. $e^{2\pi iP}=\alpha \bf 1$. Let $|p\rangle$ be a momentum
eigenstate $P|p\rangle=p|p\rangle$, then this implies that
$e^{2\pi ip}|p\rangle=\alpha |p\rangle$
and furthermore by taking $|p\rangle=|0\rangle$ we see that $\alpha =1$.
We conclude that the spectrum of $P=
L_0-\bar{L}_0$ is a subset of $\bf Z$ (i.e. it is quantized). In
CFT the eigenvalues of $L_0-\bar{L}_0$ are called the `conformal spins'.

It can be shown \cite{Nahm} that $T(X)$ can be written as
\be
T(X)=\frac{1}{2\pi}\int_{\Sigma}\; X^{\mu}T_{\mu \nu}d\sigma^{\nu}
\label{stress}
\ee
where $X=X^{\mu}\partial_{\mu}$ is a conformal Killing vector and
$\Sigma$ is some spacelike curve. In (\ref{stress}) we have
implicitly fixed the additive freedom in $T(X)$ by fixing the trace
of $T_{\mu \nu}$ to be zero, $T^{\mu}_{\mu}=0$. Furthermore varying
$\Sigma$ we find the conservation law $\partial^{\mu}T_{\mu \nu}=0$.
In light cone coordinates, and using the tracelessness of $T_{\mu \nu}$
this equation reads $\partial_-T_{++}=0$ and $\partial_+T_{--}=0$,
i.e. the only two nonzero components of $T_{\mu \nu}$ depend on
one light cone variable only.
This is an extremely strong statement for it implies that for any
$f=f(x^{+})$ one has the conservation law $\partial_-(fT_{++})=0$
(and similar equations for $T_{--}$). The charge
\be
Q_f=\int_{S^1}dx\; f(x^+)T_{++}(x^+)
\ee
is then likewise conserved. Since $f$ is arbitrary we conclude that
in 2 dimensions conformal invariance leads to an {\em infinite set
of independent conserved quantities}. Note that
from eq.(\ref{stress}) immediately follows
that $T_{++}(x^+)=\sum_n L_ne^{-inx_+}$ and $T_{--}(x^-)=\sum_n\bar{L}_n
e^{-inx^-}$. It is well known that $T_{\mu\nu}$ can be identified with the
stress energy tensor of the theory.

Let $|E,p\rangle \in \cal H$ be an eigenstate of $H$ and $P$ with energy $E$
and momentum $p$. It is easy to
check, using the commutation relations (\ref{vir}), that $L_n|E,p\rangle$
and $\bar{L}_n|E,p\rangle$ are also eigenstates of $H$ (and $P$), but with
energy $E-n$. Therefore the operators $L_n$ and $\bar{L}_n$ lower
the energy for positive $n$.
Since the energy is bounded from below (remember that
$p^0\geq 0$) there must exist a state $|E_0,p_0\rangle$ such that
$L_n|E_0,p_0\rangle=\bar{L}_n|E_0,p_0\rangle=0$ for all $n>0$. Define
$h=\frac{1}{2}(E_0+p_0)$ and $\bar{h}=\frac{1}{2}(E_0-p_0)$, called
`conformal dimensions', then from $H=L_0+\bar{L}_0$ and $P=L_0-\bar{L}_0$
follows that $L_0|E_0,p_0\rangle=h|E_0,p_0\rangle$ and
$\bar{L}_0|E_0,p_0\rangle=\bar{h}
|E_0,p_0\rangle$. Therefore denote $|E_0,p_0\rangle\equiv |h,\bar{h}\rangle$.
The linear
span of all states that one can obtain by successively acting with the
generators $\{L_{-n}\}_{n>0}$ on $|h,\bar{h}\rangle$ is an irreducible
Virasoro representation denoted by $L(h,c)$.
The same is true when we act with the generators $\{\bar{L}_{-n}\}_{
n>0}$ on $|h,\bar{h}\rangle$ but now we obtain the representation
$L(\bar{h},c)$. In this way the Hilbert space $\cal H$ decomposes
into irreducible $Vir \oplus Vir$ representations
\be
{\cal H} = \bigoplus_{i,{\bar{i}}} N_{i\bar{i}}
\, {\cal H}_i \otimes {\cal H}_{\bar{i}}  \label{dirsum}
\ee
where ${\cal H}_i=L(h_i,c)$, ${\cal H}_{\bar{i}}=L(\bar{h}_{\bar{i}}
,c)$
and $N_{i\bar{i}}$ are multiplicities (sometimes the set of these
multiplicities is called the `operator content' of the CFT).
The states $|h,\bar{h}\rangle$
which satisfy $L_n|h,\bar{h}\rangle=\bar{L}_n|h,\bar{h}\rangle=0$ for $n>0$
are called `highest weight states'. They are the states in the
Virasoro irreducible representations with lowest energy.
of course the (unique) vacuum state $|0>$ of the theory is such
a highest weight state with $h=\bar{h}=0$ and the Virasoro
representation $L(0,c)$ is called the `vacuum representation'.

Note that one only needs to know the set of highest weight states
to {\em completely diagonalize} the Hamiltonian, for applying
the Virasoro generators to them one can find a basis of the
Hilbert space consisting only of eigenvectors of the Hamiltonian.
Also note that since the energy momentum eigenvalues are restricted
to $\bar{V}^+$ (i.e. $E^2-p^2\geq 0$) it follows that $E\geq |p|$.
{}From this it is easy to show that $h_i,\bar{h}_{\bar{i}}\geq 0$
for all $i,\bar{i}$. Therefore the Virasoro representations
appearing in CFT are the so called
unitary irreducible `highest weight'
modules. In order to classify conformal field theories one
would like to find a list of all Virasoro  representations
of this type for given $c$.

It turns out \cite{vir} that the Virasoro algebra
only has unitary representations for $c>0$. For $c>0$ we have to
distinguish between two possibilities: $0<c<1$ and $c\geq 1$.
For $c\geq 1$ there exists a unique unitary irreducible Virasoro
representation $L(h,c)$ for any positive and real $h$. For $0<c<1$
the situation is drastically different however. To begin with there
only exist unitary irreducible Virasoro representations for
\be
c=c(m)=1-\frac{6}{m(m+1)}
\ee
where $m\in\{3,4,\ldots \}$. Furthermore the irreducible representation
$L(c(m),h)$ is only unitary for
\be
h=h^{(m)}(r,s)=\frac{((m+1)r-ms)^2-1}{4m(m+1)}
\ee
where $1\leq r\leq m-1$ and $1 \leq s \leq m$. Note the remarkable
fact that for $c=c(m)$ there are only a {\em finite} number of
values for $h$ that lead to unitary Virasoro representations. From
this follows that if we are given a conformally invariant
quantum field theory and by explicit calculation we find the
central value $c$ to be one of the values $c(m)$ then the
direct sum decomposition (\ref{dirsum}) is finite if
the multiplicities $N_{i \bar{i}}$ are finite (in fact for many
conformal field theories that have been studied they are either
0 or 1). For conformal field theories with $c \geq 1$
the decomposition (\ref{dirsum}) can
be infinite even if the multiplicities are all finite.

We are now in a position to roughly describe the idea of a $W$ algebra.
Suppose the Virasoro algebra is just a subalgebra of the total
symmetry algebra of the quantum field theory. With respect to
this larger algebra the Hilbert space may have a finite decomposition
even if the decomposition w.r.t. Virasoro algebra is infinite.
A conformal field theory with such a larger symmetry is called a
`rational conformal field theory'. Before we introduce $W$ algebras
more formally we still have to introduce the concept of a `primary field'.

Consider the (infinitesimal) action of the conformal Killing vectorfields
$X^{\pm}_{\pm 1},X^{\pm}_0$ on $\cal F$. Define the operators ${\cal L}_{\pm},
{\cal L}_0$ (and their barred counterparts) by
\ba
[L_{-1},\phi (x^+,x^-)] & = & e^{-ix^+}({\cal L}_{-1}\phi )(x^+,x^-)
\nonumber \\
{} [L_0,\phi (x^+,x^-)] & = & ({\cal L}_{-1}+{\cal L}_0)\phi (x^+,x^-)
\nonumber \\
{} [L_1,\phi (x^+,x^-)] & = & e^{ix^+}({\cal L}_{-1}+2{\cal L}_0+
{\cal L}_1)\phi (x^+,x^-) \label{L}
\ea
These equations are nothing but infinitesimal versions
of eq. (\ref{trans}). It can now be shown \cite{Nahm} that the
field algebra $\cal F$ decomposes into a direct sum of ${\cal L}_0$
and $\bar{\cal L}_0$ eigenspaces ${\cal F}^{(h,\bar{h})}=\{
\phi \in {\cal F} \mid {\cal L}_0\phi=h\phi \, ; \; \bar{\cal L}_0
\phi = \bar{h} \phi \}$. An element of ${\cal F}^{(h,\bar{h})}$ is
called a `field of left conformal dimension $h$ and right conformal
dimension $\bar{h}$'. Furthermore one can show that if $\phi \in
{\cal F}^{(h,\bar{h})}$ and
\be
\phi (x^+,x^-)=\sum_{k,\bar{k}}e^{-ikx^+-i\bar{k}x^-}\phi_{k,\bar{k}}
\label{four}
\ee
is its Fourier expansion, then $\phi_{k,\bar{k}}|0>=0$ for
$k>-h$ or $\bar{k}>-\bar{h}$ and also that the map $F:{\cal F}
\rightarrow {\cal H}$ defined by $F(\phi )=\phi_{-h,-\bar{h}}|0>$
is an isomorphism (to be completely precise the image of $F$ is
the common domain $\cal D$ of definition of all fields in $\cal F$).

As we have seen the Hilbertspace contains a set of states that through
the Virasoro algebra generate all other states. We called these the
highest weight states $|h,\bar{h}\rangle$. The inverse image under the
isomorphism $F$ of such a highest weight state is called a {\em
primary field}. Using the fact that ${\cal L}_0=F^{-1}L_0F$ one can
easily show that ${\cal L}_0\phi=h\phi$ which means that the
primary field $\phi
 \equiv F^{-1}(|h,\bar{h}\rangle)$ is an element of
${\cal F}^{(h,\bar{h})}$ as was to be expected. Also, expanding a
primary field into Fourier modes as in (\ref{four}) one finds that
\be      \label{pr}
[L_n,\phi_{k,\bar{k}}]=(n(h-1)-k)\phi_{k+n,\bar{k}}
\ee
and a similar equation for the commutations relation of
$\phi_{k,\bar{k}}$ with $\bar{L}_n$. Equation (\ref{pr}) expresses
the transformation properties of a primary field under infinitesimal
conformal transformations. Integrating this equation we obtain
\be
U(f)\phi (x) U(f)^{-1}=\left( \frac{df^+}{dx^+} \right) ^h
\left( \frac{df^-}{dx^-} \right)^{\bar{h}}\phi (f(x))
\ee
where $f\equiv (f^+,f^-) \in Diff_c(M)$. Fields of conformal
dimensions $(h,0)$ are called `left chiral' and do not depend
on $x^-$ while fields of conformal dimensions $(0,\bar{h})$ are
called `right chiral' and do not depend on $x^+$. Obviously a field
of conformal dimensions $(0,0)$ is constant and conversely a constant
has conformal dimensions $(0,0)$.

Consider the operators ${\cal L}_n=F^{-1}L_nF$ and their barred
counterparts $\bar{\cal L}_n=F^{-1}\bar{L}_nF$ which
map the field algebra $\cal F$ into itself. Obviously we have
${\cal L}_n\phi=0$ if $\phi$ is a primary field and $n>0$. Define
now the conformal family  $[\phi ]$
associated to the primary field $\phi$
to be the subset of $\cal F$ spanned by fields of the form
\be \label{desc}
\phi^{\{p\} \{\bar{p}\}}\equiv
{\cal L}_{-p_1}\ldots {\cal L}_{-p_k}\bar{\cal L}_{-\bar{p}_1}
\ldots \bar{\cal L}_{-\bar{p}_l} \; \phi
\ee
where $p_1\geq \ldots \geq p_k \geq 0;\; \bar{p}_1 \geq \ldots \geq
\bar{p}_l \geq 0; \; k,l\geq 0$ and $\{p\}=(p_1,\ldots ,p_k)$. The fields
(\ref{desc}) are called 'descendent fields'. Note that $[\phi ]\simeq
[\phi ]_L \otimes [\phi ]_R$ where $[\phi ]_L$  is spanned by
$\phi^{\{p\}\{0\}}$ and $[\phi ]_R$ by $\phi^{\{0\}\{\bar{p}\}}$.
Furthermore from eq.(\ref{dirsum}) follows that
\be  \label{dirsum2}
{\cal F}=\bigoplus_{i,\bar{i}}N_{i,\bar{i}} \;
[\phi^{(h_i,\bar{h}_{\bar{i}})}]_L
\otimes [\phi^{(h_i,\bar{h}_{\bar{i}})}]_R
\ee
where $\phi^{(h,\bar{h})}$ is a primary field with conformal dimensions
$(h,\bar{h})$.

The subalgebras $\cal A$ and
$\bar{\cal A}$ of $\cal F$ defined by
\be
{\cal A}=\bigoplus_{h} {\cal F}^{(h,0)} \;\;\mbox{ and } \;\;\bar{\cal A}=
\bigoplus_{\bar{h}} {\cal F}^{(0,\bar{h})}
\ee
are called the `left and right {\em chiral algebras}' of the
conformal field theory respectively. $\cal A$ and $\bar{\cal A}$
commute for if $\phi_L=\phi_L(x^+) \in {\cal A}$ and $\phi_R=\phi_R(y^-)
\in \bar{\cal A}$ then by using invariance of $\phi_L (\phi_R )$
under translations in $x^-$ (respectively $x^+$) one can always
spatially separate $(x^+,x^-)$ and $(y^+,y^-)$ which means that the
fields must commute. From this follows that the Hilbert space
decomposes into a direct sum of irreducible ${\cal A} \otimes
\bar{\cal A}$ representations
\be
{\cal H}=\bigoplus_{(a,\bar{a})} N_{a,\bar{a}}^{\cal A} \;
{\cal H}_a^{\cal A} \otimes \bar{\cal H}_{\bar{a}}^{\bar{\cal A}}
\label{chirdec}
\ee
Note that for a CFT the chiral algebras always contain the Virasoro
algebra due to the fact that, as we have seen, $T_{++}\in {\cal A}$
and $T_{--} \in \bar{\cal A}$. It can therefore happen that the
decomposition (\ref{chirdec}) is finite even if the decomposition
(\ref{dirsum}) is infinite. In that case the CFT is called a rational
conformal field theory (RCFT).
A $W$ algebra \cite{Za} is a chiral algebra that
is generated (in a way that we shall explain later) by a set of
chiral primary fields.

In the literature CFT is usually formulated within the framework
of Euclidean field theory (this is actually quite natural from the
point of view of the applications). First one can extend the fields defined
on Minkowski space $M$ to fields in two complex variables by putting
\be
\phi (\xi,\bar{\xi})=e^{iP.\Delta x}\phi (x^0,x^1)e^{-iP.\Delta x}
\ee
where $\Delta x=(iy^0,iy^1)$, $\xi=x^0-iy^0$ and $\bar{\xi}=x^1+iy^1$.
The Euclidean domain is defined as usual by $x^0=y^1=0$ and $y^0=\tau$
which effectively implies a Wick rotation $t \rightarrow -i\tau$.
It is more or less customary in CFT to work in the  so called
`radial picture' since it allows one to use the powerful theory
of complex functions and contour integrals. For this define
$z=e^{i(\xi +\bar{\xi})}$ and $\bar{z}=e^{i(\xi-\bar{\xi})}$.
In the Euclidean
domain we have $z=e^{\tau+ix^1}$ and $\bar{z}=e^{\tau-ix^1}$
which means that $z^*=\bar{z}$. Also the infinite past and future
are mapped
to $z=0$ and $z=\infty$ respectively on the plane and equal time
surfaces become circles of constant radius. The Wick rotated
spacetime $M={\bf R}\times S^1$ has become the complex plane with
the origin removed. Furthermore left and right chiral fields
depend holomorphically and antiholomorphically on $z$ respectively.

Usually however one leaves the restriction to the Euclidean domain
until after one has developed the theory. We are therefore dealing
with a theory on $\ce^2$ in which $z$ and $\bar{z}$ are considered
to be independent complex variables.
It is then often convenient to introduce slightly different fields
when working in the radial picture. Let $\phi (x^0,x^1)$ be a field
and extend it to $\ce^2$, $\phi=\phi (z,\bar{z})$. Then define
the field $\tilde{\phi}(z,\bar{z})$ by $\phi (z,\bar{z})=
z^h\bar{z}^{\bar{h}}\tilde{\phi}(z,\bar{z})$. It can now easily be
seen that the `radial field' $\tilde{\phi}$ has the expansion
\be
\tilde{\phi}(z,\bar{z})=\sum_{k,\bar{k}}z^{-k-h}\bar{z}^{-\bar{k}-
\bar{h}}\tilde{\phi}_{k,\bar{k}}
\ee
and furthermore that $\tilde{\phi}_{k,\bar{k}}|0>=0$ for $k>-h$
or $\bar{k}>-\bar{h}$. From now on we will omit the explicit
$\bar{z}$ dependence of fields where possible.

In Euclidean field theory products of operators $A(z)B(w)$ are only
defined for $|z|>|w|$ (Euclidean correlation functions are only
defined if the operators in the correlator
are time ordened. Remember that in the radial picture $|z|>|w|$
means that $z$ is later than $w$. of course in the Euclidean
functional integral definition of correlation functions
time ordening is automatic). Therefore define the radial
ordening operator $R$ as follows: $R(A(z)B(w))=A(z)B(w)$ if
$|z|>|w|$, and $R(A(z)B(w))=B(w)A(z)$ if $|z|<|w|$.

Now consider the commutator between two fields $A(z)$ and $B(z)$.
Without loss of generality we can take $A(z)$ and $B(z)$ to
be fields of left conformal dimensions $h_A$ and $h_B$
respectively.
Again using translational invariance of these fields in $\bar{z}$
it is easy to show that $[A(z),B(w)]$ is zero unless $z=w$.
Therefore the equal time
commutator can be expressed as a linear combination
of derivatives of $\delta (z-w)$, i.e.
\be
[A(z),B(w)]=\sum_{n=1}^{h_A+h_B}\{AB\}_{(n)}(w)
\frac{(-\partial)^{n-1}}{(n-1)!}\delta (z-w) \label{co}
\ee
where $\{AB\}_{(n)}(z)$ are elements of $\cal F$. The reason that
the summation runs only up to $h_A+h_B$ is that by performing a
scale transformation on this equation one finds that the field
$\{AB\}_{(n)}$ has conformal dimension $h_A+h_B-n$. Since there
exist no fields with negative conformal dimension the summation
terminates at $h_A+h_B$.
The left hand side of the equal time commutator (\ref{co})
can be written as
\be
[A(z),B(w)]=\left( \lim_{|z|>|w|}-
\lim_{|z|<|w|} \right) R(A(z)B(w))
\ee
There exists a similar expression for the delta function.
\be
\delta (z-w) =\left( \lim_{|z|>|w|}-
\lim_{|z|<|w|} \right) \frac{1}{z-w}
\ee
{}From the above follows
that the functions $A(z)B(z)-
\sum_n \{AB\}_{(n)}(w)(z-w)^{-n}$ and $B(z)A(z)-\sum_n
\{AB\}_{(n)}(w) (z-w)^{-n}$ are holomorphic for $|z|>|w|$
and $|z|<|w|$ respectively. Since they have the same
(boundary) value at $|z|=|w|$ we find that they are holomorphic for all $z$
and are the same function. One can now expand this function
in a power series. The result is
\be \label{ope}
R(A(z)B(w))=\sum_{k=h_A+h_B}^{-\infty} \frac{\{AB\}_{(k)}(w)}{(z-w)^k}
\ee
i.e. the product of two fields can be expressed as a power
series in terms of $z-w$. This is called the {\em operator
product expansion} (OPE).

Let $\{\phi_n(z,\bar{z})\}_{n\in I}$ be the set of primary fields in a
certain CFT.
{}From eq.(\ref{dirsum2}) follows that we can expand the OPE  between
$\phi_n$ and $\phi_m$ as follows
\be   \label{furu}
\phi_n(z,\bar{z})\phi_m(w,\bar{w})=\sum_{k}\sum_{\{p\}\{\bar{p}\}}
C^k_{nm;\{p\}\{\bar{p}\}}(z-w,\bar{z}-\bar{w})\, \phi_k^{\{p\}\{\bar{p}\}}
(w,\bar{w})
\ee
Define the number $N_{nm}^k$ to be 1 if $C^k_{nm;\{p\}\{\bar{p}\}}\neq 0$
for some $\{p\},\{\bar{p}\}$ and 0 otherwise.
Obviously these numbers encode whether in the OPE of $\phi_n$ and
$\phi_m$ there appear elements of the conformal family
of $\phi_k$. It can be shown that a
three point function $\langle \phi_n\phi_m\phi_k \rangle$ can only
be nonvanishing if $N_{nm}^k\neq 0$ which means that $N_{nm}^k$
encodes which interactions are present in the theory. It is usual to
denote this information in terms of the so called `fusion rules'
\be
[\phi_n]\cdot [\phi_m] =\sum_k N_{nm}^k \, [\phi_k]
\ee
which obviously is a shorthand notation for which conformal families
interact with which.
Given the unitary representations of a certain chiral algebra $\cal A$
the fusion rules provide a useful characterization of the
associated CFT. The calculation of the fusion rules of conformal field
theories with extended symmetries  is still an open problem at this time.

Using the OPE between $A$ and $B$ we can calculate the
commutator between their modes which are defined by
\be
A_n=\oint_0 \frac{dz}{2\pi i}z^{n+h_A-1}A(z)
\ee
and a similar equation for $B_n$. The commutator $[A_n,B_m]$
is the difference of two terms both
containing two contour integrals around the origin.
The difference of two of those contours can be replaced
by a contour surrounding $w$ \cite{CFT}. One then obtains
\be
[A_n,B_m]=\oint_0\frac{dw}{2\pi i}w^{n+h_B-1} \oint_w
\frac{dz}{2\pi i}z^{n+h_A-1}R(A(z)B(w))
\ee
Inserting eq.(\ref{ope}) one can calculate from this
the commutator between $A_n$ and $B_m$.

As usual the vacuum expectation value $<0|A(z)B(w)|0>$ of
a product of two fields diverges when $z\rightarrow w$. We therefore define
a normal ordened product as follows: Let the left conformal
dimensions of $A(z)$ and $B(z)$ be $h_A$ and $h_B$ respectively,
then the normal ordened product, denoted by $(AB)(z)$, is
defined by $(AB)(z)=A_-(z)B(z)+ B(z)A_+(z)$ where
$A_-(z)=\sum_{n\leq -h_a}A_nz^{-n-h_A}$ and $A_+=A-A_-$.
If
\be
A(z)B(w)=\sum_{k=h_a+h_B}^{-\infty} \frac{\{AB\}_{(k)}(w)}{(z-w)^k}
\ee
is the OPE between $A$ and $B$ then it can easily be shown that
$(AB)(z)=\{AB\}_{(0)}(z)$.

Let $\{A_i(z)\}$ be a certain set of chiral fields. The algebra
they generate is the subset of $\cal F$ that can be obtained
by taking arbitrary linear combinations, derivatives and
normal ordened products of these fields. A $W$ algebra is then
a chiral algebra which is generated by $T(z)$
and a set of chiral primary fields. These algebras play an
important role in CFT for reasons that we explained above.

of course the simplest example of a $W$ algebra is the Virasoro
algebra, i.e. the chiral algebra generated by $T(z)$ alone. Let us
now look at some extensions of this algebra.
A  very important example, is the algebra generated
by the  fields $T(z)$ and $\{J^a(z)\}$,
where the index $a$
runs over a basis of a simple Lie algebra $g$,
and the fields $J^a(z)$ are all chiral primary fields of
conformal dimension 1. The OPEs of the fields $J^a(z)$ are given
by
\be
J^a(z)J^b(w)=\frac{kg^{ab}}{(z-w)^2}+\frac{f^{ab}_cJ^c(w)}{z-w}+\ldots
\ee
where $k$ is a central value called the level and $g^{ab}$ is
the Cartan Killing form on the Lie algebra $g$.
The field algebra generated by the fields $J^a(z)$ itself contains
a stress energy tensor w.r.t. which the fields $J^a(z)$ are primary
of conformal dimension 1. It reads
\be
T(z)=\frac{1}{2(k+h)}g_{ab}(J^aJ^b)(z)  \label{su}
\ee
where $h$ is the dual coxeter number of $g$. The stress energy tensor
(\ref{su}) is called the affine Sugawara stress energy tensor and one can
show by explicit calculation that the central value associated to it
is
\be
c(g;k)=\frac{k\mbox{dim}(g)}{k+h}
\ee
The modes of the fields $J^a$ form an affine Lie algebra.
The structure and representation theory of this
algebra is well known \cite{Kac,GoOl} and it will be shown in this
thesis that many (perhaps all) $W$ algebras
are reductions of it.

The $W$ algebras considered above all have in common that
the modes of the fields that generate the $W$ algebra span
a Lie algebra, i.e. the commutator of two basis elements is
again a linear combination of basis elements. This needn't always be
true as was first realized by Zamolodchikov \cite{Za}.
He consider the  example of an algebra generated by $T(z)$ and a
chiral primary field $W(z)$ of conformal dimension 3 with
OPE
\ba
W(z)W(w) & = &\frac{c/3}{(z-w)^6}+\frac{2T(w)}{(z-w)^4}+\frac{\partial
T(w)}{(z-w)^3} + \nonumber \\
&   & +\frac{1}{(z-w)^2}\left( 2\beta \Lambda (w)+\frac{3}{10}\partial^2
T(w)\right) +\nonumber \\
&   & + \frac{1}{z-w}\left( \beta \partial \Lambda (w)+\frac{1}{15}
\partial^3T(w)\right) +\ldots
\ea
where $\Lambda=(TT)-\frac{3}{10}\partial T$ and
\be
\beta =\frac{16}{22+5c}
\ee
Since the right hand side of this OPE contains
terms quadratic in $T$ the commutator of the modes of
$W(z)$ will involve quadratic terms. The $W$ algebra generated
by $T(z)$ and $W(z)$ is called $W_3$. Like the Virasoro algebra
it has a discrete series of values for $c$ for which there exist only a
finite number of conformal dimensions that give rise to
unitary representations (note that there is an additional
label (or quantum number)
corresponding to the $W_0$ eigenvalue). These representations
have central value
\be
c=c(m)=2\left( 1-\frac{12}{m(m+1)} \right)
\ee
with $m=3,4,\ldots$. Note that the critical value $c=1$ of the
Virasoro algebra has been shifted to a higher value by the fact
that more symmetry is present.
The $W_3$ algebra has generalizations
to arbitrary $N$, i.e. algebras whose generators have conformal weights
$(2,3,\ldots ,N)$  called $W_N$. Like the Virasoro algebra
and $W_3$ all these algebras
admit a (principal)
series of ($W$) minimal models. The central charges associated
to these minimal models are (for $W_N$)
\be
c(m)=(N-1)\left( 1-\frac{N(N+1)}{m(m+1)}\right)
\ee
Again there exists only finitely many allowed conformal
dimensions for each $c$-value such that the corresponding representation
is unitary.

In the applications of CFT to string theory and statistical mechanics
it is necessary to consider CFT on higher genus Riemann surfaces. CFT on
the cilinder can be transfered to the torus as follows: Define the
torus by identifying the spatial $S^1$ at $t=0$ with the spatial $S^1$
at $t=2\pi\tau_0$. In doing so one can introduce a shift in $x$ by
$2\pi\tau_1$. Define then the partition function $Z=\mbox{Tr}_{\cal H}
(e^{2\pi
i\tau_1 P_1}e^{2\pi i\tau_0P_0})$ where as before $P_0$ and $P_1$ are
the energy and momentum operators respectively. In the Euclidean theory
we have to identify $z$ with $qz$ where $q=e^{2\pi i\tau}$ and
$\tau=\tau_1+i\tau_0$. The Torus partition function can then be shown
to be
\be
Z(q)=\mbox{Tr}_{\cal H}(q^{L_0-\frac{c}{24}}\bar{q}^{\bar{L}_0-
\frac{c}{24}})
\ee
Inserting the decompostion (\ref{chirdec}) into $Z$ one finds
\be
Z(q)=\sum_{(a,\bar{a})}N_{a\bar{a}}\,\chi_a(q)\chi_{\bar{a}}(\bar{q})
\ee
where $\chi_a(q)$ is the `character' of the representation ${\cal H}_a$
of the chiral algebra $\cal A$
\be
\chi_a(q)=\mbox{Tr}_{{\cal H}_a}(q^{L_0-\frac{c}{24}})
\ee
In order for the CFT to make sense on the torus it should not depend
on the way in which the torus is obtained from the cilinder, which means
that $Z$ should be invariant under the modular group $SL(2;{\bf Z})$
of the torus. This group is generated by the transformations
\be
T:\tau \mapsto \tau +1 \;\;\; \mbox{ and } \;\;\; S:\tau \mapsto -
\frac{1}{\tau}
\ee
and the fact that $Z$ must be invariant under these transformations
places strong constraints on the coefficients $N_{a\bar{a}}$. The
study of these constraints is called the `modular bootstrap' and has,
in case the chiral algebra is the Virasoro algebra with $c<1$ lead
to a complete classification of modular invariant partition functions
\cite{ADE}. For $c\geq 1$ modular invariance requires any CFT to
contain an infinite number of representations of the Virasoro algebra.
As we have seen
before $W$ algebras are a possible solution to this difficulty
since even though the decomposition may by infinite w.r.t. the
Virasoro algebra it can still be finite w.r.t. an extension of this
algebra. However for general $W$ algebras the modular program
is more complicated and has not been completed sofar.

The construction of $W$ algebras has proceeded roughly along three
independent lines. Many explicit results were obtained by the so
called `direct method' which boils down to solving the Jacobi identies
(for a review see \cite{BoSc}). This method however has the drawback
that it does not yield any information on the representation theory or
geometrical interpretation of $W$ algebras. It does give explicit
results on the algebraic structure however (i.e. structure constants).

Another approach to $W$ algebras is the so called coset construction
\cite{coset,BBSS}.
In this approach one constructs the generators and representations
of a $W$ algebra in terms of the representation theory of an affine
Lie algebra. The generators of the $W$ algebra become composite
fields in terms of the affine currents (or in a more mathematical
language, one embeds the $W$ algebra into the universal enveloping
algebra of an affine Lie algebra). We have already seen an example
of this in the case of the Virasoro algebra eq.(\ref{su}). One can generalize
the affine Sugawara construction as follows: Let $g$ be a simple Lie algebra,
$g'\subset g$ and let $\hat{g},\hat{g}'$ be their affinizations. The
central charge $k'$ of $\hat{g}'$ will in general be equal to $jk$
where $k$ is the central charge of $\hat{g}$ and $j$ is the index
of the embedding $g' \hookrightarrow g$. In the coset construction one
then starts with the affine Sugawara tensors $T(z)$ and $T'(z)$ associated
to $\hat{g}$ and $\hat{g}'$ respectively and considers the field
$\tilde{T}=T-T'$ which remarkably
generates a Virasoro algebra of central charge
$c(g,k)-c(g',k,)$. Now, the point is that the Virasoro algebra
$\tilde{T}$, called the coset algebra, commutes with the subalgebra
$\hat{g}'$ which means that a highest weight module $L(\Lambda )$
of $\hat{g}$ decomposes into a direct sum of
$Vir \oplus \hat{g}'$ representations. Obviously the coset space obtained
from $L(\Lambda )$ by identifying states in the same $\hat{g}'$
highest weight module is a  (not necessarily irreducible)
highest weight module of the coset algebra.
Now, for certain $g'$ it may happen that the commutant of $\hat{g}'$
in the universal enveloping algebra of $\hat{g}$ contains primary fields
of higher conformal dimension generating a $W$ algebra. In that case
we can obtain representations of this $W$ algebra in the same way as
we discussed above for the Virasoro algebra.

The coset construction has some drawbacks. For example, it
is not clear that the coset algebra (i.e. the commutant of $\hat{g}'$)
is finitely generated. In fact this is in general not the case
except, possibly, on some very specific $\hat{g}$ modules. In the case
of the algebra $W_3$ this was shown in \cite{BBSS}. This means that
the coset construction does not work as a construction between
abstract algebras but only  on a subset of
representations which is rather unappealing. Another drawback
is that the set of $W$ algebras that one has been able to treat
in this way is limited and that it does not seem
to lead to a general classification scheme. Furthermore it does
not provide any new insights into the geometrical
background of $W$ algebras. On the other hand the coset construction does yield
an explicit construction of unitary representations of $W$ algebras
and, as will be shown in the next chapter, it can be realized at the
Lagrangian level by the theory of gauged WZW models.

The most powerful method for constructing $W$ algebras,
and the one that has lead to important progress in trying to understand
them, is the
so called Drinfeld-Sokolov (DS) reduction procedure \cite{DS,BFFOW,BTV}.
This procedure is
the subject of this thesis and we shall therefore only describe it very
briefly here. Drinfeld-Sokolov reduction is in fact (classically)
simply Poisson reduction in infinitely many dimensions where as Poisson
structure one takes the Kirillov Poisson structure associated naturally
to any Lie algebra. Quantization of this reduction procedure is
performed by BRST methods \cite{FaLu,FeFr,BT}. The DS approach has lead to
many insights into the structure and meaning of $W$ algebras as
well as to the construction of field theories with
explicit $W$ symmetry.
Also DS reduction yields a functor (unlike the coset construction
where only a small number of affine Lie algebra
representations are allowed) from
the category  of  representations of the affine Lie algebra
to the category of $W$ representations
\cite{FeFr}. In particular it is possible to obtain free field
realizations of arbitrary $W$ algebras in this way \cite{FaLu,FeFr,BT}.
Furthermore
one can obtain a great deal of information on the theories
with $W$ symmetry (for example correlation functions etc.)
using the fact that they are a reduction from a theory with
affine symmetry.

This concludes our introduction into CFT and $W$ algebras.
To provide some additional motivation we shall give a quick overview of the
different areas of theoretical physics where $W$ algebras and CFT
are of interest. At the end of this first and
introductory chapter we briefly discuss the contents of the
remaining chapters.

\section{String theories of elementary particle physics}
String theory is an attempt to unify all known fundamental forces and
particles into a single consistent mathematical framework. At the microscopic
level (at distances down to $10^{-15}$ cm)
and for processes involving strong and electroweak interactions
the quantum field theory called the
'standard model' (SM) is very successful.
With it we have uncovered at least part of the truth.
On the other hand general relativity (GR) has been very
successful in describing
the large scale behaviour of the universe, in particular
all experimentally accessible phenomena involving gravitational
interactions; from
suble effects in the solar system such as the perihelion precession
of Mercury and the bending of light by the sun, to the cosmological
aspects of the universe as a whole. The interpretation
that the SM and GR provide of the fundamental
forces they describe are however
completely different. GR describes gravity purely in geometrical terms
as the curvature of spacetime, while in the SM forces are seen as the
exchange of particles (gauge bosons). Nevertheless, lead by past experience,
it is conjectured by most physicists that GR and the SM are really
two `low energy limits' of the same fundamental theory and in
order to construct
this theory it has been tried very hard to quantize GR. Unfortunately
quantum gravity is a nonrenormalizable quantum field theory essentially
due to the fact that the gravitational coupling constant (Newton's
constant) is not dimensionless. Therefore it suffers incurable
infinities which render the theory without predictive power..

String theory \cite{GSW,LuTh,Kaku,Hat}
is a new approach to the unification of the SM and GR.
The revolutionary aspect of string theory is that its basic
assumption is that matter is made up of very tiny one dimensional
objects (strings). While a particle looks like a point at
$10^{-13}$ cm, if we magnify it to scales resolving $10^{-33}$ cm
then we may see that it is in fact an extended object. Unification
of the complete particle spectrum
is achieved in string theory by interpreting  different vibrational
modes of a string as the different fundamental particles. In particular
the known particles are believed to
correspond to the massless sector of a
(super)string theory (the particle masses are assumed to be spontaneously
or dynamically generated by breaking of (super)symmetry). There is
no a priori reason to assume that particles are really
one dimensional objects, however it turns out that
(if spacetime is flat) the string spectrum contains a particle like
state which transforms as a symmetric rank 2 traceless tensor under
Lorentz transformations of spacetime,  which is massless and
transverse. This state can be identified with the graviton.
Calculating the scattering amplitudes involving external graviton
states in the low energy limit (using the string prescription)
one finds that they agree with the amplitudes of gravitons evaluated
from the Einstein-Hilbert action. Thus quantum Einstein gravity
appears as an effective low energy theory and questions like
unitarity and renormalizability must be addressed not at the level
of Einstein theory, but at the string level.

Now, string theory
is expected to have improved ultraviolet behaviour to all order
compared to point particle physics
due to the inherent fuzzyness  of the string. In point particle
physics divergencies arise because when two vertices of a diagram get
arbitrarily close the propagator connecting these vertices blows up.
In string theory the interaction between strings is pictured
as the joining and splitting of these strings. The difference with
point particle physics is that, because of the fact that strings are
extended objects, there is no Lorentz invariant notion of a spacetime
point at which the joining or splitting occured. This means that the
interaction vertex has become inherently fuzzy and there is no
analogue of the region where different vertices come
arbitrarily close. If one compares string theory with the point particle theory
of gravity on the quantum level, one learns that string theory has a
built in ultraviolet regulator provided by the tower of massive
modes of the string. These beautifully conspire to produce a finite
result for the generic higher order quantum corrections. This is
certainly one of the main achievements of string theory so far. Also
another plague that appears to kill most particle theories, known
as anomalies, are remedied in string theory.

Now, it is important to note that since any part of a
string diagram describing the joining and splitting of two
strings locally looks like the propagation of a free string (again
because of the  fuzzyness of the string diagrams)
one finds that once one
has decided upon a set of rules for free string propagation this
fixes the complete theory. In point particle physics one still has the
choice of certain factors to associate to the vertices of a Feynman
diagram.

At the present state of development the full underlying
theory of strings,
sometimes also called string field theory, is not known (there has
been progress in this area however, see for example \cite{SFT}). It {\em is}
possible to develop a perturbative approach to string
theory by expanding around a `classical solution' of the theory. This means
that in some way an indefinite number of gravitons have condensed
into a classical groundstate described by a metric $\{G_{\mu\nu}\}_{\mu\nu
=1}^D$, where $D$ is the dimension of spacetime. The coupling of the
string to this nontrivial background is described by a 2 dimensional
$\sigma$-model. This can be seen by, in analogy with
the action of a
point particle, taking the action of the string to be
proportional to the area of the worldsheet $\Sigma$
it sweeps out in spacetime
\be
S[X]=T\int_{\Sigma}d^2\sigma \; \sqrt{\mbox{det}(G_{\mu\nu}(X)
\partial_aX^{\mu}\partial_bX^{\nu})} \label{NG}
\ee
where the Roman indices $a,b,\ldots$ run over the `world sheet'
coordinates $\sigma \equiv (\sigma^0,\sigma^1 )$ and $X^{\mu}(\sigma )$
describes the position of the string.
The constant of proportionality $T$ is called the `string tension'
because it can be interpreted as the energy per unit length of the
string \cite{GSW}.

Obviouly the action (\ref{NG}) is very difficult to work with due
to the square root in the integrand. It is possible to
circumvent this difficulty by considering the theory (\ref{NG}) as
a subsector of an enlarged theory. For this introduce extra degrees
of freedom in the form of a world sheet metric $h_{ab}(\sigma )$
and consider the action
\be
S_M[X;h]=\frac{T}{2}\int_{\Sigma}d^2\sigma \sqrt{h}h^{ab}\partial_a
X^{\mu} \partial_b X^{\nu} G_{\mu\nu}(X)\label{Polac}
\ee
which reduces  to (\ref{NG}) after insertion of the equation of
motion for $h_{ab}$  into (\ref{Polac}).

At this point we would like to make a remark on why one considers
strings and not 2 or 3 dimensional objects (in principle 3 dimensional
objects would be the most appealing intuitively). It turns out
that it is easy to write down a general action for such objects, just
as we did for strings, but that the resulting quantum theory is
nonrenormalizable by power counting. Also such theories do not
have Weyl symmetry which is special to strings. There has been some
activity however in this area and  maybe some day it will
turn out to be possible after all to write down a renormalizable
theory for such higher dimensional objects.

The theory of free strings discussed above is to be considered as a first
quantized theory, which means that interactions will have to be
put in by hand. The interaction between strings is, in analogy with point
particle physics, considered to be the joining and splitting of strings.
An $N$ string scattering process will perturbatively be represented
by a sum over all Riemann surfaces connecting the $N$ external
strings. The $g^{th}$ order term in this perturbation series
corresponds to a  compact genus $g$ Riemann surface with $N$ punctures
(the $N$ tubes attached to the surface extending into the far past
and future, depending on whether the string is incoming or outgoing,
can be `scaled down' by Weyl rescalings of the world sheet metric to
punctures on a compact surface). The partition function, expressing
the vacuum to vacuum amplitude, will obviously involve a
sum over all topologies of compact Riemann sufaces (without punctures).
Since compact and orientable Riemann surfaces are classified by their
genus the sum over topologies can be implemented by a sum over genera.
In Minkowski space and after Wick
rotating both in spacetime and on the worldsheet the partition function
becomes
\be
Z=\sum_{g=0}^{\infty} \int \;[DX][Dh]\; \mbox{exp}\left(
-\int_{\Sigma_g}\, d^2\sigma \sqrt{h}h^{ab}\partial_aX^{\mu}
\partial_bX_{\mu} \right) \label{part}
\ee
where $\Sigma_g$ is a compact genus $g$ Riemann surface. Note that
this approach is necessarily perturbative.

Remember that the action (\ref{Polac}) is invariant under arbitrary
diffeomorphisms and Weyl rescalings. The measure $[DX]$
in (\ref{part}) is  defined in the usual way by the metric
\be
\|\delta X \|_{h}^2 = \int \; d^2\sigma \; \sqrt{h} \;\delta X \cdot
\delta X
\ee
which obviously depends explicitly on the world sheet metric.
It can now be shown that $DX$ is invariant under diffeomorphisms but
not under Weyl rescalings. Denoting the explicit dependence of $DX$
on $h$ by $D_hX$ we find that \cite{DiKa}
\be
[D_{e^{\phi} h}X]= e^{\frac{D}{48\pi}S_L[\phi ;h]}[D_hX]
\ee
where
\be
S_L[\phi;h]=\int \; d^2\sigma \;\sqrt{h}\left( \frac{1}{2}h^{ab}\partial_a
\phi \partial_b \phi +R\phi +\mu e^{\phi}\right)
\ee
Here $R$ is the `world sheet curvature scalar' calculated using the
metric $h$. From this follows that the Weyl invariance of (\ref{Polac})
is anomalous.

The nonanomalous gauge invariance present in the partition function
is the group of diffeomorphisms. The space of metrics on a genus $g$
surface $\Sigma_g$ is isomorphic to $Diff(\Sigma_g)\times Weyl(\Sigma_g)
\times {\cal M}_g$, where ${\cal M}_g$ is the moduli space of
metrics on $\Sigma_g$. The integration over all metrics in (\ref{part})
can therefore roughly be written as an integration over the
gauge orbit, an integration over the Weyl orbit and an integration
over modulispace (of course there will also be a Jacobian $J$ related
to the change of variables). In
order to apply the Faddeev-Popov procedure to extract the gauge volume
one has to make a choice of gauge. Let $\tau$ denote a set of moduli
(i.e. parameters on ${\cal M}_g$) and let $\hat{h}(\tau )$ be a
representative of the equivalence class of metrics determined by
$\tau$ (remember that an element $\tau$ of ${\cal M}_g$ is an
equivalence class of metrics on $\Sigma_g$). Take as gauge slice
the set of metrics of the form $e^{\phi (\sigma )}\hat{h}(\tau )$.
It is important to note that a subgroup of $Diff(\Sigma_g)$, the
subgroup of conformal diffeomorphisms $Diff_c(\Sigma_g)$, slides
along this gauge slice. There will therefore be a residual Virasoro
(gauge) symmetry left in the final answer.

Since the moduli space of metrics on a genus $g$ Riemann surface
is finite dimensional the integration over ${\cal M}_g$ is an
ordinary integration over moduli $\tau$ rather than a functional
integration.
As usual one represents the Jacobian $J$ as a functional integral
over ghosts $b,c$
\be
J=\int \; [D_hb\, D_hc]\; e^{-S_{gh}[b,c;h]}
\ee
The ghost action can be shown to be invariant under diffeomorphisms
and Weyl rescalings but like $[D_hX]$ the ghost measure suffers
a Weyl anomaly \cite{DiKa}
\be
[D_{e^{\phi}h}b][D_{e^{\phi}h}c]=e^{-\frac{26}{48\pi}S_L[\phi ;h]}
[D_hb][D_hc]
\ee
Now, inserting all these results into (\ref{part}) and extracting
the gauge volume we find that the partition function reads
\be
Z=\sum_{g=0}^{\infty}\int_{{\cal M}_g}\; d\tau \int \; [D_{\hat{h}}X]
[D_{\hat{h}}\phi ][D_{\hat{h}}b\, D_{\hat{h}}c]\; e^{-S_M[X;\hat{h}]-S_{gh}[
b,c;\hat{h}]-\frac{26-D}{48\pi}S_L[\phi ;\hat{h}]} \label{part2}
\ee
{}From this expression follows that the worldsheet metric decouples
in $D=26$ whereas it becomes popagating in $D\neq 26$
(in particular $D=4$). So even though Einstein gravity is trivial in
$D=2$, due to the fact that the Einstein-Hilbert action is proportional
to the Euler characteristic, a nontrivial 2 dimensional (non Einsteinian)
theory of gravity can be induced at the quantum level. This `induced
gravity' has turned out to be an extremely interesting but
complicated subject. In order
to see why consider the genus
0 term of (\ref{part2}). The moduli space ${\cal M}_0$ is a point and
we can take $\hat{h}_{ab}=\delta_{ab}$. Now performing the integral
over $X$ one finds
\be \label{liou}
\int [D\phi ]\; e^{\frac{D-26}{48\pi}\int d^2\sigma \;
(\frac{1}{2}(\partial_a\phi )^2+\mu^2e^{\phi})}
\ee
This functional integral is known to be difficult
due to the $e^\phi$ term in the Liouville action. For
higher genus the situation is even further complicated by the
integral over moduli space.

Note that eq.(\ref{liou}) was obtained after gauge fixing which
means that the action
\be  \label{liou2}
S[\phi]=\frac{26-D}{48 \pi}\int d^2\sigma
(\frac{1}{2}(\partial_a\phi )^2+\mu^2e^\phi )
\ee
is a gauge fixed version of some other (generally covariant) action.
One can find this other action by performing the matter integral $[DX]$
before one gauge fixes, i.e. in the genus 0 part of formula (\ref{part}).
What one finds is the following nonlocal (induced) action for the
world sheet metric $h$:
\be
S[h]=\frac{D}{48\pi}\int \, d^2\sigma \, \left(\sqrt{h}\, R
\frac{1}{\sqrt{h} \Box}\sqrt{h} R \right)
\ee
where $R$ is the 2 dimensional curvature scalar associated to the
metric $h$.
In the conformal gauge ($h_{ab} \sim \delta_{ab}$) this action indeed
reduces to (\ref{liou2}).

In the critical dimension $D=26$ the integral (\ref{part2})
simplifies considerably but remains highly nontrivial. The calculations
above were made under the assumption that spacetime is flat Minkowski
space and one might
wonder whether they remain valid when spacetime
is some nontrivial 26 dimensional manifold with metric $G_{\mu\nu}$
instead of Minkowski space. Again choosing a diffeomorphism invariant
regulator one finds that for general $G_{\mu\nu}$ Weyl invariance is
broken as a consequence of which the theory becomes unacceptable
due to the emergence of unphysical modes. We conclude that we have
to insist in this case on Weyl invariance which will obviously
restrict the possible choices for the spacetime metric $G_{\mu\nu}$.
For this note that  $G_{\mu\nu}$, describing
the graviton condensate, acts in (\ref{Polac}) as a set
of coupling constants (one for each index pair $(\mu ,\nu )$
and spacetime point $X$) and for Weyl invariance to be conserved
on the quantum level the $\beta$-function of these coupling constants
must be zero. An explicit computation yields \cite{GSW}
\be
\beta_{\mu\nu}=-\frac{1}{4\pi}\left( R_{\mu\nu}+\frac{1}{4\pi T}
R_{\mu\rho\kappa\lambda}R^{\mu\kappa\lambda}_{\nu}+\ldots \right)
\ee
Note that in the limit $T\rightarrow \infty$ (which corresponds to an
infinite string tension and therefore a pointlike string)
the vanishing of the $\beta$-function coincides with the Einstein
equations in vacuum.  The other terms are
string theoretical corrections to Einstein gravity. This
spectacular result shows that Einstein relativity and the equivalence
principle may arize as consistency conditions on the underlying 2
dimensional world sheet theory.
However, at this point it is not clear which solution of the Einstein
equations is selected by string theory or to put it differently, due to the
lack of a satisfactory string field theory the structure of spacetime
does not emerge dynamically from the theory but has to be put in by
hand as a particular background. The search for the fundamental theory
underlying string theory is therefore equivalent to the search for
a background independent formulation of string theory.

We have seen above that 2 dimensional local scale invariance
plays an important role in string theory.
It can easily be shown that any 2 dimensional quantum field  theory
with local scale invariance has a traceless
stress energy tensor. On the other hand it is well known that
a quantum field theory is invariant under conformal transformations
if and only if its stress energy tensor is traceless. We conclude
that the string groundstates are described by conformal field
theories which are associated to certain sigma models.
Having arrived at this conclusion one now turns the argument around
and interprets any CFT as a perturbative string groundstate around
which one can start a perturbation theory
essentially in the same way as we did above. From this point of view
{\em a classification of all $(c=26)$
conformal field theories would correspond
to a classification of all possible perturbative string vacua} and
as we have seen $W$ algebras play an important role in this
classification. However, since conformal field theories correspond
to groundstates, or put differently, to `classical solutions' of an
underlying theory, it is probably not the list of all CFTs that is of
fundamental importance to string theory but more the basic principles
that lead to such a classification. It is hoped that a proper formulation
of CFT probably together with a general scheme as to their construction
might lead to some clues regarding
the physical and geometrical principles underlying
string theory.

The need for a formulation of string theory that goes beyond a perturbative
expansion around a given groundstate, i.e. a background independent
formulation, has become an urgent one recently.
Namely, in \cite{Gr}
it was shown that the topological expansion of the partition function
diverges with the factorial of the genus $Z\simeq \sum_{g=0}^{\infty}(2g)!$
and is therefore not Borel summable (remember that even if a series
\be
A(\alpha )=\sum_{g=0}^\infty \alpha^gf(g)
\ee
formally diverges its Borel transform
\be
A_B(\alpha )=\sum_{\alpha =0}^\infty \alpha^g\frac{f(g)}{g!}
\ee
may still converge.
The original function can be retrieved by taking the inverse
transformation
\be
A(\alpha )=\int_0^\infty A_B(\alpha t)e^{-t}\,dt
\ee
However if $f(g)\simeq g!$ then even the Borel transformed function $A_B$
is singular).
This means that the string perturbation theory around a conformal field
theory is at best an asymptotic expansion. Usually, in gauge theories,
such behaviour is ascribed to nonperturbative effects which makes
it all the more necessary to develop the nonperturbative approach to
string theory.

In the formulation of string theory discussed above the fields
representing the coordinates of the string define a matter CFT described
by a nonlinear sigma model.
Generalizing the
theory (\ref{part2}) by taking an arbitrary CFT instead of a sigma
model we are lead to study CFTs coupled to 2-dimensional
quantum induced
gravity (i.e. CFT on a randomly fluctuating surface). In recent
years our understanding of minimal $(c<1)$ models coupled to
2 dimensional gravity has improved a great deal. A number of results
(scaling and critical dimensions) were obtained by using the fact that
2 dimensional induced gravity formulated in the light cone gauge
possesses a hidden affine $SL_2({\bf R})$  symmetry (we will come
back to this in the next chapter) \cite{Pol}. It has also been tried
to use the fact that Liouville theory (\ref{liou2}) is completely
integrable as a classical field theory to obtain results in the
quantum case \cite{Liou}. These efforts have become known under the
name `the continuum approach'.

It has turned out that 2 dimensional gravity can also  very
conveniently be studied in terms  of discretized
world sheets. In this approach the integration over metrics is replaced
by a sum over triangulations \cite{discrete}
\be
\sum_g\int [Dh] \rightarrow \sum_{\mbox{random triangulations}}
\ee
The reason that one can get exact results from this approach is that
the sum over triangulations can be represented by an integration
over (hermitean) $N\times N$ matrices.
These (matrix) models can be solved exactly and in the `double scaling
limit' (which involves taking $N\rightarrow \infty$) shown to be
equivalent to string theory at $c\leq 1$. An attractive feature of this
approach, especially in the light of the non-Borel summability
of perturbation theory, is that it provides nonperturbative
information on string theory.
A remarkable fact that was recently discovered
is that the partition function $Z$ of a matrix model can be characterized
by so called $W$-constraints which are differential equations of the
form
\be
W^{(s)}_n\cdot Z=0   \;\;\;\;\mbox{ , n}> \Delta_s
\ee
where $W^{(s)}_n$ are certain differential operators
\cite{Wconstr} that
satisfy a $W$-algebra with conformal weights $\{\Delta_s\}$
(which $W$ algebra depends on
the matrix model). Conversely it is conjectured that there is a
matrix model associated to any $W$ algebra.  Furthermore, out
of the $W$ constraints
came the realization that the partition function $Z$ can be viewed
as a $\tau$ function of a certain integrable hierarchy of KdV type
\cite{Wconstr}.

Finally we mention
another approach to 2 dimensional gravity is topological field theory
(TFT). A TFT is a quantum field theory in which all correlation functions
are independent of the metric \cite{TFT} which means that the
independence  of the world sheet metric is built into the theory
from the beginning. In the literature a large number of TFT's have been
constructed each of which seems to be equivalent to $c<1$ matter
coupled to 2D gravity. As one could have hoped the discrete and
topological approaches to string theory have turned out to be
equivalent \cite{Konts}. As with matrix models a strong relationship
with the theory of integrable hierarchies has been discovered and in
fact it has been possible, using this relation, to calculate
intersection numbers on the moduli space of Riemann surfaces for
arbitrary genus \cite{inters} which is an important mathematical
result.

{}From the string theoretical point of view
the results discussed above are unphysical due to the restriction
$c\leq 1$. Unfortunatly the methods used to obtain these results are not
applicable to the so called `strong coupling regime' $1<c<25$ due
to the appearance of tachyons rendering the whole theory ill defined.
Also note that, as we have seen in the previous section,
for $c<1$ there are only a finite number of unitary lowest energy
Virasoro representations. Since Virasoro symmetry is a
residue of the gauge symmetry present in string theory it follows,
that all Virasoro
descendent states are gauge equivalent to their highest weight
states and that the theory essentially has a finite number of
degrees of freedom. This simplifies matters drastically.
For $c>1$ however there are always infinitely many
degrees of freedom due to the fact that any CFT with $c>1$ contains
infinitely many Virasoro representations. Here again $W$ algebras
may prove a way out since  CFTs  with $W$ symmetry may be finitely
reducible even for $c>1$. This means that
the critical central charge $c=1$ can be
shifted to a higher value by considering a $W$ generalization
of ordinary 2 dimensional quantum gravity ($W$ gravity).
One is here of course faced with the
completely new concept of a gauge theory based on a nonlinear
gauge symmetry. Some progress in this area has been made however
\cite{PvN,Hull,BoGo}. Note that, like the Virasoro algebra (which arose
as a residue of 2 dimensional diffeomorphism invariance),
$W$ symmetry is here a residual (gauge) symmetry of something else.
What this larger
symmetry principle is that reduces on partial gauge fixing to
(chiral) $W$ symmetry is not known at this time.

In this section we have discussed string theory as on
of the most promising candidates for a theory of quantum gravity.
We have seen that $W$ algebras may play an important role in this
theory at different levels: first of all in the classification of the
perturbative vacua, and second in trying to generalize recent results
on $c<1$ string theories to the more realistic 'strong coupling regime'.

\section{The Quantum Hall effect}
The integer quantum Hall effect (IQHE) was discovered in 1980 by
von Klitzing, Dorda and Pepper (von Klitzing was awarded a Nobel prize
for this discovery in 1985). Since then there has been a large effort
to develop a theoretical understanding of it (see for example
\cite{Prange}) and remarkable progress has been make.
The Hall effect has drawn the interest of high
energy theorists due to certain indications
that the description of the
quantum Hall system is intimately related to conformal field
theory. In this section we shall briefly indicate how this relation
comes about.

Consider some (effectively) 2 dimensional system of electrons
in a
strong perpendicular magnetic field and at low temperatures.
The IQHE is the phenomenon that under
certain circumstances
the conductivity tensor $\sigma$ is of the form \cite{QHE}
\be \label{cond}
\sigma = \left(
\begin{array}{cc}
0 & -\frac{\nu e^2}{h} \\
\frac{\nu e^2}{h} & 0
\end{array}
\right)
\ee
where $h$ is Planck's constant and $\nu =0,1,2,\ldots$, which means
that the longitudinal or `diagonal' resistance $R_L$ is zero while
the transverse or {\em Hall} resistance $R_H$ is quantized.
We see that in such
a state the system is dissipationless which means that the QHE is
closely related to superconductivity and superfluidity.

Note that above we did not mention what material we were
actually dealing with. Acually one of the remarkable
properties of the QHE is its universality or material independence.
This indicates that quantities like the Hall conductance
$\sigma_H=\frac{\nu e^2}{h}$, which characterize the quantum Hall
state, only depend on large scale properties of the system and
do not depend on the precise microscoping details. It is truly
remarkable to see that such a directly measurable macroscopic
quantity in a rather complicated and impure medium is exactly equal
to an integer times an expression in terms of fundamental constants.

The standard picture of the quantum Hall state \cite{Prange} is that of the
`incompressible ground state' and we shall now very briefly discuss
its basic ideas. In what follows we always
assume the electron-electron interactions to be extremely weak so that
the $N$ electron states can be built by filling up one electron states
in accordance with the Pauli principle. Furthermore we take the sample
to have the topology of a disk with radius $R$ and denote it by $\Omega$.

As was shown by Landau
planar electrons in a perpendicular magnetic field can occupy
a discrete set of highly degenerate so called Landau levels which are
eigenspaces of the one particle Hamiltonian $H$. The degeneracy in the
Landau levels is labelled by the angular momentum $l$ of the energy
eigenstates. The single particle angular momentum states with
angular momentum $l$ are peaked around
a mean radius $r_l=l_B\sqrt{l}$, where $l_B$ is the so called
`magnetic length' $l_B=\sqrt{hc/eB}$.
of course $r_l$ increases as the angular momentum
increases and since
the radius of the sample is finite there exists a maximal
angular momentum $l=L$ above which there can exist no states. This means that
in a finite sample
the Landau levels are finitely degenerate. Let us denote this degeneracy by
$N_B$ and let $n_B$ be the associated particle density. It is easy to
derive an explicit expression for $n_B$ and it reads $n_B=eB/hc$.
Note that  $n_B$ is a linear function of  the magnetic field strenth.

Now suppose all occupied Landau levels are completely filled, i.e.
the particle density  $n$ is an integer multiple of $n_B$,
then
due to the lack of low lying conduction modes the diagonal conductivity
vanishes.
The semiclassical formula for the off diagonal element of the
resistivity tensor is $\rho_H=B/nec$ and inserting $n=\nu n_B$
into this formula, where $\nu$ is some positive integer,
we find $\rho_H=h/\nu e^2$ (also remember that the
longitudinal resistivity, the diagonal element of the resistivity tensor,
is zero). The conductivity tensor, which is the inverse of the
resistivity tensor, then obtains the form (\ref{cond}).
This is the standard picture of the
IQHE and in what follows we consider the
state in which the occupied
Landau levels are all completely filled in a little
more detail.

For simplicity consider the state $|\Omega \rangle$
where only the first Landau level
is filled and all higher Landau levels are empty. Clearly such a state
is incompressible since a compression would mean that at least one
electron would have to go to a state with lower angular momentum.
Since all states in the lowest Landau level
with lower angular momentum are already occupied the electron would have
to go to a lower angular momentum state in a higher Landau level. This
process however faces an energy gap which means that the state is
incompressible.
It is possible to calculate the expectation value of the particle
density operator $\rho$ in the state $|\Omega \rangle$
and what one finds is that
it is constant for radii $r<<R$
and drops rapidly to zero in the neighborhood of the boundary of the sample.
{}From these facts (i.e. the incompressibility and the uniform density)
it follows that in the state $|\Omega \rangle$
the electrons behave like a rigid droplet of (quantum)
fluid.

Another quantity of interest is the expectation value of the electron
current operator $J^i(x)$. Again
a straighforward calculation yields
\be
\langle\Omega | J^i({\bf x}) | \Omega \rangle
\;\;\sim \;\; \varepsilon^{ij}\partial_j
\langle\Omega | \rho ({\bf x})|\Omega \rangle
\ee
Obviously
the current vanishes in the interior of the sample, where the density
is constant, while it supports a transverse (chiral due to the
magnetic field)  current around
the boundary of the droplet. This is the so called {\em edge current}
and it indicates that all nontrivial dynamics of the system is
concentrated around the boundary.

Recently there has been a great deal of activity on the relation
between CFT and the QHE \cite{CFTQHE}.
We shall now briefly discuss the basic ideas underlying this relation.
As we mentioned above all
nontrivial dynamics of a Hall system is restricted to the boundary.
One might expect to be able to describe the system effectively
as a 1+1 dimensional theory. In fact in
\cite{CDTZ} it was shown
that the 2+1 dimensional Hamiltonian of the Hall sample
reduces (in the scaling limit)
to a 1+1 dimensional Hamiltonian for a system of
excitations living on the boundary of the sample.
Furthermore this theory turns out to be equivalent to
a theory of relativistic chiral scalar
fermions living on  $S^1$ \cite{FrKe}. It is known that the spectrum of
this theory can be described completely in terms of an affine $U(1)$
algebra \cite{FermU1KM}
which, as we have seen in the first section, is a particular
example of a $W$ algebra.

In \cite{FrKe} the same result was derived in a slightly different
way using classical arguments. Consider the current density
$\vec{j}$ induced by an electric field $\vec{E}$, i.e. $j^i (\vec{x},t)
=-\sigma_H\varepsilon^{ij}E_j(\vec{x},t)$, where
$\vec{x}\in \Omega$ and $\sigma_H=\frac{\nu e^2}{h}$. Using
Faraday's induction law we find $j^0=-\sigma_HB$ and we
can write these equations as follows
\be  \label{starcur}
J_{\alpha\beta}(x)=\sigma_HF_{\alpha\beta}(x)
\ee
where $F_{0i}=E_{i}$, $F_{12}=-B$ and $J_{\alpha\beta}=
\varepsilon_{\alpha \beta\gamma}j^{\gamma}$. In terms of differential
forms this equation reads $J=\sigma_H F$ while the continuity
equation for the current $j$ and Faraday's law reduce to $dJ=0$
and $dF=0$ respectively (here $d$ is the ordinary exterior differential).
Now, since $\Omega$ is contractible we find (by the Poincare lemma)
that $F=dA$ and $J=da$ where $A$ is the 3 dimensional vector potential
and $a$ is a one form called the `vector potential of the electric
current'. In terms of these potentials eqn.(\ref{starcur}) can be
written as $d(a-\sigma_H A)=0$ which can be derived from an action
principle with action
\be
S_{CS}[a-\sigma_HA]=\int_{{\bf R}\times \Omega}(a-\sigma_H A)\wedge
d(a-\sigma_H A)
\ee
which we recognize as a $U(1)$ Chern-Simons gauge theory for the
gauge field $a-\sigma_HA$. The Chern-Simons action is topological
since it does not require a metric and it is shown in \cite{FrKe}
that this explains why $\sigma_H$ does not depend on the microscopic
disorder in the sample.

We conclude that the classical electrodynamical
equations describing the quantum Hall sample are actually the
equations of motion of a Chern-Simons action. This makes it possible
to describe the large-scale properties of a Hall system in terms
of a Chern-Simons gauge field theory \cite{FrKe}.
{}From this picture the $W$
algebra arises as follows: It is well known \cite{witten}
that a Chern-Simons theory
on  a two dimensional manifold $\Omega$ is equivalent to a WZW model
(or a chiral affine current algebra) on the boundary $\partial \Omega$
of $\Omega$ so again we find that, in the scaling limit, a quantum Hall
system can be described by a chiral sector of a two dimensional
conformal field theory.

Up to now we implicitly assumed that the magnetic field $B$ is large
enough to align all electron spins in the direction of $B$. In that
case the spin wave function is totally symmetric while the orbital
wave function is (as electrons are fermions) totally antisymmetric.
In this approximation the electrons can be described as two
dimensional scalar fermions. As we have seen this approximation
leads to an affine $U(1)$  algebra. If the magnetic field is not large
enough to align all the spins then one finds, due to the fact that the
edge currents become spin carrying, that the CFT governing the
edge theory has an underlying level 1 affine$SU(2)$ symmetry \cite{FrKe}.

In the above we did not concern ourselves with the mechanism by
which the bulk of the sample creates the incompressible groundstate.
This would require taking into account the microscopic interactions
between the electrons and probably most important of all a study
of impurity effects, for it can be easily shown that there would be no
QHE without impurities \cite{Prange}. Rather we assumed that the incompressible
groundstate is indeed formed. The theory of how this state is formed
is still in its infancy.

\section{Phase transitions in 2 dimensions}
Phase transitions play an important role in many areas of theoretical
physics. Roughly a phase transition is a sudden change in
certain macroscopic properties of a many body system as a result of a
slight change in some external parameters like the temperature.
Mathematically a singularity in the free energy (or its derivatives)
of a system is indicative for a phase transition. Here we shall be
considering second order phase transitions
in two dimensional lattice models like the Ising
model.

As is well known a many body system (where we use the term `body' in
a very broad sense) is described in statistical mechanics (in the
canonical formalism) by a
map $\rho$ from the phase space ${\cal M}$ of the system
(i.e. the set of all microstates) to the interval
$[0,1]$ such that the integral (or sum) over phase space of  $\rho$ is equal
to 1. Usually the map $\rho$ has the well known Boltzman form
\be
\rho =\frac{1}{Z}e^{-\beta H}
\ee
where $H$ is the Hamiltonian of the system and $Z$ is the partition
function
\be
Z=\int_{\cal M} e^{-\beta H}\, d\mu
\ee
(here $d\mu$ is some measure on phase space). If $Q$ is some
measurable quantity (i.e. a function on phase space), then the
average value of $Q$ is given by
\be
\langle Q \rangle=\int_{\cal M} \, Q\rho \,d\mu
\ee
As the system moves through phase space the observable quantities
(observables) will fluctuate around their equilibrium values. In
general these fluctuations will not be large and a measure for their
size is given by the quantity
\be
\langle \Delta Q\rangle^2=\langle Q-\langle Q \rangle\rangle^2
\ee
It can be shown \cite{Reichl} that the size of certain
fluctuations and  thermodynamical quantities like the
specific heat are proportional. It is well known that near a phase
transition many respons functions blow up which means that one
expects to encounter abnormally large fluctuations near criticality.
These large fluctuations are a reflection of the internal adjustments
the system is making as it prepares to change to a new state.

In the case of lattice models it is possible
to calculate the correlation functions
between the  dynamical variables on the vertices, edges
or faces (depending on the model). For example, in the Ising model one can
calculate the correlation function between two arbitrary spins $s_i$ and
$s_j$ on the lattice. This correlation function, which  is given by
$\Gamma_{ij}=\langle (s_i-\langle s_i \rangle)(s_j-\langle s_j \rangle
)\rangle$ clearly measures the correlation
between fluctuations in the spins $s_i$ and $s_j$.
On a homogeneous lattice (like the
square lattice) $\Gamma_{ij}$ can only depend on the difference of the
position vectors $\vec{r}_i,\vec{r}_j$ of the spins $s_i,s_j$. It is
easy to show that $\Gamma_{ij}$ is not equal to zero even if $s_i$ and
$s_j$ are not nearest neighbours. Away from the critical point the
correlation function $\Gamma_{ij}$ decays exponentially in
$r_{ij}=|\vec{r}_i-\vec{r}_j|$ while at criticality it has power
law decay. Therefore at criticality fluctuations are correlated over
large distances, or put differently, the correlation length becomes infinite.
It is essentially this fact, i.e. the presence of
fluctuations on all lenth scales,
that made the problem of describing critical
systems so hard. The crucial ingredient which made the problem tractable
is the property of scale invariance which is at the heart of the
successfull renormalization group approach to critical phenomena.
It means in particular that
correlation functions tend to be scale invariant near criticality.
Effectively one therefore expects to be able to describe a lattice
system at criticality by a 2 dimensional quantum field theory. Indeed,
in the scaling (or continuum) limit one interprets certain lattice correlation
functions as Euclidean Green functions of some Euclidean field theory.
This field theory will be scale invariant and can be explicitly constructed
provided
the lattice system admits a self-adjoint transfer matrix
(it may be possible to do this even if there is no selfadjoint transfer
matrix but it is not known how). Remember that
a doubly periodic $N\times N$ lattice model is said to admit a
transfer matrix if there exists a $2^N \times 2^N$ matrix $T$ such
that
\be
Z=\mbox{Tr}(T^N)
\ee
If a transfer matrix exists then obviously the problem of
calculating the partition function reduces to finding the eigenvalues
of the transfer matrix.

Selfadjointness of the transfermatrix is a frequent but not fundamental
property from the point of view of statistical mechanics (for example
it fails in the theory of self avoiding walks). It turns out however
that if it is satisfied that one can use a procedure similar to
Osterwalder-Schrader reconstruction \cite{OsSc} to associate to the
set of Euclidean Green functions, obtained by taking the scaling
limit of the lattice correlation functions a unitary relativistic
quantum field theory \cite{FeFrKe}. First one constructs out of the
set of Euclidean Green functions a separable Hilbert space $\cal H$
with a unique vacuum state $\Omega$. Then it is shown that $\cal H$
carries a representation of the group of spacetime translations from
which one can construct selfadjoint operators $H$ and $P$ satisfying
the relativistic spectrum condition. Finally it is then shown that the
thus obtained quantum field theory is scale invariant.
Now, for homogeneous and isotropic systems scale invariance and the
fact that interactions are local imply conformal invariance. We
conclude that
{\em conformal field theories can be interpreted as field theories
describing statistical mechanical lattice models in the continuum limit
and at criticality}. The classification of critical phenomena (or in the
terminology of renormalization theory, all fixed points of the
renormalization group) in 2 dimensions
corresponds therefore (at least in part) to the classification of all
2D Euclidean conformal field theories. The central charge of the CFT
is interpreted as a measure of the effective number of field
degrees of freedom which have large scale fluctuations at criticality.

It has been possible to identify all minimal models
with certain critical 2 dimensional lattice models. Before giving some
examples let us note that to any CFT there is associated a whole
universality class of statistical mechanical models so a given
realization of a CFT as a lattice model is not by far unique.

The most famous identification is that of the $c=c(3)=\frac{1}{2}$
minimal model with the critical Ising model \cite{BPZ}.
In fact it was shown in \cite{Pas} that
all minimal models correspond to the critical points of exactly
solvable restricted solid on solid (RSOS) models defined
on Dynkin diagram $A_n$.

Remember that the minimal models all have $c<1$. The value $c=1$
corresponds to a free bosonic field, i.e. to a Gaussian fixed point.
In this theory there is one parameter, which can be interpreted as
the compactification radius of the free bosonic field, and the
critical exponents depend continuously on this parameter. We are therefore
dealing with a line of fixed points. This line can be shown to
correspond to the critical line the 8 vertex model \cite{Baxter}.

Statistical mechanical models with extended conformal symmetries (i.e.
$W$ symmetry) have also been found. In fact it turns out that the
3 states Potts model already mentioned above has $W_3$ symmetry
at criticality. However, a systematic study of $W$ symmetries in
statistical mechanics has not been embarked on. This appears to be
a worthwile problem for the future.

\section{Solitary wave phenomena}
Solitary waves were first discovered in 1834 by the Scottish engineer
John Scott Russell when he observed a smooth and well rounded heap of
water of approximately 10 meters length and  40 cm height that moved through
a channel of water apparently without change of form or speed. Later, in
an attempt to study this phenomenon, he successfully recreated many solitary
waves in a laboratory setting.
Partly due to scepticism and disbelief within the scientific
community it was not until 1895 before any solid theoretical description
of the phenomena observed by Russell was given. In that year D.J.Korteweg
and G.de Vries derived their famous equation for  one dimensional long,
small amplitude water waves in a shallow channel. This equation, nowadays
known as the KdV equation reads
\be
u_t=u_{xxx}+6uu_x
\ee
where $u=u(x,t)$ is the surface elevation above the equilibrium level
at $x$ at time $t$. This equation indeed has solutions of the type
described by Russell as was shown by Boussinesq and Rayleigh
\be
u(x,t)=s\kappa^2 \mbox{sech}^2(\kappa (x+\kappa^2t-x_0))
\ee
where $\kappa$ and $x_0$ are constants. A large part of the early
scepticism towards Russell's discovery was that it was believed that a
solitary wave could not be stable. With the discovery of the KdV equation
it became clear that the usual dispersive effects of wave propagation
cancel precisely against the nonlinear effects governed by the nonlinear
term $uu_x$.

Solitary wave phenomena are not restricted to hydrodynamical systems however.
In the 1960's  E.Fermi, J.Pasta and S.Ulam  accidentally ran into
some solitary waves when they were investigating
the heat conductivity of solids.
The problem was this: if a solid is modelled by a one dimensional
lattice of masses coupled by harmonic springs (i.e. springs for which the
return force is linear in the displacement from the equilibrium) then
energy is carried unhindered by the independent normal modes. This
means that the effective thermal conductivity would be infinite or put
differently that no thermal gradient is needed to push the heat through
the lattice from one end to the other.  Since this is certainly not the
case Debye suggested that the finiteness of the thermal conductivity
is caused by anharmonicity of the springs, i.e. nonlinearity in the
return force. This suggestion motivated Fermi, Pasta and Ulam to undertake
a numerical study of an anharmonic lattice. What they expected was that
an initial state in which all energy was in the lowest mode would
eventually relax into a state in which the energy would be distributed
equally among all modes due to the nonlinearity of the spring couplings.
The relaxation time would then be a measure of the diffusion coefficient.
To their astonishment what happened is that the energy flowed back and
forth among several low order modes eventually to be recollected almost
entirely into the lowest mode. Then the whole process started all over
again.

In an attempt to explain this phenomenon M.D.Kruskal and N.J.Zabusky
again ran into the KdV equation. They also found numerically that the
KdV equation has solutions in which there appear more than one solitary
waves. Furthermore they found that if two solitary waves collided, then
after some time they reappeared untouched, i.e. with the same height,
width and speed. The only evidence of a collision was a phase shift whereby
the solitary waves were not at the places they would have been  had there
not been a collision. This particle like behaviour lead to the name
{ \em solitons}.
It was only after these discoveries by Kruskal and Zabusky that the
theory of solitary waves became popular among mathematicians and physicists.

The next important step in the development of the theory of solitons was the
realization that there are infinitely many quantities $\{H_i\}_{i=1}^{\infty}$
that are conserved under the KdV flow.
This lead to the insight that the KdV equation was just part of an
infinite dimensional completely integrable system. Remember that  a
finite dimensional classical mechanical system described by some Hamiltonian
$H_1$ is called 'completely integrable' if there exist $M$ functions
$\{H_i\}_{i=1}^M$ of the coordinates and momenta (where $M$ is the number
of coordinates) that commute w.r.t. the Poisson bracket, i.e.
$\{H_i,H_j\}=0$.
This means that all the quantities $H_i$ are conserved w.r.t. the time
evolution generated (through the Hamilton equations) by $H_1$ and
furthermore that the conserved quantities Poisson commute among themselves.
A theorem by Liouville then states that one can trivialize such a system
by a canonical invertible
coordinate transformation, i.e. there exist phase space
coordinates $Q_i$ and $P_i$ (called action angle variables) such that
$Q_i$ is constant and $P_i$ is a linear function of $t$.
The transformation to the action angle coordinates is called  the action
angle transform.
Note that within the system $\{H_i\}$  the
original Hamiltonian plays no prefered role which means that in fact
we have obtained a set of $N$ physical systems, described by the
Hamiltonians $\{H_i\}$ which are completely integrable.

In infinitely many dimensions it is not so clear how to define complete
integrability but it is clear that one should at least have
infinitely many conserved quantities in involution.
The point is
now that the inverse scattering transformation used to solve many
integrable hierarchies, can be interpreted as an infinite
dimensional example of an action angle transform.

Let us now describe the KdV hierarchy, as well as a number of
other hierarchies,  as infinite dimensional
integrable systems. For this consider the space $D_N$ of differential
operators $L$ of the form
\be
L=\partial^N+\sum_{i=0}^{N-2} u_i \partial^i
\ee
where $u_i$ are smooth real valued functions of $x\in {\bf R}$ that
converge quickly to 0 when $|x|\rightarrow \infty$. On this space
the operation of taking the derivative is invertible so we can
consider the space $PD_N$ of so called `pseudodifferential operators'
\be
X=\sum_{i=-\infty}^{N-1} u_i\partial^{i}
\ee
{}From the fact that $\partial$ is a derivation it immediately follows
that
\be  \label{pseudo}
\partial^{-1}u=\sum_{i=0}^{\infty}(-1)^i(\partial^iu)\partial^{-i-1}
\ee
One usually defines the algebra of pseudodifferential operators
even when working with  a space on which $\partial$ is not invertible.
However in that case $\partial^{-1}$ is a formal generator of the
algebra satisfying the permutation relation (\ref{pseudo}).
Before we define the hierarchies let us introduce a few
more concepts that we shall need. Let $X$ be a pseudodifferential
operator, then we define the `residue' of $X$ to be the coefficient
of $\partial^{-1}$ and the `trace' of $X$ by the integral over the
residue of $X$, i.e.
\be
\mbox{Tr}(X)=\int dx\, u_{-1}(x)
\ee
The space $F(D_N)$ is defined to be the space of functionals
of the form
\be
F[L]=\int dx\, \tilde{F}(u)
\ee
where $\tilde{F}$ is a function of $\{u_i\}$ and a finite number of its
derivatives. Note that elements of $F(D_N)$ map differential operators
to real numbers.

In the inverse scattering method a crucial role is played by
the so called isospectral deformation problem. Consider the eigenvalue
equation
\be
L\psi =\lambda \psi
\ee
where $L\in D_N$. The question one asks is: suppose the functions
$u_i$ are time dependent, i.e. $u=u(x,t)$,
what time evolution is such that the
eigenvalues $\lambda$ are time independent? In turns out (as was shown
by Lax) that this is true
iff the functions $u_i$ satisfy a system of equations
that can be written as
\be
\frac{dL}{dt}=[M,L]     \label{Laxeqn}
\ee
where $M$ is again a differential operator (not necessarily of order $N$).
The pair of operators $(L,M)$ is called a Lax pair. Now, in order for
the equation (\ref{Laxeqn}) to make sense $M$ must have a very specific
form. It can be shown \cite{DS} that $M$ must be a linear combination of
\be
M_k= (L^{k/N})_+
\ee
where $L^{1/N}$ is the (unique) pseudodifferential operator such
that $(L^{1/N})^N=L$ and the index $+$ means that we only retain
terms with positive powers of $\partial$.
The hierarchy of evolution equations is now given by
\be
\frac{dL}{dt_k}=[M_k,L ] \label{lax2}
\ee
It can proved by explicit calculation \cite{DS} that the different
flows defined by (\ref{lax2}) commute which means that it is legitimate
to introduce an infinite number of times $t_k$ and to consider $u_i$ a
function of all these times. The KdV hierarchy is obtained
by taking $N=2$ while  $N=3$  leads to the Boussinesq hierarchy.

We now describe the infinite set of conserved quantities and the
(bi)Hamiltonian structures of the hierarchies constructed above.
Let $F \in F(D_N)$ be a functional on $D_N$, then we define the
pseudodifferential operator $X_F$ by
\be
X_F(\cdot )=\sum_{i=1}^{N-1} \partial^{-i}\frac{\delta F}{\delta_{u_{i-1}}}
\cdot
\ee
On $F(D_N)$ there exist two coordinated Poisson brackets, called
Gelfand-Dickii brackets given by
\ba
\{F,G\}_1(L) & = & \mbox{Tr}([L,X_{F}]_+X_{G}) \nonumber \\
\{F,G\}_2(L) & = & \mbox{Tr}(\nabla_{X_F}(L)X_G)
\ea
where
\be
\nabla_{X_F}(L)=L(X_FL)_+-(LX_F)_+L
\ee
Here 'coordinated' means that any linear combination
of these two Poisson brackets is again a Poisson bracket. Now the
point is that the Lax equation (\ref{lax2}) is Hamiltonian w.r.t.
both Poisson structures and the Hamiltonian functions are given by
\be
H_k[L]=\mbox{Tr}(L^{k/N})
\ee
These quantities are conserved w.r.t. all flows in the hierarchy
and commute among themselves w.r.t. both Gelfand-Dickii Poisson
structures. We have therefore obtained an infinite dimensional
integrable system.

Consider now the Poisson algebra $(F(D_N),\{.,.\}_2)$. Simply using
the definitions it is easy to show that for $N=2$ the `coordinate
functional' $u$ has the following Poisson bracket with itself
\be
\{u(x),u(y)\}_2= -\frac{1}{2}\delta'''(x-y)-2u(y)\delta'(x-y)+
u'(y)\delta (x-y)
\ee
which is the classical version of the Virasoro algebra. In fact
it has been shown in \cite{BaMa} that in general the algebra
$(F(D_N),\{.,.\}_2)$ is nothing but a classical version of the
$W_N$ algebra. In a similar way it can be shown that affine Lie algebras
arise as the (second) Hamiltonian structures of certain matrix
generalizations of the Lax equations.
Thus we conclude that {\em (classical) $W$ algebras arise in the theory
of solitary waves as Hamiltonian structures}.
Conversely one conjectures that to any classical $W$ algebra
there is associated a hierarchy of integrable evolution equations.
If and how the special
structure of $W$ algebras is related to the integrability of the
associated hierarchies is not known. Integrable hierarchies do
however have another algebraic structure that is explicitly linked
to the integrability, the so called `weak action angle structure'.
This is what we discuss next.

Remember that complete integrability implies the existence of so called
action-angle variables. Let $\{f_n\}$ and $\{\phi_n \}$ denote the
action and angle variables respectively, then their distinguishing
property is that the action variables are constant w.r.t. all times
while the angle variables depend linearly on all times. This implies
that $\{f_n,f_m\}=0$ and $\{f_n,\phi_m\}=\omega_{nm}(\{f_k\})$.
Transferring this algebraic structure to the Hamiltonian vectorfields
$X_i,L_n$ associated to $f_n$ and $\phi_n$ respectively we find that
$[X_n,X_m]=0$ while $[X_n,L_m]=C^k_{nm}X_k$, where $[.,.]$ denotes
the usual commutator between vectorfiels. Note that there is a great
deal of freedom in the choice for the angle variables since adding
to them a function of action variables does not destroy their
basic property, that of linear time dependence. It can be shown that
this freedom is large enough to pick the variables $\phi_n$ (and
therefore their vectorfields) such that $[L_n,L_m]=\tilde{C}_{nm}^kL_k$.
The algebra spanned by $\{X_n\}$ and $\{L_n\}$ is called the
`weak action angle structure'.

Let us now return to the case of integrable hierarchies. Consider
for simplicity hierarchies of equations in one field $u=u(x,t)$.
The equations in the hierarchy are  $u_{t_k}=X_k(u)$ where $\{X_k(u)\}_{
k=0}^{\infty}$ are (finite) polynomials in $u,u_x,u_{xx},\ldots$.
These quantities can be seen as (the components of) vectorfields on
an infinite dimensional Banach manifold and therefore we can define
their commutator \cite{Oevel}. Assuming that the hierarchy is integrable
we must have $[X_n,X_m](u)=0$ for the different flows must commute.

A crucial role is played by the so called `recursion operator', i.e.
many hierarchies (including the KdV hierarchy as we shall see later)
admit an operator $\Phi$ mapping elements of the hierarchy to higher
elements, i.e. $X_{n+1}=\Phi X_n$. Such a recursion operator may even
exist when the hierarchy doen not admit a Hamiltonian structure
(see for example the Burgers hierarchy) and is therefore not
completely integrable.

The next ingredient in the study of the weak action angle structure
of integrable hierarchies is the so called `scaling field'. The main
idea is here that integrable hierarchies are usually invariant under
scaling transformations of the type $u(x,t)\mapsto e^{\alpha \epsilon}
u(e^{\beta \epsilon}x,e^{\gamma \epsilon}t)$ where $\alpha ,\beta$
and $\gamma$ are real numbers and $\epsilon$ is some parameter. For the
KdV equation these numbers are $2,1$ and $3$ respectively. The
scaling transformations are generated infinitesimally by vectorfields
$\tau$ that have the property  $[X_n,\tau ]=\lambda X_n$. It is
a remarkable fact that we can use these scaling fields to find the
weak action angle structure of an integrable hierarchy. For this define
$L_0=\frac{1}{\lambda}\tau$ and $L_n=\Phi^nL_0$, where $\Phi$ is
again the recursion operator. It can then be shown \cite{Oevel}
that the vectorfields $X_n$ and $L_n$ form the algebra
\ba
[X_n,X_m] & = & 0 \nonumber \\ {}
[L_n,X_m] & = & (m+1)X_{n+m} \nonumber \\ {}
[L_n,L_m] & = & (m-n)L_{n+m} \label{weak}
\ea
We conclude that {\em the weak action angle structure of an integrable
hierarchy is isomorphic to half an affine $U(1)$  algebra}. Obviously it is
only {\em half} because the indices $n$ and $m$ run only over positive
numbers. This also causes the absense of central terms in the relations
(\ref{weak}).

Let us briefly give an example. The KdV hierarchy is the following set
of evolution equations: $u_{t_k}=\Phi^kX_0$ where $X_0=u_x$ and
$\Phi=\partial^2+4u+2u_x\partial^{-1}$. For $k=1$ this leads to the
KdV equation which is invariant under the scaling transformation
with $\alpha =2,\beta =1$ and $\gamma =3$. These scaling transformations
are infinitesimally
generated by the scaling field $\tau= xu_x+2u=L_0$. From this it is
easy to construct the complete weak action angle structure.

In this section we have clearly shown that $W$ algebras play an
important role in the theory of integrable hierarchies. We have
shown that nonlinear $W$ algebras appear as Hamiltonian structures
while the action angle structure of the hierarchy is intimately
related to an affine $U(1)$  algebra.

\section{Outline of the thesis}
In the previous sections we have seen that CFT and $W$ algebras play
a potentially important role in many areas of physics. Progress
towards the application of $W$ algebras in order to obtain physically
significant results is however hampered by the fact that so little
is known about them. Apart from some very special cases like
$W_3$ little is known about the representation theory of nonlinear
algebras and also it is not clear what their geometrical meaning is.
Remember that the Virasoro algebra arises through unitary representations
of the group of conformal diffeomorphisms which gives it
a clear and well understood geometrical interpretation. It is
natural to ask what the analogous interpretation is for a given
$W$ algebra. This has turned out to be a very difficult problem due
to the nonlinear nature of a generic $W$ algebra. It is only one
of the many mysteries still surrounding the concept of nonlinear
symmetries in 2 dimensions.

In this thesis we are going to present new steps towards the understanding
of $W$ algebras. Since most of the material presented in chapters 3
4 and 5 will be rather technical and algebraic we are going to ease
into this `mathematical state of mind' by providing in chapter 2 an intuitive
motivation for the constructions presented in the following chapters.
In chapter 2 we consider how Liouville theory arises as a reduction
of a WZW model and also how this translates into a reduction of the
WZW Noether current algebra. For this it will be necessary to use
a piece of mathematical machinery called `Poisson reduction' and since
this is not common knowledge among physicists we briefly outline its
main ideas. A slightly different, but equivalent, construction of
Liouville theory through `gauged WZW models' is also
discussed. The natural framework for the quantization of Poisson
reductions is the BRST formalism which will also be considered.

In chapter 3 we then start the main analysis of $W$ theory.
One of the reasons that this has been so hard is that,
unlike the theory of affine Lie algebras, the underlying finite structures
(or even the existence of these) were not known. The theory of affine Lie
algebras would have much harder to develop if finite dimensional simple
Lie algebras were not so well understood. It is with this in mind that
we introduce and develop in chapter 3 the theory of `finite $W$ algebras'.
Classically these are nonlinear
but finitely generated Poisson algebras that can
be obtained from simple Lie algebras by Poisson reduction. The
quantization of these algebras, consisting of an assignment of an
appropriate
noncommutative `operator' algebra to a given classical finite $W$
algebra,  will be performed by BRST methods.
We also develop the classical and quantum
representation theory of finite $W$ algebras. It turns out that
apart from ordinary finite dimensional representations finite $W$
algebras also admit Fock representations in terms of harmonic
oscillators. We also calculate the symplectic leaves, or what could be
called the $W$ coadjoint orbits, of arbitary finite $W$ algebras.

Having developed the theory of finite $W$ algebras in  detail
we turn to the study of their infinite dimensional counterparts. For
this we generalize the so called Drinfeld-Sokolov reduction
procedure in order to obtain a very large class of $W$ algebras.
In fact it is a conjecture that essentially
all $W$ algebras can be obtained
by (possibly repeated) use of this procedure. The classical theory will
lead to an elegant new formula for the Poisson structure of
arbitrary $W$ algebras which is, in a very special case, equivalent to
the Gelfand-Dickii Poisson brackets discussed in section 5 of the
introduction. Again the quantization is carried out by BRST methods
and an explicit algorithm is developed in order to construct the
$W$ generators and their relations. Furthermore it is shown that any
$W$ algebra can be embedded into a semisimple affine Lie algebra and also that
all $W$ algebras admit a free field realization.
All through chapter 4 we  purposely
follow the same lines as we did in chapter 3 in order to illustrate
the formal similarity between the theories of finite and
infinite $W$ algebras.

In the fifth and final chapter we consider some applications of
$W$ theory to integrable models. First it is shown that finite
dimensional completely integrable Toda theories are reductions
of a system describing free particles moving on group manifolds.
Furthermore using results obtained in chapter 3 the set of Toda theories is
extended greatly and it is shown that these models have finite $W$
symmetry. Next we consider the infinite dimensional version of this
reduction procedure. This will lead to generalizations of the
construction discussed in chapter 2 of Liouville theory as a reduction
of a WZW model. The models thus obtained are generalized Toda theories
in which the $W$ algebras of chapter 4 arise as Noether symmetry algebras.
Classical $W$ algebras can also arise as Hamiltonian structures of
integrable hierarchies however as we have seen in this introduction.
Therefore we investigate what hierarchies are associated to the
the aforementioned $W$ algebras in the last section of the final chapter..

The results of this thesis shed a new light on the theory of $W$
algebras since it becomes clear that they can be viewed as reductions
of Lie algebras. We believe that this point of view
is very promising and may lead to many interesting new
developments both in mathematics and physics.

\chapter{Classical and quantum reduction of WZW models}
In classical mechanics it is well known that some highly nontrivial
systems can be described as reductions of simpler ones. The general
scheme behind this is reduction of a mechanical system with symmetry:
suppose the system has a certain symmetry then the number of
`essential' degrees of freedom  can be reduced. However, the equations
of motion in terms of the smaller number of degrees
of freedom will in general look much more complicated and be harder
to investigate then the original equations of motion. One could now
turn this around and ask whether a given complicated system can be
described as a simpler system with  more degrees of freedom. In finite
dimensions systematic use was made of this idea to integrate
explicitly many more or less complicated systems (for a review see
\cite{OlPe}).

In field theory a similar construction was used recently in the study
of Toda field theories in general and Liouville theory in particular.
As we will show in this chapter Liouville theory is in fact nothing
but a reduction of an $SL_2({\bf R})$ Wess-Zumino-Witten (WZW) model
which is well studied and understood. Again, in the WZW variables
the Liouville eqn becomes extremely simple and the well known
Leznov-Savaliev solutions arise simply as reductions of the WZW solutions.
As one could have expected the WZW model has much more symmetry than the
Liouville model and as we will show the WZW-Liouville reduction
also manifests itself on the level of their Noether symmetry algebras.
This will lead us to the concept of Poisson reduction and, in
the quantum case, BRST quantization.

\section{Constrained WZW models}
The progress made towards solving 2 dimensional induced quantum gravity
using the so called `continuum approach' is for a large part due to
the observation by Polyakov that the induced gravity action possesses
in the chiral (or light cone) gauge a hidden affine $SL_2({\bf R})$
symmetry \cite{Pol}. Motivated by the insight that such an underlying
symmetry has to manifest itself in one form or another in all gauges
Forgac et al \cite{FWBFO} showed that Liouville theory
\be
S[\phi ]=\int d^2x\, (\frac{1}{2}(\partial_a \phi )^2+Me^{\phi}),
\ee
describing induced gravity in the conformal gauge (i.e. $ds^2=
e^\phi dx^+dx^-$), is in fact nothing but a constrained
$SL_2({\bf R})$ WZW model. This opens up the possibility to study
Liouville theory, which is very complicated, using the more
natural WZW variables. In fact it turns out that field configurations
which are singular in the Liouville variables, but which do have
regular energy-momentum densities (i.e. they are physically
regular), are regular in the
corresponding WZW variables. As we shall see the singularities
in the Liouville variables arise, in the WZW picture, from the fact
that Gauss decompositions of group elements are only defined on a
dense subset of the corresponding group (in the Liouville case
$SL_2({\bf R})$).

The WZW model is described by the following action functional
for (smooth) embeddings
$g$ of a 2 dimensional Riemann manifold $(\Sigma ,h)$ into a finite
dimensional Lie group $G$.
\be   \label{wzwaction}
S[g]=-\frac{k}{8\pi}\int d^2x \, \sqrt{h}h^{\alpha\beta}(x)\,
\mbox{Tr}(g^{-1}\partial_{\alpha}g\, g^{-1}\partial_{\beta}g)+
\frac{k}{12\pi}\int_{B_3}\mbox{Tr} (\tilde{g}^{-1}d\tilde{g})^3
\ee
where $B_3$ is a 3 manifold whose boundary $\partial B_3=\Sigma$ and
$\tilde{g}$ is an extension of $g$ from  $\Sigma$ to $B_3$. of course
this action will depend on the choice of $B_3$ and on the extension
$\tilde{g}$ but it can be shown that if $k$ is integer then
the Euclidean partition function is
independent w.r.t. these choices. From now on we shall simply take
$\Sigma ={\bf R}\times S^1$ equipped with the Minkowski metric.

The WZW action possesses left and right affine Lie symmetry, i.e.
it is invariant under arbitrary transformations of the form
\cite{KnZa,GeWi}
\be   \label{lrkm}
g(x^+,x^-)\mapsto \Omega (x^+)g(x^+,x^-)\bar{\Omega}(x^-)^{-1}
\ee
where $\Omega$ and $\bar{\Omega}$ are arbitrary group valued functions
of the lightcone coordinates $x^{\pm}$. The Noether currents associated
to this symmetry are
\be
J=k\partial_+gg^{-1} \;\;\; \mbox{ and } \;\;\; \bar{J}=kg^{-1}\partial_-g
\ee
while the field equations, coinciding with the conservation laws of these
currents, read
\be
\partial_-J=\partial_+\bar{J}=0
\ee
The currents $J$ and $\bar{J}$ are Lie algebra valued functions
on $\Sigma$ which means that if $\{t_a\}$ is a basis of the Lie
algebra of $G$ then we can write $J=J^at_a$ and $\bar{J}=\bar{J}^at_a$
and the equations of motion are equivalent to $\partial_- J^a=
\partial_+\bar{J}^a=0$ (for all $a$).

Let us now describe how Liouville theory arises from the $G=SL_2({\bf R})$
WZW model \cite{FWBFO}. For this we use the Gauss decomposition
$g=ABC$ of an arbitrary group element, where $A=\mbox{exp}(aE),
\; B=\mbox{exp}(\frac{1}{2}\phi H)$ and $C=\mbox{exp}(bF)$. Here
$\{H,E,F\}$ is the standard Cartan-Weyl basis of $SL_2({\bf R})$, i.e.
\be
H=\left(
\begin{array}{cc}
1 & 0 \\
0 & -1
\end{array}
\right)
\;\;\;\; E=\left(
\begin{array}{cc}
0 & 1 \\
0 & 0
\end{array}
\right) \;\;\;\;
F=\left(
\begin{array}{cc}
0 & 0 \\
1 & 0
\end{array}
\right)
\ee
The Gauss decomposition is local in the sense that it is only well
defined on a dense subset of $SL_2({\bf R})$. Let us for the moment
ignore this subtlety however. Also note that the arguments that
follow below are not valid for the compact group $SU(2)$ due to the
fact that it allows for no Gauss decomposition.

Inserting $g=ABC$ into the WZW action and using the well known
`Polyakov-Wiegman identity'
\ba
S[ABC] & = & S[A]+S[B]+S[C]-\frac{k}{4\pi}\int d^2x \, \mbox{Tr}(
(A^{-1}\partial_-A)(\partial_+B)B^{-1}+ \nonumber \\
&   & +(B^{-1}\partial_-B)(\partial_+C
\,C^{-1})+(A^{-1}\partial_-A)B(\partial_+C\,C^{-1})B^{-1}) \nonumber
\ea
we find that $S[g]$ reduces to the following action for the fields
$a,b$ and $\phi$
\be
S[a,b,\phi ]=\frac{k}{8\pi}\int d^2x\, \left( \frac{1}{2} \partial_+
\phi \partial_- \phi +2(\partial_-a)(\partial_+b)e^{-\phi}\right)
\ee
The equations of motion of this action are given by
\ba
\partial_-(\partial_+b\, e^{-\phi})=0 \\
\partial_+(\partial_-a \, e^{-\phi})=0 \\
\partial_+\partial_- \phi +2(\partial_-a)(\partial_+b)e^{-\phi}=0
\ea
Obviously from the first two equations follows that
\be
\partial_+b=\mu (x^+) e^\phi \;\;\; \mbox{ and } \;\;\; \partial_-a=\nu (x^-)
e^{\phi}
\ee
where $\mu$ and $\nu$ are arbitary functions of $x^+$ and $x^-$
respectively while the third equation reduces to
\be \label{bijna}
\partial_+\partial_-\phi (x) +2M(x)e^{\phi (x)} =0
\ee
where $M(x)=2\mu (x^+)\nu (x^-)$. Note that if $M$ was a constant
function then this equation would be nothing but the Liouville equation.

In all steps made above we have retained the full content of the WZW
model. Now, the phase space of any mechanical model (i.e. the space of
`positions and momenta') can be described, at least locally, as the
space of solutions of the equations of motion because through every
point $p$ in phase space goes one unique solution such that $t=0$
at the time when this solution passes through $p$ (put differently,
the space of boundary conditions is equal to the phase space). From the above
therefore follows that Liouville theory is a
reduction of the WZW model in the sense that the Liouville phase
space is equal to the subspace of the WZW phase space determined by
the constraints that $\mu$ and
$\nu$ are constant. On this reduced phase space the WZW equations
of motion reduce to the Liouville equation.

A nice byproduct of this approach to Liouville theory is that the
well known Leznov-Savaliev solutions of the Liouville equation
can be derived easily from the general classical solution of the WZW
model \cite{FWBFO}
\be
g(x^+,x^-)=g_L(x^+)g_R(x^-) \label{solwzw}
\ee
where $g_L$ and $g_R$ are arbitrary group valued functions constrained
only by boundary conditions. The Leznov-Savaliev solutions can now be
obtained from (\ref{solwzw}) by imposing the constraints and then
inserting the Gauss decomposition.

As we have discussed above a WZW model has left and right affine Lie
symmetry and one might wonder what happens to this symmetry
after reduction. For this we have to consider the Noether currents associated
to this symmetry a little bit closer. As we have seen the conservation
laws induced by the affine symmetry are given by $\partial_-J^a=\partial_+
\bar{J}^a$. A standard calculation reveals that the (equal time)
Poisson brackets between these currents read
\ba
\{J^a(x),J^b(y)\} & = &
f^{ab}_cJ^c(y)\delta (x-y)+kg^{ab}\delta '(x-y) \nonumber \\
\{\bar{J}^a(x),\bar{J}^b(y)\} & = & f^{ab}_c\bar{J}(y)^c\delta (x-y) +kg^{ab}
\delta '(x-y) \nonumber \\
\{J^a(x),\bar{J}^b(y)\} & = & 0 \label{curwzw}
\ea
where $g^{ab}$ is the inverse of $g_{ab}=\mbox{Tr}(t_at_b)$.
This current algebra will obviously be changed
by the reduction process because some currents will be constrained.
In order to see this consider again our example above. In
terms of the Cartan-Weyl basis the current $J$ can be written
as $J(x)=J^H(x)H+J^E(x)E+J^F(x)F$ (and a similar equation for the current
$\bar{J}(x)$). Inserting the Gauss decomposition into the
definition of $J$ one finds that $J^E=k(\partial_+b)\,e^{-\phi}$ and
$\bar{J}^F=-k(\partial_-a)\, e^{-\phi}$ from which follows that after
imposing the constraints we have $J^E(x)=\mu$ and $\bar{J}^F(x)=-\nu$.
{}From a mathematical point of view one therefore has a Poisson algebra
(\ref{curwzw}) subject to a set of constraints. Intuitively it is
clear that the reduced current algebra will be equal to the
Poisson algebra induced by (\ref{curwzw})
on the constrained phase space. Note that this algebra is {\em not}
obtained by simpy inserting the constraints into (\ref{curwzw}).
Let's illustrate this again with an example. Consider
the Poisson bracket between $J^H$ and
$J^E$. From (\ref{curwzw}) follows that
\be \label{he}
\{J^H(x),J^E(y)\}\sim J^E(y)\delta (x-y)
\ee
On the constraint manifold
$J^E=\mu$ and obviously a constant commutes with
everything, so we must have $\{J^H(x),J^E(y)\}=\{J^H(x),\mu \}=0$.
The right hand side however of equation (\ref{he}) is, after imposing
the constraints, equal to $\mu
\delta (x-y)$ which is nonzero. Blindly inserting the constraints
therefore leads to inconsistencies.
The proper way to deal with the construction of the reduced current
algebra (\ref{curwzw}) is by `Poisson reduction'. This is the subject
of the next section.

\section{Reduction of Poisson manifolds}
In  this section we shall briefly discuss the reduction of Poisson
manifolds \cite{poissred,Kimura}. This procedure will be used extensively
in the next chapters in order to construct the Noether current algebras
associated to certain reduced WZW models.

Let $(M,\{.,.\})$ be a Poisson manifold, i.e. $M$ is a manifold and
$\{.,.\}$ is an antisymmetric bilinear map from $C^{\infty}(M)\times
C^{\infty}(M) \rightarrow C^{\infty}(M)$ which satisfies the Jacobi
identities
\be
\{f,\{g,h\}\}+\{g,\{h,f\}\}+\{h,\{f,g\}\}=0
\ee
where $f,g,h\in C^{\infty}(M)$. It is well known that any symplectic
manifold is a Poisson manifold, but the converse need not be true.
It {\em is} true however that a Poisson manifold is the disjoint union of
symplectic submanifolds (we shall come back to this in the next chapter).

Any Poisson structure $\{.,.\}$ induces a map
\be
P:T^*M\rightarrow TM
\ee
which is defined as follows: A given vectorfield $X\in \Gamma (TM)$ ($=$
the space of smooth sections of $TM$) can be written as $X=g_idf^i$
where $g_i,f_i\in C^{\infty}(M)$. $P$ is therefore uniquely defined
by putting $P(df)=X_f$ where $X_f$ is the `Hamiltonian vectorfield
associated to $f$', i.e. $X_f=\{f,.\}$.

Now, let $M_0$ be a closed and embedded submanifold $i:M_0\hookrightarrow
M$ of $M$ of codimension $k$. The manifold $M_0$ is completely
characterized by $C^{\infty}(M_0)$, the space of smooth functions
on $M_0$ and the relation between $C^{\infty}(M)$ and $C^{\infty}(M_0)$
can be described as follows: let $I$ be the ideal in $C^{\infty}(M)$
of functions that vanish on $M_0$, i.e.
\be
I=\{f\in C^{\infty}(M) \mid
f(x)=0, \mbox{ for all } x\in M_0\},
\ee
then
\be
C^{\infty}(M_0)\simeq C^{\infty}(M)/I
\ee
A similar construction holds for the space of vectorfields $\Gamma (
TM_0)$: it is clear that any vectorfield on $M_0$ arises as the
restriction of some vectorfield on $M$. However, the restriction
of a vectorfield on $M$ to $M_0$ is not, in general, a vectorfield
on $M_0$ since it need not be tangent to $M_0$. Obviously the space of
vectorfields on $M$ which restrict to vectorfields on $M_0$
is given by
\be
{\cal N}(I)=\{X\in \Gamma (TM) \mid X(f) \in I, \mbox{ for all }f \in I\}
\ee
because an element $f$ of $I$ is constant on $M_0$ which means that
the derivative of $f$  in a direction tangent to  $M_0$ must be
zero. of course two vectorfields $X_1,X_2
\in {\cal N}(I)$ may restrict to the
same vectorfield on $M_0$. This happens when their difference vanishes
on $M_0$, or put differently when $X_1-X_2 \in I {\cal N}(I)$. We
therefore find
\be
\Gamma (TM_0)\simeq \frac{{\cal N}(I)}{I{\cal N}(I)}
\ee
It is also easy to see that the set of functions $f\in C^{\infty}(M)$
which have the property that
their Hamiltonian vectorfields $X_f$ are tangent to $M_0$ is given
by
\be
N(I)=\{f\in C^{\infty}(M) \mid \{f,g\}\in I, \mbox{ for all } g\in I\}
\ee
Note that $N(I)$ is a Poisson subalgebra of $C^{\infty}(M)$ and that
it is in fact the normalizer of $I$ ($I$ is an ideal w.r.t. the
multiplicative structure on $C^{\infty}(M)$ but not necessarily
w.r.t. the Poisson structure). Elements
of $N(I)$ are called `first class'.

It is well known in physics that first class elements generate
gauge invariances \cite{Dirac}. These gauge invariances cause
degeneracies which have the result that the Poisson structure on
$M$ does {\em not} induce a Poisson structure on $M_0$ but on a
certain coset space of $M_0$. Let us make this statement more precise.

Let $i^*TM$ and $i^*T^*M$ denote the pullbacks of the tangent
cotangent bundle to $M_0$ via the embedding $i$.
$TM_0$ and $T^*M_0$ are
subbundles of $i^*TM$ and $i^*T^*M$ respectively. Define now
another subbundle, called the annihilator bundle $\mbox{Ann}(TM_0)$,
of $i^*T^*M$ as follows: The fiber over $p\in M_0$ is given by
\be
\mbox{Ann}_p(TM_0)=\{\alpha \in i^*T^*_pM\mid \alpha (v)=0, \mbox{ for
all } v\in T_pM_0\}
\ee
The image under $P$ of the annihilator bundle is denoted by
$TM_0^{\perp}$, i.e.
\be
T_pM_0^{\perp}=\{p(\alpha ) \mid \alpha \in \mbox{Ann}_p(TM_0)\}
\ee
Now, using $TM^{\perp}$ we can define the so called {\em null distribution}
$V$ of $M_0$ which is the intersection of $TM_0$ and $TM_0^\perp$
\be
V_p=T_pM_0\bigcap T_pM_0^{\perp}
\ee

Let us explain these definitions. The annihilator bundle consists
roughly of those one forms that are exterior differentials of
elements of $I$, for let $f\in I$ and $v\in TM_0$ then $df (v)\equiv
v(f)=0$ because $f$ is constant on $M_0$. $TM_0^\perp$ is then
nothing but the span of the Hamiltonian vectorfields of elements
of $I$. Not all elements of $I$ are however first class and since
the Hamiltonian vectorfields of first class constraints are tangent
to $M_0$ we can identify the gauge directions by considering the
intersection of $TM_0^\perp$ and $TM_0$. The null distribution is therefore
nothing but an indication as to what directions on $M_0$ are gauge
directions.

It can be shown \cite{Kimura} that the null distribution $V$
is an involutive distribution which means, by Frobenius' theorem, that
$M_0$ foliates into maximal connected submanifolds or leaves of
`gauge equivalent states'. One can eliminate all gauge freedom by
identifying all points on a leaf or equivalently by `gauge fixing'.
Gauge fixing amounts to finding a submanifold of $M_0$ (called the
`gauge slice') that has exactly one point in common with every leaf.
On the gauge slice, denoted by $\tilde{M}$,
the Poisson structure $\{.,.\}$ then induces a reduced Poisson structure
$\{.,.\}^*$ \cite{AbMa,Kimura}.

Let us now make the above somewhat more concrete. Again consider a
Poisson manifold $(M,\{.,.\}$ and let the submanifold $M_0$
be given as the zero set of a collection of  `constraints'
$\{\phi^i \}\subset C^{\infty}(M)$, i.e.
\be
M_0=\{p\in M \mid \phi^i(p)=0, \mbox{ for all } i \}
\ee
$I$ is then the ideal in $C^{\infty}(M)$ generated by the constraints
\be
I=\{f_i \phi^i \mid f_i \in C^\infty (M) \mbox{ for all }
i \}
\ee
Suppose the gauge invariances generated by the first class constraints
(not all constraints need to be first class) can be completely fixed by
adding some `gauge fixing constraints', i.e. constraints that
determine the gauge slice. Denoting the {\em total} set of constraints
by $\{\phi^\mu \}$ (i.e. with Greek indices) the gauge slice is given by
\be
\tilde{M}=\{p\in M \mid \phi^\mu (p)=0, \mbox{ for all } \mu \}
\ee
and the reduced Poisson structure is
\cite{Dirac,Sunder}
\be \label{di}
\{\bar{f},\bar{g}\}^*=\overline{\{f,g\}-\{f,\phi^\mu \}\Delta_{\mu \nu}
\{\phi^\nu,g\}}
\ee
where $f,g\in C^\infty (M)$, the bar denotes the restriction to
$\tilde{M}$ (which is a submanifold of $M$) and $\Delta_{\mu \nu}$
is the inverse of $\Delta^{\mu \nu}\equiv \{\phi^\mu ,\phi^\nu \}$.
The bracket (\ref{di}) was originally discovered by Dirac and is therefore
called the Dirac bracket.

Let us now return to our example of Liouville theory.
The Poisson algebra we are dealing
with in this case is (\ref{curwzw}) for $SL_2({\bf R})$ subject
to the constraint $\phi^1 (x)\equiv J^E(x) -\mu=0$. Since the
Poisson bracket of $\phi (x)$ with itself is zero this constraint
is first class and therefore generates gauge invariances. This gauge
invariance has to be fixed and it will be shown in the next
chapters that one can fix it completely by adding the constraint
$\phi^2(x)\equiv J^H=0$. Calculating the Poisson relation between
$J^E(x)$ and $J^E(y)$ with respect to (a field theoretical version
of) the Dirac bracket one finds that they satisfy the Virasoro algebra
\be
\{J^E(x),J^E(y)\}^*=\partial J^E(y)\delta (x-y)-2J^E(y)  \delta '(x-y)
-\frac{1}{2}\delta '''(x-y)
\ee
This is then the current algebra associated to Liouvilly theory.

Above we have seen that the affine $SL_2({\bf R})$  algebra leads,
on reduction, to the Virasoro algebra. One might wonder if it is
possible to obtain more complicated $W$ algebras in this way by
reducing affine $SL_N({\bf R})$  algebras (or more generally affine Lie
algebras
over arbitrary simple Lie algebras). This
is precisely the question we address in chapter 4. It  is clear that affine
$SL_2({\bf R})$
subalgebras will play an important role in this theory because
any $W$ algebra must contain a Virasoro algebra and we have just seen
that this algebra is a reduction of the affine $SL_2({\bf R})$ algebra.
However, there are
in general
many inequivalent affine $SL_2({\bf R})$ subalgebras into a given affine Lie
algebra and it is therefore natural to conjecture
that there is a $W$ algebra associated to
every $SL_2({\bf R})$ embedding. This conjecture will be proven in
chapter 4.

\section{Gauged WZW models}
Above we have constructed Liouville theory as a constrained WZW
model. The procedure we followed was that of contraining the
WZW phase space by essentially specifying the solution space of the
reduced theory. In order to quantize Liouville theory it would
however be much more useful to have available a Langrangian realization
of the WZW-Liouville reduction. The natural framework for this is the
theory of gauged WZW models \cite{BFFOW}.

We have seen above that the WZW action (\ref{wzwaction}) is invariant
under left and right affine  transformations (\ref{lrkm}). Note that
the group valued functions $\Omega$ and $\bar{\Omega}$ are functions
of one lightcone coordinate only. Consider now two isomorphic subgroups
$H$ and $\bar{H}$ of $G$. The affinizations $\hat{H}$ and $\hat{\bar{H}}$
of these groups (consisting
of $H$ ($\bar{H}$) valued functions of the coordinates
$x^+$($x^-$)) are
subgroups of the left-right affine symmetry of the WZW lagrangian which
means that $S[g]$ certainly has $\hat{H} \times \hat{\bar{H}}$
symmetry. Gauging this symmetry means, in the WZW context, making the
action invariant under transformations
\be  \label{gauget1}
g(x^+x^-) \mapsto \alpha (x^+,x^-) \,g(x^+,x^-)\, \beta^{-1}(x^+,x^-)
\ee
where $\alpha$ and $\beta$ are $H$ respectively $\bar{H}$ valued
functions {\em of both} $x^+$ {\em and} $x^-$. For this one introduces
$\hat{h}$ respectively $\hat{\bar{h}}$ valued gauge fields $A_+$ and $A_-$
(where $\hat{h}$ ($\hat{\bar{h}}$) is the Lie algebra of $\hat{H}$
($\hat{\bar{H}}$)  which transform as usual under
gauge transformations
\ba
A_+ & \mapsto &
\alpha A_+ \alpha^{-1} +\partial_+ \alpha \, \alpha^{-1} \nonumber \\
A_- & \mapsto &
\beta  A_- \beta^{-1} +  \beta^{-1}\partial_- \beta \label{gauge2}
\ea
A `gauged' WZW action that is invariant (or more accurately, invariant
up to a total derivative) under the transformations
(\ref{gauget1}) and (\ref{gauge2}) is
\ba \label{gaugedwzw}
I[g,A_+,A_-] & = &
S[g]+\int d^2x \, ( \mbox{Tr}(A_-(\partial_+g \, g^{-1}
-M))+ \nonumber \\
&   & + \mbox{Tr}(A_+(g^{-1}\partial_-g-\tilde{M}))+\mbox{Tr}(A_-gA_+g^{-1})
)
\ea
where $M$ and $\tilde{M}$ are the constraints one wishes to impose
on the system (in the case of Liouville theory discussed above
$M=\mu E$ and $\tilde{M}=\nu F$).

It is clear that one can use the gauge invariance \ref{gauget1}
of the action (\ref{gaugedwzw}) to put $A$ and $B$ in the
Gauss decomposition $g=ABC$ equal to the unit element $e$ of $G$.
For the $SL_2({\bf R})$ case this implies $g=\mbox{exp}(\frac{1}{2}\phi H)$.
{}From the equations of motion of the fields $A_{\pm}$ one then finds
that $A_+=B^{-1}MB$ and $A_-=B\tilde{M}B^{-1}$. Substituting this,
together with the fact that $g=B$ back into the action (\ref{gaugedwzw})
one finds that it reduces to the Liouville action. What happens in
the general case (i.e. for arbitary algebras) will be examined in
chapter 5.

In the literature there are two approaches to gauged WZW models
corresponding (on the algebraic level) to the reduction approach
discussed above and the coset construction respectively. Let $H$ be
a subgroup of $G$ and let $A=A_+dx^++A_-dx^-$ be an $h$ valued
gauge field (where again $h$ is the Lie algebra of $H$). Consider
instead of (\ref{gaugedwzw}) the action
\be
\tilde{I}[g,A_-,A_+]=S[g]-\frac{k}{4\pi}\int d^2x \, \mbox{Tr}(A_-
(\partial_+g)g^{-1}+(g^{-1}\partial_-g)A_++A_-gA_+g^{-1}-A_-A_+)
\ee
This is called the action of a $G/H$ WZW model and it has been shown
in \cite{cosetwzw} that it provides a Lagrangian realization of
the Goddard-Kent-Olive coset construction. Note the difference with
the action (\ref{gaugedwzw}): the fields $A_-,A_+$ are now components
of an $h$ valued gauge field while in (\ref{gaugedwzw}) they are
completely independent.

\section{Quantum framework for constrained WZW models}
In the previous section we have considered the classical problem of
reduction of the WZW model to Liouville theory. As we have seen this
reduction has an algebraic counterpart relating the chiral Noether symmetry
algebras associated to these two theories. In the quantum case the
Hilbert space of the WZW model will carry a unitary representation
of the affine Lie algebra while the Liouville Hilbert space will
decompose into Virasoro representations. The problem is to
construct the Liouville Hilbert space as a reduction of the WZW
Hilbert space and to calculate the Liouville spectrum from the
WZW spectrum. The formalism ideally suited to handle problems like this
one is the BRST formalism \cite{BRST}.

Suppose one is given any (classical) physical system together with a set of
first class constraints $\{\phi_i\}$, i.e.
\be
\{\phi_i,\phi_j\}=C_{ij}^k\phi_k
\ee
where $C_{ij}^k$ may in principle
be arbitrary functions on the phase space of the
system. In what follows we will however always assume these `structure
constants' to be
constant. The meaning of this relation is that the Poisson bracket between
first class constraints vanishes on the constrained manifold from which
follows that the Hamiltonian vector fields of first class constraints
are tangent to it. As Dirac showed the directions on phase space
specified by these first class constraints via their Hamiltonian
vectorfields are gauge directions and have to be eliminated.  In principle
one could first reduce the system to a gauge slice and then quantize it
but in practise it is easier to do the opposite, i.e. quantize first
and reduce later. of course one has to assume that one is able to
quantize the unconstrained theory (which in the WZW case is certainly
the case \cite{GeWi}).
Suppose therefore that we have quantized our original system yielding
a Hilbert space $\cal H$. BRST quantization now involves introducing
a set of `ghosts' $c^i$ and `antighosts' $b_i$ which obey canonical
anticommutation relations $c^ib_j+b_jc^i=\delta^i_j$. Mathematically one
one has extended the phase space by adding (Grassman) degrees of
freedom. In the end one will have to remove all the Grassman variables
from the theory together with the redundant degrees of freedom
associated to the gauge directions. The way one goes about this is
as follows: Introduce the operator
\be
Q=c^i\phi_i-\frac{1}{2} C_{ij}^k c^ic^jb_k
\ee
which is known as the BRST operator. In the second term one needs,
in the infinite dimensional case, a normal ordening prescription but
we will ignore this subtlety in this chapter (we do come back to this
in chapter 4 however). Also it should be mentioned that if the structure
constants would have been structure functions (i.e. funcions on phase
space) then the BRST operator could have picked up higher order terms.
How to construct the BRST operator in the general case can be found in
\cite{Baulieu}.

The basic property of the BRST operator is that $Q^2=0$ and that it
maps the ghost number $n$ subspace ${\cal H}_n$ of the (extended) Hilbert space
(i.e. the eigenspace of the ghost number operator $N=c^ib_i$ with
eigenvalue $n$) to the ghost number $n+1$ subspace, i.e.
\be
Q:{\cal H}_n \rightarrow {\cal H}_{n+1}
\ee
A state is now called `BRST invariant' if $Q|\psi \rangle =0$. Since the
ghosts were artificially put into the theory it is clear that the
BRST invariant states of ghost number 0 are of special interest. For
such a state we have $N|\psi \rangle =0$ which means that $b_i|\psi \rangle
=0$ for all $i$ which means that the second term in the BRST operator
vanishes on the ghost number zero subspace. In fact one has
\be
Q|\psi \rangle = c^if_i |\psi \rangle
\ee
and since a state annihilated by all $b_i$ cannot be annihilated by
all $c^i$ we find that the state is BRST invariant iff
$\phi_i|\psi \rangle =0$. We conclude that a state of ghost number 0
is BRST invariant iff it is annihilated by the (quantum) constraints.
Therefore BRST invariant states are of great interest.

There are however states that are trivially BRST invariant, namely those
states $|\psi \rangle$ for which there exists a state $|\chi \rangle$
such that $|\psi \rangle = Q|\chi \rangle$ (these states are called
BRST exact). Let us therefore consider
two states to be equivalent if their difference is BRST exact. Heuristically
one could say that BRST equivalence corresponds to gauge equivalence
which means that one should not distinguish physically between BRST
equivalent states.

{}From the above we find that the true Hilbert space is the space of
BRST invariant states at ghost number zero modulo the set of
BRST exact states. Mathematically this is called the zeroth cohomology
of the BRST complex. of course similar arguments are in force with
respect to the observables in the theory, i.e. the physically interesting
observables $\cal O$ of the reduced theory are those which are BRST invariant
$[Q,{\cal O}]=0$ but not BRST exact ${\cal O}\neq [Q,{\cal O}_2]$.

Let us again apply this on Liouville theory. As we have said the WZW
Hilbert space carries a (reducible) representation of the underlying
affine Lie algebra. This algebra   is given in terms of operator
product expansions by
\be
J^a(z)J^b(w)=\frac{kg^{ab}}{(z-w)^2}+\frac{f^{ab}_cJ^c(w)}{z-w}+\ldots
\ee
and as we have seen the constraints are given in the Liouville case
by $\phi (z)=J^E(z)-1$. Introducing the ghost field $c(z)$, the
antighost field $b(z)$ satisfying the OPE
\be
c(z)b(w)=\frac{1}{z-w}+\ldots
\ee
and using the prescription outlined above we
find that the BRST operator is given by
\be
Q=(J^E(z)-1)c(z)
\ee
The total Hilbertspace is then given by
\be
{\cal H}_{tot}={\cal H}_{WZW}\otimes {\cal H}_{ghost}
\ee
and what we need to do is calculate the zeroth cohomology of $Q$.
In the case of Liouville theory this was done in \cite{BeOog}
using free field resolutions of affine and Virasoro representations.
They also showed that the cohomology space carries a
Virasoro representation
of central charge
\be
c(k)=13-6(k+2)-\frac{6}{k+2}
\ee
This Virasoro algebra is the chiral current algebra of Liouville theory.

In this chapter we have seen how Liouville theory arises as a (conformally)
reduced WZW model.
In chapter 4 we are going to generalize this construction to general
affine Lie algebras and arbitrary reductions.
The methods we are going to use for this are however essentially the
same as the ones that we explained in this chapter.

\chapter{Finite W algebras}
\section*{Introduction}
Starting with the seminal work of Weyl and Wigner
symmetry principles have been playing an ever increasing role
in theoretical physics. The simplest symmetries are those in which
the physical system is symmetric under a finite number of transformations
or when it is invariant under displacement
in a finite number of `directions'. The natural
mathematical structures describing such symmetries are discrete groups
and finite dimensional Lie groups respectively. In this century
the theory of these mathematical structures has,
motivated by physics, been brought to
a high level of development. This has lead to many new insights
into the principles of nature and in fact nowadays symmetry principles
play a role in physics that is as fundamental
as for example the action principle. A paramount example of this is invariance
under the Poincare group, which is at the heart of Einstein's special
theory of relativity and hence is a necessary criterion for any
realistic theory of nature.

Another sacred principle of fundamental physics
is locality or causality.
Combining symmetry principles with locality one is naturally lead to
the study infinite dimensional groups of smooth maps from some spacetime
manifold
$X$ to a finite dimensional Lie group $G$.
The group multiplication in such a group is just
pointwise multiplication, i.e. if $f,g\in Map(X,G)$ and $a\in G$ then
$(f.g)(a)=f(a)g(a)$.
In quantum field theory groups of the form $Map(X,G)$ and their Lie
algebras $Map(X,g)$ (where $g$ is the Lie algebra of $G$) arise
essentially in two different ways:
through the principle of local gauge invariance, which is at the heart
of modern high energy physics (i.e. the standard model of the
fundamental interactions between quarks and leptons),
and through the theory of
current groups and algebras.

Unfortunately for generic manifolds $X$ surprisingly little is known
about the group $Map(X,G)$. Especially the representation theory of these
groups is still almost unexplored. The exception to this is
the case $X=S^1$ where $Map(S^1,G)$ and $Map(S^1,g)$
are called `loop groups' and `loop algebras'. Loop groups and algebras
arise in simplified models of quantum field theory in which space is
taken to be 1-dimensional and therefore also in string models of elementary
particles. The study of loop groups and algebras is much simpler than
when $X$ is some more complicated manifold. This is caused by the
fact that they behave much like the ordinary finite dimensional
Lie groups and algebras that underly them. This remarkable fact makes
knowledge of the finite dimensional theory essential for the study
of the infinite dimensional theory.

Compared to Lie groups and algebras the theory of $W$ algebras has
evolved backwards, for up to now infinite dimensional $W$ algebras have
received the bulk of the attention. One reason for this is that it was
not entirely clear what the finite algebras underlying $W$ algebras were
and whether there was a finite version of $W$
theory at all. Considering the way in which
$W$ algebras were first introduced into physics
they really don't seem to have any relation to the theory of loop
groups and algebras. Therefore finite structures underlying
$W$ algebras have not been considered. In this chapter we shall
address the question: does the theory of $W$ algebras have a finite
counterpart? The answer to this question turns out to be affirmative.
In fact it will be shown that this finite theory is remarkably rich
and, as with loop algebras, contains already most
of the essential features of infinite dimensional $W$ algebra theory.

In this chapter we approach the problem of constructing
finite $W$ theory in the same way as we approach the theory of
infinite dimensional $W$ algebras, that is we consider $W$ algebras (finite
or infinite) to be reductions of Lie algebras. In the next chapter we
will show that infinite $W$ algebras are reductions
of affine Lie algebras (i.e. central extensions of certain loop algebras).
Our approach to constructing finite $W$ theory is by mimicing the
reduction procedure on finite dimensional Lie algebras. As we shall see
this leads to a rich theory of finitely generated nonlinear
algebras.

This chapter consists of three parts. The first part deals with the
classical theory of finite $W$ algebras. Classical finite $W$ algebras
will be constructed as Poisson reductions of the Kirillov Poisson structure
on the dual of a simple Lie algebra. In this section finite $W$ algebras
are therefore finitely generated nonlinear Poisson algebras. The classical
representation theory of these algebras is considered, i.e. we
calculate the symplectic leaves associated to arbitrary finite $W$
algebras. In the second
part of this chapter we quantize these classical finite $W$ algebras.
This is done by the well known BRST quantization procedure which is
ideally suited for situations like this. In the third and last part of
this chapter we construct the representation theory of quantum finite
$W$ algebras. As it turns out the representation theory of semisimple
Lie algebras
is in essence all one needs to construct these representations.
We also give a general procedure to obtain Fock realizations of arbitrary
finite $W$ algebras.

\section{The Classical Theory of Finite $W$ algebras}
In this section we introduce and develop the theory of finite $W$ algebras.
As we already mentioned, we will show
in the next chapter that many classical $W$ algebras can be
obtained by Poisson reduction of the Kirillov Poisson algebras naturally
associated to affine Lie algebras. Given a certain affine Lie algebra there are
several possible reductions all leading to different $W$ algebras. In fact
we will show that to any $sl_2$ embedding into the simple Lie algebra
underlying the affine Lie algebra one can associate such a Poisson
reduction. Motivated by this we consider in this chapter analogous
Poisson reductions of Kirillov Poisson algebras associated to finite
dimensional simple Lie algebras. Obviously this is the natural finite
dimensional analogue of the procedure that leads to infinite $W$ algebras.
The finitely generated Poisson algebras one obtains in this way we define
to be finite $W$ algebras. In order to make the presentation more or less
selfcontained we shall first briefly review some basic aspects of
Kirillov Poisson structures.

\subsection{Kirillov Poisson structures}

Let $(M,\{.,.\})$ be a Poisson manifold,
that is $\{.,.\}$ is a Poisson bracket on the space
$C^{\infty}(M)$ of $C^{\infty}$ functions on $M$,
and $G$ a Lie group. Also let
$\Phi :G\times M \rightarrow M$ be a smooth and proper
action of $G$ on $M$ which
preserves the Poisson structure, i.e.
\begin{equation}
\Phi_g^*\{\phi , \psi \} = \{\Phi_g^*(\phi ), \Phi_g^*(\psi )\}
\end{equation}
where $\Phi^*_g$ is the pullback of $\Phi_g:M \rightarrow M$.
Physically this implies that if $\gamma (t)$ is a solution of the
equations of motion w.r.t. some $G$-invariant Hamiltonian H (i.e.
$\Phi_g^* H=H$  where as usual $\Phi^*_gH=H \circ \Phi_g$), then
$(\Phi_g \circ \gamma )(t)$ is again a solution. If we do not want
to consider solutions that can be transformed into each other in this
way to be essentially different we are led to consider the space
$M/G$ as the 'true' phase space of the system. The only observables
(i.e. functions on $M$) that one is concerned with are those which are
themselves G-invariant and therefore descend to observables on
$M/G$. Denote the set of smooth $G$-invariant functions on $M$ by $O$,
and let then $\phi, \psi \in O$. Using the fact that $\Phi_g$ preserves
the Poisson structure for all $g \in G$ and that $\phi$ and $\psi$
are $G$-invariant we can easily deduce that $\{\phi,\psi\}$ is also
an element of $O$, i.e. $O$ is a Poisson subalgebra of $C^{\infty}(M)$.
Let $\pi : M \rightarrow M/G$ be the canonical projection. The
pullback map $\pi^* : C^{\infty}(M/G) \rightarrow O$ is in fact
an isomorphism. It assigns to a function $\hat{\phi}$ on $M/G$
the function $\hat{\phi}\circ \pi$ on $M$ which is constant along
the $G$-orbits. What one wants is now to define a Poisson structure
$\{.,.\}^*$ on $C^{\infty}(M/G)$ such that the Poisson algebras
$(C^{\infty}(M/G), \{.,.\}^*)$ and $(O,\{.,.\})$ are isomorphic.
This would mean that all ($G$-invariant) information of the original
phase space is transferred to the Poisson algebra $(C^{\infty}(M/G),
\{.,.\}^*)$. Therefore define
\begin{equation}
\{ \hat{\phi},\hat{\psi}\}^*=(\pi^*)^{-1}\{\pi^*\hat{\phi},
\pi^*\hat{\psi}\}
\end{equation}
for all $\hat{\phi},\hat{\psi} \in C^{\infty}(M/G)$.
Obviously $\pi^*$ is a Poisson algebra isomorphism between
the two Poisson algebras. So starting from a $G$-invariant theory
we have arrived at a formulation in which all essential information
is contained and all redundancy has been eliminated.

Consider now a $G$-invariant theory in which the group manifold itself
is the configuration space (for example a particle moving on $G$ with
a Hamiltonian that is invariant under $G$). The phase space of such
a system is of course $T^*G$, the cotangent bundle of $G$. The left action
of the group on itself, denoted by $L_g$,  induces a left $G-$action
$L_g^*$ on $T^*G$ which is in fact a Poisson action w.r.t. the
canonical Poisson structure on $T^*G$ ($T^*G$ carries a natural Poisson
structure because it is a cotangent bundle). Therefore we can do what
we did before and extract the $G$-invariant piece of the theory by
going to the quotient manifold
\begin{equation}
T^*G/G
\end{equation}
Obviously this space is nothing but the space of left invariant
one forms which means that it is isomorphic to $g^*$, the dual of
the Lie algebra of $G$. The canonical projection
\begin{equation}
\pi:T^*G \rightarrow g^*
\end{equation}
is therefore simply the pullback to the fiber over the identity element
of $G$ by means of left translation. As before the Poisson structure on
$T^*G$ induces a Poisson structure  on $g^*$. This Poisson structure
on $g^*$ is called the 'Kirillov' Poisson structure.

It is possible to obtain (by explicit calculation) a more concrete
formula for the Kirillov Poisson structure \cite{AbMa}.
It reads
\begin{equation}
\{F,G\}(\xi )=\xi ([\mbox{grad}_{\xi}F,\mbox{grad}_{\xi}G])
\end{equation}
where $F$ and $G$ are smooth functions on $g^*$, $\; \xi \in g^*$ and
$\mbox{grad}_{\xi}F \in g$ is determined uniquely by
\begin{equation}
\frac{d}{d\epsilon}F(\xi+ \epsilon \eta)|_{\epsilon =0}
=\eta(\mbox{grad}_{\xi}F)
\end{equation}
for all $\eta \in g^*$. In  this paper we will always take $G$ to be
a simple group which means that the Cartan-Killing form $(.,.)$
on its Lie algebra is non-degenerate. This allows us to identify
the Lie algebra $g$ with its dual, i.e. all $\xi \in g^*$ are of
the form $(\alpha,.)$ where $\alpha \in g$. The above formulas then
become
\begin{equation}
\{F,G\}(\alpha )=(\alpha,[\mbox{grad}_
{\alpha}F,\mbox{grad}_{\alpha}G])  \label{kir}
\end{equation}
and
\begin{equation} \label{df}
\frac{d}{d\epsilon}F(\alpha + \epsilon \beta) |_{\epsilon =0}=
(\beta,\mbox{grad}_{\alpha}F)
\end{equation}
where $\alpha,\beta \in g$ and $F,G \in C^{\infty}(g)$. This
is the form that we will use.

It is possible to give an even more detailed description of the
Kirillov bracket. Let $\{t_a\}$ be a basis of $g$ and let $\{J^a\}$
be the dual basis, i.e. $J^a(t_b)=\delta^a_b$. of course the functions
$J^a$ are smooth functions on $g$ and we can calculate the Kirillov
Poisson bracket between them. What one finds is
\begin{equation}
\{J^a,J^b\}=f^{ab}_{c}J^c    \label{kir2}
\end{equation}
where $f_{ab}^c$ are the structure constants of $g$ in the basis
$\{t_a\}$ and indices are raised and lowered by the Cartan-Killing
metric. Note the resemblance with the Lie bracket
\begin{equation}
[t_a,t_b]=f_{ab}^ct_c
\end{equation}
However we have to remember that the Poisson bracket
(\ref{kir2}) is defined on the space of smooth functions on $g$ which
is a Poisson algebra, i.e. it also carries a commutative multiplication
compatible with the Poisson structure. This means for example that
the expression $J^aJ^bJ^c$ makes sense, while $t_at_bt_c$ does not
(remember that if $F,G$ are two functions on $g$ then their product
is defined by $(FG)(\alpha)=F(\alpha)G(\alpha)$).

\subsection{$sl_2$-embeddings}
As we have mentioned before the inequivalent Poisson reductions of a
Kirillov Poisson algebra are labelled by the equivalence classes of
$sl_2$ embeddings into the simple Lie algebra in question. Let us
therefore take a moment to recall some basic facts on $sl_2$ embeddings
\cite{Dynkin} and prove some lemmas that we shall need later on.

Let $i: sl_2 \hookrightarrow g$ be an embedding of $sl_2$ into a
simple Lie algebra $g$
and let $L(\Lambda ;n )$ be the irreducible finite
dimensional representation of $g$
with highest weight $\Lambda$. As a representation of the
subalgebra $i(sl_2)$  of $g$,  $L(\Lambda;n )$ need not
be irreducible. In general it decomposes into a
direct sum of irreducible $sl_2$ representations
\begin{equation}
L(\Lambda ;n) \simeq \bigoplus_{j \in \frac{1}{2} \bf N} n_j(
\Lambda;i) \cdot
\underline{2j+1}_2
\end{equation}
where $\underline{2j+1}_2$ denotes the $(2j+1)$-dimensional irreducible
representation of $sl_2$ and $n_j(\Lambda ;i)$
are the multiplicities of these
representations in the decomposition.
The set of numbers $\{n_j(\Lambda;i)\}_j$ is called the
'branching rule' of the representation $L(\Lambda ;n)$
under the $sl_2$ embedding
$i$. In what follows $L(\Lambda ;n)$ will always either be the
($n-$dimensional) fundamental
or the ($n^2-1$ dimensional)
adjoint representation of $sl_n$ which we simply denote by
$\underline{n}_n$ and
$\underline{ad}_n$ respectively.

Two $sl_2$ embeddings $i$ and $i'$ are called equivalent if there
exists an inner automorphism $\psi $ of $g$ such that
\begin{equation}
i'=\psi \circ i
\end{equation}
It is easy to show that if $i \sim i'$ then $n_j(\Lambda;i)=
n_j(\Lambda;i')$
for all representations $L(\Lambda ;n)$ and $j
\in \frac{1}{2} \bf N$. The converse is
not so clear. However, for $g=sl_n$ it can
be shown  that \cite{Dynkin}
\begin{enumerate}
\item The inequivalent $sl_2$ embeddings into $sl_n$ are completely
characterized by the branching rule of the fundamental representation,
i.e. two $sl_2$ embeddings are equivalent if and only if the
branching rules of the fundamental
representation w.r.t. these two embeddings are equal.
\item There exist $P(n)$ inequivalent $sl_2$ embeddings into $sl_n$,
where $P(n)$ is the number of partitions of the number $n$.
\end{enumerate}
The set of equivalence classes of $sl_2$ embeddings
into $sl_n$ is therefore
parametrized by the branching rules of the fundamental representation,
while the different branching rules that are possible are simply given
by the partitions of $n$.\\[2mm]

{\it Example:
Let $g=sl_4$. There are 5 equivalence classes of $sl_2$ embeddings
into $sl_4$ characterized by the following decompositions
\begin{enumerate}
\item $\underline{4}_4 \simeq \underline{4}_2$
\item $\underline{4}_4 \simeq \underline{3}_2 \oplus \underline{1}_2$
\item $\underline{4}_4 \simeq 2 \cdot \underline{2}_2$
\item $\underline{4}_4 \simeq \underline{2}_2 \oplus 2 \cdot \underline{1}_2$
\item $\underline{4}_4 \simeq 4 \cdot \underline{1}_2$
\end{enumerate}
of the fundamental representation $\underline{4}_4$ of $sl_4$.
}\\[2mm]

The embeddings that are characterized
by the fact that the fundamental representation becomes an irreducible
representation of the embedded algebra
are called `principal embeddings'. If the fundamental representation
branches into singlets only the embedding is called trivial.

In what follows  we shall also need the centralizer $C(i)$ of
$i(sl_2)$ in $g=sl_n$. It corresponds to the singlets of
the adjoint action of $i(sl_2)$ in $g$.
In order to
describe $C(i)$ we have to take a look at the branching rule of the
fundamental representation
\begin{equation}
\underline{n}_n \simeq \bigoplus_{j \in \frac{1}{2}
\bf N} n_j
\cdot \underline{2j+1}_2
\end{equation}
where for convenience we simply denote $n_j \equiv n_j(\Lambda;i)$
for the fundamental representation.
Let
$q$ be the number of different values for $j$ appearing in this
decomposition for which $n_j$ is non-zero, then $C(i)$ is given by
\begin{equation}
C(i)\simeq \bigoplus_{j \in \frac{1}{2} \bf N}sl_{n_j} \; \bigoplus
(q-1)u(1)
\end{equation}
so in essence it is a direct sum of $sl_{n_j}$ algebras.

{}From the branching of the fundamental representation
one can deduce
the branching of the $(n^2-1)$-dimensional adjoint representation
$\underline{ad}_n$ of $g=sl_n$. It reads
\begin{eqnarray}
\underline{ad}_n \oplus \underline{1}_n
& \simeq & \bigoplus_{j\in \bf N} n_j^2 \cdot
(\underline{1}_2
\oplus \underline{3}_2 \oplus \ldots \oplus
\underline{4j+1}_2) \nonumber \\
& & \bigoplus_{j \neq j'} n_jn_{j'}\cdot
(\underline{2|j-j'|+1}_2 \oplus \ldots
\oplus \underline{2|j+j'|+1}_2)
\end{eqnarray}
where the singlet on the left hand side was added  to indicate
that in order to obtain the branching rule of the adjoint representation
$\underline{ad}_n$ we still have to 'subtract' a singlet from the
righthand side.

Let $\{t_0,t_+,t_-\}$ be the standard
generators of $i(sl_2)$. The Cartan element $t_0$, called the
defining vector of the embedding, can always be chosen to be
an element of the Cartan subalgebra of $g$ \cite{Dynkin}. Therefore
it defines a $\frac{1}{2}\bf Z$  gradation of $g$
given by
\be
g=\bigoplus_{m \in \frac{1}{2} \bf Z}g^{(m)}\;\; ; \;\;\;\;\
\ee
where $g^{(m)}=\{x \in g \mid [t_0,x]=mx\}$. We can choose a basis
\be
\{t_{j,m}^{(\mu )}\}_{j \in \frac{1}{2}{\bf N}; \;
-j \leq m \leq j; \; 1 \leq \mu \leq n^{ad}_j}
\ee
for $g=sl_n$ (where $n_j^{ad}$ is the multiplicity of the spin $j$
representation of $sl_2$ in the branching of the adjoint representation)
such that
\begin{eqnarray}
[t_3,t_{j,m}^{(\mu )}] & = & m t_{j,m}^{(\mu )} \nonumber \\ {}
[t_{\pm},t_{j,m}^{(\mu )}] & = & c(j,m)t_{j,m\pm 1}^{(\mu )}
\end{eqnarray}
where $c(j,m)$ are standard normalization factors.
We will always take the labeling to be such that $t^{(1)}_{1,\pm 1}\equiv
t_{\pm}$ and $t_{1,0}^{(1)} \equiv t_0$.
It is clear that
\begin{equation}
g^{(m)}= \bigoplus_{j,\mu} {\ce} t_{j,m}^{(\mu)}
\end{equation}
Note that the spaces $g^{(m)}$ and $g^{(-m)}$ are always of the same
dimension. We have the following
\begin{lemma}
The spaces $g^{(m)}$ and $g^{(n)}$ are orthogonal w.r.t. the
Cartan-Killing form on $g$, i.e.
\begin{equation}
(g^{(m)},g^{(n)})=0
\end{equation}
iff $m \neq -n$.
\end{lemma}
Proof: Let $x \in g^{(m)}$ and $y \in g^{(n)}$, then
obviously
\be
([t_0,x],y)=m(x,y)
\ee
but also
\be
([t_0,x],y)=-(x,[t_0,y])=-n(x,y)
\ee
where we used the invariance property of the Cartan
Killing form. It follows immediately that
$(n+m)(x,y)=0$. Therefore if $n\neq -m$ we must have $(x,y)=0$. This
proves the lemma \vspace{7 mm}.

For notational convenience we shall sometimes denote the basis elements
$t_{j,m}^{(\mu)}$ simply by $t_a$ where $a$ is now the multi-index
$a\equiv (j,m;\mu )$. Let $K_{ab}$ denote the matrix components of
the Cartan-Killing form in this basis, i.e. $K_{ab}=(t_a, t_b)$
and let $K^{ab}$ denote its matrix inverse.

{}From the above lemma and the fact that the Cartan-Killing form is
non-degenerate on $g$ follows immediately that $g^{(k)}$ and $g^{(-k)}$
are  non-degenerately paired. This implies that if $t_a \in g^{(k)}$
then $K^{ab}t_b \in g^{(-k)}$ (where we used summation convention).

We then have the
following lemma which we shall need later.
\begin{lemma}
If $t_a$ is a highest weight vector (i.e. $a=(j,j;\mu )$ for some
$j$ and $\mu$) then $K^{ab}t_b$ is a lowest weight vector (and
vica versa). In particular if $t_a \in C(i)$ then $K^{ab}t_b \in C(i)$.
\end{lemma}
Proof: Since $t_a$ is  a highest weight vector we have
\begin{eqnarray}
0 & = & ([t_a,t_+],x) \nonumber \\
& = & (t_a,[t_+,x])
\end{eqnarray}
for all $x \in g$ which means that $t_a$ is orthogonal to
$\mbox{Im}(ad_{t_+})$ or put differently
\begin{equation}
\mbox{Ker}(ad_{t_+}) \perp \mbox{Im}(ad_{t_+})
\end{equation}
It is easy to do the same thing for $t_-$. One therefore has the
following decomposition of $g$ into mutually orthogonal spaces
\begin{equation}
g=\left( \mbox{Ker}(ad_{t_+}) + \mbox{Ker}(ad_{t_-}) \right) \oplus
\left( \mbox{Im}(ad_{t_+}) \cap \mbox{Im}(ad_{t_-}) \right) \nonumber
\end{equation}
where
\begin{equation}
\left( \mbox{Ker}(ad_{t_+}) + \mbox{Ker}(ad_{t_-}) \right) \perp
\left( \mbox{Im}(ad_{t_+}) \cap \mbox{Im}(ad_{t_-}) \right)
\end{equation}
Also one has the following decomposition
\begin{equation}
\mbox{Ker}(ad_{t_+})+\mbox{Ker}(ad_{t_-})=
C(i) \bigoplus_{k>0}\mbox{Ker}(ad_{t_+})
^{(k)} \bigoplus_{k>0} \mbox{Ker}(ad_{t_-})^{(-k)}
\end{equation}
As we saw above $K^{ab}t_b \in g^{(-k)}$ iff $t_a \in g^{(k)}$. Therefore
\begin{eqnarray}
\mbox{Ker}(ad_{t_{\pm}})^{(\pm k)} & \perp &
\mbox{Ker}(ad_{t_{\pm}})^{(\pm l)} \;\;\;\; \mbox{for
all }k,l>0 \nonumber \\
\mbox{Ker}(ad_{t_+})^{(k)} & \perp & \mbox{Ker}(ad_{t_-})^{(-l)} \;\;\;\;
\mbox{for } k \neq l \nonumber \\
\mbox{Ker}(ad_{t_+})^{(k)} & \perp & C(i) \;\;\;\;\;\;\; \mbox{for all }k>0
\nonumber \\
\mbox{Ker}(ad_{t_-})^{(-k)} & \perp & C(i)\;\;\;\;\;\;\; \mbox{for all }k>0
\nonumber
\end{eqnarray}
from which the lemma follows \vspace{7 mm}.

Note that from the proof of this lemma follows that the spaces
$\mbox{Ker}(ad_{t_+})^{(k)}$ and $\mbox{Ker}(ad_{t_-})^{(-k)}$
are nondegenerately
paired. The same is true for the centralizer $C(i)$ with itself.
We shall use these results in the next section when we start
reduction of the Kirillov Poisson structure.

\subsection{Reductions associated to $sl_2$ embeddings}
We are now set up to describe the Poisson reductions that lead to finite
$W$ algebras. The procedure is to impose certain constraints, or in
other words to restrict oneself to a certain submanifold of a simple
Lie algebra g($=M$), then to construct the reduced phase space $\bar{M}$
and finally to calculate the Poisson structure on this reduced phase space.
This Poisson structure then corresponds to a finite $W$ algebra.
The different sets of constraints that lead to consistent reductions
are determined by the $sl_2$ embeddings. Let us now describe how precisely
this works.

Let there be given a certain $sl_2$ subalgebra $\{t_0,t_+,t_-\}$
of a simple Lie algebra $g$.
Under the adjoint action of this $sl_2$ subalgebra
$g$ branches into a direct sum of irreducible $sl_2$ multiplets.
Let $\{t_{j,m}^{(\mu )}\}$ be the basis of $g$ introduced in the
previous section. Associated with this basis is a set of $C^{\infty}$
functions $\{ J^{j,m}_{(\mu)}\}$ on $g$ (the dual basis) with the property
$J^{j,m}_{(\mu )}(t^{(\mu ' )}_{j',m'})=\delta^{\mu '}_{\mu}
\delta^j_{j'} \delta^m_{m'}$. These can be called the (global)
coordinate
functions on $g$ in the basis $\{t_{j,m}^{(\mu )}\}$
because they associate
to an element $\alpha \in g$ its $(j,m,\mu )$ component.

Let's now impose the following set of constraints
\be
\{\phi_{(\mu )}^{j,m}\equiv J^{j,m}_{(\mu )}-\delta_1^j\delta_1^m
\delta_{\mu}^1\}_{j \in \frac{1}{2}{\bf N}\; ; \; m>0} \label{constraints}
\ee
(remember that $t_{1,\pm 1}^{(1)} \equiv
t_{\pm}; \; t_{1,0}^{(1)} \equiv t_0$).
Denote the 'zero set'
in $g$ of these constraints by $g_c$.
Its elements have the form
\begin{equation}
\alpha = t_+ + \sum_j \sum_{m\leq 0}\sum_{\mu} \alpha_{(\mu )}^{j,m}
t^{(\mu )}_{j,m}
\end{equation}
where $\alpha^{j,m}_{(\mu )}$ are real or complex numbers (depending
on which case we want to consider).

As we shall see in the next chapter the
constraints postulated above are motivated in the infinite
dimensional case by the requirement that the Poisson algebra which
we obtain after reduction must be a $W$ algebra \cite{BFFOW,BTV}.
In principle, from a mathematical point of view, one could consider
more general sets of constraints, however these will, if consistent,
in general not lead to $W$ algebras. This is the reason why we define
finite $W$ algebras only to be the reductions associated to
$sl_2$ embeddings. In principle other reductions could also be handled
however by the formalism that we will develop in this thesis.

As discussed in the previous chapter
we need to find out which of the constraints
(\ref{constraints})
are first class for these will generate gauge invariances on $g_c$. This
is the subject of the next lemma.
\begin{lemma}
The constraints $\{\phi_{(\mu )}^{j,m}\}_{m \geq 1}$ are first class.
\end{lemma}
Proof: First we show that
\begin{equation}
\{J^{j,m}_{(\mu )},J^{j',m'}_{(\mu ')}\}=\sum_{j'',\mu ''}
\alpha_{j''}^{\mu ''}J^{j'',m+m'}_{(\mu '')} \label{twee}
\end{equation}
for some coefficients $\alpha_{j}^{(\mu )}$. Let $x \in g^{(m'')}$,
then
\begin{equation}
\{J^{j,m}_{(\mu )},J^{j',m'}_{(\mu ')}\}(x)=
(x,[\mbox{grad}_xJ^{j,m}_{(\mu )},\mbox{grad}_xJ^{j',m'}_{(\mu ')}])
\label{een}
\end{equation}
Now let $y$ be an element of $g^{(k)}$, then
\begin{equation}
(y,\mbox{grad}_xJ^{j,m}_{(\mu )})=\frac{d}{d \epsilon}J^{j,m}_{(\mu )}
(x+\epsilon y)|_{\epsilon = 0}
\end{equation}
which is zero except when $k=m$ (see definition of $J^{j,m}_{(\mu )}$).
{}From this and the lemmas of the
previous section one concludes that $\mbox{grad}_xJ^{j,m}_{(\mu )}
\in g^{(-m)}$ and that the Poisson bracket (\ref{een}) is
nonzero if and only if $m''=m+m'$. From this eq.(\ref{twee}) follows.

Consider now the Poisson bracket between two constraints in the
set $\{\phi^{j,m}_{(\mu )}\}_{m \geq 1}$
\begin{eqnarray}
\{ \phi^{j,m}_{(\mu )},\phi^{j',m'}_{(\mu ')} \} & = &
\{J^{j,m}_{(\mu )},J^{j',m'}_{(\mu ')} \} \nonumber \\
& = & \sum_{j'',\mu ''}\alpha_{j'',\mu ''}\,J^{j'',m+m'}_{(\mu '')}
\nonumber \\
& = & \sum_{j'',\mu ''}\alpha_{j'',\mu ''}\,\phi^{j'',m+m'}_{(\mu '')}
\nonumber
\end{eqnarray}
which is obviously equal to zero on $g_c$
Note that the fact $m,m' \geq 1$ was used in the
last equality sign. This proves the lemma \vspace{7 mm}.

Note that in general the set $\{\phi_{(\mu )}^{j,m}\}_{m\geq 1}$
is not equal to the total set of constraints since the constraints
with $m=\frac{1}{2}$ are not included. These constraints will turn
out to be second class.

Let us now determine the group
of gauge transformations on $g_c$ generated by
the first class constraints. Again we use the multi-index notation
where now Roman letters $a,b,\ldots$ run over
all $j,m$ and $\mu$, and Greek letters $\alpha,\beta\ldots$
over all $j,\mu$ but only $m>0$ (those are the indices
associated with the constraints). Let $\phi^{\alpha}$
be one of the first class
constraints (i.e. $\alpha \equiv (j,m;\mu )$
with $m\geq 1$), then the gauge transformations associated to it are
generated by its Hamiltonian vectorfields
\begin{equation}
X^{\alpha}\equiv \{\phi^{\alpha},\cdot\}
\end{equation}
Let $x=x^at_a \in g$, then the change of $x$ under a gauge transformation
generated by $\phi^{\alpha}$ is given by
\begin{eqnarray}
\delta_{\alpha}x & = & \epsilon \{\phi^{\alpha},
J^{a}\}(x)t_a \nonumber \\
& = & \epsilon \{J^{\alpha},J^a\}(x)t_a \nonumber \\
& = & \epsilon g^{\alpha a}f^c_{ab}x^bt_c \nonumber \\
& = & [\epsilon g^{\alpha a}t_a,x]  \label{ga}
\end{eqnarray}
Since $g^{\alpha a} t_a \in g^{(-k)}$ iff $t_{\alpha}
\in g^{(k)}$ (see lemmas in the
previous section) we find that the Lie algebra of the group of
gauge transformations
is given by
\begin{equation}
h=\bigoplus_{k\geq 1}g^{(-k)}
\end{equation}
This is obviously a nilpotent Lie subalgebra
of $g$ and can be exponentiated to a
group $H$. This is the gauge group generated by the first class
constraints. Note that from eq.(\ref{ga}) follows that $H$ acts on
$g_c$ in the adjoint representation, i.e the gauge orbit of a point
$x \in g_c$ is given by
\begin{equation}
{\cal O} = \{ g x g^{-1} \mid g \in H \} \nonumber
\end{equation}

Now that we have identified the gauge group we can come to the matter
of constructing the space $g_c/H$, or equivalently, gauge fixing.
Of course this can be done in many ways, however in \cite{BFFOW,BTV}
it was argued that there are certain gauges which are the most convenient
from the conformal field theoretic point of view. These are
the so called 'lowest weight gauges'. Define the subset
$g_{fix}$ of $g_c$ as follows
\begin{equation}
g_{fix}=\{t_++\sum_{j,\mu} x^j_{(\mu )} t^{(\mu )}_{j,-j} \mid
x^j_{(\mu )}\in {\ce}\}
\end{equation}
We then have the following
\begin{theorem}
\begin{equation}
H \times g_{fix} \simeq g_c
\end{equation}
\end{theorem}
Proof: Note first the obvious fact that $g^{(-l)}=g^{(-l)}_{lw} \oplus
g_0^{(-l)}$, where $g_0^{(-l+1)}=[t_+,g^{(-l)}]$ and $[t_-,g^{(-l)}_{lw}]
=0$. Let now $x \in g_c$ which means that $x=t_++ x^{(0)} + \ldots
+x^{(-p)}$ where $x^{(-k)} \in g^{(-k)}$ and $p \in \frac{1}{2} \bf N$
is the largest $j$ value in the decomposition of the adjoint
representation w.r.t. the embedding $i$.
Of course each $x^{(-k)}$ can
be written as a sum $x^{(-k)}_0 +x_{lw}^{(-k)}$ where $x^{(-k)} \in
g^{(-k)}_0$ and $x^{(-k)} \in g_{lw}^{(-k)}$.
Let also $\alpha^{(-k)}$ be an element of grade $-k$ in $h$, i.e.
$\alpha^{(-k)} \in g^{(-k)} \subset h$. Then
\begin{eqnarray}
\left( e^{\alpha^{(-k)}}x e^{-\alpha^{(-k)}} \right) ^{(-k+1)}
& = & (x^{(-k+1)}_0+x^{(-k+1)}_{lw})-ad_{t_+}(\alpha^{(-k)}) \nonumber \\
\left( e^{\alpha^{(-k)}}x e^{-\alpha^{(-k)}} \right) ^{(-k+i)}
& = & x^{(-k+i)} \;\;\;\;\;\; \mbox{for} \;\; 1 <i \leq k \nonumber
\end{eqnarray}
i.e. only the elements $x^{(-k+1)}, \ldots ,x^{(-p)}$ are changed by this
gauge transformation. Now, since $ad_{t_+}:g^{(-k)} \rightarrow
g^{(-k+1)}_0$ is bijective by definition, there exists a unique
$\alpha^{(-k)} \in g^{(-k)}$ such that $ad_{t_+}(\alpha^{(-k)})=
x_0^{(-k+1)}$, i.e. the element $x^{(-k+1)}_0$ can be gauged away by
choosing $\alpha^{(-k)}$ appropriately.
{}From this follows immediately that there exist unique
elements $\alpha^{(-l)}\in g^{(-l)}$ ($l=1, \ldots ,p$) such that
\begin{equation}
e^{\alpha^{(-p)}}\ldots e^{\alpha^{(-1)}}x \,e^{-\alpha^{(-1)}} \ldots
e^{-\alpha^{(-p)}}= y \;\;
\in g_{fix}
\end{equation}
This provides, as
one can now easily see, a
bijective map between $H \times g_{fix}$ and $g_c$ given by
\begin{equation}
\left( e^{-\alpha^{(-1)}} \ldots e^{-\alpha^{(-p)}}, y \right)
\rightarrow x
\end{equation}
This proves the theorem \vspace{7 mm}.

So starting from $g$, after imposing constraints and fixing gauge
invariances we have arrived at a submanifold $g_{fix}$ of $g$.
We now want to determine the Poisson algebra structure of $C^{\infty}
(g_{fix})$. For this we need to calculate the Dirac brackets
\begin{equation}
\{J^{j,-j}_{(\mu )},J^{j',-j'}_{(\mu ')}\}^*
\end{equation}
between the generators $\{J^{j,-j}_{(\mu )}\}$ of $C^{\infty}(g_{fix})$.
of course one could do this directly by using the formula for the
Dirac bracket, but this is rather cumbersome. We shall therefore use
another method which in the end will lead to a very elegant and general
formula for the Poisson structure on $C^{\infty}(g_{fix})$. Unfortunately
for this it will be necessary to go through some rather technical
lemmas and theorems that lay the mathematical groundwork for this
method. The reader may wish to skip right to formula (\ref{PA})
on first reading.

Consider the following general situation:
Let $\{t_i\}_{i=1}^{dim(g)}$ be a basis of the Lie algebra $g$,
$k$ be a positive integer smaller or equal to $\mbox{dim}(g)$ and
$\Lambda \equiv t_{k+1}$. Consider then the following subset
of $g$
\begin{equation}
g_f=\{\Lambda +\sum_{i=1}^{k}\alpha^i t_i \mid \alpha^i \in {\ce} \}
\end{equation}
which can be seen as the zero set of the constraints $\phi^1=J^{k+1}-1$
and $\phi^i=J^{k+i}$ for $1<i\leq \mbox{dim}(g)-k$.
Also suppose that the Kirillov
bracket on $g$ induces a Dirac bracket
$\{.,.\}^*$ on $g_f$ (i.e. all constraints
are second class).

Denote by $\cal R$ the set
of smooth functions
\begin{equation}
R:{\ce}^k \times g_f \longrightarrow g
\end{equation}
of the form
\begin{equation}
R(\vec{z};y)=\sum_{i=1}^{k} z_iR^i(y)
\end{equation}
where $\vec{z}\equiv \{z_i\}_{i=1}^k \in {\ce}^k$, $y \in g_f$ and
$R^i(y) \in g$. To any element $R \in {\cal R}$ one can associate
a map
\begin{equation}
Q_R : {\ce}^k \times g_f \longrightarrow C^{\infty}(g)
\end{equation}
defined by
\begin{equation}
Q_R(\vec{z};y)=\sum_{i=1}^{dim(g)} (R(\vec{z};y),t_i)J^i
\end{equation}
(where as before $J^i \in C^{\infty}(g)$ is defined by $J^i(t_j)=
\delta_j^i$). We are going to use certain elements of the set $\cal R$
in order to explicitly calculate the Dirac brackets on $g_f$. We have
the following theorem.
\begin{theorem}
If there exists a unique $R \in \cal R$ such that for all $\vec{z} \in
{\ce}^k$ and $y \in g_f$ we have
\be
\Lambda +[R(\vec{z};y),y] \;\;\;\; \in g_{f} \label{r1}
\ee
and
\be
Q_R(\vec{z};y)|_{g_{f}}=\sum_{i=1}^k z_iJ^i
+\left( R(\vec{z};y),\Lambda \right)   \label{r2}
\ee
(i.e. the restriction of $Q_R(\vec{z};y)$ to $g_f$ is equal to the
right hand side of (\ref{r2})) then
\begin{equation}
\sum_{ij=1}^k z_i\; \{J^i,J^j\}^*(y)
\; t_j=[R(\vec{z};y),y] \label{trick}
\end{equation}
for all $\vec{z} \in {\ce}^{k}$ and $y \in g_f$.
\end{theorem}
Note that from eq.(\ref{trick}) one can read off {\em all} Dirac brackets
between the generators $\{J^i\}_{i=1}^k$ of $C^{\infty}(g_f)$ since
the formula holds for all $\vec{z} \in {\ce}^k$ and the elements
$t_j$ are all independent. From this follows that calculating the whole
set of Dirac brackets has been brought down to determining the map $R$.
This is, in practical situations, easily done using eqs.(\ref{r1}) and
(\ref{r2}).

Also note that from equation (\ref{r2})
follows that within the Dirac bracket $\{.,.\}^*$ the function
$Q_R(\vec{z};y)$ is equal to $\sum_{i=1}^kz_iJ^i$, i.e.
\be
\{Q_R(\vec{z};y),\cdot \}^* =\sum_{i=1}^{k} z_i\{J^i,\cdot \}^*
\ee
since constants commute with everything and restriction to $g_f$
is always implied within the Dirac bracket. This explains the origin
of the formula (\ref{r2}).

The proof of the theorem is based on the following two lemmas.
\begin{lemma}
Let $M$ be a Poisson manifold, $Q \in C^{\infty}(M)$ and $X_Q$
its Hamiltonian vectorfield, i.e. $X_Q=\{Q,.\}$. Let there also be
given a set of second class constraints and let $Z$ be the submanifold
where they all vanish. If for some $p \in M$ we have $X_Q|_p(\phi_i)=0$
for all $i$ (i.e. $X_Q|_p$ is tangent to $Z$), then
\begin{equation}
\{Q,F\}(p)=\{\bar{Q},\bar{F}\}^*(p) \;\; \;\; \mbox{for all} \;\;F \in
C^{\infty}(M)
\end{equation}
where $\{.,.\}^*$ is the Dirac bracket.
\end{lemma}
Proof: Simply calculating the Dirac bracket in $p$ gives
\begin{eqnarray}
\{\bar{Q},\bar{F}\}^*(p) & \equiv & \{Q,F\}(p)-\{Q,\phi_i\}(p)
\Delta^{ij}(p)
\{\phi_j,F\}(p)  \nonumber \\
& = & \{Q,F\}(p) \nonumber
\end{eqnarray}
because by hypothesis $\{Q,\phi_i\}(p)=0$ for all $i$.
This proves the lemma \vspace{4 mm}.

The origin of equation (\ref{r1}) in the theorem becomes clear
in the following lemma.
\begin{lemma}
If for all $y \in g_f$ and $\vec{z} \in {\ce}^k$
\be
\Lambda +[R(\vec{z};y),y] \;\;\;\; \in g_f
\ee
then
\be
\{Q_R(\vec{z};y),\phi^i\}(y)=0
\ee
for all $1 \leq i \leq p-k;$ $y \in g_f$ and $\vec{z} \in {\ce}^k$.
\end{lemma}
Proof: By definition of $J^i$ we have $y=t_iJ^i(y)$ which
gives us
\begin{eqnarray}
\Lambda +[R(\vec{z};y),y] & = & \Lambda +
\sum_{ij=1}^{dim(g)} [R(\vec{z};y)^it_i,t_jJ^j](y) \nonumber \\
& = & \Lambda +
\sum_{ijlm=1}^{dim(g)} K_{ij}f^{jl}_{m}R
(\vec{z};y)^iJ^m(y)t_l \nonumber \\
& = & \Lambda +\sum_{ijl=1}^{dim(g)}\{K_{ij}R(\vec{z};y)^iJ^j,J^l\}(y)t_l
\nonumber \\
& \equiv & \Lambda +\sum_{i=1}^{dim(g)} \{Q_R(\vec{z};y),J^i\}(y)t_i
\nonumber
\end{eqnarray}
but by hypothesis this was an element of $g_{f}$. Therefore
$\{Q_R(\vec{z};y),J^i\}(y)=0$ for $i \geq k+1$.
This proves the lemma \vspace{4 mm}.

Putting together these two lemmas we find

Proof of theorem: From the previous lemmas follows that if eqns.
(\ref{r1}) and (\ref{r2})   are satisfied, then
\begin{eqnarray}
\sum_{ij=1}^{k}z_i \{J^i,J^j\}^*(y)t_j
& = & \sum_{i=1}^k \{Q_R(\vec{z};y),J^i\}^*(y)t_i
\nonumber \\
& = & \sum_{i=1}^k \{Q_R(\vec{z};y),J^i\}(y)t_i \nonumber \\
& = & \sum_{i=1}^{dim(g)}\{Q_R(\vec{z};y),J^i\}(y)t_i \nonumber \\
& = & [R(\vec{z};y),y]
\end{eqnarray}
where eqn.(\ref{r2}) was used in the first step.
This proves the theorem \vspace{7 mm}.

Note that conversely in order to show the
existence of the Dirac bracket on $g_{f}$
it is sufficient to prove that equations (\ref{r1}) and (\ref{r2}) of the
above theorem are solvable within $\cal R$.
We will now show that
this is the case when $g_f \equiv g_{fix}$ associated to an arbitrary
$sl_2$ embedding.
\begin{theorem}
Let $|i|$ be the total number of $sl_2$ multiplets in the
branching of the adjoint representation of $g$. There is a unique
$R \in {\cal R}$ such that
\be
 t_++[R(\vec{z};y),y] \;\;\;\; \in g_{fix}
\ee
and
\be
Q_R(\vec{z};y)|_{g_{fix}}= \sum_{j;\mu}
z^{(\mu )}_j J^{j,-j}_{(\mu)}+\left( R(\vec{z};y),t_+ \right)
\ee
for all $\vec{z} \in {\ce}$ and $y \in g_{fix}$.
\end{theorem}
Proof: First solve the second equation.
\begin{eqnarray}
Q_R(\vec{z};y)(y) & = & \left( R^{(p)}+ \ldots +R^{(-p)},
t_++y^{(0)}+\ldots
+y^{(-p)} \right) \nonumber \\
& = & (R^{(-1)},t_+)+(R^{(p)},y^{(-p)}) +\ldots +(R^{(0)},
y^{(0)}) \nonumber
\end{eqnarray}
Now, $y^{(-j)}\in g^{(-j)}_{lw}$ which means that $y^{(-j)}=\sum_{\mu }
y^{j,-j}_{(\mu )}t_{j,-j}^{(\mu )}$. Also $R^{(j)}=R^{(j)}_0+R^{(j)}
_{hw}$ where $R^{(j)}_0 \in \mbox{Im}(ad_{t_+})$ and $R^{(j)}_{hw} \in
\mbox{Ker}(ad_{t_+})$. Let $\{t^{j,-j}_{(\mu )}\}_{\mu } \subset g^{(j)}$
be such that $(t_{j,-j}^{(\mu )},t^{j',-j'}_{(\mu ')})=\delta^{j'}_j
\delta^{\mu }_{\mu '}$ (remember that $g^{(j)}$ and $g^{(-j)}$ are
non-degenerately paired). Then take
\begin{equation}
R_{hw}^{(j)}=\sum_{\mu} z_j^{(\mu)}t_{(\mu)}^{j,-j}
\end{equation}
Obviously the second equation of the theorem is then satisfied.

Let's now solve the first equation. For $k>0$ it reads
$[R,y]^{(k)}=0$. Writing this out gives
\begin{equation}
[R^{(k-1)}_0,t_+]=[y^{(0)},R^{(k)}]+ \ldots + [y^{(k-p)},R^{(p)}]
\label{342}
\end{equation}
Remember that $ad_{t_+}:g^{(k-1)}_0 \rightarrow g^{(k)}$ is bijective
which means that if we already know  $R^{(k)}, \ldots ,R^{(p)}$ then we
can uniquely solve this equation for $R_0^{(k-1)}$ (remember that
for $l\geq 0$ we have already determined the quantities $R_{hw}^{(l)}$).
The initial term of this sequence of equations is
obviously $R^{(-p)}=R^{(-p)}_{
hw}$ (i.e. $R^{(-p)}_0=0$). In this way we can completely determine
the positive and zero grade piece of $R(\vec{z};y)$.

For $-k \geq 0$ the situation is similar but slightly different. What
we now have to demand is that $[R,y]^{(-k)}_0=0$. This leads
to the equation
\begin{equation}
[t_+,R^{(-k-1)}]=[R^{(-k)},y^{(0)}]_0+ \ldots + [R^{(-k+p)},y^{(-p)}]_0
\label{343}
\end{equation}
where we used that $[t_+,R^{(-k-1)}]$ is already an element of
$g^{(-k)}_0$. Remember that $ad_{t_+}:g^{(-k-1)} \rightarrow
g_0^{(-k)}$ is bijective so there exists a unique $R^{(-k-1)}(z)$
such that the equation is solved, assuming we know $R^{(-k)},\ldots ,
R^{(-k+p)}$ already. In this way we can grade  by grade determine the
negative grade piece of the matrix $R$.
It is also clear
from the above construction that
$R \in \cal R$. This proves the theorem \vspace{7 mm}.

{}From theorem 2 therefore follows that
\begin{equation}
\sum_{j,j';\mu ,\mu '}z_j^{(\mu )} \;
\{J^{j,-j}_{(\mu )},J^{j',-j'}_{(\mu ')}
\}^*(y)\; t_{j',-j'}^{(\mu ')}=[R(\vec{z};y),y]  \label{redbrac}
\end{equation}
from which we can derive all relations.

It is possible to derive a general formula for $R(\vec{z};y)$ using
arguments similar to those used in \cite{BG}. Let again
\begin{equation}
g_{lw}= \bigoplus_{j \in \frac{1}{2} \bf N} g^{(-j)}_{lw}=\mbox{span}(
\{t^{(\mu )}_{j,-j}\}_{j;\mu }
\end{equation}
and let $\Pi$ be the orthogonal (w.r.t. the Cartan-Killing
form) projection onto $\mbox{Im}(ad_{t_+})$. Obviously the map
$ad_{t_+}: \mbox{Im}(ad_{t_-}) \rightarrow \mbox{Im}(ad_{t_+})$ is
invertible. Denote the inverse  of this map, extended by $0$ to
$g$ by $L$. As before what we want to do is solve the equation
\begin{equation}
[R,y]=x \; \in g_{lw}
\end{equation}
for $y \in g_{fix}$. Noting that $y=t_++w$ where $w \in g_{lw}$
and applying $\Pi$ this equation reduces to
\begin{equation}
\Pi \circ ad_{t_+}R(\epsilon )=\epsilon \Pi ([R(\epsilon ),w])
\label{master}
\end{equation}
where we introduced a parameter $\epsilon$ in the right hand side
which we want to put to 1 later. Note that the left hand side
is equal to $ad_{t_+}R(\epsilon )$ since this is already an element of
$\mbox{Im}(ad_{t_+})$. Assume now that $R(\epsilon )$
can be (perturbatively)
written as
\begin{equation}
R(\epsilon )=\sum_{k=0}^{\infty} R_k\epsilon^k
\end{equation}
(we shall have to justify this later). Consider now the zeroth order
part of the equation (\ref{master}). It reads
\begin{equation}
ad_{t_+}R_{0}=0
\end{equation}
This means that $R_0\equiv F(\vec{z})$
which is an arbitrary element of $\mbox{ker}(ad_{t_+})$.
The first order equation is equal to
\begin{equation}
ad_{t_+}R_1=\Pi ([F,w])
\end{equation}
Obviously this equation is solved by
\begin{equation}
R_1=-L\circ ad_wF
\end{equation}
Proceeding with higher orders we find
\begin{equation}
R_{k+1}(\vec{z};y)=-L\circ ad_w \left( R_k(\vec{z};y) \right)
\end{equation}
which means that
\begin{equation}
R(\vec{z};y)=\frac{1}{1+\epsilon L\circ ad_w}F(\vec{z})
\end{equation}
There are no convergence problems with this series since the
operator $L$ lowers the degree by one which means that after $2p$
steps it must cancel. We therefore find
\begin{equation}
\sum_{j,j';\mu \mu '}z^{(\mu )}_j \; \{J^{j,-j}_{(\mu )},
J^{j',-j'}_{(\mu ')}\}^*(y)\; t_{j',-j'}^{(\mu ')}=
-ad_y \left( \frac{1}{1+L ad_w}F(\vec{z}) \right) \label{blurg}
\end{equation}
Now, let $w \in g_{lw}$ and $Q \in
C^{\infty}(g_{lw})$. We then define $\mbox{grad}_wQ \in \mbox{Ker}(
ad_{t_+}) \equiv g_{hw}$ by
\begin{equation}
\left( x,\mbox{grad}_w Q \right) \equiv \frac{d}{d\epsilon}
Q(w +\epsilon x)|_{\epsilon=0} \;\;\;\mbox{for all }x \in g_{lw}
\end{equation}
Note that this uniquely defines $\mbox{grad}_wQ$ since, as we saw
before, $g_{lw}$ and $g_{hw}$ are non-degenerately paired.
Using equation (\ref{blurg}) we can now give the following general
description of the classical
finite $W$ algebra: it is nothing but the Poisson
algebra $(C^{\infty}(g_{lw}),\{.,.\}^*)$ where
\begin{equation}
\{Q_1,Q_2\}^*(w) =\left(w,[ \mbox{grad}_wQ_1, \frac{1}{1+L
\circ ad_w}\mbox{grad}_wQ_2] \right) \label{PA}
\end{equation}
for all $Q_1,Q_2 \in C^{\infty}(g_{lw})$ and $w \in g_{lw}$.
For the so called trivial embedding ($t_0=t_{\pm}=0$)
the map $L$ is equal to the zero map and the above formula reduces
to  the ordinary Kirillov bracket as it should (because in that
case $g_{fix} \equiv g$). For nontrivial $sl_2$ embeddings however
(\ref{PA}) is a new and highly nontrivial Poisson structure.

{}From now on denote the Kirillov Poisson algebra associated to a
Lie algebra $g$ by
$K(g)$ and the finite $W$ algebra $(C^{\infty}(g_{lw}),\{.,.\}^*)$
by ${\cal W}(i)$ where
$i$ is again the $sl_2$ embedding in question.

In general ${\cal W}(i)$ is a non-linear Poisson algebra
as we shall see when we consider examples. However it can happen that
${\cal W}(i)$ contains a subalgebra that is isomorphic to a Kirillov
algebra. The next theorem deals with this.
\begin{theorem}
The algebra ${\cal W}(i)$ has a Poisson subalgebra which is
isomorphic to $K(C)$ where $C$ is the centralizer of the $sl_2$ subalgebra
of $g$.
\end{theorem}
Proof: Since all elements of the centralizer are lowest
(and highest) weight vectors w.r.t.
$i(sl_2)$ the function $J^a$ is not constrained if $t_a \in C(i)$,
i.e. all the elements $J^a$ associated to the centralizer survive the
reduction. It is not obvious however that they still form the same algebra
w.r.t. the Poisson bracket (\ref{PA}).
This is what we have to show.

The part of the equation (\ref{redbrac})
that determines the Poisson relations
between the currents associated to the centralizer is
\begin{equation}
\sum_{\mu,\mu '}z^{(\mu )}_0 \; \{J^{0,0}_{(\mu )},J^{0,0}_{(\mu ')}\}^*
(y) \; t_{0,0}^{(\mu ')}=[R(\vec{z};y),y]^{(0)}
\end{equation}
The right hand side of this equation reads in more detail
\begin{equation}
[R,y]^{(0)}=[R^{(0)}_{lw},y^{(0)}]+[R^{(0)}_0,y^{(0)}]
+ \ldots + [R^{(p)},y^{(-p)}]+
[R^{(-1)},t_+]   \label{tja}
\end{equation}
Note that $R^{(0)}_{lw}$ and $y^{(0)}$ are both elements of $C(i)$
which means that the first term in the right hand side is also
an element of $C(i)$. We will now show that $R^{(0)}_0,R^{(1)}, \ldots,
R^{(p)}$ and $R^{(-1)}$ do not depend on $\{z^{(\mu )}_0 \}_{\mu}$
which means that all but the first term in the right hand side of
equation (\ref{tja}) are irrelevant for the Poisson brackets
$\{J^{0,0}_{(\mu )},J^{0,0}_{(\mu '}\}^*$.

{}From eq.(\ref{342}) follows that $R^{(k)}_0$ is only a function of
$z^{(\mu )}_p, \ldots ,z^{(\mu )}_{k+1}$ for $k\geq 0$ and therefore
$R^{(k)}=R^{(k)}(z^{(\mu )}_p, \ldots ,z^{(\mu )}_k)$. Now, $R^{(-1)}$
is determined by equation (\ref{343}) for $k=0$
\begin{equation}
[t_+,R^{(-1)}]=[R^{(0)},y^{(0)}]_0+ \ldots +[R^{(p)},y^{(-p)}]_0
\end{equation}
Note that the terms $[R^{(l)},y^{(-l)}]$ for $l>0$
certainly do not contain $z_0^{(\mu )}$ as we just saw. Also
note that
\begin{equation}
[R^{(0)},y^{(0)}]_0=[R^{(0)}_0+R^{(0)}_{lw},y^{(0)}]_0=
[R^{(0)}_0,y^{(0)}]
\end{equation}
because of the reason we mentioned earlier that $R^{(0)}_{hw}$
and $y^{(0)}$ are both in $C(i)$. However, as we have seen above
$R^{(0)}_0$ does not contain $z^{(\mu )}_0$. From this we conclude
that the only term in $[R,y]^{(0)}$ that contains $z_0^{(\mu )}$
is $[R^{(0)}_{lw},y^{(0)}]$, i.e.
\begin{eqnarray}
\sum_{\mu \mu '}z^{(\mu )}_0 \{J^{0,0}_{(\mu )},J^{0,0}_{(\mu ')}\}
^* (y) t_{0,0}^{(\mu )} & = & [R^{(0)}_{lw},y^{(0)}] \nonumber \\
& = & [R_{lw}^{(0)}, \sum_{\mu } J^{0,0}_{(\mu )}(y)t_{0,0}^{(\mu )}]
\label{hup3}
\end{eqnarray}
The generators $\{t_{0,0}^{(\mu )}\}$ are a basis of $C(i)$.
{}From this, eq.(\ref{hup3}) and theorem 2 follows
immediately that the Poisson algebra
generated by $\{J^{0,0}_{(\mu )}\}$ w.r.t. the Dirac
bracket is isomorphic to  Kirillov
algebra of $C(i)$. This proves the theorem \vspace{7mm}.

\subsection{Generalized finite Miura transformations}
In this section we present a generalized version of the Miura
transformation. Roughly this is a Poisson homomorphism of the
finite $W$ algebra ${\cal W}(i)$
in question to a certain Kirillov algebra.
In order to be able to describe this map
for arbitrary embeddings however we first have
to concern ourselves with the cases when in the decomposition of $g$
into $sl_2$ multiplets there appear half integer grades. As we have seen, in
those cases the constraints $\phi^{j,\frac{1}{2}}_{(\mu )}$ are
second class. In what follows it will be necessary to be able
to replace the usual set of constraints by an alternative set which
contains only first class constraints but which gives rise to the
same reduction \cite{FORTW}. Roughly what one does is impose only half of the
constraints that turned out to be second class in such a way that
they become first class. The other constraints that were second class
can then be obtained as gauge fixing conditions. In this way $g_{fix}$
stays the same but $g_c$ is different. Since the resulting Poisson
algebra only depends on $g_{fix}$ it is clear that we obtain the same
algebra ${\cal W}(i)$.

Let's now make all of this more precise.
We describe the $sl_n$ algebra in the standard way by traceless
$n\times n$ matrices; $E_{ij}$ denotes the matrix with a one in
its $(i,j)$ entry and zeroes everywhere else. As we said earlier
embeddings of $sl_2$ into $sl_n$ are in
one-to-one correspondence with partitions of $n$. Let $(n_1,n_2,\ldots)$ be
a partition of $n$, with $n_1\geq n_2 \geq \ldots$. Define a different
partition $(m_1,m_2,\ldots)$ of $n$, with $m_k$ equal to the
number of $i$ for which $n_i\geq k$. Furthermore let
$s_t=\sum_{i=1}^{t} m_i$. Then we have the following
\bl
An embedding of $sl_2$ in $sl_n$ under which the fundamental
representation branches according to $n\rightarrow \oplus n_i$
is given by
\ba
t_+ & = & \sum_{l\geq 1} \sum_{k=1}^{n_l-1} E_{l+s_{k-1},l+s_k},
\nonu
t_0 & = & \sum_{l\geq 1} \sum_{k=1}^{n_l} (\frac{n_l+1}{2}-k)
E_{l+s_{k-1},l+s_{k-1}}, \nonu
t_- & = & \sum_{l\geq 1} \sum_{k=1}^{n_l-1}  k
(n_l-k)
E_{l+s_k,l+s_{k-1}}.
\label{embedding}
\ea
\el
The proof is by direct computation\footnote{The commutation
relations are $[t_0,t_{\pm}]=\pm t_{\pm}$, and $[t_+,t_-]=
2t_0$.}. If the fundamental
representation of $sl_n$ is spanned by  vectors $v_1,\ldots v_n$,
on which $sl_n$ acts via $E_{ij}(v_k)=\delta_{jk}v_i$,
then the irreducible representations $n_l$ of $sl_2$
to which it reduces are spanned by $\{v_{l+s_{k-1}}\}_{1\leq k
\leq n_l}$. Previously we decomposed the lie algebra $g$ into
eigenspaces of $ad_{t_0}$. However, for our present purposes it is
convenient to introduce a different grading of the Lie algebra
$g$, that we need
to describe the generalized Miura map and
the BRST quantization. The grading is
defined by the following element of the Cartan subalgebra of
$sl_n$:
\be
\label{def:delta}
\delta=\sum_{k\geq 1} \sum_{j=1}^{m_k} \left( \frac{\sum_l
lm_l}{\sum_l m_l} -k \right) E_{s_{k-1}+j,s_{k-1}+j}.
\ee
This leads to the alternative decomposition $g=g_-\oplus g_0 \oplus g_+$ of
$g$ into spaces with negative, zero and positive eigenvalues under
the adjoint action of $\delta$ respectively (note that in
case the grading of $ad_{t_0}$ is an integral grading then
we have $t_0=\delta$ so in those cases nothing happens. In general
however $t_0\neq \delta$ and also $g^{(m)}\neq g_m$).  The subalgebra
$g_0$ consists of matrices whose nonzero entries are in square
blocks of dimensions $m_1\times m_1$, $m_2\times m_2$, etc.
along the diagonal of the matrix and is therefore
a direct sum of $sl_{m_k}$ subalgebras (modulo $u(1)$ terms).
The nilpotent subalgebra $g_+$
is spanned by
$\{E_{l+s_{k-1},r+s_k}\}_{l\geq 1;\; 1\leq k \leq
n_l-1;\; r>0}$,
and the nilpotent subalgebra $g_-$ by the transpose
of these. Let $\pi_{\pm}$ denote the projections onto $g_{\pm}$.
Then  the following theorem describes the replacement of the mixed
system of first and second class constraints by a system of first class
constraints only.
\bt
\label{firstclass}
The constraints
$\{J^{l+s_{k-1},r+s_k}-\delta^{r,l} \}_{l\geq 1;\; 1\leq k \leq
n_l-1;\; r>0}$
are first class. The gauge group
they generate is $\hat{H}=\exp(g_-)$ acting via the adjoint
representation on $g$. The resulting finite $W$ algebra is the
same as the one obtained by imposing the constraints (\ref{constraints}).
\et
Proof: decompose $g$ in eigenvalues of $\ad{\delta}$,
$g=\oplus_k g_{k}$. Note that $\ad{\delta}$ has only
integral eigenvalues. Using the explicit form of $t_+$ in
(\ref{embedding}), one easily verifies that $[\delta,t_+]=t_+$.
Thus, $t_+\in g_{1}$. Again since $[g_+,g_+]=[g_{\geq
1},g_{\geq 1}]\subset g_{\geq 2}$ it follows
(exactly like in lemma 3)  that the constraints
$\{J^{l+s_{k-1},r+s_k}-\delta^{r,l} \}_{l\geq 1;\; 1\leq k \leq
n_l-1;\; r>0}$
are first class.
The gauge group  can be determined similarly as in section 3.2,
and the analogue of theorem 1 of section 3.2 can be proven in the
same way with the same choice of $g_{fix}$, if one uses the
decomposition of $g$ in eigenspaces of $\ad{\delta}$ rather than
$\ad{t_0}$. Therefore the resulting finite $W$ algebra is the
same, because theorem 2 and 3 of section 3.2 show that it only
depends on the form of $g_{fix}$.\\

Let us explain this theorem in words. The number of second
class constraints in any system is necessarily even. If one switches
to the $ad_{\delta}$ grading the set of second class constraints is
split into half:
one half gets grade 1 w.r.t. $ad_{\delta}$
while the other half gets grade 0. Now what one does is impose only that
half that has obtained grade 1 w.r.t. the $ad_{\delta}$ grading. These
constraints are then first class. The gauge transformations they
generate can be completely fixed by imposing the constraints that
were in the other half. Having done that we are back in exactly the
same situation as before. The only difference is that we now know of
a system of first class constraints that in the end leads to the
same reduction.

Note that the number of generators of the finite $W$ algebra is equal to
$\dim(g_0)=(\sum_i m_i^2) -1$. This is indeed the same as the
number of irreducible representations of $sl_2$, minus one, one
obtains from $(\oplus_i n_i)\otimes (\oplus_i n_i)$, as the
latter number equals $(\sum_j (2j-1)n_j)-1$, and one easily
checks that $\sum_i m_i^2=\sum_j (2j-1)n_j$.

The generalized Miura transformation can now be formulated as
\begin{theorem}
There exists an injective Poisson homomorphism from ${\cal W}(i)$ to
$K(g_0)$.
\end{theorem}
Proof: First we show that
for every element $x \in t_++g_0$ there exist a unique element
$h$ in the gauge group $\hat{H}$ such that $h\cdot x \cdot h^{-1} \in
g_{fix}$.
For this note that there exists
a unique element $h' \in \hat{H}$ such that $h'.x.h'^{-1} \in g_c$.
This follows from the previous theorem and the remarks made after it.
In fact it follows from those remarks that $h'$ is an element of
that subgroup of $\hat{H}$ that is generated by the 'ex' second
class constraints (the ones that were made into first class constraints
by not imposing the other half). It was shown earlier however that
\be
g_{fix}\times H  \simeq g_c
\ee
which means that there exists a unique element $h'' \in H$ such that
\be
h''h'.x.(h''h')^{-1} \in g_{fix}   \label{nulfix}
\ee
We conclude from this that there is a surjective map from $x \in t_++g_0$
to $g_{fix}$ given by (\ref{nulfix}).
The pull back of this map $C^{\infty}(g_{fix})\equiv {\cal W}(i)
\rightarrow C^{\infty}(t_++g_0)$ (which will therefore
be injective) is then the Miura map. Of course we still have to check
whether this map is a Poisson homomorphism. This we address next.

What is the Poisson structure on $C^{\infty}(t_++g_0)$ ?
Since the constraints
of which the space $t_++g_0$ is the zero set are obviously second
class the Kirillov bracket on $g$ induces a Dirac bracket on
it. It is not difficult to see that the Dirac term in the Dirac
bracket cancels in this case which means that the Poisson algebra
$C^{\infty}(t_++g_0)$ with the induced Poisson structure is
isomorphic to the Poisson algebra $K(g_0)$. Since the transformation
from ${\cal W}(i)$ to $K(g_0)$ corresponds to a gauge transformation
(\ref{nulfix}) the map is necessarily a homomorphism.

\subsection{Examples}
The simplest examples of finite $W$ algebras are those associated
to the so called 'principal $sl_2$ embeddings'.
These embeddings are associated to the trivial partition of the
number $n$: $n=n$.
The fundamental representation of $g=sl_n$ therefore becomes
an irreducible representation of the $sl_2$ subalgebra, i.e.
\be
\underline{n}_n \simeq \underline{n}_2
\ee
The branching rule for the adjoint representation of $g$ therefore
reads
\be
\underline{ad}_n \simeq \underline{3}_2 \oplus \underline{5}_2\oplus
\ldots \oplus \underline{2n-1}_2
\ee
{}From this follows immediately that the finite $W$ algebra will have
$n-1$ generators (since there are $n-1$ $sl_2$ multiplets). Without
going into details we can immediately predict the Poisson relations
between these generators from the generalized Miura transformation
for in this case the subalgebra $g^{(0)}=g_0$ coincides with the
Cartan subalgebra of $sl_n$. Since the Cartan subalgebra is an
abelian algebra, and since the Kirillov algebra of an abelian Lie
algebra is a Poisson commutative algebra we find that a finite $W$
algebra associated to a principal embedding
must also be Poisson commutative since it is isomorphic to a Poisson
subalgebra of
$K(g_0)$. We conclude therefore that the principal $sl_2$ embedding
into $sl_n$ leads to the abelian Poisson algebras with $(n-1)$ generators
\vspace{3mm}.

The simplest nontrivial case of a finite $W$ algebra is associated to
the (only) nonprincipal embedding of $sl_2$ into $sl_3$. This embedding
is associated to the following partition of 3: $3 \rightarrow 2+1$.
The branching rule of the fundamental representation of $sl_3$ is
therefore
\be
\underline{3}_3 \simeq \underline{2}_2 \oplus \underline{1}_2
\ee
{}From this we find the following branching rule for the adjoint
representation
\be
\underline{ad}_3 \simeq \underline{3}_2 \oplus 2.\underline{2}_2 \oplus
\underline{1}_2
\ee
from which follows immediately that the finite $W$ algebra associated
to this embedding will have 4 generators. We shall go through the
construction of this finite $W$ algebra in some detail in order to illustrate
the theory discussed above.

The explicit form of the $sl_2$ embedding is
\begin{eqnarray}
t_+ & = & E_{1,3} \nonumber \\
t_0 & = & \mbox{diag}(\frac{1}{2},0,-\frac{1}{2}) \nonumber \\
t_- & = & E_{3,1}
\end{eqnarray}
where as before $E_{ij}$ denotes the matrix with a one in its $(i,j)$
entry and zeros everywhere else. The ($sl_3$ valued) function
$J=t_{j,m}^{(\mu )}J^{j,m}_{(\mu )}$ (where we used summation convention)
reads
\be
J=
\left(
\begin{array}{ccc}
\frac{1}{2}J^{1,0}_{(1)}+J^{0,0}_{(1)} & J^{\frac{1}{2},\frac{1}{2}}_{(1)}
& J^{1,1}_{(1)} \\
J^{\frac{1}{2},-\frac{1}{2}}_{(2)} & -2J^{0,0}_{(1)} & J^{\frac{1}{2},
\frac{1}{2}}_{(2)} \\
J^{1,-1}_{(1)} & J^{\frac{1}{2},-\frac{1}{2}}_{(1)} & J^{0,0}_{(1)}
-\frac{1}{2}J^{1,0}_{(1)}
\end{array} \right)
\ee
According to the general prescription the constraints are
\be
J^{1,1}_{(1)}-1=J^{\frac{1}{2},\frac{1}{2}}_{(1)}
=J^{\frac{1}{2},\frac{1}{2}}_{(2)}=0
\ee
the first one being the only first class constraint. As was shown
earlier the gauge invariance generated by this constraint can be
completely fixed by adding the 'gauge fixing condition'
\be
J^{1,0}_{(1)}=0
\ee
The Dirac brackets between the generators $\{J^{0,0}_{(1)},
J^{\frac{1}{2},-\frac{1}{2}}_{(1)},J^{\frac{1}{2},-\frac{1}{2}}_{(2)}\}$
and $J^{1,-1}_{(1)}$ can now easily be calculated. In order to
describe the final answer in a nice form introduce
\begin{eqnarray}
C & = & -\frac{4}{3}(J^{1,-1}_{(1)} + 3 (J^{0,0}_{(1)} )^2 ) \nonumber \\
E & = & J^{\frac{1}{2},-\frac{1}{2}}_{(1)} \nonumber \\
F & = & \frac{4}{3} \;
J^{\frac{1}{2},-\frac{1}{2}}_{(2)} \nonumber \\
H & = & 4 J^{0,0}_{(1)}  \label{gen}
\end{eqnarray}
(note that this is an invertible basis transformation). The Dirac
bracket algebra between these generators reads \cite{tjark}
\begin{eqnarray}
\{H,E\}^* & = & 2E \nonumber \\
\{H,F\}^* & = & -2F \nonumber \\
\{E,F\}^* & = & H^2+C   \label{rel1}
\end{eqnarray}
and $C$ Poisson commutes with everything. This algebra
which is called $\bar{W}_3^{(2)}$ was first
constructed in \cite{Ro} as a nonlinear deformation of
$su(2)$. In \cite{tjark} it was shown to be a reduction of
$sl_3$ and its representation theory was explicitly constructed.

Let's now consider the finite Miura transformation for this algebra.
Since the grading of $sl_3$  by $ad_{t_0}$ is half integer we have
to switch to the grading by $ad_{\delta}$. The explicit form of
$\delta$ is
\be
\delta = \frac{1}{3}\mbox{diag}(1,1,-2)
\ee
It is easily checked that this defines an integer grading of $sl_3$.
The crucial change is that the elements $E_{23}$ and $E_{12}$
have grade 1 and 0 w.r.t.
the $ad_{\delta}$ grading while they have grade $\frac{1}{2}$ w.r.t.
$ad_{t_0}$.
According to the prescription given in the previous section the
alternative set of constraints that one now imposes in order to
reduce the mixed system of first and second class constraints to
a system of first class constraints only is
\be
J^{1,1}_{(1)}-1=J^{\frac{1}{2},\frac{1}{2}}_{(2)}=0 \label{con}
\ee
As we already mentioned in the previous section, what has happened
here is that one has now imposed only half of the constraints that turned
out to be second class. The result is that both constraints in
(\ref{con}) are first class. The point however is that the
gauge symmetry induced by the second constraint in eq. (\ref{con})
can be completely fixed by adding the gauge fixing condition
\be
J^{\frac{1}{2},\frac{1}{2}}_{(1)}=0
\ee
which then leaves us with exactly the same set of constraints
and gauge invariances as before.

We can now describe the generalized Miura map for this case.
An arbitrary element of $t_++g_0$ is given by
\be
J_0 \equiv \left(
\begin{array}{ccc}
h+s  &  e  &  1  \\
f  &  s-h  & 0  \\
0  &  0  &  -2s
\end{array}
\right)
\ee
Note that $g_0 \simeq sl_2 \oplus u(1)$ which means that the Poisson
relations in $K(g_0)$ between the generators $h,e,f$ and $s$
(viewed as elements of $C^{\infty}(g_0)$) are given by
\begin{eqnarray}
\{h,e\} & = & e \nonumber \\
\{h,f\} & = & -f \nonumber \\
\{e,f\} & = & 2h   \label{rel2}
\end{eqnarray}
and $s$ commutes with everything.
As shown in the previous section the equation that we have to
solve in order to get explicit formulas for the Miura map is
is the following equation for $h \in \hat{H}$ (where $\hat{H}$ is
again the group of gauge transformations generated by the two
first class constraints (\ref{con})).
\be  \label{m}
h J_0 h^{-1}= \left(
\begin{array}{ccc}
J^{0,0}_{(1)} & 0 & 1 \\
J^{\frac{1}{2},-\frac{1}{2}}_{(2)} & -2J^{0,0}_{(1)} & 0 \\
J^{1,-1}_{(1)} & J^{\frac{1}{2},-\frac{1}{2}}_{(1)} & J^{0,0}_{(1)}
\end{array}
\right)
\ee
The unique solution of this equation is given by
\be
h= \left(
\begin{array}{ccc}
1 & 0 & 0 \\
0 & 1 & 0 \\
\frac{3}{2}s+\frac{1}{2}h & e & 1
\end{array}
\right)
\ee
Inserting this back into eqn.(\ref{MX}) one finds certain expressions
for $J^{0,0}_{(1)},J^{\frac{1}{2},-\frac{1}{2}}_{(1)}
,J^{\frac{1}{2},-\frac{1}{2}}_{(2)}$ and $J^{1,-1}_{(1)}$ in
terms of the functions $\{e,f,h,s\}$. In terms of the generators
(\ref{gen}) these read
\begin{eqnarray}
H & = & 2h-2s \nonumber \\
E & = & (3s-h)e \nonumber \\
F & = & \frac{4}{3}\; f \nonumber \\
C & = & -\frac{4}{3}(h^2+3s^2+fe) \label{MX}
\end{eqnarray}
It is easy to check, using the relations (\ref{rel2}) that these
satisfy the algebra (\ref{rel1}). Therefore we find that indeed
(\ref{MX}) provides an injective Poisson homomorphism from
$W_3^{(2)}$ into the Kirillov Poisson algebra of the Lie algebra
$g_0=sl_2 \oplus u(1)$ \vspace{5mm}.

In the appendix we consider more examples
of finite $W$ algebras and their quantum versions.
In particular we construct all finite $W$ algebras associated to
$sl_4$ (to get the classical versions of the algebras displayed
in the appendix just replace commutators by Poisson brackets and
forget about all terms of order $\hbar^2$ or higher).

\subsection{An algebraic description of finite $W$ algebras}
Putting together the results obtained in the previous sections we can
describe finite $W$ algebras in an elegant and purely algebraic manner.

As before latin indices will be
supposed to run over the entire basis ${t_a}$ of the Lie algebra $g$,
greek indices run over the basis of $g_+$ and
barred greek indices
(like $\bar{\alpha}$) run over the basis of
$g_-\oplus g_0$. Indices can again be raised and lowered by use of the
Cartan Killing metric.

Again let $K(g)=(C^{\infty}(g),\{.,.\})$ denote the Kirillov Poisson algebra
associated to $g$. The constraints $\{\phi^{\alpha}\}$ are of course
simply elements of $C^{\infty}(g)$. Denote by $I$ the subset
\be
I=\{f_{\alpha}\phi^{\alpha} \mid f_{\alpha} \in C^{\infty} (g) \}
\ee
of $C^{\infty}(g)$. Obviously $I$ is an ideal w.r.t. the (abelian)
multiplication map in $K(g)$ since $hf_{\alpha}\phi^{\alpha}=
\tilde{f}_{\alpha}\phi^{\alpha} \in I$ for all $h \in C^{\infty}(g)$,
where $\tilde{f}_{\alpha}=hf_{\alpha}$. In this terminology an element
$f \in C^{\infty}(g)$ is called 'first class' if
\be
\{f,\phi^{\alpha}\} \in I
\ee
It is now easy to verify that $I$ is a Poisson subalgebra of $K(g)$.
In order to see this let $f=f_{\alpha}\phi^{\alpha}$ and $h=h_{\alpha}
\phi^{\alpha}$ be elements fo $I$, then
\ba
\{f,g\} & = & \{f_{\alpha},h_{\beta}\}\phi^{\alpha}\phi^{\beta}+
\{f_{\alpha}, \phi^{\beta}\}h_{\beta}\phi^{\alpha}+\{\phi^{\alpha},h_{\beta}
\}f_{\alpha}\phi^{\beta} \nonumber \\
&   & +f_{\alpha} h_{\beta}\{\phi^{\alpha},\phi^{\beta}\} \nonumber
\ea
Obviously the first three terms are again elements of $I$. That the
last term is in $I$ follows from the fact that the constraints $\{\phi^{
\alpha}\}$ are all first class which means that there exist elements
$\Phi^{\alpha \beta}_{\gamma} \in C^{\infty}(g)$ such that
\be
\{\phi^{\alpha},\phi^{\beta}\}=\Phi^{\alpha \beta}_{\gamma}\phi^{\gamma}
\ee
The interpretation of the ideal $I$ is that it is the set of all smooth
functions on $g$ which are zero on the set
\be
C=\{x\in g \mid \phi^{\alpha}(x)=0, \;\; \mbox{for all }\alpha \}
\ee
Obviously the space $C^{\infty}(C)$ of smooth functions on $C$ is therefore
isomorphic to the set
\be
A \equiv C^{\infty}(g)/I
\ee
In the terminology that we used before this is the space of smooth
functions on the zero set of the constraints.
As we have seen the Kirillov Poisson structure on $g$ does not induce a
Poisson structure on $C$ since $C$ is plagued by gauge invariances
generated by the first class constraints $\{\phi^{\alpha}\}$. We solved
this problem before by gauge fixing. The algebraic version of this
procedure will now be described.

Define the maps
\be
X^{\alpha}:C^{\infty}(g) \rightarrow C^{\infty}(g)
\ee
by
\be
X^{\alpha}(f)=\{\phi^{\alpha},f\}
\ee
In geometric terms the $X^{\alpha}$ are of course the 'Hamiltonian
vectorfields' associated to the constraints $\{\phi^{\alpha}\}$. Now,
let $[f]$ denote the equivalence class $f+I$, i.e. $[f] \in A$, then
\ba
X^{\alpha}(f+I) & = & \{\phi^{\alpha},f\}+\{\phi^{\alpha},I\} \nonumber \\
& \subset & X^{\alpha}(f)+I
\ea
where we used the fact that $I$ is a Poisson subalgebra of $K(g)$.
Put differently we therefore have
\be
X^{\alpha}[f]=[X^{\alpha}(f)]
\ee
which means that the maps $X^{\alpha}$ descend to well defined maps
\be
X^{\alpha}:A \rightarrow A
\ee
In geometric terms this is nothing but the statement that the Hamiltonian
vectorfields of the constraints are tangent to $C$.

Now define
\be
A_{inv}=\{[f] \in A \mid X^{\alpha}[f]=0, \;\; \mbox{for all }\alpha \}
\ee
Interpretation of this space is that it is the space of smooth functions
on $C/\hat{H}$ (which we showed to be isomorphic to $g_{fix}$), where
as in the previous section $\hat{H}$ is the gauge group on $C$ generated
by the first class constraints $\{\phi^{\alpha}\}$.

The point is now that the Kirillov Poisson structure on $C^{\infty}(g)$
induces naturally a Poisson structure $\{.,.\}^*$ on $A_{inv}$.
Let $[f],[h] \in A_{inv}$ then this Poisson structure is simply given
by
\be
\{[f],[h]\}^*=[\{f,h\}] \label{alt}
\ee
of course one has to show that this Poisson structure is well defined.
For this we have to check two things: first we have to check whether
$[\{f,h\}] \in A_{inv}$ whenever $[f],[h] \in A_{inv}$, and second
we have to show that the definition does not depend on the choice
of representants $f$  and $h$. Both these checks are easily carried
out and one finds that indeed the Poisson bracket $\{.,.\}^*$ turns
$A_{inv}$ into a Poisson algebra. The whole point is now that by
construction the Poisson algebra
\be
W(g;t_0)\equiv (A_{inv},\{.,.\}^*)
\ee
is Poisson isomorphic to the finite $W$ algebra ${\cal W}(i)$
where $i$ is the $sl_2$ embedding associated to $t_0$.
The whole construction described above is
a purely algebraic description of finite $W$ algebras.

Let us now consider how to construct the Poisson relations in $W(g,t_0)$
using the algebraic formalism described above. For this one has to
construct the space $A_{inv}$. of course the algebra $C^{\infty}(g)$
is generated as an (abelian) associative algebra by the functions
$J^a$. Also the constraints $\phi^{\alpha}$ are of the form
\be
\phi^{\alpha}=J^{\alpha}-\chi (J^{\alpha})
\ee
where $\chi$ is defined by $\chi (J^{\alpha})=1$ if $t_{\alpha}=t_+$
and $\chi (J^{\alpha})=0$ otherwise. Therefore we immediately find
that $A=C^{\infty}(g)/I$ is isomorphic to the abelian associative
algebra generated by the functions $\{J^{\bar{\alpha}}\}$.

The next step is to construct $A_{inv}$. By definition this space
correponds to those elements $w$ of $A$ such that
\be
[\{\phi^{\alpha},w\}]\equiv [\{J^{\alpha},w\}]=0
\ee
or in geometric terms, the gauge invariant elements of $C$. In
order to construct a set  of generators of $A_{inv}$ we can use
theorems 1 and 5 which say that for any $C$ valued function on $g$
of the form
\be
t_++\sum_{\bar{\alpha}} J^{\bar{\alpha}}t_{\bar{\alpha}}
\ee
there exists a unique element $a:g \rightarrow \hat{H}$ such that
\be
a(t_++\sum_{\bar{\alpha}}
J^{\bar{\alpha}}t_{\bar{\alpha}})
a^{-1}=t_+ +\sum_{t_{\bar{\alpha}} \in g_{lw}}
W^{\bar{\alpha}}t_{\bar{\alpha}}
\ee
The functions $W^{\bar{\alpha}}$ will in general be polynomials in
the functions $\{J^{\bar{\alpha}}\}$ of the form
\be
W^{\bar{\alpha}}=\sum_{l=0}^p W^{\bar{\alpha}}_l
\ee
where $\mbox{deg}(W^{\bar{\alpha}}_l)=l$, $W^{\bar{\alpha}}_p=
J^{\bar{\alpha}}$ and $p=\mbox{deg}(J^{\bar{\alpha}})$.
Obviously they are gauge invariant (i.e. invariant under $\hat{H}$) and
generate the algebra $A_{inv}$. The Poisson relations in $W(g;t_0)$
can now easily be calculated using (\ref{alt})
\be
\{W^{\bar{\alpha}},W^{\bar{\beta}}\}^*=\{W^{\bar{\alpha}},W^{\bar{\beta}}\}
\ee

Let us give two examples of this construction. First we consider the
(rather trivial) case of $g=sl_2$. The basis of $sl_2$ is given by
\be
J^at_a=\left(
\begin{array}{cc}
J^2 & J^1 \\
J^3 & -J^2
\end{array}
\right)
\ee
of course $t_+=t_1$ which means that the constraint is $\phi=J^1-1$.
We therefore find that $A$ is generated by $J^2$ and $J^3$. In order
to construct the generator $W^3$ of $A_{inv}$ we solve the equation
\be
\left(
\begin{array}{cc}
1 & 0 \\ a_1 & 1
\end{array}
\right) \left(
\begin{array}{cc}
J^2 & 1 \\ J^3 & -J^2
\end{array}
\right) \left(
\begin{array}{cc}
1 & 0 \\ -a_1 & 1
\end{array}
\right) = \left(
\begin{array}{cc}
0 & 1 \\ W^3 & 0
\end{array}
\right)
\ee
which yields $a_1=J^2$ and $W^3=J^3+J^2J^2$. Note that indeed we have
\be
[\{W^3,\phi\}]\equiv [\{W^3,J^3\}]=0
\ee
verifying the gauge invariance of $W^3$. of course the relation of $W^3$
with itself is trivial $\{W^3,W^3\}=0$.

In order to  illustrate the construction in case of a slightly less
trivial case consider again the nonprincipal $sl_2$ embedding into $sl_3$.
For $sl_3$  we again choose the basis
\be
J^at_a =\left(
\begin{array}{ccc}
J^4+\frac{1}{2}J^5 & J^2 & J^1 \\
J^6 & -2J^4 & J^3 \\
J^8 & J^7 & J^4-\frac{1}{2}J^5
\end{array}
\right)
\ee
The constraints are again $\phi^3=J^3$ and $\phi^1=J^1-1$. Obviously
$A$ is generated by $\{J^2,J^4, \ldots , J^8\}$.
In order to construct the generators $\{W^4,W^6,W^7,W^8\}$  (remember
that $g_{lw}=\mbox{Span}(t_4,t_6,t_7,t_8)$) we solve the equation
\be
a (t_++\sum_{\bar{\alpha}}J^{\bar{\alpha}}t_{\bar{\alpha}})a^{-1}=
t_++\sum_{t_{\bar{\alpha}}\in g_{lw}} W^{\bar{\alpha}}t_{\bar{\alpha}}
\ee
where
\be
a= \left(
\begin{array}{ccc}
1 & 0 & 0 \\
0 & 1 & 0 \\
a_1 & a_2 & 1
\end{array}
\right)
\ee
{}From this equation follows that $a_1=\frac{1}{2}J^5$, $a_2=J^2$ and
\ba
W^4 & = & J^4 \nonumber \\
W^6 & = & J^6 \nonumber \\
W^7 & = & J^7+\frac{1}{2}J^2J^5-3J^2J^4 \nonumber \\
W^8 & = & J^8+\frac{1}{4}J^5J^5+J^2J^6
\ea
(Remember that since $\{J^{\bar{\alpha}}\}$ are functions the ordering
in the products is irrelevant. It only becomes relevant after
quantization.) Again introducing the generators
\ba
C & = & -\frac{4}{3} (W^8+3W^4W^4) \nonumber \\
E & = & W^7 \nonumber \\
F & = & \frac{4}{3}W^6 \nonumber \\
H & = & 4W^4
\ea
we find, using formula (\ref{alt}) that these form the algebra (\ref{rel1}).

It  is also possible to give a very simple description of the generalized
Miura transformation in this algebraic framework. For this note that
the Poisson bracket of a grade $l_1$ element with a grade $l_2$ element
is an element of grade $l_1+l_2$. From this it immediately follows that
the Poisson brackets between the elements $W^{\bar{\alpha}}_0$ and
$W^{\bar{\beta}}_0$ must have precisely the same form as the Poisson
brackets between the elements $W^{\bar{\alpha}}$ and $W^{\bar{\beta}}$,
i.e. the grade 0 pieces of the generators must form the same algebra.
Therefore
\be
W^{\bar{\alpha}} \mapsto W^{\bar{\alpha}}_0 \label{ma}
\ee
is a Poisson algebra isomorphism. This is again the Miura map. In order
to see this remember that the elements $W^{\bar{\alpha}}_0$ are
polynomials in the functions $\{J^{\bar{\beta}}\}_{t_{\bar{\beta}} \in
g_0}$. Also remember that these generate a Poisson algebra that is the
Kirillov Poisson algebra of a semisimple Lie algebra, namely $g_0$.
Therefore the map (\ref{ma}) is an embedding of the classical finite
$W$ algebra into a Kirillov Poisson algebra.

\subsection{The symplectic geometry of finite $W$ algebras}
Suppose we are given a Poisson manifold, that is a manifold $M$
together with a Poisson bracket $\{.,.\}$ on the space of smooth
functions on $M$. In general such a Poisson structure is not
associated to a symplectic form on $M$ since the Poisson bracket
can be degenerate, i.e. there may exist functions that Poisson
commute with all other functions. The Poisson bracket $\{.,.\}$
does however induce a symplectic form on certain submanifolds
of $M$. In order to see this consider the set of Hamiltonian
vectorfields
\be
X_f(\cdot )=\{f,.\} \;\;\;\; \mbox{ for } f\in C^{\infty}(M)
\ee
Note that the set of functions on $M$ that Poisson commute with
all other functions are in the kernel of the map $f\mapsto X_f$.
{}From this follows that in every tangent space
the span of all Hamiltonian vectorfields is only a subspace of the
whole tangentspace. We say that a Poisson structure is regular
if the span of the set of Hamiltonian vectorfields has the same
dimension in every tangentspace.
Obviously a regular Poisson structure
defines a tangent system on the manifold $M$
and from the well known relation
\be
[X_f,X_g]=X_{\{f,g\}}
\ee
it then immediately follows that this system is integrable (in the
sense of Frobenius). Therefore $M$ foliates into a disjoint union of
integral manifolds of the Hamiltonian vectorfields. Obviously the
restriction of the Poisson bracket to one of these integral manifolds
is nondegenerate and therefore associated to a symplectic form.
They are therefore called symplectic leaves.

The symplectic leaves play an important role in the representation
theory of Poisson algebras. A representation of a Poisson algebra
is a symplectic manifold $S$ together with a map $\pi$ from the
Poisson algebra to the space of smooth functions on $S$ that
is linear and  preserves both the multiplicative and Poisson
structure of the Poisson algebra. Also the Hamiltonian vectorfield
associated to $\pi (f) \in C^{\infty}(S)$ must be complete.
A representation is called {\em irreducible} if span of the set
$\{X_{\pi (f)}(s)\mid f\in C^{\infty}(M)\}$ is equal to the tangent
space of $S$ in $s$ for all $s\in S$.
The role
of the symplectic leaves becomes clear from the following theorem
\cite{NPL}: If a representation of the Poisson algebra $C^{\infty}(M)$
is irreducible
then $S$ is symplectomorphic to a covering space of a symplectic leaf
of $M$.

{}From the above it becomes clear that it is rather important to
construct the symplectic leaves associated to finite $W$ algebras.
Looking at eq.(\ref{PA}) it is clear that constructing these
symplectic orbits from scratch could be rather difficult. Luckily
we can use the fact that the symplectic orbits of the Kirillov
Poisson structure are known (by the famous Kostant-Souriau theorem
they are nothing but the coadjoint orbits and their covering spaces)
to construct these. The answer turns out to be extremely simple
and is given in the following theorem:
\bt
Let ${\cal O}$ be a coadjoint orbit of the Lie algebra $g$, then
the intersection of ${\cal O}$ and $g_{lw}$ is a symplectic orbit
of the finite $W$ algebra $W(g;t_0)$.
\et
Proof: Let again $C=\{x\in g \mid \phi^{\alpha}(x)=0\}$. The
Hamiltonian vectorfields $X^{\alpha}=\{\phi^{\alpha},.\}$ form
an involutive system tangent to  $C$ and therefore $C$ foliates.
$g_{lw}$ has one point in common with every leaf and denote the
canonical projection from $C$ to $g_{lw}$, which projects an element
$x\in C$ to the unique point $x'\in g_{lw}$, by $\pi$. This map
induces a map $\pi_*$ from the tangent bundle $TC$ of $C$ to
the tangent bundle $Tg_{lw}$ of $g_{lw}$. Let $f$ be a `gauge invariant'
function on $C$, i.e. $\{f,\phi_{\alpha}\}|_{C}=0$ then obviously
$X_f=\{f,.\}$ is a section of $TC$ (i.e. as a vectorfield it is
tangent to $C$ in every point of $C$). It need not be an element
of $Tg_{lw}$ however but using the gauge invariance we can
project it back onto the gauge slice. This projection of $X_f$
on the the gauge slice is given by $\pi_*(X_f)\in Tg_{lw}$. By
construction the Dirac bracket now has the property that
$\{f,.\}^*=\pi_*(X_f)$. What we now need to show is that $\pi_*(X_f)$
is tangent to the coadjoint orbit $\cal O$. Well, obviously we have
$\pi_*(X_f)=X_f+Y$ where $Y$ is a tangent to the gauge orbit. Since
the spaces tangent to the gauge orbits are spanned by the Hamiltonian
vectorfields $X^{\alpha}$ we find that $Y$ is a tangent vector of
$\cal O$. Since $X_f$ is by definition a tangent vector of $\cal O$ we find
that $\pi_*(X_f)$ is also tangent to the coadjoint orbit $\cal O$.
What conclude that the Hamiltonian vectorfields on $g_{lw}$ w.r.t. the
reduced Poisson bracket are all tangent to the coadjoint orbit and
therefore the symplectic orbits are submanifolds of the coadjoint
orbits. The theorem now follows immediately \vspace{5mm}.

Let us now recall some basic facts on coadjoint orbits of simple
Lie algebras. Let $f$ be an adinvariant function on $g$, i.e.
$f(axa^{-1})=f(x)$ for all $a\in G$ and $x\in g$, then we have
\ba
0 & = & \frac{d}{dt}f(e^{tx}ye^{-tx})|_{t=0}=\frac{d}{dt}
f(y+t[x,y])|_{t=0}
\nonumber \\
& = & \left( \mbox{grad}_yf,[x,y]\right) = \left( [\mbox{grad}_yf,x],y
\right)
\ea
which means that
\be
\left( [\mbox{grad}_yf,\mbox{grad}_yh],y\right) = \{f,h\}(y)=0
\ee
We conclude that any adinvariant function Poisson commutes with all
other functions. Conversely any function that Poisson commutes with all
others is adinvariant because $\{f,h\}=0$ for all $h$ implies that
$X_h(f)=0$ for all $h$, where $X_h$ is the Hamiltonian vectorfield
associated to $h$, which means that the derivative of $f$ in all
directions tangent to a symplectic orbit are zero.

Consider now the Casimir functions $\{C_i\}_{i=1}^{\mbox{rank}(g)}$
of the Lie algebra $g$. Certainly these Poisson commute with all
other functions and it can in fact be shown that the (co)adjoint (since
we have identified the Lie algebra with its dual the adjoint and
coadjoint orbits coincide)
orbits of maximal dimension of the Lie algebra are given by their
constant sets
\be
C_{\vec{\mu}}=\{x\in g \mid C_i(x)=\mu_i\, ;\; \mu_i \in {\bf R}\, ;\;
i=1,\ldots ,\mbox{rank}(g)\}
\ee
Their dimension is therefore $\mbox{dim}(g)-\mbox{rank}(g)$.

Coadjoint orbits are very important to the representation theory
of (semi)simple Lie groups. In order to see why let us recall the
Borel-Weil-Bott (BWB) theorem \cite{Bott}. Let $G$ be a compact
semisimple Lie group with maximal torus $T$ and let $R$ be a finite
dimensional irreducible representation of $G$. The space of highest
weight vectors of $R$ is 1 dimensional (call it $V$) and furnishes
a representation of $T$. The product space $G\times V$ is a
trivial line bundle over $G$ and its quotient by the action of $T$
(where $(a,v)\sim (at,t^{-1}v)$ for $a\in G$ and $t\in T$) is a
holomorphic line bundle over $G/T$ (which is a complex manifold). Now,
$L$ admits a $G$ action $(a,v)\rightarrow (a'a,v)$, where $a,a' \in G$
and $v\in V$, so the space of holomorphic sections of $L$ is naturally
a $G$ representation. The BWB theorem now states that this representation
is isomorphic to $R$. The BWB construction has however the restriction
that it applies only to compact groups. There does exist however a
generalization of the BWB construction called the coadjoint method
of Kirillov. In this method one generalizes $G/T$ to  certain
homogeneous spaces $G/H$ which can be realized as coadjoint orbits.
As we have mentioned above, a coadjoint orbit carries a natural
($G$ invariant) symplectic form $\omega$ inherited from the Kirillov
Poisson structure and can therefore be seen as a phase space of some
classical mechanical system. One then attempts to quantize this symplectic
manifold using the methods of geometric quantization \cite{geomquant}.
For this one is supposed to construct a holomorphic line bundle $L$
over the coadjoint orbit such that the first Chern class of $L$ is
equal to the (cohomology class of)
$\omega$. If such a line bundle exists the coadjoint orbit
is called quantizable. For a quantizable orbit the bundle $L$ can then
be shown to admit a hermitean metric whose curvature is equal to $\omega$.
The space of sections of $L$ thus obtains a Hilbert space structure
and is interpreted as the physical Hilbert space of the quantum system
associated to the coadjoint orbit. It carries a unitary representation
of the group $G$ and in fact one attempts to construct all unitary
representations in this way. Obviously this construction is a
generalization of the BWB method. Note that the fact that generic groups
have unitary irreducible representations only for a discrete set of
highest weights is translated into the fact that only a discrete
set of orbits is quantizable.

{}From the above it is clear that the symplectic (or $W$-coadjoint) orbits
of finite $W$ algebras are extremely important especially in the
study of global aspects of $W$ algebras. In order to see this note
that coadjoint orbits are homogeneous spaces of the Lie group $G$
in question, and it is therefore tempting to interpret the symplectic
orbits of $W$ algebras as  $W$-{\em homogeneous spaces}. These
naturally carry much information on the global aspects of $W$
transformations.
In the notation used above
the symplectic orbits of the finite $W$ algebra $W(g;t_0)$ are given
by
\be
{\cal O}_{\vec{\mu}}=C_{\vec{\mu}} \bigcap g_{lw}
\ee
which are therefore (generically) of dimension $\mbox{dim}(g_{lw})-
\mbox{rank}(g)$.

Let us give some  examples. As we have seen the finite $W$ algebras
associated to principal $sl_2$ embeddings are trivial in the sense
that they are Poisson abelian. This means that all Hamiltonian
vectorfields are zero and that all symplectic orbits are points.
Again the simplest nontrivial case is the algebra $W_3^{(2)}$.
The 2 independent Casimir functions of $sl_3$ are
\be
C_1=\mbox{Tr}(J^2) \;\;\; \mbox{ and } C_2 =\mbox{Tr}(J^3)
\ee
where $J=J^at_a$. The spaces ${\cal O}_{\vec{\mu}}$ can be
constructed by taking for $J$ the constrained and gauge fixed matrix
\be
J= \left(
\begin{array}{ccc}
j & 0 & 1 \\
G_+ & -2j & 0 \\
T & G_- & j
\end{array}
\right)
\ee
One then easily finds
\ba
C_1 & = & 3j^2+T \nonumber \\
C_2 & = & -6j^3 + 3G_+G_-+6jT
\ea
Obviously $\mbox{dim}(g_{lw})=4$ and introducing the variables
$x_1=j; \; x_2=G_+ ;\; x_3=G_-;\; x_4=T$ we find that
\be
{\cal O}_{\mu_1,\mu_2}=\{\vec{x}\in {\bf R}^4 \mid 3x_1^2+x_4=\mu_1
\; \mbox{ and } \; 2x_1x_4+x_2x_3-2x_1^3=\mu_2 \}
\ee
We can investigate the topological nature of these symplectic orbits
by inserting $x_4=\mu_1-3x_1^2$ into the second equation. We find that
topologically ${\cal O}_{\mu_1,\mu_2}$ is equivalent to the 2 dimensional
surface in ${\bf R}^3$ determined by the equation
\be
x_2x_3=\mu_2-2\mu_1x_1+8x_1^3
\ee
In the plane $x_1=\alpha$ this is generically
a hyperbola but it degenerates if
$x_1$ is a zero of the polynomial $\mu_2-2\mu_1x_1+8x_1^3$. For example
the orbit ${\cal O}_{0,0}$ degenerates into the two lines
$x_1=x_3=0$ and $x_1=x_2=0$ for $x\rightarrow 0$ and has a
singular point in the origin. In fact this phenomenon also occurs
in ordinary Lie group coadjoint orbits. Consider for example the
case of $SL_2({\bf R})$ where the coadjoint orbits are given by
the following hyperplanes in ${\bf R}^3$ \cite{Wittencoad}
\be
x_2x_3=x^2_1+\mu
\ee
where $\mu \in {\bf R}$. Here also for $\mu =0$ the coadjoint orbit
degenerates if $x_1\rightarrow 0$. This coadjoint orbit is in fact
conelike and for arbitrary non-compact semisimple Lie groups there
are always just a finite number of them (infinite dimensional
Lie groups may have infinitely many of them). The group representations
that one can associate to them using the Kirillov method are called
unipotent representations. We therefore interpret the $W(g;t_0)$
symplectic orbits that are singular as unipotent finite $W$
representations. Also it should be noted that, in the $SL_2({\bf R})$ case,
the origin itself
is also a coadjoint orbit, i.e. it is a fixed point of the group
action. In the same way the singular points in the unipotent $W$
orbits are fixed points of global $W$ transformations.

This concludes our presentation of the theory of classical finite
$W$ algebras.

\section{Quantum finite $W$ Algebras}

Having developed the classical theory of finite $W$ algebras we now
turn to their quantization.
In quantum mechanics, quantization amounts to replacing Poisson
brackets by commutators, or in a mathematically more sophisticated
language, replacing a Poisson algebra
$({\cal A}_0,\{.,.\} )$ by an
associative algebra ${\cal A}$ depending on a parameter $\hbar$
such that
\begin{itemize}
\item  ${\cal A}$ is a free $\ce[[\hbar]]$ module,
\item ${\cal A}/\hbar {\cal A}\simeq {\cal A}_0$
\item if $\pi$ denotes the natural map $\pi : {\cal A} \rightarrow {\cal
A}/\hbar {\cal A}\simeq {\cal A}_0$, then $\{\pi(X),\pi(Y)\}=
\pi((XY-YX)/\hbar)$.
\end{itemize}
In most cases one has a set of generators
for ${\cal A}_0$, and ${\cal A}$ is completely fixed by giving
the commutation relations of these generators.

For example, let ${\cal A}_0$ be the Kirillov Poisson algebra $K(g)$
associated to $g$, then a quantization of
this Poisson algebra is the algebra
${\cal A}$
generated by the $J^a$ and $\hbar$, subject to the commutation relations
$[J^a,J^b]=\hbar f^{ab}_{\,\,\, c} J^c$ (note that here the $J^a$ are
no longer functions on $g$ but the quantum objects associated
to them). Obviously, the Jacobi
identities are satisfied. Specializing to
$\hbar=1$, this algebra is precisely the universal enveloping
algebra ${\cal U}g$ of  $g$.

To find quantizations of finite $W$ algebras, one can first
reduce the $sl_n$ Kirillov Poisson algebra, and then try to
quantize the resulting algebras that we studied in the previous
sections. On the other hand, on can also first quantize and then
constrain. We will follow the latter approach, and thus study
the reductions of the quantum Kirillov algebra
\be \label{eq:qkir}
[J^a,J^b]=\hbar f^{ab}_{\,\,\, c} J^c.
\ee
We want to impose the same constraints on this algebra as we
imposed previously on the Kirillov Poisson algebra, to obtain
the quantum versions of the finite $W$ algebras related to
$sl_2$ embeddings. Imposing constraints on quantum algebras can
be done using the BRST formalism \cite{KoSt}.

\subsection{BRST quantization of finite $W$ algebras}
Consider again the map $\chi$ defined by $\chi (J^{\alpha})=1$ if
$t_{\alpha}=t_+$ and $\chi (J^{\alpha})=0$ otherwise.
In terms of $\chi$, the
constraints can be written as $\pi_+(J)=
\chi(\pi_+(J))$, where $\pi_+$ denotes the
projection $g\rightarrow g_+$ and $J=J^at_a$
It is
this form of the constraints that we will use. Furthermore we
will take $\hbar=1$ for simplicity; the explicit $\hbar$
dependence can be determined afterwards.

To set up the BRST framework
we need to introduce anticommuting
ghosts and antighosts $c_{\alpha}$ and $b^{\alpha}$, associated
to the constraints that we want to impose \cite{KoSt}. They satisfy
$b^{\alpha}c_{\beta}+c_{\beta}b^{\alpha}=
\delta^{\alpha}_{\beta}$ and generate
the Clifford algebra $Cl(g_+\oplus g_+^*)$.
The quantum Kirillov algebra is
just the  universal enveloping algebra ${\cal U}g$,
and the total
space on which the BRST operator acts is $\Omega={\cal
U}g\otimes Cl(g_+\oplus g_+^*)$. A ${\bf Z}$
grading on $\Omega$ is defined by
$\deg(J^a)=0$, $\deg(c_{\alpha})=+1$ and $\deg(b^{\alpha})=-1$,
and we can decompose $\Omega=\oplus_k \Omega^k $ accordingly.
The BRST differential on $\Omega$ is given by $d(X)=[Q,X]$,
where $Q$ is the BRST charge
\be \label{def:q}
Q=(J^{\alpha}-\chi(J^{\alpha}))c_{\alpha}-\hf
f^{\alpha\beta}_{\,\,\, \gamma} b^{\gamma}c_{\alpha}c_{\beta}.
\ee
and $[.,.]$
denotes the graded commutator (as it always will from now on)
\be
[A,B]=AB- (-1)^{\mbox{deg}(A).\mbox{deg}(B)}BA
\ee
Note that $\mbox{deg}(Q)=1$.

This is the standard BRST complex associated to the first class
constraints of the previous section. Of interest are the
cohomology groups of this complex, $H^k(\Omega;d)$. The zeroth
cohomology group is the quantization of the classical finite $W$ algebra.
Because the gauge group $\hat{H}$ in (39) acts
properly on $C$, we expect the higher cohomologies of the BRST
complex to vanish, as they are generically related to
singularities in the quotient $C/\hat{H}$. In the mathematics
literature the cohomology of the BRST complex is called the Hecke
algebra ${\cal H}(g,g_+,\chi)$ associated to
$g,g_+,\chi$.
Hecke algebras related to arbitrary $sl_2$
embeddings have not been computed, apart from those
related to the principal $sl_2$ embeddings. In that case it was
shown by Kostant \cite{kost2} that the only nonvanishing
cohomology is $H^0(\Omega;d)$ and that it is isomorphic to the
center of the universal enveloping algebra. Recall that the
center of the ${\cal U}g$ is generated by the set of independent
Casimirs of $g$. This set is closely related to
the generators of standard infinite $W_n$-algebras; in that case
there is one $W$ field for each Casimir which
form a highly nontrivial algebra \cite{BBSS}. We see that for
finite $W$ algebras
the same generators survive, but that they form a trivial abelian
algebra. For non principal $sl_2$ embeddings however quantum finite $W$
algebras are non-trivial.

We now want to compute the cohomology of $(\Omega;d)$.
Unfortunately this is  a difficult problem to approach directly. We
therefore use a well known trick of cohomology theory, namely we split
the complex $(\Omega ;d)$ into a double complex and calculate the
cohomology via a spectral sequence argument. For this note
that the operator $d$ can be
decomposed into two anti commuting pieces. Write $Q=Q_0+Q_1$, with
\ba \label{def:q01}
Q_1 & = & J^{\alpha}c_{\alpha}-\hf
f^{\alpha\beta}_{\,\,\, \gamma} b^{\gamma}c_{\alpha}c_{\beta},
\nonu
Q_0 & = & -\chi(J^{\alpha})c_{\alpha},
\ea
and define $d_0(X)=[Q_0,X]$, $d_1(X)=[Q_1,X]$, then one can
verify by explicit computation that
$d_0^2=d_0d_1=d_1d_0=d_1^2=0$. Associated to this decomposition
is a bigrading of $\Omega=\oplus_{k,l} \Omega^{k,l}$ defined by
\ba \label{def:bigrade}
\deg(J^a)=(-k,k), & & \mbox{ if }t_a\in g_{k} \nonu
\deg(c_{\alpha})=(k,1-k), & & \mbox{ if }t_{\alpha} \in g_{k} \nonu
\deg(b^{\alpha})=(-k,k-1), & & \mbox{ if }t_{\alpha} \in g_{k},
\ea
with respect to which $d_0$ has degree $(1,0)$ and $d_1$ has
degree $(0,1)$. Thus $(\Omega^{k,l};d_0;d_1)$ has the structure
of a double complex. Explicitly, the action of $d_0$ and $d_1$
is given by
\ba \label{eq:q01}
d_1(J^a) & = & f^{\alpha a}_{\,\,\,b}J^bc_{\alpha},
\nonu
d_1(c_{\alpha}) & = & -\hf f^{\beta\gamma}_{\,\,\, \alpha}c_{\beta}
c_{\gamma}, \nonu
d_1(b^{\alpha}) & = & J^{\alpha} + f^{\alpha \beta}_{\,\,\,
\gamma}b^{\gamma}c_{\beta}, \nonu
d_0(J^a)=d_0(c_{\alpha}) & = & 0, \nonu
d_0(b^{\alpha}) & = & -\chi(J^{\alpha}).
\ea
To simplify the algebra, it is advantageous to introduce
\be \label{def:hj}
\hj^a=J^a+f^{a\beta}_{\,\,\, \gamma}b^{\gamma} c_{\beta}.
\ee
Our motivation to introduce these new elements $\hj^a$ is
twofold: first, similar expressions were encountered
in a study of the effective action for $W_3$ gravity \cite{BG},
where it turned out that the BRST cohomology for the infinite $W_3$
algebra case could conveniently be expressed in terms of
$\hj$'s; second, such expressions were introduced for the $J^a$'s
that live on the Cartan subalgebra of $g$ in \cite{FeFr}, and
simplified their analysis considerably. In terms of $\hj$ we
have
\ba \label{eq:q01hat}
d_1(\hj^a) & = & f^{\alpha a}_{\,\,\,\bar{\gamma}}
\hj^{\bar{\gamma}} c_{\alpha}, \nonu
d_1(c_{\alpha}) & = & -\hf f^{\beta\gamma}_{\,\,\, \alpha}c_{\beta}
c_{\gamma}, \nonu
d_1(b^{\alpha}) & = & \hj^{\alpha}, \nonu
d_0(\hj^a) & = & -f^{a \beta}_{\,\,\, \gamma} \chi(J^{\gamma})
c_{\beta}, \nonu
d_0(c_{\alpha}) & = & 0, \nonu
d_0(b^{\alpha}) & = & -\chi(J^{\alpha}).
\ea

The advantage of having a double complex, is that we can apply
the techniques of spectral sequences \cite{botttu}
to it, in order to compute the cohomology of $(\Omega;d)$. The
results from the theory of spectral sequences that we need are
gathered in the next section.

\subsection{Spectral Sequences for Double Complexes}
Let $(\Omega^{p,q};d_0;d_1)$ denote a double complex, where $d_0$
has degree $(1,0)$ and $d_1$ has degree $(0,1)$. The standard
spectral sequence for this double complex is a sequence of
complexes $(E_r^{p,q};D_r)_{r\geq 0}$, where $D_r$ is a
differential of degree $(1-r,r)$, and is defined as follows:
$E_0^{p,q}=\Omega^{p,q}$, $D_0=d_0$, $D_1=d_1$, and for $r\geq 0$
\be \label{def:ss}
E^{p,q}_{r+1} = H^{(p,q)}(E_r;D_r)=
\frac{\ker\, D_r:E^{p,q}_r \rightarrow
E_r^{p+1-r,q+r}}{\im\, D_r:E^{p-1+r,q-r}_r\rightarrow E_r^{p,q}}.
\ee
The differential $D_{r+1}$ for $r>0$
is given by $D_{r+1}(\alpha)=d_1 (\beta)$,
where $\beta$ is chosen such that $d_0 \beta=D_r \alpha$. Such a
$\beta$ always exists and $D_{r+1}$ is uniquely defined in this
way. The usefulness of this spectral sequence is provided by the
following \cite{specseq}
\bt
If $\cap_{q}\oplus_{p,q\geq s}\Omega^{p,q}=\{0\}$,
then $E_{\infty}^{p,q}=\cap_r
E_r^{p,q}$ exists, and
 \be
\label{isom}
E_{\infty}^{p,q}\simeq\frac{F^q H^{(p+q)}(\Omega;d)}{F^{q+1}
H^{(p+q)}(\Omega;d)},
\ee
where
\be
\label{filter}
F^q H^{(p+q)}(\Omega;d)=\frac{\ker\ d:\oplus_{s\geq q}
\Omega^{p+q-s,s} \rightarrow \oplus_{s\geq q}
\Omega^{p+q+1-s,s}}{\im\ d:\oplus_{s\geq q} \Omega^{p+q-1-s,s}
\rightarrow \oplus_{s\geq q} \Omega^{p+q-s,s}}
\ee
\et
This spectral sequence is especially useful if it
collapses at the $n^{th}$ term for some $n$, \ie\ $D_r=0$ for
$r\geq n$, because then $E_n\simeq E_{\infty}$ and one needs only
to compute the first $n$ terms of the spectral sequence.

If the double complex is also an algebra, i.e. there is a
multiplication operator $m: \Omega^{p,q} \otimes_{\ce} \Omega^{p',q'}
\rightarrow \Omega^{p+p',q+q'}$, and $d$ satisfies the Leibniz
rule with respect to this multiplication, then (\ref{isom}) is
also an equality of algebras. It is in general  nontrivial
to reconstruct the full algebra structure of
$H^{\ast}(\Omega;d)$ from $E^{p,q}_{\infty}$, due to the
quotient in the right hand side of (\ref{isom}).

Another useful tool in the computation of cohomology is the
following version of the K\"unneth theorem
\bl
\label{kunneth}
Let $A=\oplus_k A^k$
be a graded differential algebra over $\ce$ with a
differential $d$ of degree 1 that satisfies the Leibniz rule, and
assume that $A$ has two graded subalgebras $A_1=\oplus_k A_1^k$
and $A_2=\oplus_k A_2^k$ such
that $d(A_1)\subset A_1$ and $d(A_2)\subset A_2$, that
$A_2^k=\{0\}$ for $k$ sufficiently large, and that
$m:A_1\otimes_{\ce} A_2 \rightarrow A$ given by $m(a_1\otimes
a_2)=a_1 a_2$ is an isomorphism of vector spaces. Then
\be \label{kunneth2}
H^{\ast}(A;d)\simeq \{a_1a_2 | a_1\in H^{\ast}(A_1;d), a_2\in
H^{\ast}(A_2;d) \}.
\ee
\el
Proof: form the double complex $(\Omega^{p,q};d_0;d_1)$ with
$\Omega^{p,q}\simeq m(A_1^p \otimes A_2^q)$, $d_0(a_1
a_2)=d(a_1)a_2$, and $d_1(a_1
a_2)=(-1)^{\deg(a_1)} a_1 d(a_2)$. The spectral sequence
for this double complex collapses at the $E_2$ term, and one
finds $E^{p,q}_{\infty}=\{a_1a_2|a_1 \in H^p(A_1;d), a_2\in
H^q(A_2;d)\}$. The condition $A_2^k=0$ for $k$ sufficiently
large guarantees that $F^q H^{p+q}=0$ for $q$ sufficiently
large, and one can assemble $H^{\ast}(\Omega;d)$ from
$E^{p,q}_{\infty}$ using (\ref{isom}). This leads to
(\ref{kunneth2}) on the level of vector spaces. Because $a_1a_2$
is really a representative of an element of $H^{\ast}(A;d)$, it
follows that (\ref{kunneth2}) is also an isomorphism of
algebras.\\

By induction, one can easily prove that the theorem still holds
if instead of two subalgebras $n$ subalgebras $A_1,\ldots A_n$
are given, with $A\simeq A_1 \otimes \ldots \otimes A_n$. The
condition $A_2^k=\{0\}$ for sufficiently large $k$ is replaced by
$A_2^k=\ldots =A_n^k=\{0\}$ for sufficiently large $k$. Without
such a condition, it may not be so easy to reassemble $H^{\ast}$
from $E_{\infty}$. To illustrate some of the difficulties that
can arise, let us give an example where $H^{\ast}$
cannot be recovered directly from $E_{\infty}$. This example is
not related to the above theorem, but it represents a situation
we will encounter in the computation of the BRST cohomology.

Consider the algebra $\Omega=\ce[x,y]/(y^2=0)$, where $x$ is an even
generator of bidegree $(1,-1)$, and $y$ is an odd generator of
bidegree $(0,-1)$. The differentials $d_0,d_1$ are given by
$d_{0,1}(x)=0$, $d_0(y)=x$ and $d_1(y)=-1$. One immediately
computes $H^k(\Omega;d)=\ce[x]/(x-1)\ce[x]\simeq \ce$
for $k=0$, and $H^k=0$
otherwise. The spectral sequence associated to the double
complex collapses at the first term, and one finds
$E^{p,q}_{\infty}=\ce \delta_{p,0} \delta_{q,0}$. Because $F^1
H^{\ast}=0$, one deduces that $H^k(\Omega;d)\simeq \ce \delta_{k,0}$.
On the other hand, we could also have started with the mirror
double complex obtained by interchanging $d_0$ and $d_1$ and the
bigrading. Thus, we assign bidegree $(-1,1)$ to $x$ and bidegree
$(-1,0)$ to $y$. The spectral sequence associated to the mirror
double complex also collapses at the first term, but now one
finds $E^{p,q}_{\infty}=0$. This is not in conflict with the
previous computation, because we cannot a priori find a $q$ for
which $F^q H^{p+q}=0$, and we can only conclude that $F^q
H^{p+q}\simeq F^{q+1} H^{p+q}$. If we compute explicitly with
respect to this bigrading what $F^q H^{p+q}$ is, we find that
it is only nontrivial for $p+q=0$, and then $F^q
H^0=x^q\ce[x]/(x-1)x^q\ce[x]\simeq\ce$ for $q\geq0$, and $F^q
H^0=\ce[x]/(x-1)\ce[x]\simeq\ce$ for $q<0$. This indeed yields
$E^{p,q}_{\infty}=0$. The lesson is that one should be careful
in deriving $H^{\ast}(\Omega;d)$ from $E^{p,q}_{\infty}$.

Finally, let us present another fact that will be useful later.
\bl \label{filter2}
Suppose $A$ is a differential graded algebra, $A=\oplus_{n\geq
0} A^n$, with a differential of degree $1$. Assume furthermore
that $A$ has a filtration
\be \label{filt}
\{0\}=F^0 A \subset F^1 A \subset F^2 A \subset \cdots \subset A,
\ee
such that $F^pA F^qA\subset F^{p+q}A$, and that $d$ preserves the
filtration, $d(F^pA)\subset F^pA$. If $H^k(F^{p+1}A/F^pA;d)=0$
unless $k=0$, then we have the following isomorphism of vector
spaces
\be \label{isom2}
H^0(A;d)\simeq \oplus_{p\geq 0} H^0(F^{p+1}A/F^pA;d).
\ee
\el
Proof: One can assign a spectral sequence to such a filtered
graded algebra \cite{specseq}, whose first term contains the
cohomologies $H^k(F^{p+1}A/F^pA;d)$. If only $H^0\neq 0$, then
the spectral sequence collapses at the first term, and because
the filtration is bounded from below ($\{0\}=F^0A$), one can
collect the vector spaces that make up $E^{\ast,\ast}_{\infty}$,
to get the isomorphism (\ref{isom2}).

\subsection{The BRST Cohomology}
The first people to propose using the theory of spectral sequences
in the setting of $W$ algebras were
Feigin and Frenkel \cite{FeFr}.  In fact, in this way they computed the
BRST cohomology in the infinite dimensional case
but only for the special example of the principal $sl_2$
embedding (which are known to lead to the $W_N$ algebras).
Their calculation however has the drawback that it is
very difficult to generalize to arbitrary embeddings and that it
constructs the cohomology in an indirect way (via commutants of
screening operators).
Now, to any double complex there are canonically associated two spectral
sequences which can be considered to be each others mirror image.
In \cite{BT} it was first proposed to
calculate the BRST cohomology using a spectral sequence that is the
mirror spectral sequence of the one used by Feigin and Frenkel.
As it turns out this has drastic simplifying consequences for the
calculation of the cohomology.

The following theorem \cite{BT} gives the BRST cohomology on the level of
vectorspaces.
\bt  \label{final}
As before let $g_{lw}\subset g$ be the kernel of the map
$\ad{t_-}:g\rightarrow g$. Then the BRST cohomology is
given by the following isomorphisms of vector spaces
\be \label{coho}
H^{k}(\Omega;d)\simeq ({\cal U}g_{lw})\delta_{k,0}.
\ee
\et
Proof: The computation of the BRST cohomology is simplified
considerably due to the introduction of the new set of generators
$\hj^a$. The simplification arises due to the fact that
$H^{\ast}(\Omega;d)\simeq H^{\ast}(\Omega_{red};d)$, where $\Omega_{red}$ is
the subalgebra of $\Omega$ generated by
$\hj^{\bar{\alpha}}$ and $c_{\alpha}$.
In order to show this
apply the K\"unneth theorem \ref{kunneth} to
$\Omega_{red}\otimes (\otimes_{\alpha} \Omega_{\alpha})$, where
$\Omega_{\alpha}$ is the algebra generated by $\hj^{\alpha}$ and
$b^{\alpha}$. Note that $[\hj^{\alpha},b^{\alpha}]=0$ and that
the conditions of the K\"unneth theorem are satisfied. Therefore,
$H^{\ast}(\Omega;d)\simeq H^{\ast}(\Omega_{red};d)\otimes
(\otimes_{\alpha} H^{\ast}(\Omega_{\alpha};d))$. Now
$(\Omega_{\alpha};d)$ is essentially the same complex as the
one we examined in the last part of the previous section, and one
easily proves that $H^k(\Omega_{\alpha};d)\simeq
\ce\delta_{k,0}$. This shows $H^{\ast}(\Omega;d)\simeq
H^{\ast}(\Omega_{red};d)$
The reduced complex $(\Omega_{red};d)$ is described by the
following set of relations:
\ba \label{eq:q01hat2}
d_1(\hj^{\bar{\alpha}}) & = & f^{\alpha \bar{\alpha}}_{\,\,\,
\bar{\gamma}} \hj^{\bar{\gamma}} c_{\alpha}, \nonu
d_1(c_{\alpha}) & = & -\hf f^{\beta\gamma}_{\,\,\, \alpha}c_{\beta}
c_{\gamma}, \nonu
d_0(\hj^{\bar{\alpha}}) & = & -f^{{\bar{\alpha}}
\beta}_{\,\,\, \gamma} \chi(J^{\gamma})
c_{\beta}, \nonu
d_0(c_{\alpha}) & = & 0, \nonu
[\hj^{\bar{\alpha}},\hj^{\bar{\beta}}]
& = & f^{\bar{\alpha}\bar{\beta}}_{
\,\,\,\bar{\gamma}} \hj^{\bar{\gamma}}, \nonu
[\hj^{\bar{\alpha}},c_{\beta}] & = &
-f^{\bar{\alpha}\gamma}_{\,\,\,\beta} c_{\gamma}, \nonu
[c_{\alpha},c_{\beta}] & = & 0.
\ea
The $E_1$ term of the spectral sequence is now given
by the $d_0$ cohomology of $\Omega_{red}$. To compute the
cohomology we use lemma \ref{filter2}. Define the filtration on
$\Omega_{red}$ by declaring $F^p\Omega_{red}$ to be spanned as a vector
space by $\{ \hj^{\bar{\alpha}_1} \hj^{\bar{\alpha}_2} \cdots
\hj^{\bar{\alpha}_r} c_{\beta_1}
 c_{\beta_2} \cdots c_{\beta_s} | r+s\leq p\}$. Thus
 $F^p\Omega_{red}/F^{p-1}\Omega_{red}$ is spanned by the
 products of precisely $p$ $\hj$'s and $c$'s, and in this
 quotient $\hj$ and $c$ (anti)commute with each other. Now let
 us rewrite $d_0(\hj^{\bar{\alpha}})$ as
 \ba
 \label{rewri}
 d_0(\hj^{\bar{\alpha}}) & = &
 -\tr([\chi(J^{\gamma})t_{\gamma},t^{\bar{\alpha}}] t^{\beta}
 c_{\beta} ) \nonu
 & =  & -\tr([t_+,t^{\bar{\alpha}}]t^{\beta}c_{\beta} ).
 \ea
 From this it is clear that $d_0(\hj^{\bar{\alpha}})=0$ for
 $t^{\bar{\alpha}}\in g_{hw}$. Furthermore, since $t_{\bar{\alpha}}\in
 g_0\oplus g_-$ and $\dim(g_{lw})=\dim(g_0)$, it
 follows that for each $\beta$ there is a linear combination
$a(\beta)_{\bar{\alpha}}\hj^{\bar{\alpha}}$ with
 $d_0(a(\beta)_{\bar{\alpha }}\hj^{\bar{\alpha}})=c_{\beta}$.
 This proves that
 \be \label{symalg}
 \bigoplus_{p>0}\frac{F^p\Omega_{red}}{F^{p-1}\Omega_{red}} \simeq
 \bigotimes_{t_{\bar{\alpha}}\in g_{lw}} \ce[\hj^{\bar{\alpha}}]
 \bigotimes_{t_{\bar{\alpha}}\not\in g_{lw}}
 (\ce[\hj^{\bar{\alpha}}] \oplus d_0(\hj^{\bar{\alpha}})
 \ce[\hj^{\bar{\alpha}}]).
 \ee
Using the K\"unneth theorem (lemma \ref{kunneth}) for
(\ref{symalg}), we find that
 \be \label{E1}
 H^k(\Omega_{red};d_1) =
 \bigotimes_{t_{\bar{\alpha}}\in g_{lw}}
\ce[\hj^{\bar{\alpha}}]\delta_{k,0} =
 ({\cal U}g_{lw})\delta_{k,0}.
 \ee
 Because there is only cohomology of degree 0, the spectral
 sequence collapses, and $E_{\infty}= E_1$. Because
 $\Omega_{red}^{k,l}=0$ for $l>0$, we can find
 $H^{\ast}(\Omega_{red};d)$ from $E_{\infty}$. This proves the theorem.

As expected, there is only cohomology of degree
zero, and furthermore, the elements of $g_{lw}$ are in one-to-one
correspondence with the components of $g$ that made up the
lowest weight gauge in section 1. Therefore
$H^{\ast}(\Omega;d)$ really is a quantization of the finite $W$
algebra. What remains to be done is to compute the algebraic
structure of $H^{\ast}(\Omega;d)$. The only thing that
(\ref{coho}) tells us is that the product of two elements $a$
and $b$ of bidegree $(-p,p)$ and $(-q,q)$ is given by the
product structure on ${\cal U}g_{lw}$, modulo terms of bidegree
$(-r,r)$ with $r<p+q$. To find these lower terms we need
explicit representatives of the generators of
$H^0(\Omega;d)$ in $\Omega$. Such representatives can be
constructed using the so-called tic-tac-toe construction
\cite{botttu}: take some $\phi_0\in g_{lw}$, of bidegree $(-p,p)$.
Then $d_0(\phi_0)$ is of bidegree $(1-p,p)$. Since
$d_1d_0(\phi_0)=-d_0d_1(\phi_0)=0$, and there is no $d_1$ cohomology of
bidegree $(1-p,p)$, $d_0(\phi_0)=d_1(\phi_1)$ for some $\phi_1$ of
bidegree $(1-p,p-1)$. Now repeat the same steps for $\phi_1$,
giving a $\phi_2$ of bidegree $(2-p,p-2)$, such that
$d_0(\phi_1)=d_1(\phi_2)$. Note that $d_1d_0(\phi_1)=
-d_0d_1(\phi_1)=-d_0^2(\phi)=0$. In this way we find a sequence
of elements $\phi_l$ of bidegree $(l-p,p-l)$. The process stops
at $l=p$. Let
\be
W(\phi)=\sum_{l=0}^{p} (-1)^l \phi_l
\ee
Then
$dW(\phi)=0$, and $W(\phi)$ is a representative of $\phi_0$ in
$H^0(\Omega;d)$. The algebraic structure of $H^0(\Omega;d)$ is
then determined by calculating the commutation
relations of $W(\phi)$ in $\Omega$, where $\phi_0$ runs over a
basis of $g_{lw}$. This is the quantum finite $W$ algebra.

Let us now give an example of the construction described above.

\subsection{Example}
Consider again the nonprincipal $sl_2$ embedding into $sl_3$
associated to the following partition
of the number 3: $3=2+1$. We constructed the classical $W$ associated
to this embedding earlier. We shall now quantize this Poisson algebra
by the methods developed above.
Take the following basis of $sl_3$:
\be \label{basis21}
r_at_a=\mats{\frac{r_4}{6}-\frac{r_5}{2}}{r_2}{r_1}{r_6}{-
       \frac{r_4}{3}}{r_3}{r_8}{r_7}{\frac{r_4}{6}+\frac{r_5}{2}}.
\ee
Remember that
(in the present notation)
the $sl_2$ embedding is given by $t_+=t_1$, $t_0=-t_5$ and
$t_-=t_8$. The nilpotent subalgebra $g_+$ is spanned by
$\{t_1,t_3\}$, $g_0$ by $\{t_2,t_4,t_5,t_6\}$ and $g_-$ by
$\{t_7,t_8\}$. The $d_1$ cohomology of $\Omega_{red}$ is
generated by $\{\hj^4,\hj^7,\hj^6,\hj^8\}$, and using the
tic-tac-toe construction one finds representatives for these
generators in $H^0(\Omega_{red};d)$:
\ba \label{reps21}
W(\hj^4) & = & \hj^4, \nonu
W(\hj^6) & = & \hj^6, \nonu
W(\hj^7) & = & \hj^7-\hf\hj^2\hj^5-\hf\hj^4\hj^2+\hf\hj^2, \nonu
W(\hj^8) & = & \hj^8+\deel{1}{4}\hj^5\hj^5+\hj^2\hj^6-\hj^5.
\ea
Let us introduce another set of generators
\ba \label{newgens21}
C & = & -\deel{4}{3} W(\hj^8)-\deel{1}{9}W(\hj^4)W(\hj^4)-1,
\nonu
H & = & -\deel{2}{3}W(\hj^4)-1, \nonu
E & = &  W(\hj^7), \nonu
F & = & \deel{4}{3} W(\hj^6).
\ea
The commutation relations between these generators are given by
\ba \label{alg21}
[H,E] & = & 2E, \nonu
[H,F] & = & -2F, \nonu
[E,F] & = & H^2+C, \nonu
[C,E]=[C,F]=[C,H] & = & 0.
\ea
These are precisely the same as the relations for the
finite $W_3^{(2)}$ algebra given in \cite{tjark}.
Notice that in this case the quantum relations are identical to the
classical ones.
The explicit
$\hbar$ dependence can be recovered simply by multiplying the
right hand sides of (\ref{alg21}) by $\hbar$.

In \cite{Ro} Rocek studied nonlinear deformations of $su(2)$ by
imposing Jacobi identities on a general deformation. He also
gave a general discussion of the existence of unitary representations
and some other features of their representation theory. As a solution
of the Jacobi identities he found $\bar{W}^{(2)}_{3}$. In \cite{SSN}
an attempt was made to formulate a gauge theory based on this
nonlinear algebra by using the Noether method. However due to the
nonlinearity of $\bar{W}^{(2)}_3$ the action they obtained has
infinitely many terms. It is therefore an interesting fact that this
algebra is a reduction of $sl_3$ as we have shown above. It might be
possible to construct a manageable action for $\bar{W}_3^{(2)}$
using this knowledge.

In the appendix we also discuss the explicit quantization of all
the finite $W$ algebras that can be obtained from $sl_4$.
There one does encounter certain quantum effects, i.e the quantum
relations will contain terms of order $\hbar^2$ or higher.

In this section we have seen that it is possible to quantize the
classical finite $W$ algebras introduced in the first section of this
chapter using the BRST formalism.
What we have obtained are nonabelian and in general nonlinear
associative algebras. These algebras are in principle still abstract.
For applications to physics we are interested in the
representation theory of quantum finite $W$ algebras. This is the
subject of the final section of this chapter.

\section{The Representation Theory of Finite $W$ algebras}
In physics symmetry groups are always represented in terms of unitary
linear operators. It follows that knowledge of the
representation theory of groups and algebras is as important to
physics as knowledge of their structure.
For any physical application of finite $W$ algebras
it is therefore crucial to develop their representation theory.

Even if the representation theory of finite $W$ algebras turns out
to be of no particular interest to physics on its own it certainly
will be important to the theory of infinite $W$ algebras. As we said
in the introduction of this chapter the structure and representation
theory of affine Lie algebras is determined for a large part by the
structure and representation theory of simple Lie algebras. For example,
the space of singular or highest weight vectors of an affine Lie algebra
representation will carry a representation of the simple Lie algebra
underlying the affine algebra. We can therefore sweep out the set
of singular vectors by acting with this underlying finite algebra
on the vacuum state of the theory (which is of course also a
singular vector).

Answering questions like 'what are the $W$ singular vectors' have turned
out to be rather difficult even though some progress has been made in
particular cases. Here finite $W$ algebras may provide an important
now tool for the same reasons as described above.
So also seen from this point of view it is very important to develop
the representation theory of finite $W$ algebras.

Let us therefore begin with the simplest nontrivial case, the finite
$W$ algebra $\bar{W}^{(2)}_3$. The finite dimensional representation
theory of this algebra is the subject of the next section.

\subsection{The representation theory of $\bar{W}_3^{(2)}$}
Consider again the associative algebra generated by $\{H,E,F,C\}$ subject
to the relations (\ref{alg21}). We shall now construct the finite
dimensional representation theory of this algebra. For this we are going
to need the following
\bl
The following identities are true
\begin{eqnarray}
[ H,E^k ]  & = & 2kE^k \nonumber \\ {}
[ H,F^k ]  & = & -2kF^k  \nonumber \\ {}
[ E,F^k ] & = & F^{k-1} (kH^2 +kC -2k(k-1)H +
\frac{2}{3}k(k-1)(2k-1))
\end{eqnarray}
\el
Proof: One can easily
prove these identities by induction using the defining
relations of $\bar{W}^{(2)}_3$ and the formula
\be
\sum_{i=1}^{p-1}i^2 =\frac{1}{6}p(p-1)(2p-1) \nonumber
\ee
\vspace{6mm}

Consider the real span of $H$ and $C$ to be the Cartan subalgebra. The
next theorem then identifies the highest weight representations
of $\bar{W}_3^{(2)}$.
\bt
Let $p$ be a positive integer and $x$ a real number.
\begin{enumerate}
\item For every pair $(p,x)$ the algebra $\bar{W}_3^{(2)}$ has	a
unique highest weight representation $W(p,x)$ of dimension p.
\item $W(p,x)$ is spanned by eigenvectors (weight vectors) of $H$.
The weights are given by
\be
\{j(p,x)-2k \}_{k=0}^{p-1}
\ee
where the highest weight $j(p,x)$ is equal to
\be
j(p,x)=p+x-1
\ee
\item The value of the central element $C$ on $W(p,x)$ is given by
\be
c(p,x)=\frac{1}{3}(1-p^2)-x^2
\ee
\end{enumerate}
\et
Proof: Define the $p$-dimensional representation $W$ as follows: Let
$v$ be a (formal) element and define
\begin{eqnarray}
H\cdot v & = & jv \nonumber \\
E\cdot v & = & 0  \nonumber \\
C\cdot v & = & cv \nonumber
\end{eqnarray}
Define the vectorspace $W$ as the span of the set
\be
\{ v_k \equiv F^k \cdot v \}_{k=0,1,2,...} \nonumber
\ee
$W$ is then a representation of $W_3^{(2)}$ and by the previous
lemma we find
\begin{eqnarray}
H\cdot v_k & = & (j-2k)v_k  \nonumber \\
E\cdot v_k & = & (k(j^2+c)-2k(k-1)j+\frac{2}{3}k(k-1)(2k-1))v_{k-1}
\nonumber \\
C\cdot v_k & = & cv_k \nonumber
\end{eqnarray}
Since we are looking for finite dimensional representations there
must be a positive integer $p$ such that $F^pv=0$. From
\begin{eqnarray}
0 & = & EF^p \cdot v \nonumber \\
  & = & (pj^2-2p(p-1)j+pc+\frac{2}{3}p(p-1)(2p-1))v_{p-1} \nonumber
\end{eqnarray}
follows that
\be
pj^2-2p(p-1)j+pc+\frac{2}{3}p(p-1)(2p-1)=0   \nonumber
\ee
Solving this equation for $j$ and demanding that it be real we find
that
\begin{eqnarray}
c(p,x) & = & \frac{1}{3}(1-p^2)-x^2  \nonumber \\
j(p,x) & = & p+x-1 \nonumber
\end{eqnarray}
where $x$ is an arbitrary real number. This proves the theorem
\vspace{6mm}.

So we get infinitely many highest weight representations of a
given dimension.
Not all these representations are irreducible however. In fact there are
$(p-1)$ reducible representations of dimension $p$.
This is the
subject of the next theorem.
\bt
Let $k \in \{1, ... ,p-1\}$ then
\begin{enumerate}
\item $W(p;\frac{2}{3}k-\frac{1}{3}p)$
has a $p-k$ dimensional sub-representation isomorphic to \\
$\; W(p-k;-\frac{1}{3}(k + p))$.
\item The quotient representation is isomorphic to $W(k;\frac{2}{3}p
-\frac{1}{3}k)$, i.e.
\be
W(p;\frac{2}{3}k-\frac{1}{3}p)/W(p-k;-\frac{1}{3}(k+p)) \simeq
W(k;\frac{2}{3}p-\frac{1}{3}k)
\ee
\end{enumerate}
\et
Proof: Consider the representation $W(p,x)$. It might happen that
$E.v_k=0$ for some $k<p$. The subspace of $W(p,x)$ spanned by
$F^l.v_k$ is then an invariant subspace. Let us now consider when
this happens. Again we have to solve the equation $EF^k.v=0$. This
leads (as in the proof of the previous theorem) to the equation
\be
k(j^2+c-2(k-1)j+\frac{2}{3}(k-1)(2k-1))=0 \nonumber
\ee
This equation has three solutions for $k$, namely two we already
know
\begin{eqnarray}
k & = & 0 \nonumber \\
k & = & p \nonumber
\end{eqnarray}
and a new one
\be
k=\frac{1}{2}p+\frac{3}{2}x  \nonumber
\ee
This means that if $p$ and $x$ are such that $\frac{1}{2}p+
\frac{3}{2}x$ is an integer between 1 and $p-1$ then
\be
E.v_{\frac{1}{2}p+\frac{3}{2}x}=0 \nonumber
\ee
The sub-representation that can be built out of $v_{\frac{1}{2}p+
\frac{3}{2}x}$ is obviously a $p-k$ dimensional highest weight
representation of dimension $p-k$. It is therefore equal to
$W(p-k;\tilde{x})$ for some $\tilde{x}$. Since $C$ acts on the
sub-representation $W(p-k;\tilde{x})$ as the same multiple of
unity as on $W(p;\frac{2}{3}k-\frac{1}{3}p)$
and since $H(F^kv)=j(p;\frac{2}{3}k-\frac{1}{3}p)-2k$
we see that in order
to determine $\tilde{x}$ we have to solve the equations
\begin{eqnarray}
c(p;\frac{2}{3}k-\frac{1}{3}p) & = & c(p-k;\tilde{x}) \nonumber \\
j(p;\frac{2}{3}k-\frac{1}{3}p)-2k & = & j(p-k;\tilde{x}) \nonumber
\end{eqnarray}
Inserting the expression for $c(p,x)$ and $j(p,x)$
given in the previous
theorem we find $\tilde{x}=-\frac{1}{3}(k+p)$.
For the second part a similar argument shows that we have
to solve $\tilde{x}$ from the equations
\begin{eqnarray}
c(p;\frac{2}{3}k-\frac{1}{3}p) & = & c(k;\tilde{x}) \nonumber \\
j(p;\frac{2}{3}k-\frac{1}{3}p) & = & j(k;\tilde{x}) \nonumber
\end{eqnarray}
Again using the explicit expressions for $c$ and $j$ we find
$\tilde{x}=\frac{2}{3}p-\frac{1}{3}k$. This proves the theorem
\vspace{7mm}.

The most important representations from a physical point of
view are the unitary ones. The existence of these unitary
representations is our next subject.

Define an anti-involution $\omega$ on $W_3^{(2)}$ by
\begin{eqnarray}
\omega (E) & = & F \nonumber \\
\omega (F) & = & E \nonumber \\
\omega (H) & = & H \nonumber \\
\omega (C) & = & C
\end{eqnarray}
It is easy to check that $\omega$ is an algebra endomorphism, i.e.
that it preserves the relations of $W^{(2)}_3$.
\bl
For all positive integers $p$ and real numbers $x$ the representation
$W(p;x)$ carries a unique bilinear symmetric form $\langle .,.
\rangle$ such that
\begin{enumerate}
\item $\langle v,v \rangle =1$
\item $\langle .,.\rangle$ is contravariant w.r.t. $\omega$, i.e.
$\langle A \cdot v_l,v_k \rangle = \langle v_l,\omega (A)\cdot v_k
\rangle$ for all $A \in \bar{W}^{(2)}_3$.
\end{enumerate}
\el
Proof: Define the bilinear symmetric form $\langle.,.\rangle$
by putting
$\langle v,v \rangle =1$ and
\be
\langle v_l,v_k \rangle =\delta_{lk}\prod_{i=1}^{k}A(i)
\ee
for $l,k \in \{0,1,...,p-1\}$, where
\be
A(k)=k(j^2+c)-2k(k-1)j+\frac{2}{3}k(k-1)2k-1) \nonumber
\ee
We have to check contravariance. Obviously it suffices to
check this for the generators. For example take $A=F$, then
\be
\langle F \cdot v_l , v_k \rangle =\delta_{l+1,k}\prod_{i=1}^{k}
A(i) \nonumber
\ee
On the other hand $\langle v_l, \omega (F)\cdot v_k \rangle =
\langle v_l, E \cdot v_k \rangle$ which by lemma 1 is equal to
\begin{eqnarray}
\langle v_l,E \cdot v_k \rangle & = & \langle v_l,v_{k-1} \rangle
A(k) \nonumber \\
& = & \delta_{l+1,k}\prod_{i=1}^{k-1}A(i)A(k) \nonumber \\
& = & \delta_{l+1,k}\prod_{i=1}^{k}A(i) \nonumber
\end{eqnarray}
which proves contravariance for $F$. The proof for the other
generators is similar. Uniqueness follows by
induction on $k$ \vspace{7mm}.

So the representations $W(p;x)$ all carry bilinear symmetric
forms such that the basis vectors $v_k$ are orthogonal. This
is not enough for unitarity however for we still need to check if
the bilinear form is positive definite. This is the subject of
the next theorem.
\bt
The representation $W(p,x)$ is unitary if and only if
\be
x>\frac{1}{3}p-\frac{2}{3}
\ee
\et
Proof: We have to find out which
$W(p,x)$ have the property that the vectors $v_l$ all
have positive norm $\langle v_l , v_l \rangle >0$
for $l=1,...,p-1$. Now, by definition
\be
\langle v_l , v_l \rangle  =
\prod_{k=1}^{l} (k(j^2 +c)-2k(k-1)j+\frac{2}{3}k(k-1)(2k-1))
\nonumber
\ee
For unitarity we have to demand that this is $>0$ for all
$l=1,2,...,p-1$. This can only be the case if $A(k)>0$ for all
$k=1,...,p-1$ separately. Inserting the expressions for
$j(p;x)$ and $c(p;x)$ into this equation we find that it reduces
to
\be
x>\frac{2}{3}k-\frac{1}{3}p \nonumber
\ee
This must be true for all $k=1,...,p-1$ which is only the case if
\be
x>\frac{1}{3}p-\frac{2}{3} \nonumber
\ee
This proves the theorem \vspace{6mm}.

\begin{figure}[htbp]
\vspace{3.3cm}
\begin{picture}(100,55)(-81,0)
\setlength{\unitlength}{1mm}
\put (50,5){\line (0,1){52}}
\put (1,5){\line (1,0){98}}
\put (20,5){\line (0,1){1}}
\put (80,5){\line (0,1){1}}
\put (96,0){x}
\put (46,53){p}
\put (20,0){-1}
\put (80,0){1}
\put (46,13){2}
\put (46,23){3}
\put (46,33){4}
\put (46,43){5}
\put (50,15){\circle*{1.5}}
\put (40,25){\circle*{1.5}}
\put (60,25){\circle*{1.5}}
\put (30,35){\circle*{1.5}}
\put (50,35){\circle*{1.5}}
\put (70,35){\circle*{1.5}}
\put (20,45){\circle*{1.5}}
\put (40,45){\circle*{1.5}}
\put (60,45){\circle*{1.5}}
\put (80,45){\circle*{1.5}}
\put (70,25){unitary}
\put (5,25){non-unitary}
\end{picture}
\caption{The location of the reducible (depicted by dots)
and unitary representations.}
\end{figure}
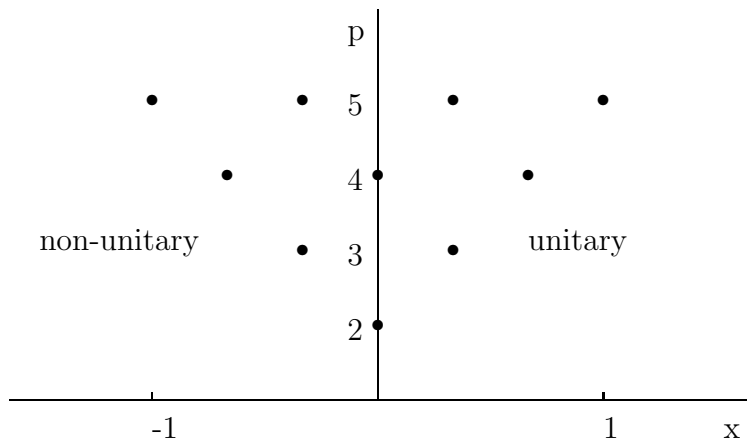

Note that from theorems 2 and 3 follows that the quotient
representation of a reducible rep w.r.t. its invariant
subspace is a unitary irreducible representation. This concludes
our presentation of the representation theory of $\bar{W}^{(2)}_3$.

\subsection{Induced representions}
In the previous section we constructed the representation theory of
$\bar{W}_3^{(2)}$ in a way that is in principle completely analogous
to the way in which one constructs the representation theory of $sl_2$.
This of course was possible due to the special structure of this algebra.
However, one look at the finite $W$ algebras listed in the appendix of
this chapter reveals that such an analysis would be extremely cumbersome
for other finite $W$ algebras because of their very complicated (nonlinear)
structure. Also one would hardly be able to make any general statements
which means that one would have to construct the representation theory case
by case.

In the previous section however we did not use one crucial piece of
information, namely that finite $W$ algebras are reductions of
simple Lie algebras. It might by possible to let representations of
these Lie algebras 'descend' to finite $W$ representations. Let
us explain how this works in general.
Let $g$ again be a simple Lie algebra. We have seen that finite $W$
algebras are the zeroth cohomology of the complex $(\Omega;d)$,
where $\Omega={\cal U}g \otimes Cl(g_+ \oplus g_+^*)$, $d(X)=[Q,X]$
and $Q$ is given by (\ref{def:q}). The Clifford algebra
$Cl(g_+ \oplus g_+^*)$ is finite dimensional and has a unique irreducible
module given by
\be
L_{gh} = Cl(g_+ \oplus g_+^*) / Cl(g_+ \oplus g_+^*).g_+
\ee
Essentially we can describe $L_{gh}$ as follows: it is spanned by
\be
\{ [c_{\alpha_{i_1}}\ldots c_{\alpha_{i_k}}]\}_{\alpha_{i_1} \leq \ldots
\leq \alpha_{i_k}}
\ee
where we have defined some ordening on the basis $\{t_{\alpha}\}$ of
$g_+$. Obviously the dimension of $L_{gh}$ is $2^{|g_+|}$, where
$|g_+|=\mbox{dim}(g_+)$.
We can introduce a grading on  $L_{gh}$ by putting
\be
\mbox{deg}([c_{\alpha_{i_1}} \ldots c_{\alpha_{i_k}}])=k
\ee
so we get
\be
L_{gh}=\bigoplus_{k\geq 0}L_{gh}^{(k)}
\ee
We find therefore that any representation $L_{\Lambda}$ of $g$ induces
a unique representation $L_{\Lambda}(\Omega )=L_{\Lambda} \otimes L_{gh}$
of $\Omega$ and furthermore that this representation is graded
\be
L_{\Lambda }(\Omega )= \bigoplus_{k \geq 0} L_{\Lambda}(\Omega )^{(k)}
\ee
where $L_{\Lambda}(\Omega )^{(k)} =L_{\Lambda} \otimes L_{gh}^{(k)}$.

Now the point is that $Q$ turns $L_{\Lambda}(\Omega )$ into a complex
\be
Q: L_{\Lambda}(\Omega )^{(n)} \longrightarrow L_{\Lambda}(\Omega )^{(n+1)}
\ee
and that the zeroth cohomology $H^0(\Omega;d)$ of the complex $(\Omega;d)$
has a natural action on the cohomology spaces $H^n(L_{\Lambda}(\Omega );Q)$
that turns them into $H^0(\Omega ;d)$ representations.
We can describe this action as follows: Let $[x] \in H^0(\Omega;d)$ and
$[v] \in H^n(L_{\Lambda}(\Omega );Q)$ then $[x].[v]\equiv [x.v]$. It is
a standard excercise to verify that this definition does not depend on
the representatives $x$ and $v$ of $[x]$ and $[v]$ respectively, and
also that $x.v \in \mbox{ker}(Q) \cap L_{\Lambda}(\Omega )^{(n)}$
if $v \in \mbox{ker}(Q) \cap L_{\Lambda}(\Omega )^{(n)}$.
Now, since $H^0(\Omega ;d)$
is nothing but the finite $W$ algebra in question we conclude that the
spaces $H^n(L_{\Lambda}(\Omega );Q)$ are finite $W$ representations.
This provides us with a functor from the category of $g$ representations
to the category of representations of the finite $W$ algebra in question.

Even though  this procedure is quite general and elegant it is
not the one that we shall persue. The reason for this is that it is
rather nontrivial to find $g$ representations $L_{\Lambda}$ that reduce
to interesting finite $W$ representations. Let us illustrate this in the
example of $W_3^{(2)}$. Again choosing the basis $\{t_a\}$ as specified by
(\ref{basis21}) $Q$ reads
\be
Q=J^3c_3+(J^1-1)c_1
\ee
Also $L_{gh}^{(0)}={\ce}$, $L_{gh}^{(1)}={\ce}[c_1] \oplus
{\ce}[c_3]$ and $L_{gh}^{(2)}={\ce}[c_1c_3]$. Let us take $L_{\Lambda}$
to be the fundamental representation  of $sl_3$, then one can easily verify
that $H^*(L_{\Lambda}(\Omega );Q)=0$ which means that the fundamental
representation of $sl_3$ reduces to the trivial representation of
$W_3^{(2)}$. This in fact holds true for all irreducible representation
$L_{\Lambda}$ of $sl_3$ which is rather disappointing. In view of this
we shall take a slightly different approach towards the construction
of the representation
theory of finite $W$ algebras.  In essence what we will show is that
there exists an injective algebra homomorphism from any finite $W$
algebra to the universal enveloping algebra of the grade 0 piece
$g_0$ of $g$ (of course this is a quantum version of the Miura map
presented earlier). This means that any $g_0$ representation induces
a representation of the finite $W$ algebra. Since $g_0$ is just a
semisimple Lie algebra its representation theory is well known.

Let us now make all this more precise. Denote by
$X^{0,0}$ the component of an element $X\in \Omega_{red}$ of
bidegree $(0,0)$, so that $W^{0,0}(\phi)=(-1)^p\phi_p$, then
\be
[W^{0,0}(\phi),W^{0,0}(\phi')]=[W(\phi),W(\phi')]^{0,0}
\ee
and
therefore $W(\phi) \rightarrow W(\phi)^{0,0}$ is a homomorphism of
algebras. We now have the following important theorem \cite{BT} which is a
quantum version of the generalized Miura map.
\bt \label{qmiura} (Quantum Miura Transformation)
The map $W(\phi) \rightarrow W(\phi)^{0,0}$, or, equivalently,
the map $H^0(\Omega;d) \rightarrow H^0(\Omega;d)^{0,0}$, is an
isomorphism of algebras.
\et
Proof: Now that we know the cohomology of $(\Omega;d)$, let us
go back to the original double complex $(\Omega_{red};d_0;d_1)$.
Because there is only cohomology of degree $0$, we know that the
$E^{p,q}_{\infty}$ term of the spectral sequence associated to
$(\Omega_{red};d_0;d_1)$ must vanish unless $p+q=0$. If we look
at $d_0(\hj^{\bar{\alpha}})=f^{\beta\bar{\alpha}}_{\,\,\,
\bar{\gamma}}\hj^{\bar{\gamma}}c_{\beta}$, we see that
$d_0(\hj^{\bar{\alpha}})=0$ if and only if $\bar{\alpha}\in
g_0^{\ast}$. From this it is not difficult, repeating arguments
similar as those in the proof of theorem \ref{final},
to prove that the only nonvanishing piece of $\oplus_r
E^{r,-r}_1$ is in $E^{0,0}_1$. This implies that
$E^{p,q}_{\infty}$ is only nonzero for $p=q=0$. Because
$H^0(\Omega;d)=E^{0,0}_{\infty}=H^0(\Omega;d)^{0,0}$ as vector
spaces, it follows that the map $H^0(\Omega;d)\rightarrow
H^0(\Omega;d)^{0,0}$ can have no kernel, and is an isomorphism.

The quantum Miura map is an injective algebra homomorphism into the
subalgebra $\Omega_{red}^0$
of $\Omega_{red}$ generated by $\{J^{\bar{\alpha}}\}_{
t_{\bar{\alpha}}\in g_0}$. Since
\be
[\hat{J}^{\bar{\alpha}},\hat{J}^{\bar{\beta}}]=f^{\bar{\alpha}
\bar{\beta}}_{\bar{\gamma}} \hat{J}^{\bar{\gamma}}
\ee
we find that $\Omega^0_{red}$ is isomorphic as an algebra to
${\cal U}g_0$. We have therefore indeed found the result that any
finite $W$ algebra can be embedded into ${\cal U}g_0$.

Since ${\cal U}(g_0)$
is abelian for the principal $sl_2$ embeddings, this
implies that in those cases the quantum finite $W$ algebras are
also abelian, something which was already proven by Kostant \cite{kost2}.
To get some interesting novel structure, one should therefore
consider nonprincipal $sl_2$
embeddings.

Again let us consider the example of $3=2+1$.
The expressions for the quantum Miura transformation of this
algebra
are obtained from (\ref{reps21}) by restricting these
expressions to the bidegree $(0,0)$ part. If we introduce
$s=(\hj^4+3\hj^5)/4$, $h=(\hj^5-\hj^4)/4$, $f=2\hj^6$ and
$e=\hj^2/2$, then $h,e,f$ form an $sl_2$ Lie
algebra, $[h,e]=e$ and $[h,f]=-f$ and $[e,f]=2h$ while $s$ commutes with
everything.
In terms of $s$ and $h,e,f$, the quantum Miura
transformation reads
\ba \label{miura21}
C & = & -\deel{4}{3}(h^2+\frac{1}{2}ef+\frac{1}{2}fe)-\deel{4}{9}s^2+
\deel{4}{3}s-1, \nonu
H & = & 2h-\deel{2}{3}s+1, \nonu
E & = & -2(s-h-1)e, \nonu
F & = & \deel{2}{3}f.
\ea
Notice that $C$ contains the second Casimir of $sl_2$, which is
what one would expect because it commutes with everything.
Every $g_0=sl_2\oplus u(1)$ module gives, using
the expressions (\ref{miura21}), a module for the finite quantum
algebra $W_3^{(2)}$.  So if we have a representation of $sl_2 \oplus u(1)$
in terms of $n \times n$ matrices, we immediately get a representation
of $\bar{W}^{(2)}_3$ in terms of $n \times n$ matrices.

An irreducible representation of $sl_2 \oplus u(1)$ is characterized
by two numbers: the 'spin' $j$ of the $sl_2$ representation and a
$u(1)$ 'charge' $s$. It is then easy to verify that the $(j,s)$
representation of $sl_2 \oplus u(1)$ gives rise to the
representation $W(2j+1; 1-\frac{2}{3}s)$ of $W_3^{(2)}$. Note also
that the map from $sl_2 \oplus u(1)$ representations to $W_3^{(2)}$
representations is surjective.

In the appendix we give explicit formulas for the quantum Miura
transformations associated to the finite $W$ algebras that can be
obtained from $sl_4$.

\subsection{Fock realizations}
Using the quantum Miura transformation we can turn any Fock realization
of $g_0$ into a Fock realization of the corresponding finite $W$
algebra. In this section we briefly discuss  how we can obtain
Fock realizations for simple Lie algebras.
We will then show in the
example of $W_3^{(2)}$ that its finite dimensional representations
can be realized as submodules of certain Fock modules.

As usual define an oscillator algebra $O$ to be an
algebra generated by operators $\{a_i,a^{\dagger}_i\}_{i=1}^{n}$
with the following commutation relations
\be
[a_i,a_j]=[a_i^{\dagger},a_j^{\dagger}]=0 \;\; ; \;\;
[a_i,a_j^{\dagger}]=\delta_{ij}
\ee
In physics terminology this algebra describes a collection of $n$
uncoupled harmonic oscillators.
The Hilbert space
of such a system
is isomorphic to the space
${\ce}[z_1, \ldots ,z_n]$ of polynomials in $n$ variables,
and (the quantum operators) $a_i$ and $a_j^{\dagger}$ can be
realized as
\be
a_i=\frac{\partial}{\partial z_i}\;\;\;\; \mbox{and} \;\;\;\;
a_i^{\dagger}=z_i
\ee
This is called the 'Fock realization' of the (quantum) harmonic
oscillator algebra.
Since any polynomial $P$ can be reduced to a multiple of 1 by successive
application of $\{a_i\}$ we can construct any polynomial $Q$ out
of $P$ by the action of the universal enveloping algebra of the
oscillator algebra. This means that the Fock realization is irreducible.
A Fock realization of a Lie algebra $g$ is an injective homomorphism from
$g$ to the universal enveloping algebra of the oscillator
algebra.

As we shall need this later
we now discuss how to obtain a Fock realization for an arbitrary
simple Lie algebra $g$
(see for details \cite{Kostant}).

Let $G$ be a simple Lie group with Lie algebra $g$. Furthermore
let $B_-$ be its lower Borel subgroup and $N_+$ the group
associated to the positive roots. Consider the coset space
\be
B_-\backslash G=\{B_-g \mid g\in G\}
\ee
The triple $(G,B_-\backslash G,B_-)$ is a principal bundle with
total space $G$, basespace $B_-\backslash G$ and structure group
$B_-$. Let now $\chi:B_- \rightarrow \ce$ be a 1-dimensional
representation  (a character) of the Lie group $B_-$, then we can
construct the associated complex line bundle ${ \cal L}_{\chi}$ over
$B_-\backslash G$
\be
{ \cal L}_{\chi}=G\times {\ce}/ \sim
\ee
where $(b.g,\chi (b^{-1})z) \sim (g,z)$ for all $b \in B_-$.
Denote the set of holomorphic global sections of this bundle by
$\Gamma_{\chi}$. The group $G$ acts from the right on ${\cal L}_{\chi}$
as follows
\be
[(g,z)]\cdot g' =[(gg',z)]
\ee
and this induces of course an action of $G$ (and therefore of $g$)
on $\Gamma_{\chi}$.

Now let $H$ and $b_-$ be the Cartan subalgebra and
negative Borel subalgebra of $g$ respectively. The derivative
of $\chi$ in the unit element
\be
\phi_{\chi}(b)=\frac{d}{dt}\chi (e^{tx})|_{t=0} \;\;\;\;\; x \in g
\ee
is a one dimensional representation of $b_-$. However
this implies that
\be
0=[\phi_{\chi}(h),\phi_{\chi}(e_{-\alpha})]=
-\langle \alpha , h \rangle \phi_{\chi} (e_{\alpha})
\ee
where
$h \in H$ and $e_{-\alpha}$ is a rootvector of $g$, which means
that $\phi_{\chi}(e_{-\alpha})=0$ for all $-\alpha \in \Delta_+$.
Therefore $\phi_{\chi}$ determines really an element $\Lambda_{\chi}$
of $H^*$. It is also clear that all elements of $H^*$ can be
obtained in this way.

It was already said above that $\Gamma_{\chi}$ is a representation
space of $g$. In fact the Borel-Weil
theorem states that for
$\Lambda_{\chi}$ an dominant weight of $g$ the space
of holomorphic sections $\Gamma_{\chi}$ of ${\cal L}_{\chi}$ is isomorphic
to the
finite dimensional irreducible representation of $g$ with highest weight
$\Lambda_{\chi}$. If $\Lambda_{\chi}$ is not a dominant weight
then the complex line bundle ${\cal L}_{\chi}$ has no non-trivial
holomorphic
sections.

The construction that leads to the Fock realization of the Lie
algebra $g$ is very similar to the Borel-Weil construction. First
one makes use of the fact that $B_-\backslash G$ possesses a
canonical decomposition, called the Bruhat decomposition
\be
B_-\backslash G= \bigcup_{w \in W} \; C_w
\ee
where $W$ is the Weyl group of $g$. The spaces $C_w$ are called
'Schubert cells' and are isomorphic to ${\ce}^n$. Denote
from now on the Schubert cel corresponding to the unit element
of $W$ by $Y$. Since $Y \subset B_-\backslash G$
the group $G$ acts on $Y$ by right multiplication.
It can also be shown that $Y$ is diffeomorphic to $N_+$. This
diffeomorphism is defined by assigning to $[g] \in Y$ the unique
element $x \in N_+$ such that $[x]=[g]$ (i.e. such that
$x=bg$ for some $b \in B_-$). The right $G$-action on $Y$ induces a right
$G$-action on $N_+$ via this diffeomorphism. Let's from now
on  identify $Y$ and $N_+$.

The set $B_-N_+$ is a submanifold of $G$ and the triple $(B_-N_+,Y,B_-)$
is again a principle bundle (in fact it is nothing but the restriction
of the bundle $(G,B_-\backslash G,B_-)$ to the cell $Y\subset B_-\backslash
G$. Let again $\chi$ be a character of $B_-$ and let $\bar{{\cal L}}_{\chi}$
be the associated complex line bundle over $Y$. Denote the space of
sections of this line bundle by $\bar{\Gamma}_{\chi}$. It is a standard
result (and easy to prove) that there exists a 1-1 correspondence between
$\bar{\Gamma}_{\chi}$ and the set $\bar{R}_{\chi}$ of
holomorphic functions on $B_-N_+$ such that
\be
f(b.g)=\chi^{-1}(b)f(g)    \label{prop}
\ee
where $b\in B_-$ and $g \in N_+$. Since obviously the space
of holomorphic functions on $N_+$
and $\bar{R}_{\chi}$ are isomorphic (a holomorphic
function $f$ on $N_+$ uniquely determines a holomorphic function $\hat{f}$ on
$B_-N_+$ with the property  (\ref{prop}) if we set $\hat{f}(b.g)=
\chi^{-1}(b)f(g)$) we find that
\be
\bar{\Gamma}_{\chi}\simeq Hol(Y)
\ee
In \cite{Kostant} Kostant showed that there exists a representation
of the Lie algebra $g$ of $G$ in terms of first order smooth differential
operators on the manifold $Y$. Since the space $Y$
is isomorphic to ${\ce}^{|\Delta_+|}$ where $|\Delta_+|$ is the
number of positive roots of $g$ we can choose a set
of global coordinates $z\equiv \{z_{\alpha}\}_{\alpha \in \Delta_+}$ on $Y$.
The representation
\be
\sigma_{\Lambda}:g \rightarrow \mbox{Diff}(Y)
\ee
of $g$ on the set of differential operators
$Diff(Y)$ on $Y$
is then given by
\be
\sigma_{\Lambda}(x)=\xi (x)+h_{\Lambda }(x) \;\;\;\; x\in g
\ee
where $\xi (x)$ is a Killing vector of the right $G$-action on $Y$
\be
\left( \xi (x)f \right) (z)=\frac{d}{dt}f(z.e^{tx})|_{t=0}
\;\;\;\; z \in Y
\ee
and
\be
\left( h_{\Lambda}(x)f \right)(z)=\langle \Lambda_* , Ad_z x \rangle
f(z) \label{hlab}
\ee
Here $\Lambda_*$ is the trivial extension of $\Lambda \in H^*$
to $g$
and $f$ is some holomorphic function of $Y$.
Note that in (\ref{hlab}) we made use of the fact we have
identified $Y$ and $N_+$ or else $Ad_z$ would not make sense.

The subspace ${\ce}[\{z_{\alpha}\}]$ of $Hol(Y)$ is
obviously invariant  under  $\sigma_{\Lambda}(g)$ which means
that we have obtained Fock realizations of the Lie algebra $g$
(one for each $\Lambda \in H^*$). Let us denote by $F_{\Lambda}$
the pair $({\ce}[\{z_{\alpha}\}],\sigma_{\lambda})$.

For $sl_2$ the Fock realizations $\sigma_{\Lambda}$ have the form
\begin{eqnarray}
\sigma_{\Lambda}(e) & = & \frac{d}{dz} \nonumber \\
\sigma_{\Lambda}(f) & = & (\Lambda,\alpha )z-z^2\frac{d}{dz} \nonumber \\
\sigma_{\Lambda}(h) & = & \frac{1}{2}(\Lambda ,\alpha )-z\frac{d}{dz}
\label{F}
\end{eqnarray}
where $\alpha$ is the root of $sl_2$. The expressions for arbitrary
$sl_n$ can be found in \cite{BMP} and will not be given here.

Using the results described above one is always able to immediately
write down a Fock realization of the algebra $g_0$ (since it is
essentially a direct sum of $sl_k$ algebras). Then using the quantum
Miura transformation one thus arrives at a Fock realization of the
finite $W$ algebra in question. Let us now explicitly do this in
the example of $\bar{W}_3^{(2)}$. Inserting the expressions
(\ref{F}) into the Miura map (\ref{miura21}) one finds
\begin{eqnarray}
\sigma_{\Lambda ,s}(H) & = & (\Lambda,\alpha )-\frac{2}{3}s+1-2z\frac{d}{dz}
\nonumber \\
\sigma_{\Lambda ,s}(E) & = & 2(1-s+\frac{1}{2}(\Lambda ,\alpha ))
\frac{d}{dz}-2z\frac{d^2}{dz^2} \nonumber \\
\sigma_{\Lambda ,s}(F) & = & \frac{2}{3}(\Lambda,\alpha )z-\frac{2}{3}z^2
\frac{d}{dz} \nonumber \\
\sigma_{\Lambda ,s}(C) & = & -\frac{1}{3}(\Lambda,\alpha )^2-
\frac{2}{3}(\Lambda,\alpha )-\frac{4}{9}s^2+\frac{4}{3}s-1 \label{FW}
\end{eqnarray}
(where we consider $s$ to be a number). This realization is equal to
the zero mode structure of the free field realization of the infinite
$W_3^{(2)}$ algebra constructed in \cite{B}. Note however that the
derivation is completely different since in \cite{B} (using standard
methods) the expressions in terms of free fields were obtained by
constructing the generators of the commutant of certain screening
charges. Constructing this commutant is in general however rather
cumbersome. The method we presented above is more direct and
works for arbitrary embeddings (and realizations).

The Fock realizations of $g$ are not all irreducible. In fact all
finite dimensional irreducible representations of simple Lie
algebras can be realized as submodules fo certain Fock realizations.
Let us briefly describe how this works before we show that the
same statement is true for finite $W$ algebras.

Suppose $D \in Diff(Y)$ is a differential operator such that
$[\sigma_{\Lambda}(x),D]=0$ for all $x \in g$ then the subspace
$V=\{v \in F_{\Lambda} \mid Dv=0\}$ of $F_{\Lambda}$ is an invariant
subspace, i.e. $\sigma_{\Lambda}(g) V \subset V$. A `complete'
set of such operators can be constructed as follows: Let $x$ be an
element of $n_+$, then define the differential operator $\rho (x)$
by
\be
(\rho (x)f)(z) = \frac{d}{dt}f(e^{-tx}z)|_{t=0}
\ee
where $f \in C^{\infty}(Y)$. Here we again used the isomorphism
$Y \simeq N_+$ in order to define the left action of $N_+$ on $Y$.
Now let $\{e_i\}_{i=1}^l$ be the simple rootvectors of $g$ corresponding
to the simple roots $\{\alpha_i\}_{i-1}^l$, then  the differential
operators
\be
\rho (e_i)^{( \Lambda , \alpha_i )}
\ee
(called intertwiners) commute with $\sigma_{\Lambda}(x)$ for all $x \in g$
if $\Lambda$ is an integral dominant weight (i.e. if $( \Lambda ,
\alpha_i)$ is a positive integer). Furthermore the submodule
\be
V=\{v \in F_{\Lambda} \mid \rho (e_i)^{( \Lambda ,\alpha_i
) +1}v=0, \mbox{ for all }i\}  \label{findim}
\ee
is isomorphic to the (unique) finite dimensional irreducible $g$
representation $L_{\Lambda}$ with highest weight $\Lambda$. The
operators $s_i\equiv \rho (e_i)$ are usually called `sreening charges'
in physics literature.

Let us illustrate all this with an example. For $sl_2$ we have
$\rho (e)=-\frac{d}{dz}$. It is easy to check that for $( \Lambda,
\alpha ) =m \in {\bf Z}_+$
the space $V=\{v \in {\ce}[z] \mid
(\frac{d}{dz})^{m+1}v=0\}$ carries the $m+1$ dimensional irreducible
multiplet of $sl_2$.

Since the intertwiners commute with all elements of ${\cal U}g_0$
they also commute with the Fock representants of the finite $W$
algebra in question. Therefore the finite dimensional spaces
(\ref{findim}) are also finite dimensional representation spaces
of the finite $W$ algebra. This provides us with Fock realizations
of the finite dimensional representations of $W$ algebras.

Let us again consider the case of $W_3^{(2)}$. Let $m=( \Lambda ,
\alpha )$ then the subspace $V=\{v \in {\ce}[z] \mid
(\frac{d}{dz})^{m+1}v=0\}$ of the Fock representation $F_{\Lambda ,s}=
({\ce}[z],\sigma_{\Lambda ,s})$ of $W_3^{(2)}$ then it is easy to see
that $(V,\sigma_{\Lambda ,s})$ is isomorphic as a $W_3^{(2)}$
representation to $W(m+1;1-\frac{2}{3}s)$.

\section{Relation with Lie algebra cohomology}
According to the general construction in this chapter one associates
a finite $W$ algebra to any triple $(g,g_+,\chi )$ where $g$ is a
simple Lie algebra, $g_+$ is the positive grade subspace w.r.t. some
grading element $\delta$ and $\chi$ is a certain one dimensional representation
(character) of $g_+$ (note that in principle one could consider the case
where $g_+$ and $\chi$ are {\em any} nilpotent subalgebra and character
respectively, not necessarily related to an $sl_2$ embedding. The
analysis given in the previous sections still remains valid without
alterations). We shall now briefly indicate the relation of these
constructions with Lie algebra cohomology.

Recall the definition of Lie algebra cohomology: Let $h$ be a Lie algebra
and let $(V, \phi )$ be a representation of $h$ (i.e. $V$ is a
vectorspace and $\phi$ is a Lie algebra homomorphism from $h$
to the set of linear transformations on $V$).
Consider the spaces
\be
C^p=\mbox{Hom}(\bigwedge\nolimits^ph,V)
\ee
(where of course $C^0=V$) and define
\be
C= \bigoplus_{p=0}^{\infty} C^p
\ee
The Lie algebra $h$ has a natural action on $C^p$ defined as follows:
Let $x\in h$ and $\alpha \in C^p$, then $x.\alpha \equiv \phi (x)
\circ \alpha \in C^p$. One then defines a coboundary operator
$d_p:C^p \rightarrow C^{p+1}$ by putting
\ba
(d_p\alpha )(x_1 \wedge \ldots \wedge x_{p+1}) & = & \sum_{j=1}^{p+1}
(-1)^{j+1}(x_j.\alpha )(x_1 \wedge \ldots \wedge \hat{x}_j \wedge
\ldots \wedge x_{p+1}) \nonumber \\
&   & + \sum_{1\leq i < j p+1} (-1)^{i+j} \alpha ([x_i,x_j]\wedge
\ldots \wedge \hat{x}_i \wedge \ldots \wedge \hat{x}_j \wedge \ldots
\wedge x_{p+1})     \nonumber
\ea
where $\alpha \in C^p$ and $x_i\in h$ for all $i=1, \ldots , p+1$.
It is now easy to check that $d_p \circ d_{p-1}=0$ for all $p$
and the cohomology derived from this complex
\be
H^p(h;V) = \frac{\mbox{Ker}(d_p:C^p\rightarrow C^{p+1})}{\mbox{Im}(d_{p-1}:
C^{p-1} \rightarrow C^p)}
\ee
is called the $p^{th}$ Lie algebra cohomology of $h$ with coefficients
in $V$.

We will now slightly reformulate the above to make contact with BRST
cohomology. For this note that
\be
\mbox{Hom}(\bigwedge\nolimits^ph,V) \simeq \bigwedge\nolimits^p h^* \otimes V
\ee
where the isomorphism can be described as follows:
Let $\alpha \in \mbox{Hom}(\bigwedge^ph,V)$ then
\be \label{liecoho}
\alpha \mapsto \sum_{i_1<\ldots i_p}< (e_{i_1}^* \wedge \ldots \wedge
e^*_{i_p})\otimes \alpha (e_{i_1}\wedge \ldots \wedge e_{i_p})
\ee
where $\{e_i\}$ and $\{e_i^*\}$ are dual bases of $h$ and $h^*$
respectively, i.e. $e_i^*(e_j)=\delta_{ij}$. It is clear that the
map (\ref{liecoho}) is independent of the chosen basis and that it
defines an isomorphism.

It is now straightforward to calculate the action of $d_p$ on the
element $e_{i_1}^* \wedge \ldots \wedge e_{i_p}^*)\otimes v \in
\bigwedge^ph\otimes V$. It reads
\ba \label{dact}
d((e^*_{i_1}\wedge \ldots \wedge e^*_{i_p})\otimes v) & = &
\hat{d}(e_{i_1}^*\wedge \ldots \wedge e^*_{i_p})\otimes v+ \nonumber \\
&   & + \sum_i (e_i^*\wedge e^*_{i_1} \wedge \ldots \wedge e^*_{i_p})
\otimes \phi (e_i)v
\ea
where the derivation
$\hat{d}:\bigwedge^ph^* \rightarrow \bigwedge^{p+1}h^*$ is determined
by
\be
\hat{d}(e^*_i) = -\frac{1}{2}\sum_{jk}C^i_{jk}e^*_j \wedge e^*_k
\ee
and $C^i_{jk}$ are the structure constants of $h$ in the basis
$\{e_i\}$.

Now, having chosen a certain basis $\{e_i\}$ in $h$ we can interpret
elements of $\bigwedge^ph$ as (formal) polynomials of anticommuting variables
$\{\theta_i\}$ with coefficients in $V$, i.e. if
\be
\alpha = \sum_{i_1< \ldots < i_p} e_{i_1}^* \wedge
\ldots \wedge e^*_{i_p}\otimes v^{i_1\ldots i_p}
\ee
(where $v^{i_1\ldots i_p} \in V$)
then the polynomial in terms of the Grassman variables $\theta_i$
associated to it is given by
\be
\alpha = \sum_{i_1 < \ldots < i_p} v^{i_1 \ldots i_p} \theta_{i_1}
\ldots \theta_{i_p}
\ee
The advantage of the description of Lie algebra cohomology in terms of
Grassman variables is that the operator
\be
d=\bigoplus_p d_p: C\rightarrow C
\ee
takes on the following simple form
\be
d=\sum_i\phi (e_i)  \theta_i-\frac{1}{2}\sum_{ijk}C^i_{jk}\theta_j\theta_k
\frac{\partial}{\partial \theta}
\ee
as can easy be checked using formula (\ref{dact}). Note that $\theta_i$
and $\frac{\partial}{\partial \theta_j}$ satisfy canonical anti
commutation relations. The relation between Lie algebra cohomology
and BRST cohomology is now clear: Letting $h=g_+$ and
denoting
\be
c_i=\theta_i \;\;\; \mbox{ and } \;\;\; b^i=\frac{\partial}{\partial
\theta_i}
\ee
we find that the Lie algebra cohomology of $g_+$ with coefficients
in the universal enveloping algebra of $g$ is equal to the
cohomology of the operator $d_1$ (in the notation of the previous sections).
Finite $W$ algebras therefore do not correspond precisely to Lie
algebra cohomology due to the fact that the BRST operator involves
another term, i.e. $d=d_0+d_1$.  At this point we can only speculate
as to the relevance of finite $W$ algebras to the theory of Lie algebras
but it may be that this role is an important one. In order to see this
recall that Lie algebra cohomology arises in the theory of Lie groups
as de Rham cohomology of left invariant vectorfields
and also through the theory of group extensions \cite{Knapp}. Finite
$W$ algebras may therefore correspond globally (i.e. on the group level)
with a `deformation' of de Rham theory. Whether this is true and
what this new geometric theory is is not clear at this time but we
will come back to this in future publications.

\section*{Conclusion}
We have seen in this chapter that the theory of $W$ algebras has a
natural finite counterpart. The classical theory, which is a theory
of Poisson reductions of the very important Kirillov Poisson structure,
lead to a set of Poisson algebras (eqn. (\ref{PA})) which are in
general nonlinear but finitely generated. In the second part of the
chapter we have seen that all these algebras can be quantized by the
BRST formalism. A completely algorithmic method has been presented
to construct the generators of the quantum finite $W$ algebra and to
calculate their relations. In the last section we have developed the
representation theory of quantum finite $W$ algebras. It was shown that
this can be done easily by the so called 'quantum Miura transformation'.
In this way it was also possible to obtain realizations of all finite
$W$ algebras in terms a finite collection of harmonic oscillators.

\section*{Appendix}
In this appendix we discuss the finite $W$ algebras that can be obtained
from $sl_4$. Using the standard constructions developed in this
chapter we calculate their relations and the quantum Miura maps.

\subsection*{$4= 2+ 1 + 1$}
The basis of $sl_4$ we use to study this quantum algebra is
\be \label{basis211}
r_at_a=\left(\begin{array}{cccc}
\frac{1}{2}r_{10}-\frac{1}{8}r_8 -\frac{1}{8}r_9 & r_{11} &
r_{12} & r_{15} \\
r_5 & \frac{3}{8}r_8-\frac{1}{8}r_9 & r_7 & r_{14} \\
r_4 & r_6 & -\frac{1}{8}r_8+\frac{3}{8}r_9 & r_{13} \\
r_1 & r_2 & r_3 &
-\frac{1}{2}r_{10}-\frac{1}{8}r_8 -\frac{1}{8}r_9
\end{array} \right).
\ee
The $sl_2$ embedding is given by $t_+=t_{15}$, $t_0=t_{10}$ and
$t_-=\hf t_1$. The nilpotent algebra $g_+$ is spanned by
$\{t_{13},t_{14},t_{15}\}$, $g_0$ by $\{t_4,\ldots,t_{12}\}$ and
$g_-$ by $\{t_1,t_2,t_3\}$. The $d_1$ cohomology of
$\Omega_{red}$ is generated by $\hj^a$ for $a=1, \ldots ,9$.
Representatives that are exactly $d$-closed, are given by
$W(\hj^a)=\hj^a$ for $a=4,\ldots,9$, and by
\ba \label{reps211}
W(\hj^1) & = & \hj^1 + \hj^4\hj^{12}+\hj^5\hj^{11}+\deel{1}{4}
\hj^{10}\hj^{10}+\deel{\hbar}{2}\hj^{10}, \nonu
W(\hj^2) & = & \hj^2 + \hj^6\hj^{12} + \hf\hj^8\hj^{11} +
\hf\hj^{10}\hj^{11}+\hbar\hj^{11}, \nonu
W(\hj^3) & = & \hj^3 + \hj^7\hj^{11} + \hf\hj^9\hj^{12} +
\hf\hj^{12}\hj^{10}+\deel{\hbar}{2}\hj^{12}.
\ea
Introduce a new basis of fields as follows
\ba \label{newgens}
U & = & \deel{1}{4} (\whj{8}+\whj{9}), \nonu
H & = & \deel{1}{4} (\whj{8}-\whj{9}), \nonu
F & = & -\whj{7}, \nonu
E & = & -\whj{6}, \nonu
G_1^- & = & -\whj{3},  \nonu
G_1^+ & = & \whj{2}, \nonu
G_2^- & = & \whj{5}, \nonu
G_2^+ & = & \whj{4}, \nonu
C & = & \whj{1}+\hf EF+\hf FE + H^2+\hf U^2 + 2\hbar U.
\ea
If we compute the commutators of these expressions, we find that
$C$ commutes with everything, $\{E,F,H\}$ form a $sl_2$
subalgebra and $G_i^{\pm}$ are spin $\hf$ representations for
this $sl_2$ subalgebra. $U$ represents an extra $u(1)$ charge.
The nonvanishing commutators, with $\hbar$ dependence,  are
\ba \label{alg211}
[E,F] & = & 2\hbar H , \nonu
[H,E] & = & \hbar E , \nonu
[H,F] & = & -\hbar F , \nonu
[U,G_1^{\pm}] & = & \hbar G_1^{\pm}, \nonu
[U,G_2^{\pm}] & = & -\hbar G_2^{\pm}, \nonu
[H,G_i^{\pm}] & = & \pm\deel{\hbar}{2} G_i^{\pm}, \nonu
[E,G_i^-] & = & \hbar G_i^+, \nonu
[F,G_i^+] & = & \hbar G_i^-, \nonu
[G_1^+,G_2^+] & = & -2\hbar E(U+\hbar), \nonu
[G_1^-,G_2^-] & = & 2\hbar F(U+\hbar), \nonu
[G_1^+,G_2^-] & = & \hbar(-C+EF+FE+2H^2+\deel{3}{2}U^2+2HU)+\hbar^2
(2H+3U), \nonu
[G_1^-,G_2^+] & = & \hbar(C-EF-FE-2H^2-\deel{3}{2}U^2+2HU)+\hbar^2
(2H-3U).
\ea
Let us also present the quantum Miura transformation for this
algebra. In this case, $g_0=sl_3\oplus u(1)$. Standard
generators of $g_0$ can be easily identified. A generator of
the $u(1)$ is $s=\hf\hj^8+\hf\hj^9+2\hj^{10}$, and the $sl_3$
generators are $e_1=\hj^5$, $e_2=\hj^6$, $e_3=\hj^4$,
$f_1=\hj^{11}$, $f_2=\hj^7$, $f_3=\hj^{12}$,
$h_1=-\hf\hj^8+\hf\hj^{10}$ and $h_2=\hf\hj^8-\hf\hj^9$.
The convention is such that the commutation relations between
$\{e_i,f_i,h_i\}$ are the same as those of corresponding
matrices defined by
\be \label{conv211}
a_i e_i + b_i f_i + c_i
h_i=\hbar\mats{c_1}{a_1}{a_3}{b_1}{c_2-c_1}{a_2}{b_3}{b_2}{-c_2}.
\ee
The quantum Miura transformation reads
\ba \label{miura211}
U & = & \deel{1}{6}(s-2h_2-4h_1), \nonu
H & = & \hf h_2, \nonu
F & = & -f_2, \nonu
E & = & -e_2, \nonu
G_1^- & =  & -f_2f_1-\deel{1}{3}(s-2h_2-h_1+3\hbar)f_3, \nonu
G_1^+ & = & e_2f_3+\deel{1}{3}(s+h_2-h_1+3\hbar)f_1, \nonu
G_2^- & = & e_1, \nonu
G_2^+ & = & e_3, \nonu
C & = & (\deel{1}{24}s^2+\deel{\hbar}{2}s) +
\hf(e_1f_1+f_1e_1+e_2f_2+f_2e_2+e_3f_3+f_3e_3) \nonu
& & +
\deel{1}{3} (h_1^2+h_1h_2+h_2^2).
\ea
In $C$ we again recognize the second casimir of $sl_3$. It is a
general feature of finite $W$ algebras that they contain a
central element $C$, whose Miura transform contains the second
casimir of $g_0$. $C$ is the finite counterpart of the energy
momentum tensor that every infinite $W$ algebra possesses.

\subsection*{$4= 2+2$}

A convenient basis to study this case is
\be \label{basis22}
r_at_a=\left(\begin{array}{cccc} \frac{r_6}{4}+\frac{r_{10}}{2}
& \frac{r_5}{2}+\frac{r_8}{2} & r_{12} & r_{14} \\
\frac{r_7}{2}+\frac{r_9}{2} & -\frac{r_6}{4}-\frac{r_{11}}{2} &
r_{13} & r_{15} \\ r_1 & r_2 & \frac{r_6}{4}-\frac{r_{10}}{2} &
\frac{r_5}{2}-\frac{r_8}{2} \\ r_3 & r_4 &
\frac{r_7}{2}-\frac{r_9}{2} & -\frac{r_6}{4}+\frac{r_{11}}{2}
\end{array} \right).
\ee
The $sl_2$ embedding is given by $t_+=t_{12}+t_{15}$,
$t_0=t_{10}-t_{11}$ and $t_-=\hf(t_1+t_4)$. The subalgebra $g_+$
is spanned by $\{t_{12},\ldots,t_{15}\}$, $g_0$ by
$\{t_5,\ldots,t_{11}\}$ and $g_-$ by $\{t_1,\ldots,t_4\}$. The
$d_1$ cohomology of $\Omega_{red}$ is generated by
$\hj^1,\ldots,\hj^7$. The $d$-closed representatives are
$W(\hj^a)=\hj^a$ for $a=5,6,7$, and
\ba \label{reps22}
W(\hj^1) & = & \hj^1-\hv\hj^5\hj^9+\hv\hj^7\hj^8+\hv\hj^8\hj^9+
 \hv\hj^{10}\hj^{10}+\deel{3\hbar}{4}\hj^{10}-\deel{\hbar}{4}
 \hj^{11}, \nonu
 W(\hj^2) & = &
 \hj^2+\hv\hj^5\hj^{10}+\hv\hj^5\hj^{11}-\hv\hj^6\hj^8+
 \hv\hj^8\hj^{10}-\hv\hj^{10}\hj^{11}+\deel{\hbar}{2}\hj^8,
 \nonu
 W(\hj^3) & = &
 \hj^3+\hv\hj^6\hj^9-\hv\hj^7\hj^{10}-\hv\hj^7\hj^{11}+
 \hv\hj^9\hj^{10}-\hv\hj^9\hj^{11}+\deel{\hbar}{2}\hj^9,
 \nonu
W(\hj^4) & = &
\hj^4+\hv\hj^5\hj^9-\hv\hj^7\hj^8+\hv\hj^8\hj^9+
\hv\hj^{11}\hj^{11}+\deel{\hbar}{4}\hj^{10}-
\deel{3\hbar}{4}\hj^{11}.
\ea
To display the properties of the algebra as clearly as possible,
we introduce a new basis of fields
\ba \label{newgens22}
H & = & \hf\whj{6}, \nonu
E & = & -\whj{7}, \nonu
F & = & -\whj{5}, \nonu
G^+ & = & \whj{3}, \nonu
G^0 & = & \whj{1}-\whj{4}, \nonu
G^- & = & -\whj{2}, \nonu
C & = & \whj{1}+\whj{4}+\deel{1}{8}\whj{6}\whj{6}+
 \hv\whj{5}\whj{7} \nonu
 & & +\hv\whj{7}\whj{5}+\deel{\hbar}{4}\whj{6}.
\ea
Here, $C$ is the by now familiar central element, $\{E,H,F\}$
form an $sl_2$ algebra and $\{G^+,G^0,G^-\}$ form a spin $1$
representation with respect to this $sl_2$ algebra. The
nonvanishing commutators are
\ba \label{alg22}
[E,F] & = & 2\hbar H, \nonu
[H,E] & = & \hbar E, \nonu
[H,F] & = & -\hbar F, \nonu
[E,G^0] & = & 2\hbar G^+, \nonu
[E,G^-] & = & \hbar G^0, \nonu
[F,G^+] & = & \hbar G^0, \nonu
[F,G^0] & = & 2\hbar G^-, \nonu
[H,G^+] & = & \hbar G^+, \nonu
[H,G^-] & = & -\hbar G^-, \nonu
[G^0,G^+] & = & \hbar(-CE+EH^2+\hf EEF+\hf EFE)-2\hbar^3 E,
\nonu
[G^0,G^-] & = & \hbar(-CF+FH^2+\hf FFE+\hf FEF)-2\hbar^3 F,
\nonu
[G^+,G^-] & = & \hbar(-CH+H^3+\hf HEF+\hf HFE)-2\hbar^3 H.
\ea
Since $g_0=sl_2\oplus sl_2 \oplus u(1)$, the quantum Miura
transformation expresses this algebra in term of generators
$\{e_1,h_1,f_1\}$, $\{e_2,h_2,f_2\}$, $s$ of $g_0$. The relation
between these generators and the $\hj^a$ are:
$s=\hj^{10}-\hj^{11}$, $h_1=\hf(\hj^6+\hj^{10}+\hj^{11})$,
$h_2=\hf(\hj^6-\hj^{10}-\hj^{11})$, $e_1=\hf(\hj^7+\hj^9)$,
$e_2=\hf(\hj^7-\hj^9)$, $f_1=\hf(\hj^5+\hj^8)$ and
$f_2=\hf(\hj^5-\hj^8)$. The commutation relations for these are
$[e_1,f_1]=\hbar h_1$, $[h_1,e_1]=2\hbar e_1$,
$[h_1,f_1]=-2\hbar f_1$, and similar for $\{e_2,h_2,f_2\}$.
For the quantum Miura transformation one
then finds
\ba \label{miura22}
H & = & \hf(h_1+h_2), \nonu
E & = & -e_1-e_2, \nonu
F & = & -f_1-f_2, \nonu
G^+ & = & \hf e_1h_2-\hf e_2h_1 +\hv s(e_1-e_2)+\hbar(e_1-e_2),
\nonu
G^0 & = & f_1e_2-f_2e_1+\hv s (h_1-h_2)+\hbar(h_1-h_2), \nonu
G^- & = & \hf f_1h_2 -\hf f_2 h_1 -\hv s (f_1-f_2)-\hbar
(f_1-f_2), \nonu
C & = & (\deel{1}{8}s^2+\hbar s) +
\hf(e_1f_1+f_1e_1+e_2f_2+f_2e_2) +
\hv(h_1^2 +h_2^2).
\ea
The infinite dimensional version of this algebra is one of the
`covariantly coupled' algebras that have been studied in
\cite{tjark2}. The finite algebra (\ref{alg22})
is almost a Lie algebra. If we assign particular values to $C$
and to the second casimir $C_2=(H^2+\hf EF+\hf FE)$ of the $sl_2$
subalgebra spanned by $\{E,H,F\}$, then (\ref{alg22}) reduces to
a Lie algebra. For a generic choice of the values of $C$ and
$C_2$ this Lie algebra is isomorphic to $sl_2 \oplus sl_2$. An
interesting question is, whether a similar phenomena occurs
for different covariantly coupled algebras.

\subsection*{$4=3+1$}

The last nontrivial nonprincipal $sl_2$ embedding we consider is
$\underline{4}_4\simeq \underline{3}_2\oplus \underline{1}_2$. We choose yet
another basis
\be \label{basis31}
r_at_a = \left( \begin{array}{cccc}
\frac{r_{5}}{12}-\frac{r_{6}}{3} & r_8 & r_{11} & r_{12} \\
r_4 & -\frac{r_5}{4} & r_{13} & r_{14} \\
\frac{r_3}{2}-\frac{r_9}{2} & r_{10} &
\frac{r_5}{12}+\frac{r_6}{6}+\frac{r_7}{2} & r_{15} \\
r_1 & r_2 & \frac{r_3}{2}+\frac{r_9}{2} &
\frac{r_5}{12}+\frac{r_6}{6}-\frac{r_7}{2}
\end{array} \right),
\ee
in terms of which the $sl_2$ embedding is $t_+=t_{11}+t_{15}$,
$t^0=-3t_6+t_7$ and $t_-=2t_3$. The subalgebra $g_+$ is
generated by $\{t_{11},\ldots,t_{15}\}$, $g_-$ is generated by
$\{t_1,t_2,t_3,t_9,t_{10}\}$, and $g_0$ is generated by
$\{t_4,\ldots,t_8\}$. The $d_1$ cohomology is generated by
$\hj^1,\ldots,\hj^5$, and $d$-closed representatives are given
by $\whj{4}=\hj^4$, $\whj{5}=\hj^5$, and
\ba \label{reps31}
\whj{1} & = & \hj^1+\www{1}{6}{3}{6}-\www{1}{12}{6}{3}
+\www{1}{4}{7}{3}+\hj^4\hj^{10}+\www{1}{4}{6}{9}-\www{1}{4}{7}{9}-
\www{1}{3}{4}{5}\hj^8 \nonu
& &
-\www{1}{3}{4}{6}\hj^8-\www{1}{108}{6}{6}
\hj^6+\www{1}{12}{6}{7}\hj^7+\www{31\hbar}{12}{4}{8}-
\deel{3\hbar}{4}\hj^9+\www{5\hbar}{48}{6}{6}
\nonu
& &
+\www{\hbar}{6}{7}{6}-\www{3\hbar}{16}{7}{7}-
\deel{3\hbar^2}{8}\hj^7-\deel{7\hbar^2}{24}\hj^6,
\nonu
\whj{2} & = & \hj^2-\www{1}{2}{3}{8}-\www{1}{3}{5}{10}-
\www{1}{6}{6}{10}+\www{1}{2}{7}{10}-\www{1}{2}{8}{9}+
\deel{3\hbar}{2}\hj^{10}+
\www{1}{9}{5}{5}\hj^8 \nonu
& &
+\www{1}{9}{5}{6}\hj^8+
\www{1}{36}{6}{6}\hj^8-\www{1}{4}{7}{7}\hj^8-
\hbar\hj^5\hj^8-\www{\hbar}{2}{6}{8}-\www{\hbar}{2}{7}{8}+
2\hbar^2\hj^8, \nonu
\whj{3} & = & \hj^3 +\hj^4\hj^8+\www{1}{12}{6}{6}+
\www{1}{4}{7}{7}+\deel{\hbar}{2}\hj^7-\deel{\hbar}{2}\hj^6.
\ea
We introduce a new basis
\ba \label{newgens31}
U & = & \deel{1}{4}\whj{5}, \nonu
G^+ & = & \whj{4}, \nonu
G^- & = & \whj{2}, \nonu
S & = & \whj{1}, \nonu
C & = &
\whj{3}+\deel{1}{24}\whj{5}\whj{5}-\deel{\hbar}{2}\whj{5}.
\ea
In this case, the fields are not organized according to $sl_2$
representations, because the centralizer of this $sl_2$
embedding in $sl_4$ does not contain an $sl_2$. Again $C$ is a
central element, and the nonvanishing commutators are
\ba \label{alg31}
[U,G^+] & = & \hbar G^+ , \nonu
[U,G^-] & = & -\hbar G^- , \nonu
[S,G^+] & = & \hbar G^+ (-\deel{2}{3} C+\deel{20}{9}U^2-
\deel{43\hbar}{9}U+\deel{29\hbar^2}{27}), \nonu
[S,G^-] & = & \hbar  (\deel{2}{3} C-\deel{20}{9}U^2+
\deel{43\hbar}{9}U-\deel{29\hbar^2}{27})G^-, \nonu
[G^+,G^-] & = & \hbar S-\deel{4\hbar}{3}CU+\deel{3\hbar^2}{4}C
    +\deel{88\hbar}{27}U^3-\deel{17\hbar^2}{2}U^2+
    \deel{25\hbar^2}{6}U.
\ea
This is the first example where the brackets are no longer
quadratic, but contain third order terms.
For the sake of completeness, let us also give the quantum Miura
transformation for this algebra. We identify generators
$\{e,f,h\}$,$s_1$,$s_2$ of $sl_2\oplus u(1) \oplus u(1)=g_0$ via
$f=\hj^8$, $e=\hj^4$, $h=\deel{1}{3}(\hj^5-\hj^6)$,
$s_1=\deel{1}{3}(2\hj^6+\hj^5)$ and $s_2=\hj^7$. The only
nontrivial commutators between these five generators are
$[e,f]=\hbar h$, $[h,e]=2\hbar e$ and $[h,f]=-2\hbar f$.
The quantum
Miura transformation now reads
\ba \label{miura31}
U & = & \deel{1}{4}s_1+\deel{1}{2}h, \nonu
G^+ & = & e, \nonu
G^- & = & (\deel{1}{4}s_1^2+\deel{1}{2}s_1 h + \deel{1}{4}h^2
-\deel{1}{4}s_2^2-\deel{\hbar}{2}(3s_1+3h+s_2)+2\hbar^2)f,
\nonu
S & = & -\deel{1}{12} e(8s_1+4h-31\hbar)f-\deel{1}{108}
(s_1-h)^3+\deel{1}{12}(s_1-h)s_2^2+\deel{5\hbar}{48}(s_1-h)^2
\nonu
& & + \deel{\hbar}{6}s_2(s_1-h)-\deel{3\hbar}{16}s_2^2-
\deel{3\hbar^2}{8}s_2-\deel{7\hbar^2}{24}(s_1-h),
\nonu
C & = & (\deel{1}{2}ef+\deel{1}{2}fe+\deel{1}{4}h^2)+
(\deel{1}{4}s_2^2+\deel{\hbar}{2}s_2)+
(\deel{1}{8}s_1^2-\hbar s_1).
\ea
This completes our list of finite quantum $W$ algebras from
$sl_4$.

\chapter{W algebras from affine Lie algebras}
In this chapter we develop, both classically and quantum mechanically,
the point of view that CFT's with $W$ symmetry are reductions of theories
with affine symmetry. Examples of such  theories are the
well known Wess-Zumino-Witten models describing strings moving on
group manifolds \cite{GeWi}. Due to this correspondence theories with
affine  symmetry obtain a universal status and many properties of quantum
field theories with $W$ symmetry can be derived by pulling down the
properties of the theory with affine symmetry.
A classic example of this was the
discovery by Polyakov \cite{Pol} that the 2 dimensional induced gravity
action which appears in noncritical string theory has, in the chiral
(or light cone) gauge, a `hidden' affine $sl(2)$ symmetry. Motivated by
this it was shown \cite{dublin} that the induced gravity action,
sometimes also called `Liouville gravity', is nothing but a gauged
version of the WZW model associated to $sl(2;{\bf R})$. As we have seen
in chapter 2 this corresponds on an algebraic level to a
realization of the Virasoro algebra
(which is the symmetry algebra of the Liouville theory) as a
reduction of the affine $sl(2;{\bf R})$  algebra. The consequences of this
correspondence between the WZW model and the Liouville  action are
quite far reaching. For example it relates the correlation functions
in chiral 2D gravity to those in an $sl(2;{\bf R})$ WZW model and it
also gives a handle on the study of renormalization effects in 2D
gravity \cite{KPZ}.

Recently there has also been a great deal of interest in theories which
are called `higher spin generalizations of 2D gravity'. These higher spin
gravities, which are generalizations of ordinary gravity in 2 dimensions,
arise when one reduces an $sl(N;{\bf R})$ WZW model instead of an
$sl(2;{\bf R})$ model. The theories which one obtains after reduction
turn out to be Toda field theories which have been studied extensively
within the context of integrable systems. Motivated by the
$sl(2;{\bf R})$ case one interprets Toda field theory as a
generalization of 2D  induced gravity in the chiral gauge.

The classical versions of the covariant gravity theories that
reduce in the chiral gauge to Toda theories have been constructed
recently in \cite{BG}. The physical and geometrical interpretation
of these generalized 2D gravity theories remains a mistery however.

On the algebraic level the reduction from $sl(N;{\bf R})$ WZW to
Toda field theory is a reduction from an affine $sl(N;{\bf R})$
algebra to $W_N$. This lead to the remarkable insight that Toda field
theories, which have been around for a long time, have as their
symmetry algebras the nonlinear $W_N$ algebras (this was realized
independently in \cite{Gervais}). of course it was well known that Toda
field theories are conformally invariant for this follows immediately from
their description as dimensional reductions of selfdual Yang-Mills
theories \cite{dimred}. In this description one shows that the SDYM
equations together with the constraint of spherical symmetry are
equivalent to the Toda field equations of motion. However, it was not
known that $sl(N;{\bf R})$ Toda field theory was in fact invariant
under the higher spin extension of conformal symmetry known as $W_N$.
We see that $W$ algebras may also have applications in Yang-Mills theory.

In principle the generators of rotations with respect to which one
defines sperical symmetry are not uniquely defined. In fact every
$SU(2)$ embedding into the gauge group of the Yang-Mills system
can be seen as generating rotations. The standard Toda field theories
with $W_N$ symmetry are related to the so called `principal $SU(2)$
embeddings'. In \cite{BaVe} the problem of constructing the field
theories associated to the other $SU(2)$ embeddings was addressed.
This lead to a whole new class of generalized Toda field theories all
of which are conformally invariant.

In view of the fact that ordinary Toda field theories are invariant
under $W$ extensions of conformal symmetry it is natural to ask
whether this is also true for the generalized Toda theories. The answer
to this question is affirmative and in this chapter we shall construct
these $W$ algebras. Again they will be reductions of affine Lie algebras and
in analogy with the reductions that lead generalized Toda theories the
different reductions are associated to the different $sl(2)$ embeddings
into the gauge group. The class of $W$ algebras we thus obtain is very
large and contains many known examples.

This chapter consists of 2 parts. The first part deals with the classical
theory. The classical $W$ algebras thus obtained have, apart from the
applications described above, also great relevance to the theory of
integrable systems and solitary wave physics. In fact they are nothing
but the second Hamiltonian structures of new and highly complicated
but integrable hierarchies of evolution equations that generalize the
KdV hierarchy. We shall be taking a look at some of these in the next
chapter.

The second part of this chapter deals with the quantization of the
classical $W$ algebras constructed in the first part. For this we
again use the BRST formalism which is ideally suited for jobs like
this. In the entire chapter we continually emphasize the formal
similarity between the theory of finite $W$ algebras presented
in the previous chapter and the theory of infinite $W$ algebras.
As we mentioned at the beginning of the previous chapter this is
completely analogous to the theory of loop groups and algebras over
finite dimensional simple Lie algebras.

\section{Classical W algebras}
In this section we develop the classical theory of $W$ algebras. Within
the context of integrable systems classical $W$ algebras have a longer
history than in conformal field theory. There they were known as
Gelfand-Dickii Poisson brackets on  spaces of
pseudodifferential operators \cite{GD}. An example of a Gelfand-Dickii bracket
is the well known second Hamiltonian structure of the KdV hierarchy.
This bracket, which is the GD bracket associated to $sl_2$ is known
to be nothing but the Virasoro algebra. The classical
$W_N$ algebras are just the GD bracket algebras associated to $sl_N$.

It was shown by Drinfeld and Sokolov \cite{DS} that the GD Poisson
structure associated to a Lie algebra $g$ is a reduction of the
Kirillov Poisson algebra associated to the affine Lie algebra  over
$g$. This reduction, which is just Poisson reduction in infinitely
many dimensions, is nowadays called Drinfeld-Sokolov reduction. It was
the first evidence supporting the claim that many $W$ algebras
(perhaps all) can be viewed as reductions of some sort of Lie algebras.
In this first section we generalize the (classical) construction by
Drinfeld and Sokolov to include a large number of $W$ algebras other
than $W_N$. We also generalize the well known Miura map and present an
elegant algebraic description of classical $W$ algebras.
In doing this we shall benefit greatly from the
theory developed in the previous chapter since many of the statements
that we have to prove here have already been proved in the finite case.
We shall therefore refer heavily to the previous chapter.

\subsection{Generalized Drinfeld-Sokolov reduction}
Let $\{t_a\}$ be a basis of the simple Lie algebra $g$.
Furthermore let $\cal F$ be the algebra of (classical) fields generated
by $\{J^a(z)\}$ (where the parameter $z$ is thought of as parametrizing
$S^1$, or equivalently $U(1)$), i.e. $\cal F$ is defined inductively as
follows:
\begin{enumerate}
\item $J^a(z)$ is an element of $\cal F$ for all $a$.
\item If $A(z),B(z)\in {\cal F}$ then $\alpha A(z)+\beta B(z) \in
{\cal F}$.
\item If $A(z) \in {\cal F}$ then $\frac{dA}{dz}(z)\in {\cal F}$.
\item If $A(z),B(z) \in {\cal F}$ then $(AB)(z)=A(z)B(z) \in {\cal F}$.
\end{enumerate}
Note that since we are still dealing with classical fields the
multiplication is abelian, i.e. $(AB)(z)=(BA)(z)$. The fields
$J^a(z)$ are to be considered as functionals on the affine Lie algebra
$\bar{g}=Map(S^1;g)\oplus {\ce}c$ over $g$: Any element $y\in Map(S^1;
g)$ can be written as $y=y^at_a$ where $y^a$ are smooth  maps from
$S^1$ to the real numbers. The functional $J^a(z)$ is then defined by
\be
J^a(z)(y)=y^a(z)
\ee
Define $Fun(\bar{g})$ to be the set of functionals on $\bar{g}$ of the
form
\be
F=\int_{S^1}dz \; A(z)
\ee
where $A(z) \in {\cal F}$. Obviously $Fun(\bar{g})$ is a subset of the
space of smooth functions on $\bar{g}$.

$Fun(\bar{g})$ is a Poisson algebra w.r.t. the Kirillov Poisson structure
on $\bar{g}$ which is given by
\be
\{F,G\}(y)=((\; [\partial+y,\mbox{grad}_yF],\mbox{grad}_yG\; ))
\label{infkir}
\ee
where for $y_1,y_2\in \bar{g}$
\be
((y_1,y_2))=\int_{S^1} dz \; (y_1(z),y_2(z))
\ee
is the Killing form on $\bar{g}$ and $\mbox{grad}_yF \in \bar{g}$ is
defined by
\be
\frac{d}{d\epsilon} F(y+\epsilon y_2)|_{\epsilon =0} =((\mbox{grad}_yF,
y_2))
\ee
for all $y_2\in \bar{g}$. The commutator bracket $[.,.]$ in (\ref{infkir})
denotes the Lie bracket of $g$, {\em not} of $\bar{g}$ which involves an extra
central term. This is why there appears a $\partial \equiv \frac{d}{dz}$ in
(\ref{infdir}). Had we taken $[.,.]$ to be the Lie bracket in $\bar{g}$
then the right hand side of (\ref{infkir}) would have been the usual
expression for the Kirillov bracket (i.e. without the $\frac{d}{dz}$).
In terms of the functionals $J^a(z)$ the bracket (\ref{infkir}) reads
\be   \label{KMkir}
\{J^a(z),J^b(w)\}=f^{ab}_cJ^c(w)\delta (z-w) +kg^{ab}\delta '(z-w)
\ee
where $g^{ab}$ is the inverse of $g_{ab}=(t_a,t_b)$ (as in
the previous chapter $(.,.)$ denotes the Cartan-Killing form on
$g$), $[t_a,t_b]=f^{c}_{ab}t_c$ and $k$ is the `central element'.
Note that strictly speaking the right hand side of this equation is
no longer an element of $\cal F$  due to the delta functions, i.e. the
Poisson bracket of two elements of $\cal F$ takes one outside of $\cal F$.
This however is a well known phenomenon and the usual trick to avoid
it is by considering so called `smeared fields'. This however will
nowhere be essential to us and we shall therefore ignore this subtlety
from now on.

The algebra (\ref{KMkir}) is well known to arise
as the current algebra of the so called WZW model.
It should be compared to the Poisson relation
\be
\{J^a,J^b\}=f^{ab}_cJ^c
\ee
which we used as the starting point for the theory of finite $W$
algebras in the previous chapter.

The algebra $\cal F$ contains fields which form Virasoro algebras.
An example of such a Virasoro field is the so called (classical)
`affine Sugawara stress
energy tensor' given by
\be
T^s(z)=\frac{1}{2k}g_{ab}J^a(z)J^b(z)
\ee
It is easy to verify using the relations (\ref{KMkir}) that the
Poisson bracket of this field with itself reads
\be
\{T^s(z),T^s(w)\}=2T^s(w)\delta '(z-w) -\partial T^s(w)\delta (z-w)
\ee
Its Fourier modes therefore satisfy a Virasoro algebra with
central charge 0. In fact, in the WZW realization of the algebra
(\ref{KMkir}) the
affine Sugawara stress energy tensor is really the
true stress energy tensor of the system.

There are however many other elements of $\cal F$
whose modes form a Virasoro
algebra. Let $\lambda^a$ be a set of real numbers then the
fields
\be
T^I(z)=T^s(z)+\lambda_a \partial J^a (z)     \label{imp}
\ee
form Virasoro algebras with nonzero central extensions (even classically)
\be
\{T^I(z),T^I(w)\}=2T^I(w)\delta ' -\partial T^I (w) \delta -k^2
\lambda_a \lambda^a \delta '''
\ee
One could think that all these Virasoro fields have the same status.
This is however not the case since the fields $\{J^a(z)\}$
are only {\em all}
primary fields w.r.t. the affine Sugawara stress energy tensor. This
follows from
\be
\{T^I(z),J^a(w)\}=(J^a(w)-f^{ab}_c\lambda_bJ^c(w))\delta '
-\partial J^a(w)\delta+k\lambda^a\delta ''       \label{impcur}
\ee
One can see from this equation that if we require all
fields $J^a(z)$ to be primary then all $\lambda^b$ must be zero.
In that case $T^I(z)$ reduces to $T^s(z)$ and all fields $J^a(z)$
become primary fields of conformal weight 1.

Suppose we want to apply the Poisson reduction scheme that we used in
the previous chapter to the case at hand. We have seen that this can
lead to nonlinear algebras which, in this case, will be generated by
a finite number of fields. However, we do want to make the restriction
that the algebra that we obtain after reduction is a $W$
algebra. Other reductions may be interesting from a mathematical
point of view, but in this thesis we are only interested in reductions
that lead to
$W$ algebras since these appear to be the most interesting from a
physical point of view.
Let us therefore heuristically try to find a set of constraints
that may lead to interesting reductions (of course these constraints
will turn out to be just the constraints that we used in the
previous chapter to construct finite $W$ algebras).
We shall then show that these constraints indeed lead to $W$ algebras.

As we mentioned above all fields $J^a(z)$ are primary fields of
conformal weight 1 w.r.t. the affine Sugawara stress energy tensor. Therefore
a constraint of the form
\be
J^a(z)=\mbox{constant}   \label{c}
\ee
is not in agreement with conformal invariance
since constants
have conformal weight 0. To put it differently, since the affine Sugawara
stress energy tensor does not Poisson
commute, even on the constrained manifold,
with the constraint (i.e. it is not a gauge invariant quantity)
it does not descend to a generator of the reduced algebra.
At this moment we recall that the affine Sugawara stress energy tensor
was by far not the only Virasoro generator. In fact, from
equation (\ref{impcur}) follows that, even though not all currents
$J^a(z)$ are primary, most of them
are and some may have conformal weight 0 w.r.t.
an improved stress energy tensor. This opens up the possibility
that if we impose a set of constraints like (\ref{c}) that perhaps
one or more improved stress energy tensors descend to become
generators in the reduced algebra. of course we can turn this around,
fix a certain improved stress energy tensor and then impose a
set of constraints compatible to it, i.e. one constrains only
currents that are primary and have conformal dimension 0.
The choice we make for the improvement term will be determined
by an $sl_2$ embedding.  Given the improved stress energy tensor
we shall postulate a set of constraints and then motivate these
afterwards by showing that they lead to a $W$ algebra.
The role of the $sl(2)$ subalgebra will become clear later.

Let $\{t_0,t_+,t_-\}$ be an $sl_2$ subalgebra of $g$, then we can
again choose a basis $\{t_{j,m}^{(\mu )}\}$ compatible with the
$sl_2$ multiplet structure of $g$ (see previous chapter). An arbitrary
smooth map from $S^1$ into $g$ can then be written as
\be
J(z)=\sum_{j,m;\mu} J^{j,m}_{(\mu )}(z) t_{j,m}^{(\mu )}
\ee
Here again we take $t^{(1)}_{1,\pm 1}\equiv t_{\pm}$ and $t_{1,0}^{(1)}
\equiv t_0$. We now take the improved affine Sugawara stress energy tensor to
be
\be
T^I(z)=T^s(z)+ \partial J^{1,0}_{(\mu )}(z)
\ee
i.e. we let $\lambda^at_a=t_0$. The nice thing about this improved
stress energy tensor is that all currents $\{J^{j,1}_{(\mu )}\}_{j;\mu}$
are primary of conformal dimension 0.
They can therefore in principle be constrained to nonzero constants
without breaking conformal invariance. Not any choice will lead
to consistent reductions however (we will come to this later). A
choice that leads to consistent and physically interesting reductions is
\be
J^{j,1}_{(\mu )}(z)=\delta^j_1 \delta^1_{\mu}
\ee
i.e. one constrains the current associated to $t_+$ to 1 and all
other currents of grade 1 are constrained to 0.
The currents $\{J^{j,m}_{(\mu )}(z)\}$
for $m>1$ are primary fields of negative conformal dimension.
This we should not allow since negative conformal dimensions lead to
correlations in the quantum field theory that increase with distance.
Therefore these fields have to be eliminated from the theory and the
only way to do this consistently with conformal invariance
is by constraining them to be zero.

When the gradation of $g$ by $t_0$ is half integer there will also
be currents $\{J^{j,\frac{1}{2}}\}_{j;\mu}$ which are primary fields
of conformal weight $\frac{1}{2}$ w.r.t. the improved stress energy
tensor. In principle there is no reason to constrain these currents
and in fact the constraints specified above do lead to consistent
reductions. However, the algebras that one thus obtains are {\em not}
$W$ algebras (we shall comment further on this in a moment). We therefore
also constrain these currents to zero.

Putting it all together we arrive at the following set of constraints
\be   \label{fconstr}
\{\phi^{j,m}_{(\mu )}(z) \equiv J^{j,m}_{(\mu )}-\delta^j_1 \delta^m_1
\delta^1_{\mu } \}_{\mu; \; j \in \frac{1}{2} {\bf N} ; \; m \geq 0}
\ee
Note that the constraints (\ref{constraints}) are the direct finite
dimensional analogue of the constraints (\ref{fconstr}) thus
finally explaining the term 'finite $W$ algebra'.

Now we have the following
\bl
The constraint $\phi^{j,m}_{(\mu )}(z)$  is first class if $m\geq 1$.
\el
The proof of this lemma in the finite case has been given in the
previous chapter. In the infinite dimensional case the proof is
completely analogous.

As in the finite dimensional case we can determine the gauge group
generated by these first class constraints. Following exactly the
same arguments we find that it is given by
\be
\bar{H}\equiv Map(S^1,H)
\ee
where $H$ is the group associated to the nilpotent Lie subalgebra
\be
h=\bigoplus_{k \geq 1}g^{(-k)}
\ee
of $g$. Here, as before, $g^{(l)}=\mbox{Span}(\{t_{j,l}^{(\mu )}\}_{j;\mu}$.
The action of an element $g(z) \in Map(S^1,H)$
on $y(z) \in Map(S^1,g)$ is given by
\be
y \mapsto gyg^{-1}+\partial g g^{-1}
\ee
Note that all these formulas are the natural affine
versions of the formulas encountered in the theory of finite $W$
algebras.

The constrained surface in $\bar{g}$
will be denoted by $\bar{g}_c$, i.e. $\bar{g}_c$ consists of
elements of the form
\be   \label{cc}
y(z)=t_++\sum_j\sum_{m\leq 0}\sum_{\mu}
 y^{j,m}_{(\mu )}(z)t_{j,m}^{(\mu )}
\ee

One would again like to eliminate the gauge invariance by fixing it.
of course there are many ways to do this. On the level of the
reduced algebra a change in gauge means just an invertible change
of generators. Note that a relatively simple algebra can be made to look
very complicated when written in a nontrivial basis. Choosing
the right gauge is therefore very important. For $W$ algebras there
is a specific type of basis which is very natural. In such a basis
the algebra is generated by the Virasoro field (the stress energy
tensor of the conformal field theory) and a number of $W$ fields
that are {\em primary} w.r.t. this stress energy tensor.
As we will show there exists a natural gauge choice that
automatically puts the $W$ algebra in this preferred basis.
In fact it is precisely the `lowest weight'  gauge that we used
in the previous chapter. In order to see this introduce the space
$\bar{g}_{fix}$ consisting of currents of the form
\be  \label{fixcur}
y(z)=t_++\sum_{j;\mu } y^{j,-j}_{(\mu )}(z)t_{j,-j}^{(\mu )}
\ee
It can be shown that
\be
\bar{H} \times \bar{g}_{fix} \simeq \bar{g}_c     \label{gfix}
\ee
The proof of this statement uses precisely the same `counting by
grade' argument that was used in the finite case.  From (\ref{gfix})
follows immediately that
\be
\bar{g}_c/\bar{H} \simeq \bar{g}_{fix}
\ee
Note that since the improved stress energy tensor $T^I(z)$ weakly
commutes with the first class constraints (i.e. it commutes after
imposing the constraints) it is a gauge invariant quantity and
descends to a generator of the reduced algebra. In fact it will
be the stress energy tensor of the reduced theory.

The Poisson bracket (\ref{KMkir}) induces a Dirac bracket on
$\bar{g}_{fix}$. This Dirac bracket reads in the infinite dimensional
case
\be  \label{Dir}
\{\bar{F},\bar{G}\}^*=\overline{\{F,G\}-\int \; dzdw \{F,
\phi^{j,m}_{(\mu )}(z)\} \Delta^{(\mu \mu ')}_{j,m;j',m'}(z,w)
\{\phi_{(\mu ')}^{j',m'}(w),G\}}
\ee
where
\be
\int \; dz \, \Delta^{(\mu \mu ')}_{j,m;j'm'}(z,w)
\Delta_{(\mu '\mu '')}^{j',m';j'',m''}(w,z')
=\delta^{j''}_{j}\delta^{m''}_{m}\delta^{\mu}_{\mu ''} \delta (z-z')
\ee
and
\be
\Delta^{j,m;j'm'}_{(\mu \mu' )}(z-w)=\{\phi^{j,m}_{(\mu )}(z),
\phi^{j',m'}_{(\mu ')}(w)\}
\ee
In these formulas the bar denotes restriction to $\bar{g}_{fix}$.
The algebra that we are interested in is then the subalgebra
${\cal F}_{lw}$ of $\cal F$
generated by the fields $\{J^{j,-j}_{(\mu )}(z)\}_{j;\mu }$ and with
Poisson relations
\be
\{J^{j,-j}_{(\mu )}(z),J^{j',-j'}_{(\mu ')}(w)\}^*
\ee
{}From now on we denote this algebra by $W(\bar{g};t_0)$.
We now have the following important theorem
\bt
The algebra $W(\bar{g};t_0)$ is a $W$ algebra.
\et
Proof: we will show that the fields $J^{j,-j}_{(\mu )}(z)$ are
all primary fields w.r.t. the stress energy tensor $T^I(z)$.
Consider the generator $Q^I_{\epsilon}$
of conformal transformations defined by
\be
Q^I=\int \; dz \; \epsilon (z) T^I(z)
\ee
The action of this element on an arbitrary current $J^a(z)$ can
easily be calculated. Denoting $J(z)=J^a(z)t_a$ we find
\be  \label{transf}
\delta_{\epsilon} J(z) \equiv \{Q^I_{\epsilon},J(z)\} =
(\epsilon J)'-k\epsilon ''t_0 -\epsilon '[t_0,J]
\ee
Now, note that $\bar{g}_c$ has been so constructed that
it is stable under the transformations generated by $Q^I_{\epsilon}$.
This follows immediately from the fact that $Q^I_{\epsilon}$ commutes
weakly with the constraints. A current of the form (\ref{cc}) will
therefore be form invariant. All this does not mean however that the
submanifold $\bar{g}_{fix}$ of $\bar{g}_c$ will be stable.
In fact let $J(z)$ be a current of the form (\ref{fixcur}) then
from (\ref{transf}) one finds that
\be
\delta_{\epsilon}J=\sum_{j;\mu}\; (\epsilon \, \partial J^{j,-j}_{(\mu )}
+(j+1)\epsilon 'J^{j,-j}_{(\mu )}\, )\; t_{j,-j}^{(\mu )}
-k\epsilon ''t_0
\ee
Due to the last term in this equation it follows
that if $J \in \bar{g}_{fix}$ then $J+\delta_{\epsilon}
J$ is {\em not} an element of $\bar{g}_{fix}$. Note however that
$J+\delta_{\epsilon}J$ {\em is} still an element of $\bar{g}_c$. From
eq.(\ref{gfix}) it therefore immediately follows that there exists
a unique gauge transformation that brings us back onto $\bar{g}_{fix}$.
The combined action of $Q^I_{\epsilon}$ and $\bar{H}$ therefore
induces an action $\delta^*_{\epsilon}$ on $\bar{g}_{fix}$.
It is well known that this action is still generated by $Q^I_{\epsilon}$
but now by means of the Dirac bracket instead of the ordinary bracket.
Let $h(z)=k\epsilon (z) ''t_- \; \in \bar{h}$, where $\bar{h}$
is the Lie algebra of the gauge group $\bar{H}$, then the infinitesimal
gauge transformation generated by this element is given by
\be
\delta_{h}J=[h(z),J(z)]+kh'(z)
\ee
and it can easily be verified that
\ba
\delta^*_{\epsilon}J & \equiv & \delta_{\epsilon}J+\delta_hJ \nonumber \\
& = & \sum_{j;\mu}\; (\, \epsilon \partial J^{j,-j}_{
(\mu )}+(j+1)\epsilon 'J^{j,-j}_{(\mu )}-k^2\epsilon '''
\delta^j_1 \delta^{\mu}_1)t_{j,-j}^{(\mu )} \label{prim} \\
& \equiv & \{Q^I_{\epsilon},J(z)\}^*
\ea
{}From this equation it follows immediately that the current
$J^{j,-j}_{(\mu )}(z)$ is a primary field of conformal dimension
$j+1$ w.r.t. the stress energy tensor $T^I(z)$, i.e.
\be
\delta_{\epsilon}^*J^{j,-j}_{(\mu )}=\epsilon \partial J^{j,-j}_{(\mu )}
+(j+1)\epsilon 'J^{j,-j}_{(\mu )}  \label{primary}
\ee
This proves the theorem \vspace{7mm}.

In the calculations leading up to equation (\ref{prim}) it is
crucial that $[t_-,t_{j,-j}^{(\mu )}]=0$. If we had not constrained the
elements $t_{j,\frac{1}{2}}^{(\mu )}$ we could still have performed all the
calculations, i.e. we could have found a compensating
gauge transformation $h$ and we would have found an equation similar
to equation (\ref{prim}). However, this equation would not have had
this specific form, i.e. the currents would not have been primary
fields and therefore the algebra would not have been a $W$ algebra.

Note that from the proof of the theorem follows that we can predict
the field content of the $W$ algebra $W(\bar{g};t_0)$ without having
to actually calculate any Poisson brackets. Namely, the number of
fields that generate the $W$ algebra is equal to the number of $sl_2$
multiplets $|i|$ into which $g$ decomposes under the adjoint action of
$sl_2$. Also from eq.(\ref{primary}) follows that the conformal
dimensions are simply given by $\{\Delta = j+1\}$ where $j$ are the spins
of the $sl_2$ multiplets.

Having proved that the algebras that one obtains after reduction
w.r.t. the constraints (\ref{fconstr}) are $W$ algebras we now want
to calculate the Dirac brackets between the fields $J^{j,-j}_{(\mu )}(z)$.
These are the brackets that will involve the nonlinearities. However,
one look at the formula (\ref{Dir}) for the Dirac bracket makes it clear
that a direct calculation of these relations would be extremely
cumbersome. Fortunately the method used in the previous chapter to
summarise all Dirac brackets within one formula involving a matrix
$R$ generalizes to the infinite dimensional case. Again the proofs of
the statements that we will make below are exact analogues of the
proofs given in the previous chapter and will therefore be omitted.

Let $\cal S$ be the set of smooth test functions on $S^1$ and
denote by $\bar{\cal R}$ the set of smooth maps
\be
R:{\cal S}^{|i|} \times \bar{g}_{fix} \rightarrow \bar{g}
\ee
(where $|i|$ again denotes the number of $sl_2$ multiplets in the
branching of the adjoint representation of $g$) of the form
\be
R[\vec{a};y](z)=\sum_{j;\mu}\sum_k (\partial^k a_j^{(\mu )})(z)
R^j_{(\mu ;k)}[y](z)
\ee
where $\vec{a} \in {\cal S}^{|i|}$,
$y\in \bar{g}_{fix}$ and $R^j_{(\mu ;k)}[y] \in \bar{g}$. The
crucial thing about elements of $\bar{\cal R}$ is that they are
linear in the test fuctions $a_i(z)$.

{}From now on denote by $J_{fix}(z)$ the $\bar{g}_{fix}$ valued function
on $\bar{g}$ given by
\be
J_{fix}(z)=t_++\sum_{j;\mu } J^{j,-j}_{(\mu )}(z) t_{j,-j}^{(\mu )}
\ee
we then have the following theorem which is the direct analogue of
theorems 2 and 3 of the previous chapter.
\bt
There exists a unique $R \in \bar{\cal R}$ such that
\be    \label{that}
t_++[R[\vec{a};y],y]+k\partial R[\vec{a};y] \;\;\;  \in \bar{g}_{fix}
\ee
and
\be       \label{and}
(J_{fix}(z),R[\vec{a};y](z))=\sum_{j;\mu}  a^{(\mu )}_j(z)
J^{j,-j}_{(\mu )}(z) + (R[\vec{a};y](z),t_+)
\ee
for all $\vec{a} \in {\cal S}^{|i|}$ and $y \in \bar{g}_{fix}$.
Furthermore this $R$ has the property that
\be
\int dz \, a_j^{(\mu )}(z)\;
\{J^{j,-j}_{(\mu )}(z),J^{j',-j'}_{(\mu ')}(w)\}^*
(y)\; t_{j',-j'}^{(\mu ')}=
\left[ R[\vec{a};y],y \right] +k\partial R[\vec{a};y]  \label{infdir}
\ee
\et
Eq.(\ref{infdir}) makes it possible to
calculate all the Dirac brackets. All one has to do is solve $R$ from
equations (\ref{that}) and (\ref{and}) after which one can read off
the desired Poisson relations from eq.(\ref{infdir}).

Again it is possible to derive a general formula for $R$ \cite{BG}.
It reads
\be
R[\vec{a};y]=\frac{1}{1+L(\partial +ad_w)}F[\vec{a}]
\ee
where $w(z)=y(z)-t_+$ for $y \in \bar{g}_{fix}$ and
\be
F[\vec{a}]=a_{j,-j}^{(\mu )}t^{j,-j}_{(\mu )}
\ee
Let $Q$ be a smooth
function on $\bar{g}_{lw}=Map(S^1,g_{lw})$ then define
$\mbox{grad}_wQ \in \bar{g}_{hw}$ by
\be
\frac{d}{d\epsilon}Q(w+\epsilon w')|_{\epsilon =0}=((\mbox{grad}_wQ,
w'))
\ee
for all $w'\in \bar{g}_{lw}$. From eq.(\ref{infdir}) follows that
the Dirac bracket between two arbitrary smooth functions $F$ and $G$ on
$\bar{g}_{lw}$ is  then given by
\be
\{F,G\}^*(w)=(([\partial +w,\frac{1}{1+L(\partial +ad_w)}\mbox{grad}_wF],
\mbox{grad}_wG))
\ee
Note that for the trivial embedding (L=0) this formula reduces to the
ordinary Kirillov bracket.

Let us now give some examples of the formalism developed above.
In  order to illustrate the rather formal construction developed
above we consider the simplest example in some detail. Let
$g=sl_2$, then the only nontrivial $sl_2$ embedding into $g$ is the
identity map. An arbitrary element $y\in Map(S^1,g)$ can then be
written as
\be
y=\left(
\begin{array}{cc}
y^{1,0} & y^{1,1} \\
y^{1,-1} & -y^{1,0}
\end{array}
\right)
\ee
while $\bar{g}_{fix}$ consists of currents of the form
\be
y_{fix}=\left(
\begin{array}{cc}
0 & 1 \\
y^{1,-1} & 0
\end{array}
\right)
\ee
Let us now try to find the map $R\equiv R_{1,-1}t^{1,-1}+R_{1,0}t^{1,0}+
R_{1,1}t^{1,1}$. From eq.(\ref{and}) follows
immediately that $R[a;y]_{1,-1}=a$. The matrix elements $R_{1,0}$ and
$R_{1,1}$ can be derived from eq.(\ref{that}). Doing this one easily
finds that
\ba
R[a;y]_{1,-1} & = & a \nonumber \\
R[a;y]_{1,0} & = & -k\partial a \nonumber \\
R[a;y]_{1,1} & = & ay^{1,-1}+\frac{1}{2}k^2\partial^2a \nonumber
\ea
{}From eq.(\ref{infdir}) now follows that
\be
\int dz \; a(z)\{J^{1,-1}(z),J^{1,-1}(w)\}^*(y) \; t_{1,-1} =
2k\partial (ay^{1,-1})-ka\partial y^{1,-1}+\frac{1}{2}k^3\partial^3a
\ee
which means that if we define $T(z)=-\frac{1}{k}J^{1,-1}(z)$ then
\be
\{T(z),T(w)\}^*=2T(w)\delta ' -\partial T(w)\delta -k\delta '''
\ee
which is the Virasoro algebra. We conclude that the Virasoro algebra,
which is the simplest $W$ algebra, is a reduction of the affine
$sl_2$  algebra. This now explains our choice of constraints. Since
an affine $sl_2$  algebra leads to a Virasoro algebra and since any $W$
algebra {\em must} contain the Virasoro algebra it seems natural to
reduce an affine Lie algebra such that some affine $sl_2$  subalgebra reduces
to the required Virasoro algebra. This is precisely what we did.

It was shown in \cite{BFFOW} that the reduction of the affine $sl_N$
algebra w.r.t. the principal $sl_2$ embedding leads to $W_N$. Let us
therefore illustrate our formalism by considering a case in which
the $sl_2$ embedding is nonprincipal. As in the previous chapter
we take for this the nonprincipal $sl_2$ embedding into $g=sl_3$.
An arbitrary element $y\in Map(S^1,g)$ can then be written as
(in the basis compatible with the $sl_2$ embedding)
\be
y=\left(
\begin{array}{ccc}
\frac{1}{2}y^{1,0}+y^{0,0} & y^{\frac{1}{2},\frac{1}{2}}_{(1)} & y^{1,1} \\
y^{\frac{1}{2},-\frac{1}{2}}_{(2)} & -2y^{0,0} & y^{\frac{1}{2},\frac{1}{2}}
_{(2)} \\
y^{1,-1} & y^{\frac{1}{2},-\frac{1}{2}}_{(1)} & y^{0,0}-\frac{1}{2}y^{1,0}
\end{array}
\right)
\ee
while $\bar{g}_{fix}$ consists of currents of the form
\be
y_{fix}=\left(
\begin{array}{ccc}
y^{0,0} & 0 & 1 \\
y^{\frac{1}{2},-\frac{1}{2}}_{(2)} & -2y^{0,0} & 0 \\
y^{1,-1} & y^{\frac{1}{2},-\frac{1}{2}}_{(1)} & y^{0,0}
\end{array}
\right)
\ee
Again it is easy to calculate $R$ using equations (\ref{that}) and
(\ref{and}). Applying equation (\ref{infdir}) oncemore we find
that in the Dirac brackets between the generators
\ba
T(z) & = & -\frac{4}{3}(J^{1,-1}+3(J^{0,0})^2) \nonumber \\
G^+(z) & = & J^{\frac{1}{2},-\frac{1}{2}}_{(1)}\nonumber \\
G^-(z) & = & \frac{4}{3}J^{\frac{1}{2},-\frac{1}{2}}_{(2)} \nonumber \\
H(z) & = & 4 J^{0,0} \nonumber
\ea
read
\ba
\{T(z),T(w)\}^* & = & 2T(w)\delta '-\partial T(w)\delta -\frac{1}{2}
\delta''' \nonumber \\
\{T(z),G^{\pm}(w)\}^* & = & \frac{3}{2}G^{\pm}(w)\delta '-
\partial G^{\pm}(w)\delta  \nonumber \\
\{T(z),H(w)\}^* & = & H(w)\delta '-\partial H(w) \delta \nonumber \\
\{H(z),G^{\pm}(w)\}^* & = & \pm 2G^{\pm}\delta \nonumber \\
\{H(z),H(w)\}^* & = & \delta ' \nonumber \\
\{G^+(z),G^-(w)\}^* & = & (H^2(w)+T(w)+\partial H(w))\delta-
2H(z)\delta ' +\delta ''' \nonumber
\ea
This algebra is  known in the literature as $W_3^{(2)}$. The
quadratic version of $sl_2$ studied in the previous chapter is
the finite $W$ algebra underlying this algebra.

Using the formalism developed above it is easy to construct many
other examples.

\subsection{An algebraic description of $W$ algebras}
It is also possible to give an elegant algebraic description of
infinite $W$ algebras. For this remember that we can replace the mixed
set of first and second class constraints (\ref{constraints}) by an
equivalent (in the sense that they lead to the same $W$ algebra) set
of constraints that is entirely first class. The point was roughly to split
into half the space $g^{(\frac{1}{2})}$, associated to the second
class constraints, and to add one half of it to $g^{(0)}$ and the other
half to $g^{(1)}$. All this was accomplished by grading the algebra
by the element $\delta$ of the Cartan subalgebra instead of $t_0$.
Consider again the triangular decomposition
\be
g=g_-\oplus g_0 \oplus g_+
\ee
of the algebra $g$ into a negative, zero and positive grade piece w.r.t.
the operator $ad_{\delta}$.
As before let $\{t_a\}\equiv \{t_{j,m}^{(\mu )}\}$ be a basis of $g$
and let $\{t_{\alpha}\}$ and $\{t_{\bar{\alpha}}\}$ be the subbases
of $g_+$ and $g_0 \oplus g_-$ respectively. Remember that the
alternative set of constraints is then given by
\be
\phi^{\alpha}(z) \equiv J^{\alpha}(z)-\chi (J^{\alpha}(z))
\ee
where $\chi (J^{\alpha} (z) )=1$ if $t_{\alpha}=t_+$ and zero
otherwise. Let $\cal C$ be the algebra of fields generated by the
constraints $\{\phi^{\alpha}(z)\}$, i.e. as $\cal C$ contains all
derivatives products and their linear combinations of the constraints.
Denote by $I$ the subalgebra
\be
I \equiv {\cal F}.{\cal C}
\ee
of $\cal F$. Obviously $I$ is an ideal in $\cal F$ and one can
easily verify that it also is a Poisson subalgebra w.r.t. the Kirillov
Poisson structure. Consider the quotient algebra
\be
{\cal F}/I \equiv {\cal F}_{red}
\ee
It is an easy exercise to show that ${\cal F}_{red}$ is isomorphic
to the field algebra generated by $\{J^{\bar{\alpha}}(z)\}$. Now,
define ${\cal F}^{inv}_{red}$ by
\be
{\cal F}^{inv}_{red}=\{[A(z)] \in {\cal F}_{red}\mid [\{\phi^{\alpha},
A(w)\}]=0, \mbox{  for all }\alpha\}
\ee
then the Kirillov bracket on $\cal F$ naturally defines a Kirillov
bracket on ${\cal F}^{inv}_{red}$ as follows
\be
\{[A(z)],[B(w)]\}^*=[\{A(z),B(w)\}]
\ee
By construction this algebra is isomorphic to the $W$ algebra
$W(\bar{g};t_0)$.

Let us now explicitly construct ${\cal F}_{red}^{inv}$. Its elements
are fields $A(z)$ of ${\cal F}_{red}$ such
that
\be
[\{\phi^{\alpha}(z),A(w)\}]==0
\ee
As in the finite case we can construct a set of generators of
${\cal F}_{red}^{inv}$ as follows: Let
\be
J_c(z)=t_++\sum_{\bar{\alpha}} J^{\bar{\alpha}}(z)t_{\bar{\alpha}}
\ee
then there exists a unique element $a:\bar{g} \rightarrow \bar{\hat{H}}$,
where $\hat{H}$ is the Lie group associated to $g_+$, such that
\be
a(z)J_c(z)a(z)^{-1}+\partial a(z)a(z)^{-1}=t_++\sum_{t_{\bar{\alpha}}\in
g_{lw}}W^{\bar{\alpha}}(z)t_{\bar{\alpha}}   \label{bla}
\ee
The fields $W^{\bar{\alpha}}(z)$ are differential polynomials in
the fields $J^{\bar{\alpha}}(z)$ of the form
\be
W^{\bar{\alpha}}(z)=\sum_{l=0}^{p} W_l^{\bar{\alpha}}(z)
\ee
where the degree of $W^{\bar{\alpha}}(z)$ is $l$, $W^{\bar{\alpha}}_p(z)
=J^{\bar{\alpha}}(z)$ and $p=\mbox{deg}(J^{\bar{\alpha}}(z))$.
Obviously these fields are gauge invariant. The Poisson relations
in $W(\bar{g};t_0)$ are now given by
\be
\{W^{\bar{\alpha}}(z),W^{\bar{\beta}}(w)\}^*=\{W^{\bar{\alpha}}(z),
W^{\bar{\beta}}(w)\}  \label{algre}
\ee

Let us give two examples of the construction described above.
Obviously we have
\be
J_c(z)=\left(
\begin{array}{cc}
J^{1,0}(z) & 1 \\
J^{1,-1}(z) & -J^{1,0}(z)
\end{array}
\right)
\ee
{}From eq.(\ref{bla}) we then find that
\be
W(z)=J^{1,-1}(z)+(J^{1,0})^2(z)+\partial J^{1,0}(z)
\ee
Using eq.(\ref{algre}) one can now easily calculate the relation
of this field with itself and one finds that it is a Virasoro algebra
as expected.

It is now easy to also do the $W_3^{(2)}$ case. One finds
\ba
W^{0,0}(z) & = & J^{0,0}(z) \nonumber \\
W^{\frac{1}{2},-\frac{1}{2}}_{(2)}(z)  & = & J^{\frac{1}{2},-\frac{1}{2}}
_{(2)} \nonumber \\
W^{\frac{1}{2},-\frac{1}{2}}_{(1)}(z) & = &
J^{\frac{1}{2},-\frac{1}{2}}_{(1)}(z) +\frac{1}{2}J^{1,0}
J^{\frac{1}{2},\frac{1}{2}}_{(1)}-3J^{\frac{1}{2},\frac{1}{2}}_{(1)}
J^{0,0}+\partial J^{\frac{1}{2},\frac{1}{2}}_{(1)} \nonumber \\
W^{1,-1}(z) & = & J^{1,-1}+\frac{1}{4}(J^{1,0})^2+J^{\frac{1}{2},
\frac{1}{2}}_{(1)}J^{\frac{1}{2},-\frac{1}{2}}_{(2)}+\frac{1}{2}
\partial J^{1,0} \nonumber \\
\ea
Introducing the generators
\ba
T(z) & = & \frac{4}{3}(W^{1,-1}+3W^{0,0}W^{0,0}) \nonumber \\
G^+(z) & = & W^{\frac{1}{2},-\frac{1}{2}}_{(1)} \nonumber \\
G^-(z) & = & \frac{4}{3} \; W^{\frac{1}{2},-\frac{1}{2}}_{(2)} \nonumber \\
H(z) & = & 4 W^{0,0} \nonumber
\ea
we find, using formula (\ref{algre}) that these form the algebra
$W_3^{(2)}$.

As in the finite case the $W$ algebra $W(\bar{g};t_0)$ has in general
a Kirillov subalgebra. Here, in the infinite dimensional case,
this subalgebra is isomorphic to the
Kirillov algebra associated to the affine Lie algebra over the
centralizer of the $sl_2$ embedding. In short we have
\bt
The $W$ algebra $W(g;t_0)$ has a Poisson subalgebra which is isomorphic
to $K(\bar{C})$ where $\bar{C}$ is the affinization of the centralizer
$C$ of the $sl_2$ subalgebra $\{t_0,t_{\pm}\}$ of $g$.
\et
Proof: This follows immediately from the fact that
\be
W^{0,0}_{(\mu )}(z) = J^{0,0}_{(\mu )}
\ee
for all $\mu$ \vspace{5mm}.

We now describe a generalization of the Miura transformation.
For this note that the Poisson bracket of a grade $l_1$ field
and a grade $l_2$ field is a field of grade $l_1+l_2$. Therefore
the Poisson brackets between the fields $W_0^{\bar{\alpha}}(z)$
and $W_0^{\bar{\beta}}(z)$ must be of precisely the same form
as the Poisson brackets between the fields $W^{\bar{\alpha}}(z)$
and $W^{\bar{\beta}}(z)$. This means that the map
\be
W^{\bar{\alpha}}(z) \mapsto W^{\bar{\alpha}}_0(z) \label{mi}
\ee
is a Poisson algebra homomorphism from the algebra generated by
the fields $\{W^{\bar{\alpha}}(z)\}$ to the algebra generated
by the fields $\{W^{\bar{\alpha}}_0(z)\}$. Since these fields
are themselves elements of the algebra generated by the fields
$\{J^{\bar{\alpha}}(z)\}_{t_{\bar{\alpha}}\in g_0}$ we find
that the map (\ref{mi}) is an embedding of the algebra $W(\bar{g};t_0)$
into the Kirillov algebra associated to $\bar{g}_0$. This is
the Miura map. We postpone the discussion of some examples of this
construction until we have quantized the $W$ algebras $W(\bar{g};t_0)$.

In this section we have generalized the classical Drinfeld-Sokolov
reduction scheme. What we have obtained is a large class of $W$ algebras.
It is a notable fact that the development of the theory of infinite
$W$ algebras was completely analogous to the development of the theory of
finite $W$ algebras. This is in fact very much the same situation as
with the theories of finite Lie algebras and loop algebras. The way we have
presented the material was to emphasize the similarity between the
finite and the infinite theory.

In the next section we are going to quantize the DS reduction scheme
for arbitrary $sl_2$ embeddings. Apart from some formal changes
related to the fact that we will be dealing with algebras of field
operators the quantization will again follow closely the lines
which were presented in the previous chapter.

\section{Quantum $W$ algebras}
Having developed the theory of generalized DS reductions in the
previous section we now turn to their quantization. The first attempt
to quantize the standard DS reduction leading to $W_N$ was made in
\cite{FaLu}. First they represented the classical $W_N$ algebras in
terms of $N-1$ free bosonic fields and then quantized them by
quantizing the free fields and normal ordening the expressions of the
$W$ generators. In general this procedure does not lead to a quantum
algebra that closes, i.e. one will not be able to represent the
commutator of two $W$ generators in terms of $W$ generators. It is
easy to see this even if one has only a finite number of degrees of
freedom (like with finite $W$ algebras). However, the procedure {\em does}
work, more or less by accident, in the case of $sl_N$ (but not
for other algebras) and leads to exactly the quantum relations
originally given by Zamolodchikov.

A proper way to quantize DS reductions is by the BRST procedure. This
was poineered in \cite{BeOog,FeFr} where the BRST cohomology for
$W_N$ was calculated by a spectral sequence argument. Unfortunately
the arguments used in \cite{FeFr} are very difficult to generalize to
DS reductions associated to arbitrary $sl_2$ embeddings. Also
it makes use of a free field realization of the original affine Lie algebra
which means that in the end one obtains the $W$ algebra in its free
field form. If the $W$ algebra is representable in terms of affine
currents, as we will show to be the case, then this will be obscured
by the too explicit free field form of the $W$ generators.

Another disadvantage of the methods developed in \cite{BeOog,FeFr}
is that the $W$ algebra is in the end described as the commutant of
certain screening operators. This is very inconvenient since calculating
such a commutant is in general very complicated.

In this section we shall quantize the generalized DS reductions by
a method that resembles the methods of \cite{FeFr} but which works
for arbitrary embeddings and which also lead to a completely
algorithmic procedure for constructing the $W$ algebra generators
and their relations. This will also lead to a quantization of the
generalized Miura transformation.

\subsection{Quantization of generalized DS reductions}
Let $\{t_a\}$ be a basis of the Lie algebra $g \equiv sl_n$.
The affine Lie
algebra $\bar{g}$ associated to $g$ is the span of $\{J^a_n\}$
and the central element $K$. The commutation relations are given by
\ba
[J^a_n,J^b_m] & = & f^{ab}_cJ^c_{n+m}+ng^{ab}K\delta_{n+m,0} \nonumber \\ {}
[K,J^a_n] & = & 0  \label{comm}
\ea
where $g^{ab}$ is the inverse of
$g_{ab}=\mbox{Tr}(t_at_b)$ and $[t_a,t_b]=f_{ab}^ct_c$.
As usual we use $g_{ab}$ to raise and lower indices.
Let ${\cal U}_k\bar{g}$ $({k \in \ce})$
be the universal enveloping algebra
of $g$ quotiented by the ideal generated by $K-k$.
The algebra ${\cal U}_k\bar{g}$ is the natural quantization of the
Kirillov Poisson algebra associated to $\bar{g}$.

Let there again be given an $sl_2$ subalgebra $\{t_0,t_{\pm}\}$ of $g$
and consider the grading defined by $\delta$
\be
g=\bigoplus_{k \in {\bf Z}} g_k
\ee
where $g_k=\{x \in g \mid [\delta ,x]=kx \}$. Remember that the basis
elements $\{t_a\}$ of $g$ are chosen such that they all are basis vectors
of some $sl_2$ multiplet. Since the $sl_2$ subalgebra
$\{t_0,t_{\pm}\}$ is a triplet under its own adjoint action  there
must be some $\alpha_+$ such that
$t_+=A\, t_{\alpha_+}$.
Define the character $\chi$ of $g_+$  (where $\bar{g}_+$ is the affinization
of $g_+$) by putting $\chi (J^{\alpha}_n) = A\delta^{\alpha ,\alpha_+}
\delta_{n+1,0}$
The constraints are then $J^{\alpha}_n=\chi (J^{\alpha}_n)$.
As we have seen these constraints are first class
which means that we can
use the BRST formalism. Thereto introduce the fermionic ghost variables
$c_{\alpha}^n,b^{\alpha}_n$ with ghost numbers 1 and $-1$
respectively and relations
\be
c_{\alpha}^nb^{\beta}_m+b^{\beta}_mc_{\alpha}^n =\delta_{\alpha \beta}
\delta_{n+m,0}
\ee
The algebra generated by these ghost variables is the Clifford algebra
$Cl(g_+ \oplus g_+^*)$. As usual one then considers the algebra
\be
\Omega_k = {\cal U}_kg \otimes Cl(g_+ \oplus g_+^*)
\ee

For calculational purposes it is convenient (as is standard
practice in conformal field theory) to introduce the following 'basic fields'
\ba
J^a(z) & = & \sum_n J^a_n \,z^{-n-1}  \nonumber \\
c_{\alpha}(z) & = & \sum_n c^n_{\alpha}\,z^{-n} \nonumber \\
b^{\alpha}(z) & = & \sum_n b^{\alpha}_n \,z^{-n-1}
\ea
It is well known that the commutation relations in $\Omega_k$ can then be
expressed in terms of the operator product expansions (OPE)
\begin{eqnarray}
J^a(z) J^b(w) & = & \frac{kg^{ab}}{(z-w)^2} + \frac{f^{ab}_c}{z-w}J^c(w)
+ \ldots \\
c_{\alpha}(z)b^{\beta}(w) & = & \frac{\delta^{\beta}_{\alpha}}{z-w}
\end{eqnarray}
Now inductively define the {\em algebra of fields} $F(\Omega_k )$
as follows:
\begin{enumerate}
\item $J^a(z),c_{\alpha}(z),b^{\alpha}(z)$
are elements of $F(\Omega_k )$
with 'conformal dimensions' $\Delta =1,0,1$ respectively.
\item if $A(z),B(z)\in F(\Omega_k )$ then $\alpha A(z)+\beta B(z) \in F
(\Omega_k )$
\item if $A(z)$ is an element of $F(\Omega_k)$
of conformal dimension $\Delta$ then $\frac{dA}{dz}(z)$ is also an
element of $F(\Omega_k )$ and has conformal dimension $\Delta+1$
\item if $A(z),B(z)$ are elements of $F(\Omega_k )$ of conformal dimensions
$\Delta_A$ and $\Delta_B$ respectively then
the normal ordened product $(AB)(z)\equiv A_-(z)B(z)\pm
B(z)A_+(z)$ (where one has the minus sign if $A$ and  $B$ are
fermionic) is also an element of
$F(\Omega_k )$ and has conformal dimension $\Delta_A + \Delta_B$.
Here $A_-(z)=\sum_{n\leq -\Delta_A}A_n\, z^{-n-\Delta_A}$ and $A_+(z)=
A(z)-A_-(z)$.
\end{enumerate}
We say that $F(\Omega_k )$ is 'generated' by the basic fields. Note that
$F(\Omega_k ) \subset {\cal U}_kg[[z,z^{-1}]]$.
The algebra $F(\Omega_k)$ is graded by ghost number, i.e. $J^a(z)$,
$c_{\alpha}(z)$ and $b^{\alpha}(z)$ have degrees 0,1 and -1 respectively
and we have the decomposition
\be
F(\Omega_k)=\bigoplus_n F(\Omega_k)^{(n)}
\ee

The algebra of fields $F(\Omega_k)$ is not simply the set of
`words' in the fields that can be made using the rules given
above, there are also relations. If we denote the operator
product expansion of $A$ and $B$ by $A(z)B(w)\sim\sum_r\frac{
\{AB\}_r}{(z-w)^r}$, then the relations valid in
$F(\Omega_k)$ are \cite{BBSS}
\ba
(AB)(z)-(BA)(z)\equiv [A,B](z)
& = & \sum_{r>0} (-1)^{r+1}\frac{\partial^r}{r!}
\{AB\}_r \nonu
(A(BC))(z)-(B(AC))(z) & = & ([A,B]C)(z) \nonu
\dif(AB)(z) & = & (\dif A B)(z) + (A\dif B)(z). \label{rel}
\ea

The BRST operator is then \cite{KoSt} $d(.)=[Q,.]$
where $Q=\oint \frac{dz}{2\pi i}Q(z)$ and
\be  \label{brstoper}
Q(z)=(J^{\alpha}(z)-\chi (J^{\alpha}(z)))c_{\alpha}(z)-\frac{1}{2}
f^{\alpha \beta}_{\gamma}(b^{\gamma}(c_{\alpha}c_{\beta}))(z)
\ee
As one immediately sees from eq. (\ref{brstoper}) we have
\be
d(F(\Omega_k)^{(l)}) \subset F(\Omega_k)^{(l+1)})
\ee
which means that $d$ is of degree 1. Furthermore, by explicit calculation
one finds that $d^2=0$ and the space $F(\Omega_k)$ is turned into
a complex
\be \label{complex}
\ldots \rightarrow F(\Omega_k)^{(n-1)} \rightarrow F(\Omega_k)^{(n)}
\rightarrow F(\Omega_k)^{(n+1)} \rightarrow \ldots
\ee
We want to calculate the zeroth cohomology of this complex.
This calculation is completely analogous to the one we did for finite
$W$ algebras \cite{BT2} and it will therefore not be necessary to go
into it in detail. Again one splits the BRST operator into two anticommuting
pieces $d_0$ and $d_1$ (\ref{def:q01})  and turns the complex (\ref{complex})
into a double complex by the double grading (\ref{def:bigrade}). The
action of $d_0$ and $d_1$ on the generators
$\hat{J}^a(z)=J^a(z)+f^{a\beta}_{\gamma}(b^{\gamma}c_{\beta})(z), \;
c_{\alpha}(z)$ and $b^{\alpha}(z)$ of $F(\Omega_k)$ is the same as in the
finite case (\ref{eq:q01hat}) except for
\be
D_1(\hat{J}^a(z))  =  f^{\alpha a}_{\bar{\beta}}\hat{J}^{\bar{\beta}}(z)
c_{\alpha}(z)+(kg^{a\alpha}-f^{\alpha
e}_{\beta} f^{\beta a}_{e}) \partial
c_{\alpha}(z)
\ee
Obviously the last two terms, which involve derivatives,  were not present
earlier. These terms do not spoil however the fact that the
subalgebra $F^{\alpha}(\Omega_k)$ of $F(\Omega_k)$ generated by
$J^{\alpha}(z)$ and $b^{\alpha}(z)$ is a subcomplex and that the
cohomology of this complex is equal to $\ce$ as in the finite case.
{}From here on the calculation is completely analogous, i.e. again one
shows that $H^*(F(\Omega_k);d)\simeq H^*(F_{red}(\Omega_k);d)$
where $F_{red}(\Omega_k)$ is the subcomplex of $F(\Omega_k)$ generated
by $\hat{J}^{\bar{\alpha}}(z)$ and $c_{\alpha}(z)$, and then one
calculates $H^*(F_{red}(\Omega_k);d)$ by a spectral sequence
argument. It will be no surprise that the final answer reads
\be \label{finalcoho}
H^n(F(\Omega_k);D) \simeq F_{lw}(\Omega_k)\delta_{n,0}
\ee
where $F_{lw}(\Omega_k)$ is the subspace of $F(\Omega_k)$ generated
by the fields $\{J^{\bar{\alpha}}(z)\}_{t_{\bar{\alpha}} \in g_{lw}}$.
Equation (\ref{finalcoho}) should  be understood to hold purely
on the level of vectorspaces. In order to explicitly construct the
generators of cohomology we again need to apply the tic-tac-toe construction.
It is however possible to immediately write down one generator, namely
the one that will correspond with the stress energy tensor of the
reduced theory. For this consider the field
\be
T=T_{I}+T_{gh}
\ee
where $T_{I}$  is the improved affine Sugawara stress energy tensor
\be
T_{I}=\frac{1}{2(k+h)}g_{ab}(J^aJ^b)-s_a\partial J^a,
\ee
$s_a$ are determined by $t_0=s^at_a$  and $T_{gh}$ is the `ghost
stress energy tensor'
\be
T_{gh}=(\delta_{\alpha}-1)b^{\alpha}\partial c_\alpha + \delta_{\alpha}
\partial b^{\alpha}
\ee
The coefficients $\delta_{\alpha}$ are determined by $[t_0,t_{\alpha}]=
\delta_\alpha t_{\alpha}$ while $h$ is the dual Coxeter number of the
Lie algebra $g$.
Obviously $T\in F(\Omega_k)$ and the operator product expansion
of $T$ with itself reveals that it satisfies a Virasoro algebra
with central charge
\be \label{central}
c(k;t_0)=\mbox{dim}(g^{(0)})-\frac{1}{2}\mbox{dim}(g^{(\frac{1}{2})})-
12\mbox{Tr}\left( \frac{\rho}{\sqrt{k+h}}-t_0\sqrt{k+h}\right)^2
\ee
where $\rho$ is the half sum of the positive roots. (Remember that
the grade $n$ subspace of $g$  w.r.t $t_0$ is denoted by $g^{(n)}$
while the grade $n$ subspace w.r.t. $\delta$ is denoted by $g_n$.)
The point is that one can show by explicit calculation that $T(z)$ is
a nontrivially BRST closed field, i.e.
\be
d(T(z))=0
\ee
and that it therefore is a generator of cohomology. Since the product
of two cohomology classes is simply the cohomology class of the product of
two representatives we find that in cohomology the field $T$ also generates
a Virasoro algebra with central charge (\ref{central}).

To obtain the other generators of $H^*(F(\Omega_k);d)$ one again applies
the tic-tac-toe construction in the same way as we did in the finite
case. For this consider a
generator $\hat{J}^{\bar{\alpha}}(z)$ of degree $(p,-p)$
of the field algebra
$F_{lw}(\Omega_k)$ (i.e. $t_{\bar{\alpha}}\in g_{lw}$) then
the generator of cohomology associated to this element
is given by
\be \label{tttc}
W^{\bar{\alpha}}(z)=\sum_{l=0}^p (-1)^l W^{\bar{\alpha}}_l(z)
\ee
where $W^{\bar{\alpha}}_0(z)\equiv J^{\bar{\alpha}}(z)$ and
$W^{\bar{\alpha}}_l(z)$ can be determined inductively by
\be
d_1(W^{\bar{\alpha}}_l(z))=d_0(W^{\bar{\alpha}}_{l+1}(z)) \label{ind}
\ee
It is easy to
check, using the fact that $d_0(J^{\bar{\alpha}}(z))=0$ for
$t_{\bar{\alpha}}\in g_{lw}$ that indeed
$d(W^{\bar{\alpha}}(z))=0$. These then are the generators of the $W$
algebra. The relations in the $W$ algebra can be found by
calculating the OPE's between the fields
$\{W^{\bar{\alpha}}(z)\}_{t_{\bar{\alpha}}\in g_{lw}}$
using the OPEs in  $F(\Omega_k)$.
In principle this
algebra closes only modulo $d$-exact terms. But since we
computed the $d$ cohomology on a reduced comlex generated by
$\hj^{\bar{\alpha}}(z)$ and $c_{\alpha}(z)$, and this reduced complex
is zero at negative ghost number, there simply aren't any $D$
exact terms at ghost number zero. Thus the algebra generated by
$\{W^{\bar{\alpha}}(z)\}_{
t_{\bar{\alpha}}\in g_{lw}}$ closes in itself.

As was shown in the previous chapter for finite $W$ algebras,
the operator product algebra
generated by the fields $W^{\bar{\alpha}}(z)$ is isomorphic to the operator
product algebra generated by
their (bi)grade (0,0) components $W^{\bar{\alpha}}_p(z)$ (the proof in the
infinite dimensional case is completely analogous and will not be repeated
here).
The fields $W^{\bar{\alpha}}_p(z)$ are of course elements of the field
algebra generated by the currents $\{\hat{J}^{\bar{\alpha}}(z)\}_{
t_{\bar{\alpha}} \in g_0}$. The relations (i.e. the OPEs)
satisfied by these currents are almost identical to the relations satisfied
by the unhatted currents
\be
\hat{J}^{\bar{\alpha}}(z)\hat{J}^{\bar{\beta}}(w)=\frac{kg^{\bar{\alpha}
\bar{\beta}}+ k^{\bar{\alpha}\bar{\beta}}}{(z-w)^2}+\frac{f^{\bar{\alpha}
\bar{\beta}}_{\bar{\gamma}}\hat{J}^{\bar{\gamma}}(w)}{z-w}+ \ldots
\label{hat}
\ee
where $k^{\bar{\alpha}\bar{\beta}}=f^{\bar{\alpha}\lambda}_{\gamma}
f^{\bar{\beta}\gamma}_{\lambda}$. Now, it is easy to see that
$g_0$ is just a direct sum
of $sl_{p_j}$ and $u(1)$ algebras, i.e. forgetting for a moment about the
$u(1)$ algebras one can write
\be
g_0 \sim \bigoplus_j sl_{p_j}
\ee
Within the $sl_{p_j}$ component of $g_0$ we have the identity
\be
k^{\bar{\alpha}\bar{\beta}}=
g^{\bar{\alpha}\bar{\beta}}(h-h_j) \label{shift}
\ee
where $h$ is the dual coxeter number of $g$ and $h_j$ is
the dual coxeter number of $sl_{p_j}$. We therefore find that the
field algebra generated by the currents $\{\hat{J}^{\bar{\alpha}}(z)\}_{
t_{\bar{\alpha}} \in g_0}$, denoted from now on by
$\hat{F}_0$, is nothing but the field algebra
associated to a semisimple affine Lie algebra
the components of which are affine $sl_{p_j}$ and $u(1)$
Lie algebras. This semisimple affine
Lie algebra is not simply $\bar{g}_0$ (whose field algebra is generated
by the unhatted currents) however
because in $\bar{g}_0$ all components have the same level while in $\hat{F}_0$
the level varies from component to component as follows from
equation (\ref{shift}). This is just a result
of the ghost contributions $k^{\bar{\alpha}\bar{\beta}}$ in  the OPEs
of the hatted currents.

{}From the above we find that the  map
\be
W^{\bar{\alpha}}(z) \mapsto (-1)^p W^{\bar{\alpha}}_p(z)
\ee
is an embedding of the $W$ algebra into $\hat{F}_0$.
This provides a quantization
and generalization to arbitrary $sl_2$ embeddings of the well known
Miura map.
In \cite{FaLu} the standard Miura map for $W_N$
algebras was naively quantised by simply
normal ordening the classical expressions. This is known to work only
for certain algebras \cite{BoSc}. Our construction gives a rigorous
derivation of the quantum Miura transformations for arbitrary affine Lie
algebras and $sl_2$ embeddings (the generalized Miura transformations
for a certain special class of $sl_2$ embeddings were also recently
given in \cite{De}).

As a result of the generalized quantum Miura transformation {\em any}
representation or realization of $\hat{F}_0$ gives rise to a representation
or realization of the $W$ algebra. In particular one obtains a free field
realization of the $W$ algebra by choosing the Wakimoto free field realization
of $\hat{F}_0$. Given our formalism it is therefore straightforward to
construct
free field realizations for any $W$ algebra that can be obtained by
Drinfeld-Sokolov reduction.

\subsection{Examples}
In this section we consider some representative
examples of the construction described
above. For notational convenience we shall omit the explicit $z$
dependence  of the fields where possible.

\subsubsection{The Virasoro algebra}
The Virasoro algebra is the simplest $W$ algebra and it is well
known to arise from the affine $sl_2$  algebra by
quantum Drinfeld-Sokolov reduction \cite{BeOog}. It is the
$W$ algebra associated to the only nontrivial embedding of $sl_2$ into itself,
namely the identity map. We consider this example here to contrast our
methods to the ones used by Bershadsky and Ooguri.

Choose the following basis of
$sl_2$
\be \label{basis1}
J^at_a=\mat{J^2/2}{J^3}{J^1}{-J^2/2}.
\ee
where $t_0=-t_2, \; t_+=t_1$ and $t_-=t_3$.
The positive grade piece of the Lie algebra $g=sl_2$ is generated by
$t_1$, and the constraint to be imposed is $J^1=1$. The BRST
current $d(z)$ is given simply by
\be \label{brst1}
d(z)=(J^1 c_1)-c_1.
\ee
The `hatted' currents are $\hj^1=J^1,\hj^2=J^2+2(b^1c_1)$ and
$\hj^3=J^3$. The
actions of $d_0$ and $d_1$ are given by
\be \label{d0d1}
\begin{array}{lclclcl}
d_0(\hj^2) & = & -2c_1 & \qquad & d_1(\hj^3) & = & (\hj^2c_2)+
(k+2)\dif c_1 \\
d_0(b^1) & = & -1 & \qquad & d_1(b^1) & = & \hj^1 .
\end{array}
\ee
On the other fields $d_0$ and $d_1$ vanish. From (\ref{d0d1}) it
is immediately clear that \\ $H(F_{red}(\Omega_k);d_0)$ is generated by
$W^3_0\equiv\hj^3$, in accordance with the general arguments in
section 2. To find the generator of the $d$-cohomology,
we apply the tic-tac-toe construction. We are looking for an
element $W^3_1(z)\in F_{red}(\Omega_k)$ such that
$d_0(W^3_1(z))=d_1(W^3_0)$. The strategy is to
write down the most general form of $W^3_1(z)$, and then to fix
the coefficients. In general, $W^{\bar{\alpha}}_l$ must satisfy
the following two requirements
\begin{enumerate}
\item if $W^{\bar{\alpha}}_l$ has bidegree $(-k,k)$, then
$W^{\bar{\alpha}}_{l+1}$ must have bidegree $(-k-1,k+1)$
\item if we define inductively the {\it weight} $h$
of a monomial in the
$J^{\bar{\alpha}}$ by $h(J^{\bar{\alpha}})=1-k$ if
$t_{\bar{\alpha}}\in g_k$,
$h((AB))=h(A)+h(B)$ and $h(\dif A)=h(A)+1$\footnote{
The weight $h$ is similar to the conformal weight, but not
always equal to it.
It is independent of the way in which the
$\hj$ are ordered.}, then $h(W^{\bar{
\alpha}}_l)=h(W^{\bar{\alpha}}_{l+1})$
\end{enumerate}
These two conditions guarantee that the most general form of
$W^{\bar{\alpha}}_l$ will contain only a finite number of
parameters, so that in a sense
the tic-tac-toe construction is a finite
algorithm. In the case under hand, the most general form of
$W^3_1$ is $a_1 (\hj^2\hj^2) + a_2 \dif \hj^2$, and the $D_0$ of
this equals $-4a_1(\hj^2c_1)-(4a_1+2a_2)\dif c_1$. Thus,
$a_1=-1/4$ and $a_2=-(k+1)/2$. Since $d_1(W^3_1)=0$, the
tic-tac-toeing stops here and the generator of the $D$
cohomology reads
\be \label{deft}
W^3=W^3_0-W^3_1=\hj^3+\frac{1}{4} (\hj^2\hj^2)
+\frac{(k+1)}{2}\dif\hj^2.
\ee
As one can easily verify, $T=W^3/(k+2)$ generates a Virasoro algebra
with central charge
\be
c(k)=13-6(k+2)-\frac{6}{k+2}
\ee
a result first found
by Bershadsky and Ooguri \cite{BeOog}.

Let's now consider the quantum Miura transformation.
In the case at hand $\bar{g}_0$ is the Cartan subalgebra
spanned by $t_2$ and $\hat{F}_0$ is an affine $u(1)$ field algebra at
level $k+2$
generated by $\hat{J}^2$. Indeed defining the field $\partial \phi \equiv
\nu \hat{J}^2$ where $\nu = \sqrt{2(k+2)}$, it is easy to check that
\be
\partial \phi (z) \, \partial \phi (w) = \frac{1}{(z-w)^2}+\ldots
\ee
In terms of the field $\phi$ the (bi)grade (0,0)
piece of $T$ is given by
\be
T^{(0,0)}= \frac{1}{2}(\partial \phi \partial \phi) + \alpha_0
\partial^2 \phi   \label{V}
\ee
where $\alpha_0=\frac{1}{2}\nu-\frac{1}{\nu}$. This is the usual
expression for the Virasoro algebra in terms of a free field $\phi$.

\subsubsection{The Zamolodchikov $W_3$ algebra}
Having illustrated the construction in some detail for the Virasoro
algebra, we will now briefly discuss two other examples. We start with
the Zamolodchikov $W_3$ algebra \cite{Za}. This algebra is associated
to the principal $sl_2$ embedding into $sl_3$. In terms of the
following basis of $sl_3$
\be
J^at_a = \left(
\begin{array}{ccc}
\frac{J^4}{6}+\frac{J^5}{2} & \frac{1}{2}(J^6+J^7) & J^8 \\
\frac{1}{2}(J^2+J^3) & -\frac{J^4}{3} & \frac{1}{2}(J^7-J^6) \\
J^1 & \frac{1}{2}(J^2-J^3) & \frac{J^4}{6}-\frac{J^5}{2}
\end{array}
\right)
\ee
The $sl_2$ subalgebra is given in this case by $t_+=4t_2,\; t_0=-2t_5$
and $t_-=2t_7$. The constraints are therefore $J^1=J^3=0$ and $J^2=1$
according to the general prescription. The BRST current associated
to these (first class) constraints read
\be
Q(z)=J^1c_1+(J^2-1)c_2+J^3c_3+2(b^1(c_2c_3)).
\ee
The BRST cohomology is generated by $\hat{J}^7$ and $\hat{J}^8$ since
$t_7$ and $t_8$ are the lowest weight vectors w.r.t. the $sl_2$ subalgebra.
The tic-tac-toe construction gives as generators of $H^*(F(\Omega_k);d)$
the fields $W^7=W^7_0-W^7_1$ and $W^8=W^8_0-W^8_1+W^8_2$ where
\ba
W^7_0 & = & \hat{J}^7 \nonumber \\
W^7_1 & = & -\frac{1}{6}(\hat{J}^4\hat{J}^4)-\frac{1}{2}(\hat{J}^4
\hat{J}^5)-2(k+2)\partial \hat{J}^5 \nonumber \\
W^8_0 & = & \hat{J}^8 \nonumber \\
W^8_1 & = & -(\hat{J}^5\hat{J}^6)+\frac{1}{3}(\hat{J}^4\hat{J}^7)-
(k+2)\partial \hat{J}^6 \nonumber \\
W^8_2 & = & -\frac{1}{27}(\hat{J}^4\hat{J}^4\hat{J}^4)+\frac{1}{3}
(\hat{J}^4(\hat{J}^5\hat{J}^5))+\frac{(k+2)}{3}(\hat{J}^4\partial
\hat{J}^4) +(k+2)(\hat{J}^5\partial \hat{J}^4) \nonumber \\
&   & +\frac{2(k+2)^2}{3}\partial^2 \hat{J}^4  \label{axpl}
\ea
with some work, for instance by using the program for computing
OPE's by Thielemans \cite{Thiele}, one finds that
\ba
T & = & \frac{1}{2(k+3)}W^7 \nonumber \\
W & = & \left( \frac{3}{(5c+22)(k+3)^3} \right)^3W^8
\ea
generate the $W_3$ algebra with central charge
\be     \label{cenw3}
c(k)=50-24(k+3)-\frac{24}{k+3}
\ee

For any principal embedding the grade zero subalgebra $\bar{g}_0$ is
just the Cartan subalgebra. In the case at hand $\hat{F}_0$ is therefore
a direct sum of two affine $u(1)$ field algebras, both of level $k+3$,
generated by $\hat{J}^4$ and $\hat{J}^5$. Defining $\partial \phi_1
\equiv \nu_1\hat{J}^4$ and $\partial \phi_2 \equiv \nu_2\hat{J}^5$,
where $\nu_1=\sqrt{6(k+3)}$ and $\nu_2=\sqrt{2(k+3)}$ it is easy to
check that
\be
\partial \phi_i(z)\partial \phi_j(w) =\frac{\delta_{ij}}{(z-w)^2}+\ldots
\ee
According to the general prescription the Miura transformation reads
in this case $W^7\mapsto -W^7_1$ and $W^8\mapsto W^8_2$, and the
fields
\ba
T^{(0,0)} & = & -\nu_2^{-2}W_1^7 \nonumber \\
W^{(0,0)} & = & \left( \frac{3}{(5c+22)(k+3)^3}\right)^{\frac{1}{2}}
W_2^8
\ea
also generate a $W_3$ algebra with central charge (\ref{cenw3}). Note
that according to (\ref{axpl}) $T^{(0,0)}$ and $W^{(0,0)}$ only
depend on $\phi_1$ and $\phi_2$. This is the well known free field
realization of $W_3$.

The two examples given above are associated to principal embeddings.
The construction presented above also works for nonprincipal
embeddings however as we shall illustrate in the next example.

\subsubsection{The $W_3^{(2)}$ algebra}
To describe the $W_3^{(2)}$ algebra, we pick the following
basis of $sl_3$
\be \label{basis4}
J^at_a=\mats{\frac{J^4}{6}-\frac{J^5}{2}}{J^6}{
J^8}{J^2}{-\frac{J^4}{3}}{J^7}{
J^1}{J^3}{\frac{J^4}{6}+\frac{J^5}{2}}.
\ee
The $sl_2$ embedding reads $t_+= t_1$, $t_0=
t_5$ and $t_- = t_8$. The
gradation of $\bar{g}$ with respect to $ad_{t_0}$ is half-integer
which means that there will be second class constraints \cite{BTV}
corresponding to the fields with grade -1/2. If one wants to use the
BRST formalism all constraints should be first class. One way to get
around this problem is to introduce auxiliary fields \cite{B}. This is not
necessary however as was shown in \cite{FORTW,BT} since it is always
possible to replace the half integer grading and the constraints
associated to it by an integer grading and
a set of first class constraints that nevertheless
lead to the same Drinfeld-Sokolov reduction. As mentioned earlier we have
to replace the grading by $t_0$ by a grading w.r.t. an element $\delta$.
\footnote{Essentially what one does is split the set of second class
constraints in two halves. The constraints in the
first half, corresponding to positive
grades w.r.t. $\delta$, are still imposed but have now become first class.
The other half can then be obtained as gauge fixing conditions of the
gauge invariance generated by the first half.}
In this specific case  $\delta = \frac{1}{3}\mbox{diag}(1,1,-2)$.
With respect to $ad_{\delta}$,
$t_1$ and $t_3$ have degree $1$ and span $g_+$. The BRST
current is
\be \label{brst32}
d(z)=((J^1-1)c_1)+J^3c_3.
\ee
Notice that there is no need for auxiliary fields, since the
constraints $J^1=1$ and $J^3=0$ are first class. The
cohomology $H(F_{red}(\Omega_k);D_0)$ is generated by $\{\hj^4,
\hj^6,\hj^7,\hj^8\}$. Again using the tic-tac-toe construction one finds
that $H(F(\Omega_k);D)$ is generated by $W^4=W^4_0$; $W^6=W^6_0$;
$W^7=W^7_0-W^7_1$ and $W^8=W^8_0-W^8_1$, where
\ba
W^4_0 & = & \hj^4, \nonu
W^6_0 & = & \hj^6, \nonu
W^7_0 & = & \hj^7, \nonu
W^7_1 & = & \frac{1}{2}(\hj^2\hj^5) + \frac{1}{2}(\hj^2\hj^4) -
(k+1) \dif \hj^2, \nonu
W^8_0 & = & \hj^8, \nonu
W^8_1 & = & -\frac{1}{4} (\hj^5 \hj^5) - (\hj^2\hj^6) +
\frac{(k+1)}{2}\dif\hj^5.
\ea
The OPEs of the hatted currents involving shifts in level are in this case
\ba \label{hjope}
\hj^4(z)\hj^4(w) & \sim & \frac{6(k+\frac{3}{2})}{(z-w)^2} + \cdots \nonu
\hj^5(z)\hj^5(w) & \sim & \frac{2(k+\frac{5}{2})}{(z-w)^2} + \cdots \nonu
\hj^4(z)\hj^5(w) & \sim & \frac{3}{(z-w)^2}+\cdots \nonu
\hj^2(z)\hj^6(w) & \sim & \frac{(k+1)}{(z-w)^2} +
\frac{\frac{1}{2}(\hj^4-\hj^5)}{(z-w)} +
\cdots.
\ea
If we now define the following generators
\ba
H & = & -W^4/3, \nonu
G^+ & = & W^6, \nonu
G^- & = & W^7, \nonu
T & = & \frac{1}{k+3}(W^8+\frac{1}{12}(W^4 W^4)),
\ea
we find that these generate the $W_3^{(2)}$ algebra
\cite{B}
\ba
T(z)T(w) & = & \frac{c(k)}{(z-w)^4}+\frac{2T(w)}{(z-w)^2}+\frac{\partial
T(w)}{z-w} \ldots \nonumber \\
T(z)G^{\pm}(w) & = & \frac{\frac{3}{2}G^{\pm}(w)}{(z-w)^2}+\frac{
\partial G^{\pm}(w)}{z-w} \ldots \nonumber \\
T(z)H(w) & = & \frac{H(w)}{(z-w)^2}+\frac{\partial H(w)}{z-w} \ldots
\nonumber \\
G^+(z)G^-(w) & = & \frac{(k+1)(2k+3)}{(z-w)^3}+\frac{3(k+1)H(w)}{(z-w)^2}
+\nonumber \\
&   & +\frac{1}{z-w}(3H^2(w)-(k+3)T(w)+\frac{3(k+1)}{2}\partial H(w)) +\ldots
\nonumber \\
H(z)H(w) & = & \frac{\frac{1}{3}(2k+3)}{(z-w)^2}
\ea
with
\be
c(k)=25-6(k+3)-\frac{24}{k+3}
\ee
a formula that was found in \cite{B} by a counting argument.

In the case at hand the subalgebra $g_0$ is spanned by the elements
$t_4,\; t_5,\; t_6$ and $t_2$. Obviously $g_0 \simeq sl_2 \oplus u(1)$.
Therefore $\hat{F}_0$ is the direct sum of an affine $sl_2$ and an affine
$u(1)$ field algebra, and using equation (\ref{shift}) the levels of these
can be calculated to be $k+1$ and $k+3$ respectively.
Indeed if we introduce the currents
\ba
\partial \phi & = & \frac{1}{4} (\hat{J}^4+3\hat{J}^5) \nonumber \\
J^0 & = & \frac{1}{4} (\hat{J}^5 - \hat{J}^4) \nonumber \\
J^- & = & \frac{1}{2} \; \hat{J}^2 \nonumber \\
J^+ & = & 2 \; \hat{J}^6 \nonumber
\ea
then these satisfy the following OPEs
\ba
J^0(z)J^{\pm}(w) & = & \frac{\pm J^{\pm}(w)}{z-w}+\ldots
\nonumber \\
J^+(z)J^-(w) & = & \frac{(k+1)}{(z-w)^2}+\frac{2J^0(w)}{z-w}+
\ldots \nonumber \\
J^0(z)J^0(w) & = & \frac{\frac{1}{2}(k+1)}{(z-w)^2}+\ldots
\nonumber \\
\partial \phi (z)\, \partial \phi (w) & = & \frac{\frac{3}{2}(k+3)}{
(z-w)^2}+ \ldots  \nonumber
\ea
and all other OPEs are regular. As stated before the shifts in the
levels that one can see in these OPEs are a result of the ghost
contributions.

The quantum Miura transformation in this case reads: $W^4 \mapsto W^4_0,\;
W^6 \mapsto W^6_0,$ \\ $W^7 \mapsto -W^7_1, \; W^8 \mapsto -W^8_1$. This means
that in terms of the currents $J^{\pm},\; J^0$ and $\phi$ the grade (0,0)
components of the $W$ generators read (let's for notational convenience
denote the (0,0) components of $H,G^+,G^-$ and $T$ again simply by
the same letters since they generate an isomorphic algebra anyway)
\ba
H & = & J^0-\frac{1}{3}\partial \phi \nonumber \\
G^+ & = & \frac{1}{2}J^+ \nonumber \\
G^- & = & (J^-J^0)+(J^0J^-)+2(k+3)\partial J^--2 \partial \phi J^-
\nonumber \\
T & = & \frac{1}{2(k+3)} \left( 2(J^0J^0)+(J^-J^+)+
(J^+J^-) +(k+3)\partial J^0
+ \frac{2}{3}(\partial \phi \partial \phi ) +
Q_0\partial^2 \phi \right) \nonumber
\ea
where $Q_0=-(k+1)$. We recognise
in the expression for $T$ the improved $sl_2$ affine Sugawara stress energy
tensor
and the free boson in a background charge.
Note that these formulas provide an embedding
of $W_3^{(2)}$ into $\hat{F}_0$.
In \cite{BTV} a realization of this type was called a
'hybrid field realization' since it represents the $W$ algebra partly
in terms of affine currents and partly in terms of free fields.

It is
now easy to obtain a realization of $W_3^{(2)}$ completely in terms
of free fields by inserting for the affine $sl_2$  currents $J^{\pm},\; J^0$
their Wakimoto free field form
\ba
J^- & = & \beta \nonumber \\
J^+ & = & -(\gamma^2 \beta )-k\partial \gamma - \sqrt{2k+6}
\,\gamma \,i \partial \varphi
\nonumber \\
J^0 & = & -2(\gamma \beta )-\sqrt{2k+6} \,i \,\partial \varphi
\ea
where as usual $\beta, \gamma$ and $\varphi$ are bosonic fields with the
following OPEs
\ba
\gamma (z) \beta (w) & = & \frac{1}{z-w}+ \ldots \nonumber \\
i\partial \varphi (z)\, i \partial \varphi (w) & = & \frac{1}{(z-w)^2}+\ldots
\ea

This example gives a nice taste of the general case. By the Miura
transformation one can write down for any $W$ algebra a hybrid field
realization, i.e. a realization partly in terms of free fields and partly
in affine currents. When required one can then insert for the affine currents
the Wakimoto free field realization giving you a realization of the $W$
algebra completely in terms of free fields.

\section*{Summary and conclusions}
In this chapter we have shown that a large class of $W$ algebras can be
obtained by generalizing and quantizing the Drinfeld-Sokolov
reduction scheme. The inequivalent reductions are related to the
inequivalent embeddings of $sl_2$. The formalism yields completely
algorithmic methods for calculating the $W$ algebra generators
and relations in both the classical and quantum case. We also
generalized and quantized the Miura map. This lead to the statement
that any $W$ algebra can be embedded into a semisimple affine Lie algebra.
Using this result it was possible to give a
general and explicit method for constructing free field
realizations for arbitrary $W$ algebras.

The question naturally arises whether all $W$ algebras can be
described as BRST cohomologies of affine Lie algebras. The answer
to this question is not known at this time but there are some
indications that it is true. No matter how it turns out, the description
of $W$ algebras as reductions of Lie algebras is extremely powerful.
It is in fact the only approach that, up to date, has lead to any
results w.r.t. the physical models behind $W$ algebras or their
geometry.

\chapter{W symmetry in integrable systems}
The intersection between topics related to string theory
like conformal field theory,
$W$ gravity, matrix models, topological field theory
etc. and the theory of integrable systems is substantial.
Consider for example the facts that the second Hamiltonian
stucture of the KdV hierarchy is equal to the Virasoro algebra
and also that the partition function of the one matrix model
is (after taking the double scaling limit)
a KdV tau function. Also remember that the theory describing
2 dimensional induced gravity in the chiral gauge is nothing
but Toda field theory which is a well known integrable system.
It appears that string theory is connected on a fundamental
level with the theory of integrable systems. Certainly one reason
for this is conformal invariance which, as is well known, is
infinite dimensional in 2 dimensions.

In this chapter we shall consider some applications to
integrable systems of the theory developed in the previous
chapters. In the first section show that finite dimensional
Toda systems are reductions of systems describing free particles
moving on group manifolds. Using this picture of finite Toda
systems we shall also be able to generalize them by again
considering nonprincipal $sl_2$ embeddings.
This is of course a close analogue of the reduction
of WZW models to Toda field theories which is also
briefly reviewed in the second section.
The (finite and infinite) Toda systems thus obtained have as
their symmetry algebras the $W$ algebras constructed in the
previous chapters.

In the last section we construct new hierarchies
of integrable evolution equations that generalize the KdV
type hierarchies introduced by Gelfand and Dickii. The
difference between the Gelfand-Dickii hierarchies and the
hierarchies we shall be considering is that the GD hierarchies
contain evolution equations for scalar fields while our
hierarchies will in general be for matrix fields. Also the
spatial derivatives in the equations will be replaced by
covariant derivatives w.r.t. some static gauge field.

\section{Generalized finite Toda systems}
Let $G$ be a simple Lie group of rank $l$ and let $g$ be its
Lie algebra. Consider the following coupled set of equations
of motion for the coordinates $\{q_i\}_{i=1}^l$
\be
\frac{d^2q_i}{dt^2}+\mbox{exp}(\sum_{j=1}^l K_{ij}q_j)=0 \label{To}
\ee
where $K_{ij}$ is the Cartan matrix of $g$. Systems of this form
are called finite dimensional Toda systems and are well known in
the theory of integrable systems (for a review see \cite{OlPe}).
In  this section we will show, using the formalism developed in
the previous chapters, that the Toda system (\ref{To}) is a
reduction of a system describing a free particle moving on the
group $G$. As we will show this reduction is again associated to
a principal $sl_2$ embedding. By also considering the other
$sl_2$ embeddings we will also be able to generalize the class
of finite dimensional Toda systems substantially.

The action of a 'free' particle moving on a  Riemann manifold $M$  with
local coordinates $\{x^{\mu}\}$ and metric
(in these coordinates) $h_{\mu \nu}(x)$ is
\be     \label{freerie}
S=\frac{1}{2}\int dt \; h_{\mu \nu}(x)
\frac{dx^{\mu}}{dt}\frac{dx^{\nu}}{dt}
\ee
An example of such a manifold is a simple Lie group. The Lie algebra $g$
of a simple Lie group $G$ is simple which means that the Cartan-Killing
form on it is nondegenerate. Since the Lie algebra of $G$ is nothing
but the tangent space in the unit element we can take the Cartan-Killing
form to be the 'metric in the unit element'. Because of the group structure
however we can extend this metric to the whole group manifold by left
translation. What we thus obtain is the unique left and right invariant
metric on $G$. This allows us to write down the following
coordinate free formula for the action of a free particle moving on
the group $G$:
\be \label{freeac}
S[g]=\frac{1}{2} \int dt \tr \left(
g^{-1}\frac{dg}{dt} g^{-1}\frac{dg}{dt} \right).
\ee
where $g=g(t)$ is the world line of the particle.
Indeed let $\{x^{\mu}\}$ be a set of local coordinates on $G$ then we can
write
\be
g^{-1}\frac{dg}{dt}=R^a_{\mu}(x)\frac{dx^{\mu}}{dt}t_a
\ee
where as before $\{t_a\}$ is a basis of the Lie algebra $g$.
Inserting this into eq.(\ref{freeac}) we find that it reduces
to an action of the form (\ref{freerie}) with
\be
h_{\mu \nu}(x)= R^a_{\mu}(x)R^b_{\nu}(x)\mbox{Tr}(t_at_b)
\ee
The group manifold $G$ together with the metric $h$ is a Riemann
manifold of constant curvature. This system will be our starting
point.

The action (\ref{freeac}) satisfies the following identity which
is a finite version of the Polyakov-Wiegman identity
\be \label{polwie}
S[gh]=S[g]+S[h]+\int dt \tr \left(
g^{-1} \frac{dg}{dt} \frac{dh}{dt} h^{-1} \right),
\ee
from which one immediately deduces the equations of motion,
\be \label{eqnmot}
\frac{d}{dt}
\left( g^{-1} \frac{dg}{dt} \right)=
\frac{d}{dt}
\left(  \frac{dg}{dt} g^{-1} \right)=0.
\ee
The action (\ref{freeac}) is invariant under
the group $G \times G$ where if $(h_1,h_2) \in G \times G$
then
\be
g(t) \mapsto h_1 g(t) h_2
\ee
Also note that the phase space of a particle moving on $G$  is
the cotangent bundle $T^*G$ of $G$ which means that the true
phase space of the free particle is
\be
T^*G/(G\times G)
\ee
The invariance of the system under $G\times G$ leads to the
conserved quantities $J^a,\bar{J}^a$ defined by
\be \label{conscurr}
J \equiv \frac{dg}{dt}g^{-1} \equiv J^at_a
\,\,\,\, \mbox{\rm and} \,\,\,\,
\bar{J} \equiv g^{-1}\frac{dg}{dt} \equiv \bar{J}^at_a.
\ee
The equations that express the conservation of these currents in
time coincide with the equations of motion of the system, so
fixing the values of these conserved quantities completely fixes
the orbit of the particle once its position on $t=0$ is
specified. The conserved quantities form  a
Poisson algebra \cite{papad}
\ba
\{J^a,J^b\} & = & f^{ab}_{\,\,\, c}J^c \nonumber \\
\{\bar{J}^a,\bar{J}^b\} & = & f^{ab}_{\,\,\, c}\bar{J}^c \nonumber \\
\{J^a,\bar{J}^b\} & = & 0 \label{pois}
\ea
This is precisely the
Kirillov Poisson algebra we used as a starting point for the
construction of finite $W$ algebras in chapter 3. These were obtained by
imposing constraints on the Poisson algebra (\ref{pois}), and we
want to do the same here to get systems with finite $W$
symmetry.
In terms of the decomposition $g=g_-\oplus
g_0 \oplus g_+$, these constraints were $\pi_{+}(J)=t_+$,
where $\pi_{\pm}$ are the projections on $g_{\pm}$. Here we want
to impose the same constraints, together with
similar constraints on
$\bar{J}$,
\be \label{constr}
\pi_+(J)=t_+\,\,\,\,\, \mbox{\rm and} \,\,\,\,\,
\pi_-(\bar{J})=t_-.
\ee
Let $G_{\pm}$ denote the subgroups of $G$
with Lie algebra $g_{\pm}$, and $G_0$ the subgroup with Lie
algebra $g_0$, then almost every element $g$ of $G$ can be
decomposed as $g_-g_0g_+$, where $g_{\pm,0}$ are elements of the
corresponding subgroups, because $G$ admits a generalized Gauss
decomposition $G=G_-G_0G_+$\footnote{Strictly speaking
$G_-G_0G_+$ is only dense in $G$ but we will ignore this subtlety
in the remainder}. Inserting $g=g_-g_0g_+$ into (\ref{constr})
we find
\ba \label{solco}
g_0^{-1} t_+ g_0 & = & \frac{dg_+}{dt} g_+^{-1} \nonu
g_0 t_- g_0^{-1} & = & g_-^{-1} \frac{dg_-}{dt}.
\ea
In the derivation of these equations one uses that
$\pi_+(g_-t_+g_-^{-1})=t_+$, and a similar equation with
$\pi_-$ and $t_-$, which follow from the fact that $t_{\pm}$
have degree $\pm 1$. The constrained currents look like
\ba \label{jconstr}
J & = & g_-\left(t_+ + \frac{dg_0}{dt} g_0^{-1} +
g_0 t_- g_0^{-1} \right) g_-^{-1}, \nonu
\bar{J} & = & g_+^{-1}\left(t_- + g_0^{-1} \frac{dg_0}{dt} +
 g_0^{-1} t_- g_0 \right) g_+.
\ea
The equations of motion now become
\ba \label{eqnm2}
0=g_-^{-1} \frac{dJ}{dt} g_- & = & \frac{d}{dt}
\left( \frac{dg_0}{dt} g_0^{-1} \right) + [g_0t_-g_0^{-1},t_+],
\nonu
0=g_+ \frac{d\bar{J}}{dt} g_+^{-1} & = & \frac{d}{dt}
\left( g_0^{-1} \frac{dg_0}{dt} \right) +
[t_-,g_0^{-1}t_+g_0].
\ea
Note that these equations describe a particle moving on $G_0$ in some
background potential, i.e. the system is no longer free.

If we consider again the principal $sl_2$ embedding then $g_0$ is the
Cartan subalgebra. Let $\{H_i\}$ be a basis of the CSA then any
element of $g_0$ can be written as
\be
\mbox{exp}(q^iH_i)
\ee
Inserting this into (\ref{eqnm2}) we find that they reduce to
\be
\frac{d^2q_i}{dt^2}+\mbox{exp}(K_{ij}q_j)=0
\ee
which are exactly the Toda equations (\ref{To}). For nonprincipal
embeddings the equations (\ref{eqnm2}) are more complicated matrix
equations. For obvious reasons we call these `generalized finite
Toda systems'.

It is easy to find an action that has (\ref{eqnm2}) as its equations
of motion. It reads
\be \label{finalact2}
S[g_0]  =  \frac{1}{2} \int dt \tr \left(
g_0^{-1}\frac{dg_0}{dt} g_0^{-1}\frac{dg_0}{dt} \right)
 - \int dt \tr
\left( g_0t_-g_0^{-1} t_+ \right).
\ee
This generalized Toda action describes a particle moving on
$G_0$ in some background potential. Two commuting copies of the
finite $W$ algebra leave the action (\ref{finalact2}) invariant
and act on the space of solutions of the
equations of motion (\ref{eqnm2}).

The general solution of the the equations of motion
of the free particle is given by
\be
g(t)=h_1e^{tX}h_2
\ee
where $h_1,h_2 \in G$ and $X \in g$.
{}From this follows that the general solution of the
equations (\ref{eqnm2}) can be constructed as follows: Let
$h_0^{(1)},h_0^{(2)}$ be elements of $G_0$.
Let $X_0$ be an arbitrary element of $g_0$.
If $g_0(t)$ is defined by the Gauss decomposition
\be \label{allsol}
g_-(t)g_0(t)g_+(t) = h_0^{(1)} \exp t(X_0+(h_0^{(1)})^{-1}t_+
h_0^{(1)}+h_0^{(2)}t_-(h_0^{(2)})^{-1}) \,\,h_0^{(2)},
\ee
then $g_0(t)$ is the most general solution of (\ref{eqnm2}).

This concludes our presentation of generalized finite Toda systems.

\section{Generalized Toda field theory}
It is also possible to obtain Toda field theories by the same
kind of arguments that we used in the previous section.  For
completeness we shall briefly describe this reduction in this
section.

Consider the Wess-Zumino-Witten action
\be
S[g]=-\frac{1}{4\lambda}\int_{\Sigma}
d^2\xi\; \mbox{Tr}(g^{-1}\partial_{\mu}g
g^{-1}\partial^{\mu}g)+k\Gamma (g) \label{WZW}
\ee
where $g$ is a $G$ valued matter field, $\lambda ,k$ are
dimensionless coupling constants and $\sigma =S^1 \times {\bf R}$.
The first term of the action
(\ref{WZW}) is a nonlinear sigma model and the second term is
called the Wess-Zumino term
\be
\Gamma [g]=\frac{1}{24\pi}\int_B \; \mbox{Tr}((g^{-1}dg)^3)
\ee
where $B$ is a three dimensional manifold such that $\partial B=\Sigma$.
The Wess-Zumino term $\Gamma$ is only well defined if $k$ is an
integer. The action $S[g]$ is conformally invariant for $k$
positive and
\be
\lambda^2 = \frac{4\pi}{k}
\ee
In fact the action (\ref{WZW}) is only called the WZW action for
this value of $\lambda$.

The WZW action possesses a left and right affine  symmetry. As in the
case of the free particle moving on $G$ the field equations coincide
with the conservation laws of the Noether currents associated to
these symmetries. They read
\be
\partial_-J=0 \; \mbox{ and }\; \partial_+\bar{J}=0
\ee
where the conserved currents $J$ and $\bar{J}$ are given by
\be
J=\partial_+gg^{-1} \; \mbox{ and } \; \bar{J}=g^{-1}\partial_-g
\ee
In order to describe the reduction \cite{dublin,LS} to Toda theory
we again introduce (as in the previous chapter) the constraints
\be
\pi_-(J)=t_- \; \mbox{ and } \; \pi_+(\bar{J})=t_+
\ee
Here again $\pi_{\pm,0}$ are the projection operators projecting
onto $g_{\pm,0}$ respectively.
Again using the generalized Gauss decomposition
\be
g(x^+,x^-)=g_+(x^+,x^-)g_0(x^+,x^-)g_-(x^+,x^-)
\ee
we find that the constraints are equivalent to
\ba
g_+^{-1}\partial_-g_+ & = & g_0t_+g_0^{-1} \nonumber \\
\partial_+g\,g^{-1}_- & = &  g_0^{-1}t_-g_0 \nonumber
\ea
and the constrained currents can be written as
\ba
J & = & g_+(\partial_+g_0\, g_0^{-1} + t_-)
g_+^{-1}+\partial_+g_+\, g_+^{-1}\nonumber \\
\bar{J} & = & g_-^{-1}(g_0^{-1}\partial_-g_0+t_+)g_-+g_-^{-1}\partial_-
g_- \nonumber
\ea
Now writing the WZW equation of motion as
\be
[\partial_+-J,\partial_--0]=0   \label{wzweqn}
\ee
and conjugating this equation by $g_+^{-1}(x^+,x^-)$ we see
that (\ref{wzweqn}) is equivalent to
\be
\partial_-(\partial_+g_0 \, g_0^{-1})=[t_-,g_0t_+g_0^{-1}]
\ee
which is the equation of motion of a generalized Toda field theory
\cite{LS,BaVe,dublin}.

A particular case is again when the $sl_2$ embedding is principal.
We then have
\be
g_0(x^+,x^-)=\mbox{exp}(\sum_i\phi_i(x^+,x^-)H_i) \label{exp}
\ee
where $H_i$ is a basis of the Cartan subalgebra of $g$.
Substituting (\ref{exp}) into (\ref{wzweqn}) we find that
it reduces to
\be
\partial_+\partial_-\phi_i+\mbox{exp}(\sum_j\, K_{ij}\phi_j)=0
\ee
which is the standard Toda field equation.

\section{Covariant KdV type hierarchies}
The nonlinear  Poisson algebras $W(\bar{g};t_0)$ constructed in the
previous chapter can be interpreted in two ways: as current algebras
of certain classical field theories and as the second Hamiltonian
structures of certain integrable hierarchies of evolution equations.
The former picture we developed in the previous section and it is
the purpose of this section to construct the integrable hierarchies
associated to an interesting subclass of $W$ algebras. For this
we shall closely follow the approach of Drinfeld and Sokolov (see
chapter 2).

Consider the case $g=sl_{2N}$ where $N=1,2,\ldots$ and take
the defining vector of the $sl_2$ embedding to be
\be           \label{defvec}
t_0=\frac{1}{2}\mbox{diag}(1, \ldots ,1,-1,\ldots ,-1)
\ee
where there are $N$ entries 1 and $N$ entries $-1$.
With respect to the $sl_2$ subalgebra determined by this defining vector
the fundamental representation of $sl_{2N}$ branches into
a direct sum of $N$ two dimensional representations of $sl_2$ (doublets)
while the adjoint representation branches as follows (in the notation
of chapter 3)
\be
\underline{ad}_{2N} \rightarrow (N^2-1)\underline{1}_2 \oplus N^2
\underline{3}_2                \label{422}
\ee
{}From this follows that the centralizer $C$ of the $sl_2$ embedding
is isomorphic to $sl_N$ and that
the resulting $W$ algebra has an affine $sl_N$
subalgebra. Furthermore, in (\ref{422}) we see that there are
$N^2$ triplets (i.e. $j=1$ representations of $sl_2$) from which
follows that the $W$ algebra will contain $N^2$ field of conformal
weight 2. of course one of those will be the stress energy tensor.

The `step up' operator $t_+$ of the $sl_2$ embedding determined by
(\ref{defvec}) is given by
\be
t_+=\left(
\begin{array}{cc}
0 & {\bf 1} \\
0 & 0
\end{array}
\right)
\ee
where $\bf 1$ is the $N \times N$ unit matrix. According to the
general prescription given in the previous chapter an general element
$y \in \bar{g}_{fix}$ will be of the form
\be
y(z)=\left(
\begin{array}{cc}
A(z) & {\bf 1} \\
T(z) & A(z)
\end{array}
\right)
\ee
where $A$ and $T$ are $N \times N$ matrices. of course $A$ is
traceless, i.e. $A \in Map(S^1,sl_N)$ while $T \in Map(S^1,gl_N)$.

Let us now, in analogy with the standard DS description, construct
the Lax operator. For this consider again the differential
equation
\be
(\partial - y)\Psi =0    \label{Lax}
\ee
where $\Psi$ is a 2-vector whose entries $\psi_i$
are $N\times N$  matrices.
Here $A$ acts on $\psi_i$ in the adjoint representation, i.e.
$A.\psi_i=[A,\psi_i]\equiv -ad_A\psi$. Eliminating the component
$\psi_2$ equation (\ref{Lax}) reduces to
\be
(D^2-T)\psi_1=0
\ee
where $D=\partial +ad_A$. From this we find that the Lax operator
is given by
\be
L=D^2-T
\ee

The covariant derivative $D$ is a derivation, i.e.
\be
D(PQ)=(DP)Q+PDQ
\ee
and has a formal inverse $D^{-1}$ within the algebra of pseudo
differential operators with matrix coefficients. The first few
terms of $D^{-1}$ are
\be \label{Din}
D^{-1}=\partial^{-1}-ad_A\partial^{-2}+(ad_{\partial A}+ad^2_A)
\partial^{-3}+\ldots
\ee
It is easily checked that $DD^{_1}=D^{-1}D=1$. In particular
this means that $D^{-1}D(PQ)=PQ$. Using the derivation property
of $D$ and on defining $R=DQ$ we find
\be
D^{-1}(PR)=PD^{-1}R-D^{-1}((DP)D^{-1}R)
\ee
Iterating this equation we find the following permutation rule
\be
D^{-1}Q=\sum_{i=0}^{\infty} (-1)^i(D^iQ)D^{-i-1}
\ee
which is identical to the one for $\partial^{-1}$. From these
considerations we find that we might as well work with the algebra
of covariant pseudodifferential operators instead of with ordinary
pseudodifferential operators.

Having found the Lax operator $L$ we can now easily determine its formal
root. It reads
\be
L^{\frac{1}{2}}=D-\frac{1}{2}TD^{-1}+\frac{1}{4}(DT)D^{-2}+\ldots
\ee
Using (\ref{Din}) one can easily check that in terms of ordinary
pseudodifferential operators $L^{\frac{1}{2}}$ is given by
\be
L^{\frac{1}{2}}=\partial +ad_A-\frac{1}{2}T\partial^{-1}+
\frac{1}{4}(\partial T + \{ad_A,T\})\partial^{-2}+\ldots
\ee
where $\{.,.\}$ here denotes the anticommutator.

The hierarchy of evolution equations is given by (see last section
of chapter 2)
\be             \label{hier}
\frac{dL}{dt_{2k+1}}=[(L^{\frac{2k+1}{2}})_+,L] \;\;\;\;\;\;
(k=0,1,2,\ldots )
\ee
where the $+$ means that we are to consider only the positive power
part (w.r.t. the covariant derivative). For the hierarchy (\ref{hier})
to be anywhere near completely integrable the different flows
must commute, i.e they must correspond to commuting vector fields
on the (infinite dimensional) phase space of the system.
This is proved in the following
\bl
The different flows in the hierarchy (\ref{hier}) commute, i.e.
\be
\frac{d^2}{dt_idt_j}L=\frac{d^2}{dt_jdt_i}L
\ee
\el
Proof: It is not difficult to show \cite{DS} that if equation
(\ref{hier}) holds then also
\be  \label{24}
\frac{dL^{\frac{i}{2}}}{dt_{2k+1}}=[(L^{\frac{2k+1}{2}})_+,L^{\frac{i}{2}}]
\ee
Using this equation we find
\be
\frac{d^2}{dt_idt_j}=[[(L^{\frac{i}{2}})_+,L^{\frac{j}{2}}]_+,L]+
[(L^{\frac{j}{2}})_+,[(L^{\frac{i}{2}})_+,L]]
\ee
Therefore
\be    \label{tussen}
(\frac{d^2}{dt_idt_j}-\frac{d^2}{dt_jdt_i})L=[[(L^{\frac{i}{2}})_+,
L^{\frac{j}{2}}]_+-[(L^{\frac{j}{2}})_+,L^{\frac{i}{2}}]_++
[(L^{\frac{j}{2}})_+,(L^{\frac{i}{2}})_+],L]
\ee
where we used the Jacobi identity. From
\be
[L^{\frac{i}{2}},L^{\frac{j}{2}}]=0
\ee
it follows that
\be
[(L^{\frac{i}{k}})_+,L^{\frac{j}{k}}]_+=[(L^{\frac{j}{k}})_+,
(L^{i}{k})_-]_+
\ee
Inserting this into eq. (\ref{tussen}) we find the desired result.
This proves the lemma \vspace{5mm}.

The first two equations of the hierarchy (\ref{hier}) are
\ba
\frac{dT}{dt_1} & = & DT  \label{cK1}\\
\frac{dT}{dt_3} & = & \frac{1}{4}(D^3T)-\frac{3}{4}\{DT,T\} \label{cK2}
\ea
while the `connection' components are constant w.r.t. all time
variables, i.e.
\be
\frac{dA}{dt_{2k+1}}=0
\ee
This can be seen as follows. The Lax equation was so constructed
\cite{DS} that its right hand side is a differential operator of order
0. However, the left hand side has in the covariant case order 1 which
means that the coefficient of the order 1 term must be zero. This
coefficient is exactly
\be
2\frac{d}{dt_{2k+1}}ad_A
\ee

Obviously the hierarchy constructed above is a covariant generalization
of the KdV hierarchy. If we write $T=u +T^aI_a$
(where $\{I_a\}$ is a basis of $sl_N$) then equation (\ref{cK1})
becomes
\ba
\frac{du}{dt_1} & = & \partial u \nonumber \\
\frac{dT^a}{dt_1} & = & \partial T^a + f^a_{bc}T^bA^c
\ea
while eqn.(\ref{cK2}) reads
\ba
\frac{du}{dt_3} & = & \frac{1}{4} \partial^3u -\frac{3}{2}u\partial u
-\frac{3}{4N}\partial (g_{ab}T^aT^b) \nonumber \\
\frac{dT^a}{dt_3} & = & \frac{1}{4}\partial T^a-\frac{3}{2} \partial
(T^0T^a)
\ea
Obviously the first of these equations is the ordinary KdV equation
plus a term that couples the field $u$ to the nondiagonal fields
$T^a$. Using eq.(\ref{hier}) and the explicit form of the Lax operator
it is easy to construct the other elements of the covariant KdV
hierarchy.

Integrable evolution equations like the KdV equation
are characterized by the fact that they have
infinitely many conserved quantities. This property, which
signals soliton like solutions, is of fundamental importance.
We will now show that the covariant KdV hierarchy constructed
above also admits infinitely many conserved quantities.
\bt
Let $Res(\sum_iA_iD^i)=A_{-1}$, then
the quantities
\be
H_{2k+1}=\int \; \mbox{Tr}\, Res(L^{\frac{2k+1}{2}}) \; dz
\ee
are conserved w.r.t. all flows in the hierarchy.
\et
Proof: Consider
\be
Res[AD^k,BD^l]
\ee
This quantity is 0 if $l,k>0$ or $l,k<0$. The only interesting case
is $k>0,l<0$. Take $l=-p$ where $p>0$ then $Res[AD^k,BD^{-p}]=0$
if $p>k+1$ and
\be
(\frac{k}{p+1})(A(D^{k-p+1}B)+(-1)^{k-p}(D^{k-p+1}A)B)
\ee
Also note that if
\be
g=(\frac{k}{p-1})\sum_{i=0}^{k-p}(-1)^i(D^iA)D^{k-p-i}B
\ee
then
\be
Dg=(\frac{k}{p-1}(AD^{k-p+1}B+(-1)^{k-p}(D^{k-p+1}A)B)
\ee
which means that
\be
\mbox{Tr}(Dg)=\mbox{Tr}Res[ADk,BD^{-p}]
\ee
However, $\mbox{Tr}(Dg)=\partial \mbox{Tr}(g)$ from which it follows
that $\mbox{Tr}Res[AD^k,BD^{-p}]$ is a total derivative. Using
this and eq.(\ref{24}) the theorem follows vspace{5mm}.

The first two Hamiltonians are given by
\ba
H_1 & = & \frac{1}{2}\int \; u(z) \, dz \nonumber \\
H_2 & = & \frac{1}{2}\int \; (u^2+\frac{1}{N}g_{ab}T^aT^b)\, dz
\nonumber
\ea

Note that for $N=1$ the hierarchy discussed above reduces to the
ordinary KdV hierarchy. We have therefore obtained a generalization
of this hierarchy and the $W$ algebra $W(\bar{g};t_0)$ is its second
Hamiltonian structure.

It is possible to repeat the whole procedure oulined above for the
the Boussinesq
hierarchy. For this consider the $sl_2$
embedding into $sl_{3N}$ determined by the following defining  vector
\be
t_0=\mbox{diag}(1, \ldots ,1,0, \ldots ,0,-1, \ldots ,-1)
\ee
where there are $N$ entries 1,0 and $-1$. With respect to the $sl_2$
subalgebra determined by this defining vector the fundamental
representation of $sl_{3N}$ banches into a direct sum of $N$ three
dimensional $sl_2$ representations (triplets) while the adjoint
representation branches as follows
\be
\underline{ad}_{3N} \rightarrow (N^2-1)\underline{1}_2 \oplus
N^2 \underline{3}_2 \oplus N^2 \underline{5}_2
\ee
Again from this we see immediately that the $W$ algebra $W(\bar{g};t_0)$
will contain an affine $sl_N$ subalgebra, $N^2$ fields of conformal dimension
2 and $N^2$ fields of conformal dimension $3$. According to the
general prescription given in the previous chapter a general element
$y \in \bar{y}_{fix}$ will be of the form
\be
y(z)=\left(
\begin{array}{ccc}
A & T & W \\
{\bf 1} & A & T \\
0 & {\bf 1} & A
\end{array}
\right)
\ee
where $A,T$ and $W$ are again $N \times N$ matrices.

Again we can easily construct the Lax operator. It reads
\be
L=D^3-2TD-(DT)-W
\ee
where as before $D=\partial +ad_A$. The formal cube root of this
operator is given by
\be
L^{1/3}=D-\frac{2}{3}TD^{-1}+\frac{1}{3}(\partial T -W)D^{-2}+\ldots
\ee
As before the hierarchy is then given by
\be     \label{Bou}
\frac{dL}{dt_{2k+1}}=[(L^{{2k+1}/3})_+,L]
\ee
Again let us give the first few elements of this hierachy. The first
one reads
\ba
\frac{dT}{dt_1} & = & DT \nonumber \\
\frac{dW}{dt_1} & = & DW \nonumber
\ea
and the second one is given by
\ba
\frac{dT}{dt_3} & = & DW \nonumber \\
\frac{dW}{dt_3} & = & -\frac{1}{3}D^3T+\frac{4}{3}\{T,DT\}-\frac{4}{3}
[T,W]
\ea
which is a covariant version of the Boussinesq equation. As in the case
of the covariant KdV hierarchy the connection components $A$ are constant
w.r.t. all times.

Again it is possible, in precisely the same way, to show that all
the flows commute and also that the quantities
\be
H_{2k+1} = \int \; \mbox{Tr}Res (L^{2k+1}/3)
\ee
are conserved.

It is easy to construct covariant versions of all the
elements of the Gelfand-Dickii hierarchy of hierarchies of
which the KdV and Boussinesq hierarchies are the first two. For this
one has to consider the $sl_2$ embedding into $sl_{MN}$ with respect to
which the fundamental representation of $sl_{MN}$ branches into
$N$ $M$-dimensional $sl_2$ representations.

\section{Discussion}
In this chapter we have considered some applications of the theory
developed in the previous chapters to the theory of integrable systems.
This lead to generalized Toda theories in finite and infinite dimensions
and also to new integrable hierarchies. The hierarchies we presented
above are however related to a very special kind of $sl_2$ embedding
and one might wonder as to the general structure.  A possible answer
to this question has been given in \cite{GHM} where it was shown that
to any conjugation class of the Weyl group associated a given
simple Lie algebra one can associate an integrable hierarchy. This
approach seems quite different from our approach in which $sl_2$
embeddings are the fundamental objects, but this is only superficial
since $sl_2$ embeddings and conjugation classes of the Weyl group
are in 1-1 correspondence for almost all simple Lie algebras. We
therefore conjecture that the $W$ algebras constructed in this thesis
are the second Hamiltonian structures of the hierarchies of \cite{GHM}.
It should be noted here that in the approach of \cite{GHM}
the covariant derivative structure (which is of course related to
affine subalgebras of the $W$ algebra) is not clear.

\end{document}